\documentclass[11pt,a4paper,openany]{book}
\usepackage[english]{babel}
\usepackage{amsmath,amsthm,amssymb,epsfig,latexsym}
\usepackage[numbers,square,sort&compress]{natbib}
\usepackage[toc,page]{appendix}
\newcommand{\be}[1]{\begin{equation}\label{#1}}
\newcommand{\ba}[1]{\begin{multline}\label{#1}}
\newcommand{\ee}{\end{equation}}
\newcommand{\ea}{\end{multline}}
\newcommand{\non}{\nonumber\\ \rule{0pt}{30pt}}

\newcommand{\num}{\\ \rule{0pt}{20pt}}
\newcommand{\numa}[1]{\\ \rule{0pt}{#1pt}}
\newcommand{\dis}{\displaystyle}

\newcommand{\tr}{\mathop{\rm tr}}

\newcommand{\Res}{\mathop{\rm Res}}
\renewcommand{\Im}{\mathop{\text{Im}}}
\renewcommand{\Re}{\mathop{\text{Re}}}

\newcommand{\Sym}{\mathop{\rm Sym}\limits}
\newcommand{\diag}{\mathop{\rm diag}}
\newcommand{\bs}[1]{\boldsymbol{#1}}
\newcommand{\bu}{\bar u}
\newcommand{\bv}{\bar v}

\newcommand{\bxi}{\bar{\xi}}
\newcommand{\bet}{\bar{\eta}}

\newcommand{\so}{\scriptscriptstyle \rm I}
\newcommand{\st}{\scriptscriptstyle \rm I\hspace{-1pt}I}
\newcommand{\sth}{\scriptscriptstyle \rm I\hspace{-1pt}I\hspace{-1pt}I}
\newcommand{\stf}{\scriptscriptstyle \rm I\hspace{-1pt}V}
\newtheorem{prop}{Proposition}[section]
\newtheorem{lemma}{Lemma}[section]
\newtheorem{cor}{Corollary}[section]

\numberwithin{equation}{chapter}
\numberwithin{equation}{section}

\DeclareRobustCommand\maybe[1]{\ifnum#1=\value{chapter}\relax\else{#1}.\fi}
\begin{document}

\thispagestyle{empty}
\renewcommand{\chaptername}{Lecture}
\renewcommand{\appendixname}{Appendices}
\begin{center}
\vspace*{5cm}
{\Large\bf Algebraic Bethe ansatz\\}
\vspace*{2cm}
{\large\bf N.A. Slavnov\footnote{nslavnov@mi.ras.ru}}\\
\vspace*{1cm}
{\it Steklov Mathematical Institute of Russian Academy of Sciences, Moscow, Russia}\\
\vspace*{2cm}
{\bf Abstract}\\
\end{center}
\vspace*{5mm}
\begin{small}
This course of lectures on the algebraic Bethe ansatz was given in the Scientific and Educational Center of
Steklov Mathematical Institute in Moscow. The course includes both classical well known results and very recent ones.
\end{small}

\newpage
\tableofcontents
\newpage


\chapter*{Introduction}

This text is written on the basis of lectures that the author gave in 2014--2015 in the Scientific and Educational Center of the
Steklov Mathematical Institute in Moscow. As the name suggests, these lectures tell about algebraic Bethe  ansatz.
Let me give a brief historical comment for those readers who first hear these words.

In 1931 H.~Bethe proposed an original method for constructing the eigenfunctions of the quantum Hamiltonian of a Heisenberg spin chain \cite{Bet31}.
This method was called the Bethe ansatz\footnote{From the German {\it ansatz}, that is, approach, method.} and gave rise to a new approach to the
study of a wide class of quantum systems. Despite the fact that the models solved by the Bethe ansatz are
$(1+1)$-dimensional, they find a rather wide application in different areas of quantum physics, for example, in condensed matter physics,
models of superconductivity and nonlinear optics. Moreover, at the beginning of the 21st century, it was unexpectedly found that this
method is very effective in solving a number of problems in theories of higher dimensions, in particular, in supersymmetric gauge
field theories and string theory.

At the turn of the 70-80s of the 20th century in the works of
Leningrad School under the leadership of L.~D.~Faddeev a Quantum  Inverse Scattering Method (QISM) was developed \cite{FadST79,FadT79,Skl80}. In the framework of this method
it was found that many quantum models solved by Bethe ansatz and having a completely different physical interpretation,
can be described by the same algebra of operators, being in essence
different representations of this algebra. In this case, many important properties of physical systems can be established already
at the level of algebra, without using its concrete representation. This approach was called
algebraic Bethe ansatz. If very briefly, then the algebraic Bethe is a method of working with a special operator algebra
describing a rather wide class of quantum systems. The lectures that are offered
attention of the reader tell just about this.

Despite the considerable volume, these lectures in no way claim to be complete. In particular, one cannot find here
the original method proposed by H.~Bethe, which, after the appearance of the algebraic ansatz
became known as the coordinate Bethe ansatz. I did not do this, because the description of the coordinate  Beta ansatz already exists in
numerous and beautiful  literature (see e.g. \cite{Gaud83,KBIr}). The lectures also do not say anything about the nested (hierarchical) algebraic Bethe ansatz
\cite{KulRes81,KulRes82,KulRes83}, but for another
reason: this approach is still largely in the development stage. I also did not mention related approaches, such as the separation of variables method
by Sklyanin \cite{Skl85}  and the method of the Baxter $Q$-operator \cite{Bax72,Bax82}. I did not try to cover the largest number of topics and approaches. Instead, I focused
on one method, but I tried to cover it in detail. I also put in these lecture notes a series of results that were obtained
in recent years and which have not yet been described in monographs.

The list of bibliography placed at the end of the lecture notes is also not intended to be complete. I believe that such a list
is unrealistic, since the literature on the algebraic Bethe ansatz and related problems is extremely extensive. Let me mention,
however, a number of review articles and books \cite{Takh85,FadLH96,Ize99,Plun09,Fra11,WanYCS15,Lev16} in which the reader
can find material that is not included in these lectures. It should be noted, however, that the above works
are mainly dealing with the application of the algebraic Bethe ansatz to the calculation of the spectrum of quantum-mechanical Hamiltonians.
The peculiarity of this lecture notes is that besides the spectral problem I tried to pay more attention  to the application
of this method to the calculation of correlation functions. At the same time, I touched only on the part of the problem that is quite
common to a large class of physical models. Detailed and comprehensive coverage of issues related to the calculation of correlation functions,
is beyond the scope of this course. The reader is referred to the book \cite{KBIr} in which he can get acquainted with some of these issues and methods of their solution.

To understand the lectures it is necessary to know the linear algebra, analysis, the theory of functions of a complex variable, some information from the functional analysis. Knowledge of quantum mechanics is welcome, however, as a rule, this is required only for the physical interpretation of the results obtained. In the text, for the most part, the double numeration of formulas is adopted: the first number is the section number in the lecture, the second number is the number of the formula in this section. When referring to formulas from other lectures, an additional number is introduced, showing the number of the lecture.

In conclusion, I apologize in advance for misprints  that are unavoidable in a text of such a volume.
I also want to thank S.~Belliard, N.~Kitanine, K.~K.~Kozlowski, J.~M.~Maillet, S.~Z.~Pakuliak, E.~Ragoucy, and V.~Terras in co-authorship with
whom a number of new results, included in the
these lectures, were obtained. I am also grateful to A.~A.~Hutsalyuk, A.~V.~Ilina, A.~N.~Liashyk, D.~S.~Rudneva, and V.~S.~Fanaskov who were the first
to take the blow, listened to these lectures and made many valuable comments.


%
%

\chapter{Matrix tensor product\label{CHA-TPM}}

Most of the content in this lecture is introductory. It mainly concerns the tensor
product of linear spaces and  matrices acting in these spaces. A reader who is familiar with these issues
can look through this lecture, paying attention only to the notation that will be used further.
In particular, we use the superscripts to denote matrix elements (for example, $A^{jk}$), while
the subscripts are reserved for designating spaces in which the matrices act nontrivially (for example, matrix $R_ {23}$
acts nontrivially in spaces with numbers $2$ and $3$). Only section~\ref{01-Sec22} has a direct relation to quantum integrable models.
There we introduce the Yang--Baxter equation and find one of its solutions.

\section{Tensor product of linear spaces\label{01-Sec1}}

As a rule, we  deal with tensor products of linear spaces
$\mathbb{C}^2$, therefore the main description is given for this case. However, a generalization to
other finite-dimensional spaces  is quite obvious.

Let us have two finite-dimensional linear spaces $V_1$ and $V_2$ with bases $\{e^k\}$ and $\{f^k\}$, respectively. Their tensor product $V_1\otimes V_2$ is the space generated by the elements $e^{ik} = e^{i} \otimes f^{k}$, which in turn are called tensor products of basis vectors. Thus, if the dimensions of the spaces $V_1$ and $V_2$  respectively are $N_1$ and $N_2$, then the space $V_1\otimes V_2$ has dimension $N_1N_2$.

The operation of the tensor product is bilinear, that is,
\be{01-bilin}
 \begin{aligned}\
&(\lambda x_{1}+\mu x_{2})\otimes y=\lambda x_{1}\otimes y+\mu x_{2}\otimes y, \qquad \{x_1,x_2\}\in V_1,\quad y\in V_2,\\
&x\otimes (\lambda y_{1}+\mu y_{2})=\lambda x\otimes y_{1}+\mu x\otimes y_{2}, \qquad x\in V_1,\quad \{y_1,y_2\}\in V_2.
\end{aligned}
\ee
Then, for any two vectors $x\in V_1$ and  $y\in V_2$, their tensor product can be defined as a decomposition with respect to the basis $e^{ik}$.
Namely, if
\be{01-razloah}
x=\sum_{i=1}^{N_1} \lambda_i e^i,\qquad\qquad
y=\sum_{k=1}^{N_2} \mu_k f^k,
\ee
then
\be{01-tenzprodxy}
x\otimes y = \sum_{i=1}^{N_1}\sum_{k=1}^{N_2} \lambda_i\mu_k \;e^{ik}.
\ee

Let $V_k\sim \mathbb{C}^2$, $k=1,2$.
Then a basis in the space $V_1\otimes V_2$ consists of four vectors, for example,
\be{01-eij}
e^{11}=\begin{pmatrix} 1\\0\end{pmatrix}\otimes\begin{pmatrix} 1\\0\end{pmatrix},
\quad
e^{12}=\begin{pmatrix} 1\\0\end{pmatrix}\otimes\begin{pmatrix} 0\\1\end{pmatrix},
\quad
e^{21}=\begin{pmatrix} 0\\1\end{pmatrix}\otimes\begin{pmatrix} 1\\0\end{pmatrix},
\quad
e^{22}=\begin{pmatrix} 0\\1\end{pmatrix}\otimes\begin{pmatrix} 0\\1\end{pmatrix}.
\ee
This implies that  $V_1\otimes V_2$ is isomorphic to the space  $\mathbb{C}^4$.

Consider two vectors $x$ and $y$ belonging to the different spaces $\mathbb{C}^2$. We can understand them as two-component columns
\be{01-xy1}
x=\begin{pmatrix} x^1\\x^2\end{pmatrix},
\qquad
y=\begin{pmatrix} y^1\\y^2\end{pmatrix}.
\ee
Then we can write down
their tensor product $x \otimes y$ as a two-component block-column (that is, $2 \times 1$ block-matrix), each block in turn being a two-component column:
\be{01-xoy}
x\otimes y=\begin{pmatrix} x^1y^1\\x^1y^2\numa{18}x^2y^1\\x^2y^2\end{pmatrix}\hspace{-13.5mm}\overline{\phantom{..........}}\hspace{2mm}.
\ee
The first block is the column $y$ multiplied by the component $x^1$, the second block  is the column
$y$ multiplied by the component $x^2$. In general, $x\otimes y$ is a $4$-component column. According
to this rule the basis vectors $e^{ik}$ \eqref{01-eij} take the form
\be{01-eijd}
e^{11}=\left(\begin{smallmatrix} 1\\0\\{}\\0\\0\end{smallmatrix}\right),\hspace{-10mm}\overline{\phantom{......}}\hspace{8.5mm}
\qquad
e^{12}=\left(\begin{smallmatrix} 0\\1\\{}\\0\\0\end{smallmatrix}\right),\hspace{-10mm}\overline{\phantom{......}}\hspace{8.5mm}
\qquad
e^{21}=\left(\begin{smallmatrix} 0\\0\\{}\\1\\0\end{smallmatrix}\right),\hspace{-10mm}\overline{\phantom{......}}\hspace{8.5mm}
\qquad
e^{22}=\left(\begin{smallmatrix} 0\\0\\{}\\0\\1\end{smallmatrix}\right).\hspace{-10mm}\overline{\phantom{......}}\hspace{8.5mm}
\ee
Generically, the components of the tensor product are numbered by two indices, namely, if $z =x\otimes y$, then
$z^{aj} = x^{a} y^{j}$:
\be{01-Comp-xy}
\begin{pmatrix} z^{11}\\z^{12}\numa{18} z^{21}\\z^{22}\end{pmatrix}\hspace{-10.5mm}\overline{\phantom{........}}\hspace{2mm}
=\begin{pmatrix} x^1y^1\\x^1y^2\numa{18}x^2y^1\\x^2y^2\end{pmatrix}\hspace{-13.5mm}\overline{\phantom{..........}}\hspace{2mm}.
\ee
Here the first index $a=1,2$ shows the number of the block, while the second index $j=1,2$ shows the number of the element in the block.

Of course, we could enumerate the components of the vector $z$ with a single index running through the values $1, 2, 3, 4$,
however, the two-index enumeration emphasizes that we  deal with the tensor product of two spaces.

Consider now matrices acting in the space $V_1\otimes V_2$. Clearly, these are $4\times 4$ matrices,
whose matrix elements could be enumerated by a pair of indices taking the values $1,2,3,4$. It is also clear that in this case it
is more convenient to determine the matrix elements by two pairs of indices $A^{ab,jk}$, each of which takes values $1,2$:
\be{01-A}
A=\begin{pmatrix}
A^{11,11}&A^{11,12}&A^{12,11}&A^{12,12}\\
A^{11,21}&A^{11,22}&A^{12,21}&A^{12,22}\numa{20}
A^{21,11}&A^{21,12}&A^{22,11}&A^{22,12}\\
A^{21,21}&A^{21,22}&A^{22,21}&A^{22,22}
\end{pmatrix}
\hspace{-29mm}\vrule \hspace{-26.5mm}\overline{\phantom{.................................................}}
\hspace{5mm}.
\ee
Thus, the matrix $A$ has block structure. It can be understood as a $2\times 2$ block-matrix, where any block in its turn
is a $2\times 2$ matrix. The first pair of superscripts $a$ and $b$ specifies the block number and corresponds to the first space $V_1$, the second pair
$j$ and $k$ is the element number in the block and corresponds to the second space $V_2$. Therefore, if $z\in V_1\otimes V_2$,
then the action of the matrix $A$ on the vector $z$ is given by the formula
\be{01-act-A-comp}
(Az)^{aj}=A^{ab,jk}z^{bk}.
\ee
Here and below, summation over repeated indices is assumed.
It is easy to check that this action is equivalent to the usual action of $4\times 4$ matrices on $4$-component vectors
\be{01-act-Az}
\begin{pmatrix}
A^{11,11}&A^{11,12}\;&A^{12,11}&A^{12,12}\\
A^{11,21}&A^{11,22}\;&A^{12,21}&A^{12,22}\\
A^{21,11}&A^{21,12}\;&A^{22,11}&A^{22,12}\\
A^{21,21}&A^{21,22}\;&A^{22,21}&A^{22,22}
\end{pmatrix}\!\!
\begin{pmatrix} z^{11}\\z^{12}\\z^{21}\\z^{22}\end{pmatrix}
\!=\!\begin{pmatrix}
A^{11,11}z^{11}+A^{11,12}z^{12}+A^{12,11}z^{21}+A^{12,12}z^{22}\\
A^{11,21}z^{11}+A^{11,22}z^{12}+A^{12,21}z^{21}+A^{12,22}z^{22}\\
A^{21,11}z^{11}+A^{21,12}z^{12}+A^{22,11}z^{21}+A^{22,12}z^{22}\\
A^{21,21}z^{11}+A^{21,22}z^{12}+A^{22,21}z^{21}+A^{22,22}z^{22}
\end{pmatrix}\!\!.
\ee
(Here and further we do not specify explicitly the block structure
of matrices and vectors). This action on vectors determines the rules of the product of matrices acting in
$V_1\otimes V_2$. Let $A$, $B$, and $C$ act in $V_1\otimes V_2$. Then
\be{01-Prod}
\text{if}\qquad C=AB,\qquad\text{then}\qquad C^{ab,jk}=A^{ac,j\ell}B^{cb,\ell k}.
\ee
Again, it is easy to check that this rule
is equivalent to the standard rule  for the product of $4\times 4$ matrices.

An important special case of matrices acting in $V_1\otimes V_2$ are matrices that act nontrivially
only in one of the two spaces, while in the other space they act as the identity operator.
We denote such the matrices by the subscripts: $A_1$ or $A_2$. The latter
show in which space ($V_1$ or $V_2$) the given matrix acts nontrivially
as a $2\times 2 $ matrix $A$:
\be{01-act-A1A2}
A_1(x\otimes y)=(Ax)\otimes y,\qquad A_2(x\otimes y)=x\otimes (Ay).
\ee
Clearly, the entries of the matrices $A_1$ and $A_2$ have the form
\be{01-El-A1A2}
A_1^{ab,jk}=A^{ab}\delta^{jk}, \qquad A_2^{ab,jk}=\delta^{ab}A^{jk}.
\ee
It is important to understand that in this case we are actually dealing with one $2 \times 2$ matrix $A$. We just
extend the action of this matrix from the space $\mathbb{C}^2$ to the space $V_1\otimes V_2$, and we can do this in two different ways. For clarity, we give the explicit form of the matrices $A_1$ and $A_2$:
\be{01-AIIA}
A_1=\begin{pmatrix}
A^{11}&0&A^{12}&0\\
0&A^{11}&0&A^{12}\\
A^{21}&0&A^{22}&0\\
0&A^{21}&0&A^{22}
\end{pmatrix},\qquad
A_2=\begin{pmatrix}
A^{11}&A^{12}&0&0\\
A^{21}&A^{22}&0&0\\
0&0&A^{11}&A^{12}\\
0&0&A^{21}&A^{22}
\end{pmatrix}.
\ee
In both cases, we have the same matrix elements $A^{\alpha\beta}$ ($\alpha,\beta=1,2$),
but they are placed in different ways.

Let us agree upon terminology that will be used below. Let
matrices $A_1$ and $B_2$ be given. We say that these matrices act in different spaces.
Strictly speaking, this is not true, because both matrices act in the tensor product $V_1\otimes V_2$,
however, their non-trivial action indeed applies to different spaces: $A_1$ on $V_1$, and $B_2$ on $V_2$.
For brevity, we omit the word {\it nontrivially}.

Now we can introduce the notion of a tensor product of matrices. Let $A$ and $B$ be $2\times 2$ matrices.
It is natural to call their tensor product $A\otimes B $ a matrix acting in $V_1 \otimes V_2$ as follows:
\be{01-act-AB1}
(A\otimes B)(x\otimes y)=(Ax)\otimes( By).
\ee
Let us introduce matrices $A_1$ and $B_2$ according to the definition above. Then
\begin{align}\label{01-act-AB12}
A_1B_2(x\otimes y)&=A_1\bigl(x\otimes( By)\bigr)=(Ax)\otimes( By),\\
\label{01-act-AB21}
B_2A_1(x\otimes y)&=B_2\bigl((Ax)\otimes y\bigr)=(Ax)\otimes( By).
\end{align}
Hence,
\be{01-A-ten-B}
(A\otimes B)=A_1B_2=B_2A_1.
\ee
The commutativity of the matrices $A_1$ and $B_2$  follows from the fact that they act in different spaces.
This can be easily checked by the direct calculation
\begin{align}\label{01-AB12}
(A_1B_2)^{ab,jk}&=A_1^{ac,j\ell}B_2^{cb,\ell k}=A^{ac}\delta^{j\ell}\delta^{cb}B^{\ell k}=A^{ab}B^{jk},\\
\label{01-BA21}
(B_2A_1)^{ab,jk}&=B_2^{ac,j\ell}A_1^{cb,\ell k}=\delta^{ac}B^{j\ell}A^{cb}\delta^{\ell k}=B^{jk}A^{ab}.
\end{align}

Thus, the tensor product of the matrices
$A$ and $B$ is a  $4\times4$ matrix having a block structure.
Every block is the $2\times2$ matrix $B$ multiplied by the corresponding
element $A^{ab}$ of the matrix $A$:
\be{01-AB2}
A\otimes B=\begin{pmatrix}
A^{11}B^{11}&A^{11}B^{12}&A^{12}B^{11}&A^{12}B^{12}\\
A^{11}B^{21}&A^{11}B^{22}&A^{12}B^{21}&A^{12}B^{22}\\
A^{21}B^{11}&A^{21}B^{12}&A^{22}B^{11}&A^{22}B^{12}\\
A^{21}B^{21}&A^{21}B^{22}&A^{22}B^{21}&A^{22}B^{22}
\end{pmatrix}
%
\ee

{\sl Remark.} Generally, we can tensor multiply matrices of any size.
Let we have an $N_1\times M_1$ matrix $A$  and an $N_2\times M_2$ matrix $B$. Their tensor product
is a matrix $C=A\otimes B$ of the size $N_1N_2\times M_1M_2$ and having a block structure.
One should understand this matrix as  $N_1\times M_1$ block-matrix, where every block in its turn
is an $N_2\times M_2$ matrix. The entries of the matrix $C$ are numerated by two pairs of
superscripts $C^{ab,jk}$. The first pair  $a=1,\dots,N_1$ and $b=1,\dots,M_1$   determines the number of the block,
the second pair $j=1,\dots,N_2$ and $k=1,\dots,M_2$ corresponds to the number of the element in the block. The entries
are $C^{ab,jk}=A^{ab}B^{jk}$. In particular, we used just this rule for the tensor product
of two $2\times 1$ matrices  $x$ and $y$ in \eqref{01-xoy}.

Similarly to the notation $A_1$ and $A_2$, we could denote by $A_ {12}$ a matrix  that acts nontrivially in both components of the tensor product
$V_1\otimes V_2$. However, as long as we are dealing with the tensor product of
only two spaces, such the notation is often superfluous. Exceptions are cases when we are dealing with
a permutation of spaces $V_1$ and $V_2$. Let us consider this case.

An important operator acting in the space $V_1\otimes V_2$ is a permutation matrix $P$.
The main property of this matrix is
\be{01-Perm-xy}
P(x\otimes y)=y\otimes x,
\ee
for arbitrary $x$ and $y$, that is
\be{01-Perm-VV}
P(V_1\otimes V_2)=V_2\otimes V_1.
\ee
It is easy to find from \eqref{01-Perm-xy} that the entries
of the permutation matrix have the form
\be{01-Permut-mel}
P^{ab,jk}=\delta^{ak}\delta^{bj},
\ee
and hence,
\be{01-Permut}
P=\begin{pmatrix}
1&0&0&0\\
0&0&1&0\\
0&1&0&0\\
0&0&0&1
\end{pmatrix}.
\ee
It is easy to check that $P^2=\mathbf{1}$ and
\be{01-Permut-AB}
P(A\otimes B)P=B\otimes A,
\ee
for all $A$ and $B$. If a matrix $C$ acts in $V_1\otimes V_2$, then the
permutation matrix swaps the pairs of the indices
\be{01-Permut-C}
(PCP)^{ab,jk}=P^{ac,j\ell}C^{cc',\ell\ell'}P^{c' b,\ell'k}=
\delta^{a\ell}\delta^{cj}C^{cc',\ell\ell'} \delta^{c'k}\delta^{\ell' b}=
C^{jk,ab}.
\ee
We will write this equation in the form
\be{01-Permut-C1}
PC_{12}P=C_{21},\qquad \text{where}\qquad C_{21}^{ab,jk}=C_{12}^{jk,ab}.
\ee
In this case, the  subscripts play an important role. They show which pair of superscripts refers to the first
space $V_1$, and which pair refers to the second space $V_2$. It is easy to check that the matrices $C_ {12}$ and $C_ {21}$
differ from each other by the permutation of the second and the third rows, as well as the second and the third columns.

However, with the exception of such cases, we will always assume that the first pair of the superscripts refers to $V_1$, and the second pair to $V_2$.
Then the subscripts are not important. In particular, it is not necessary to write the subscripts of the permutation matrix $P_ {12}$. This is all the more justified, because due to $P^2=\mathbf{1}$ we have $P_ {12} = P_ {21}$
(since by definition $P_{21}=P_{12}P_{12}P_{12}$).

Concluding this section, we introduce a notion of the trace. As usual, we call the trace of the matrix the sum of its
diagonal elements, that is, $\tr A=A^{aa,jj}$  (we recall that summation over  repeated indices is assumed).
However, in the tensor product $V_1 \otimes V_2$, one can take a partial trace over one of the two spaces:
$\tr_1$ is the trace over the space $V_1$; $\tr_2$ is the trace over the space $V_2$. It is easy to guess that
$\tr_1A=A^{aa,jk}$ and $\tr_2A=A^{ab,jj}$. Thus, the trace of the matrix $A$ over the first space is a matrix
acting in the second space, while the trace of the matrix $A$ with respect to the second space is a matrix
acting in the first space.

Let us calculate, for example, the trace of the permutation matrix over every of the two spaces.
We have
\begin{align}\label{01-trP1}
{\tr}_1P=P^{aa,jk}=\delta^{ak}\delta^{aj}=\delta^{jk}=\mathbf{1}_2,\\
\label{01-trP2}
{\tr}_2P=P^{ab,jj}=\delta^{aj}\delta^{bj}=\delta^{ab}=\mathbf{1}_1,
\end{align}
where  $\mathbf{1}_1$ and $\mathbf{1}_2$ denote the identity matrices acting respectively in
$V_1$  and $V_2$. It is clear, however, that $\mathbf{1}_1=\mathbf{1}_2=\mathbf{1}$, because both matrices $\mathbf{1}_1$ and $\mathbf{1}_2$
act as the identity operator in both spaces. Thus, the trace of the permutation matrix over any of the two spaces
is equal to the identity operator.

\section{Tensor product of several spaces\label{01-Sec2}}

Consider now a tensor product of several spaces $V_1\otimes \cdots \otimes V_N$,
where every $V_k$ is isomorphic to $\mathbb{C}^2$. The elements of this space are linear combinations of vectors $z=x_1\otimes \cdots \otimes x_N$.
Each vector has $2^N$ components, which are numbered by a sequence of superscripts $z^{a_1\dots a_N}$.
Every index $a_\alpha$ corresponds to its space $V_\alpha$ and takes two values $a_\alpha=1,2$.

Matrices acting in this space have $N$ pairs of superscripts: $A^{a_1b_1,\dots,a_Nb_N}$.
They  have the size $2^N\times 2^N$. Their action on vectors is completely analogous to the action in the
tensor product of two spaces, namely, if $w = Az$, then
\be{01-actN}
w^{a_1\dots a_N}=A^{a_1b_1,\dots,a_Nb_N}z^{b_1\dots b_N}.
\ee
The product of two matrices is defined in a similar way: if  $C=AB$, then
\be{01-prod-AB-N}
C^{a_1b_1,\dots,a_Nb_N}=A^{a_1c_1,\dots,a_Nc_N}B^{c_1b_1,\dots,c_Nb_N}.
\ee

One can understand these matrices as a system of nested $2\times 2$ blocks-matrices.
First, a $2^N\times 2^N$ matrix is divided into four blocks of size $2^{N-1}\times 2^{N-1}$,
then each block in its turn is divided into four blocks of the size $2^{N-2}\times 2^{N-2}$
and so on. The first pair of superscripts $a_1$ and $b_1$ numerates the largest blocks $2^{N-1}\times 2^{N-1}$,
the second pair numerates the  blocks of the size $2^{N-2}\times 2^{N-2}$  and so on.
However these formulas are almost never used, because of their cumbersomeness.

One can also introduce a permutation matrix acting in two of $N$ spaces. Now we should provide it with subscripts, in order to indicate these spaces.
For instance,
\be{01-Per-kn}
P_{kn}V_1\otimes \cdots \otimes V_k\otimes \cdots \otimes V_n \otimes \cdots\otimes V_N=
V_1\otimes \cdots \otimes V_n\otimes \cdots \otimes V_k \otimes \cdots\otimes V_N.
\ee
The permutation matrix has entries
\be{01-Per-kn-comp}
P_{kn}^{a_1b_1,\dots,a_Nb_N}=\delta^{a_kb_n}\delta^{a_nb_k}
\prod_{\substack{j=1\\j\ne k,n}}^N\delta^{a_jb_j},
\ee
that is, this matrix acts in the product $V_k\otimes V_n$ as the usual permutation matrix $P$ considered above,
while in other spaces it acts as the identity matrix.

Below, we will often deal with matrices that, like the permutation matrix, act only
in two spaces. We will supply such matrices with two subscripts. For instance,
\be{01-A-kn-comp}
A_{kn}^{a_1b_1,\dots,a_kb_k,\dots,a_nb_n,\dots,a_Nb_N}=A^{a_kb_k,a_nb_n}
\prod_{\substack{j=1\\j\ne k,n}}^N\delta^{a_jb_j}.
\ee
As an example, we give an explicit form of matrices acting in two components of the tensor product of three spaces
$V_1\otimes V_2\otimes V_3$ in section~\ref{01-Sec3}.

Sometimes we use the  subscripts in order to stress, that a matrix $W$ acts nontrivially in all $N$ spaces: $W_{1\dots N}$.
This notation is convenient, for example, when writing the main property
of the permutation matrix
\be{01-Per-kn-prop}
P_{kn}W_{1\dots k\dots n\dots N}P_{kn}=W_{1\dots n\dots k\dots N},
\ee
that is, the permutation matrix $P_{kn}$ exchange the positions of the indices $k$ and $n$.

Finally, just as in the case of the tensor product of two spaces,
we can define partial traces of matrices with respect to some component of the tensor product:
\be{01-trVk}
{\tr}_k A =A^{a_1b_1,\dots,a_ka_k,\dots,a_Nb_N}.
\ee
It is easy to check, for example, that the trace of the permutation matrix $P_{kn}$ over any of two spaces $V_k$ or $V_n$
results to the identity operator.

\subsection{Example.\label{01-Sec21}}

Let us calculate in the space $V_1\otimes V_2\otimes V_3\otimes V_4$
\be{01-U1}
{\tr}_1 (P_{12}A_{13}A_{14}).
\ee

{\sl First way.} Using $P_{12}^2=\mathbf{1}$ we have
\be{01-Sp1-U1}
{\tr}_1 (P_{12}A_{13}A_{14})={\tr}_1 (P_{12}\underline{A_{13}A_{14}}P_{12}P_{12}).
\ee
The underlined matrices are  between two permutation matrices $P_{12}$. Hence, the actions of $P_{12}$
on these matrices leads to the replacement of the subscript $1$ with the subscript $2$. We obtain
\be{01-Sp1-U2}
{\tr}_1 (P_{12}A_{13}A_{14})={\tr}_1 (A_{23}A_{24}P_{12}).
\ee
Now there is only one matrix $P_{12}$ who acts in the space $V_1$, while $A_{23}$ and $A_{24}$
act in $V_1$ as the identity operator. Therefore, taking the trace ${\tr}_1$, we actually should take the trace of the
permutation matrix only. The latter gives the identity operator, what leads to
\be{01-Sp1-U3}
{\tr}_1 (P_{12}A_{13}A_{14})=A_{23}A_{24}.
\ee

{\sl Second way.} Let us compute the trace directly. We have
\begin{multline}\label{01-Sp2-U1}
{\tr}_1 (P_{12}A_{13}A_{14})={\tr}_1\left(P_{12}^{a_1c_1,a_2c_2,a_3c_3,a_4c_4}
A_{13}^{c_1c'_1,c_2c'_2,c_3c'_3,c_4c'_4}A_{14}^{c'_1b_1,c'_2b_2,c'_3b_3,c'_4b_4} \right)\\
=P_{12}^{a_1c_1,a_2c_2,a_3c_3,a_4c_4}
A_{13}^{c_1c'_1,c_2c'_2,c_3c'_3,c_4c'_4}A^{c'_1a_1,c'_2b_2,c'_3b_3,c'_4b_4}.
\end{multline}
Substituting here \eqref{01-Per-kn-comp},  \eqref{01-A-kn-comp} we find
\be{01-Sp2-U2}
{\tr}_1 (P_{12}A_{13}A_{14})=\delta^{a_1c_2}\delta^{a_2c_1}\delta^{a_3c_3}\delta^{a_4c_4}
\delta^{c_2c'_2}\delta^{c_4c'_4}\delta^{c'_2b_2}\delta^{c'_3b_3}A^{c_1c'_1,c_3c'_3}A_{14}^{c'_1a_1,c'_4b_4}.
\ee
Summing up the $\delta$-symbols we obtain
\be{01-Sp2-U3}
{\tr}_1 (P_{12}A_{13}A_{14})=A^{a_2c'_1,a_3b_3}A^{c'_1b_2,a_4b_4}=A_{23}A_{24}.
\ee
We see that the second method, being formally applicable, is more cumbersome than the first one.

\section{Yang--Baxter equation\label{01-Sec22}}

Let a $R(u_1,u_2)$ act $V_1\otimes V_2$ and depend on two complex variables $u_1$ and $u_2$. Consider a space $V_1\otimes V_2\otimes V_3$ and introduce
matrices $R_{12}(u_1,u_2)$,  $R_{13}(u_1,u_3)$, and $R_{23}(u_2,u_3)$ (their explicit form
is given in section~\ref{01-Sec3}). Let us study an equation
\be{01-Yangian}
R_{12}(u_1,u_2)R_{13}(u_1,u_3)R_{23}(u_2,u_3)=
R_{23}(u_2,u_3)R_{13}(u_1,u_3)R_{12}(u_1,u_2).
\ee

Equation \eqref{01-Yangian} is a matrix functional equation. It is called the Yang--Baxter equation and plays an exceptionally important
role in the theory of exactly solvable models (see e.g. \cite{Jim90,PerA06}). The solutions of this equation are called
$R$-matrices. In this exercise, we do not set out to find
all solutions of the Yang--Baxter equation. Instead  we will try to find at least one
nontrivial solution.

Observe that if $R(u_1,u_2)$ is a solution of \eqref{01-Yangian}, then $\varphi(u_1,u_2)R_{12}(u_1,u_2)$ also is a solution, where
$\varphi(u_1,u_2)$ is an arbitrary scalar function of two variables. Therefore, we can look for the solutions of equation \eqref{01-Yangian}
up to multiplication by a scalar function.

One obvious solution to \eqref{01-Yangian} is the identity matrix. The second slightly less obvious solution is the permutation matrix $P$. Indeed, multiplying the identity
\begin{equation}\label{01-c33}
P_{23}P_{13}=P_{23}P_{13}
\end{equation}
from the right by $P_{12}$ we obtain
\begin{equation}\label{01-c3}
P_{23}P_{13}P_{12}=P_{12}P_{12}P_{23}P_{13}P_{12}=P_{12}P_{13}P_{23},
\end{equation}
where we have used $P_{12}^2=\mathbf{1}$ and
$P_{12}P_{13}P_{23}P_{12}=P_{23}P_{13}$.

Let us try to find a solution in the form of a linear combination of the identity and the permutation matrices and depending on the difference $u_1-u_2$.
Let $u_1-u_2=u$ and $u_2-u_3=v$.
Then $u_1-u_3=u+v$, and the Yang--Baxter equation takes the form
\be{01-Yangian2}
R_{12}(u)R_{13}(u+v)R_{23}(v)= R_{23}(v)R_{13}(u+v)R_{12}(u).
\ee
We are looking for the solution to this equation in the form $R(u)=\phi(u)\mathbf{1}+cP$, where $\phi(u)$ is a smooth function
is $c$ is a constant. The matrix equation \eqref{01-Yangian2}
yields a functional equation for the function $\phi(u)$. Let us derive this equation. For this we substitute $R(u)=\phi(u)\mathbf{1}+cP$ into \eqref{01-Yangian2}, and expand the lhs and the rhs of \eqref{01-Yangian2} over powers of $c$. At $c^0$ and $c^3$ we obtain obvious identities.
At $c^1$ we also have an identity:
\begin{multline}\label{01-c1}
\phi(v)\phi(u+v)P_{12}+\phi(v)\phi(u)P_{13}+\phi(u)\phi(u+v)P_{23}\\
=\phi(u)\phi(u+v)P_{23}+\phi(v)\phi(u)P_{13}+\phi(v)\phi(u+v)P_{12}.
\end{multline}

A nontrivial equation arises at $c^2$ only:
\be{01-c2}
\phi(u)P_{13}P_{23}+\phi(v)P_{12}P_{13}+\phi(u+v)P_{12}P_{23}=
\phi(u)P_{23}P_{13}+\phi(v)P_{13}P_{12}+\phi(u+v)P_{23}P_{12}.
\ee
Multiplying \eqref{01-c2} from the left by $P_{12}$ we obtain
\be{01-c2-2}
\phi(u)P_{12}P_{13}P_{23}+\phi(v)P_{13}+\phi(u+v)P_{23}=
\phi(u)P_{12}P_{23}P_{13}+\phi(v)P_{23}+\phi(u+v)P_{13}.
\ee
Now we multiply this equation from the right by $P_{13}$:
\be{01-c2-3}
\phi(u)\mathbf{1}+\phi(v)\mathbf{1}+\phi(u+v)P_{23}P_{13}=
\phi(u)P_{12}P_{23}+\phi(v)P_{23}P_{13}+\phi(u+v)\mathbf{1}.
\ee
Finally, multiplying \eqref{01-c2-3} from the right by $P_{23}$, we eventually obtain
\be{01-c2-23}
\phi(u)P_{23}+\phi(v)P_{23}+\phi(u+v)P_{12}=
\phi(u)P_{12}+\phi(v)P_{12}+\phi(u+v)P_{23},
\ee
or equivalently,
\be{01-c2-4}
\bigl(\phi(u)+\phi(v)-\phi(u+v)\bigr)P_{23}=\bigl(\phi(u)+\phi(v)-\phi(u+v)\bigr)P_{12}.
\ee
Since $P_{12}$ and $P_{23}$ are different matrices, the equality is possible if and only if
\be{01-fuct-eq}
\phi(u+v)=\phi(u)+\phi(v)
\ee
for any $u$ and $v$. It is well known that the linear function $\phi(u) = au$ (where $a$ is an arbitrary constant)
is the only smooth  solution of the functional equation \eqref{01-fuct-eq}. The constant $a$ can be included
in the constant $c$, and thus we find that the matrix
\be{01-sol-YB}
R(u_1,u_2)=(u_1-u_2)\mathbf{1}+cP
\ee
solves the Yang--Baxter equation  \eqref{01-Yangian}. This is the only solution that depends on the difference of
the arguments and is a linear combination of the permutation matrix and the identity matrix. We also note that in
the derivation of this solution we did not use the fact that the spaces $V_k$ were isomorphic to $\mathbb{C}^2$.
Therefore, the solution obtained is also valid for matrices $R$ acting in the space $\mathbb{C}^N\otimes \mathbb{C}^N$
for arbitrary $N$.


\section{Explicit form of matrices in the space $V_1\otimes V_2\otimes V_3$ \label{01-Sec3}}

As an illustration, we present the explicit form of  matrices acting in the space $V_1\otimes V_2\otimes V_3$.
Let a matrix $R$ act $V_1\otimes V_2$, where $V_k\sim\mathbb{C}^2$. It has an explicit form
\be{01-R-single}
R=\begin{pmatrix}
R^{11,11}&R^{11,12}&R^{12,11}&R^{12,12}\\
R^{11,21}&R^{11,22}&R^{12,21}&R^{12,22}\\
R^{21,11}&R^{21,12}&R^{22,11}&R^{22,12}\\
R^{21,21}&R^{21,22}&R^{22,21}&R^{22,22}
\end{pmatrix}
\ee

Let us introduce matrices $R_{12}$, $R_{13}$, and $R_{23}$, acting $V_1\otimes V_2\otimes V_3$.
Each of them acts only in a pair of spaces as a matrix $R$ with elements $ R^{ab,jk} $. In the remaining space,
it acts as the identity matrix. All matrices have the size  $8\times 8$. Below is their explicit form.
\be{01-R12}
R_{12}=\begin{pmatrix}
R^{11,11}&0&R^{11,12}&0&R^{12,11}&0&R^{12,12}&0\\
0&R^{11,11}&0&R^{11,12}&0&R^{12,11}&0&R^{12,12}\\
R^{11,21}&0&R^{11,22}&0&R^{12,21}&0&R^{12,22}&0\\
0&R^{11,21}&0&R^{11,22}&0&R^{12,21}&0&R^{12,22}\\
R^{21,11}&0&R^{21,12}&0&R^{22,11}&0&R^{22,12}&0\\
0&R^{21,11}&0&R^{21,12}&0&R^{22,11}&0&R^{22,12}\\
R^{21,21}&0&R^{21,22}&0&R^{22,21}&0&R^{22,22}&0\\
0&R^{21,21}&0&R^{21,22}&0&R^{22,21}&0&R^{22,22}
\end{pmatrix}
\ee
\vspace{5mm}
\be{01-R13}
R_{13}=\begin{pmatrix}
R^{11,11}&R^{11,12}&0&0&R^{12,11}&R^{12,12}&0&0\\
R^{11,21}&R^{11,22}&0&0&R^{12,21}&R^{12,22}&0&0\\
0&0&R^{11,11}&R^{11,12}&0&0&R^{12,11}&R^{12,12}\\
0&0&R^{11,21}&R^{11,22}&0&0&R^{12,21}&R^{12,22}\\
R^{21,11}&R^{21,12}&0&0&R^{22,11}&R^{22,12}&0&0\\
R^{21,21}&R^{21,22}&0&0&R^{22,21}&R^{22,22}&0&0\\
0&0&R^{21,11}&R^{21,12}&0&0&R^{22,11}&R^{22,12}\\
0&0&R^{21,21}&R^{21,22}&0&0&R^{22,21}&R^{22,22}
\end{pmatrix}
\ee

\vspace{5mm}

\be{01-R23}
R_{23}=\begin{pmatrix}
R^{11,11}&R^{11,12}&R^{12,11}&R^{12,12}&0&0&0&0\\
R^{11,21}&R^{11,22}&R^{12,21}&R^{12,22}&0&0&0&0\\
R^{21,11}&R^{21,12}&R^{22,11}&R^{22,12}&0&0&0&0\\
R^{21,21}&R^{21,22}&R^{22,21}&R^{22,22}&0&0&0&0\\
0&0&0&0&R^{11,11}&R^{11,12}&R^{12,11}&R^{12,12}\\
0&0&0&0&R^{11,21}&R^{11,22}&R^{12,21}&R^{12,22}\\
0&0&0&0&R^{21,11}&R^{21,12}&R^{22,11}&R^{22,12}\\
0&0&0&0&R^{21,21}&R^{21,22}&R^{22,21}&R^{22,22}
\end{pmatrix}
\ee
%


%
%
\chapter{Quantum integrable systems\label{CHA-IS}}

In this lecture we describe the general scheme for constructing quantum integrable models in the framework of the Quantum Inverse Scattering
Method (QISM). We also show how this scheme works in a specific example of the $XXX$ Heisenberg chain \cite{Hei28}.
It was to this model that H.~Bethe applied his new method \cite{Bet31}, which later became one of the starting points of QISM.
The physical content of this model can be found in \cite{Gaud83}.

\section{$XXX$  Heisenberg chain \label{02-sec0}}

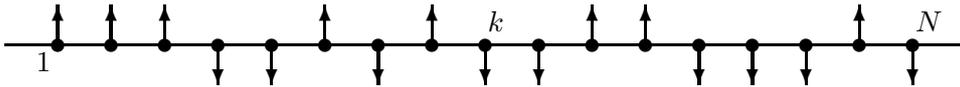
\begin{figure}[h!]
\begin{picture}(400,70)
\put(10,0){\begin{picture}(380,70)
\thicklines
\put(20,30){\line(1,0){360}}
\multiput(40,30)(20,0){17}{\circle*{5}}
\multiput(40,30)(20,0){3}{\vector(0,1){15}}
\put(140,30){\vector(0,1){15}}
\put(180,30){\vector(0,1){15}}
\put(240,30){\vector(0,1){15}}
\put(260,30){\vector(0,1){15}}
\put(340,30){\vector(0,1){15}}
\multiput(100,30)(20,0){2}{\vector(0,-1){15}}
\multiput(200,30)(20,0){2}{\vector(0,-1){15}}
\put(160,30){\vector(0,-1){15}}
\multiput(280,30)(20,0){3}{\vector(0,-1){15}}
\put(360,30){\vector(0,-1){15}}
\put(32,20){$1$}
\put(361,35){$N$}
\put(201,35){$k$}
\end{picture}}
\end{picture}
\caption{\label{02-QH-XXZchain} $XXX$ Heisenberg chain.}
\end{figure}

Consider a chain of $N$ spin-$1/2$ particles  (see. Fig.~\ref{02-QH-XXZchain}) and assume that
the nearest neighbors in this chain interact with each other.

Then the Hamiltonian of this quantum system can be written in the form
\be{02-QH-Ham-XYZ}
H=\sum_{k=1}^N\Bigl(J_x\sigma_k^x\sigma_{k+1}^x+J_y\sigma_k^y\sigma_{k+1}^y+J_z\sigma_k^z\sigma_{k+1}^z\Bigr).
\ee
Here $J_x$, $J_y$, and $J_z$ respectively are interaction constants along the axis $x$, $y$, and $z$, while  $\sigma_{k}^{x,y,z}$ are the projections
of the spin operators of  $k$th particle in these axis. Such the model is  called the $XYZ$ Heisenberg chain (magnet). If two of the interaction constants
are equal, for example $J_x = J_y$, but $J_x \ne J_z$, then the model is called the $XXZ$ Heisenberg chain. Finally, in the case $J_x = J_y = J_z$
the model is known as the $XXX$ Heisenberg chain, and its Hamiltonian has the form
\be{02-QH-Ham-XXZ}
H=\sum_{k=1}^N\Bigl(\sigma_k^x\sigma_{k+1}^x+\sigma_k^y\sigma_{k+1}^y+\sigma_k^z\sigma_{k+1}^z\Bigr),
\ee
where, without any loss of generality, we set all interaction constants equal to $1$.

{\sl Remark}.  We implicitly assumed the  periodic boundary
conditions in \eqref{02-QH-Ham-XYZ},  \eqref{02-QH-Ham-XXZ}, that is, in fact the chain is wrapped in a ring. Indeed, in the sum over $k$ we have the projections of the operator $\sigma_ {N+1}$ for $k = N$, however, there is no such particle in the chain. The situation can be corrected if we set $\sigma_{N+1}^{x,y,z}=\sigma_{1}^{x,y,z}$,  and
then the chain turns into the ring. In this case, the model is still one-dimensional in the sense that the coordinate of the particle on the chain
is given by one integer $k$. Generally, the requirement of periodicity is not necessary. One can also consider open chains \cite{Skl88,Gaud83}.
In this case, we simply replace the upper summation limit in the formulas \eqref{02-QH-Ham-XYZ}, \eqref{02-QH-Ham-XXZ} by $N-1$. However, due to several
reasons we prefer to deal with closed chains. First, these models are simpler from a mathematical point of view. Secondly,
in closed chains, the condition of translational invariance is ensured from the very beginning. Finally, thirdly, the most interesting case is
the case of long chains  (formally we send $N \to \infty$). If we are interested in the processes that occur
far from the boundaries of such a long chain, then for physical reasons they should not depend on the boundary conditions. Therefore, for
descriptions of these physical processes, we can choose the model that deals with the simplest mathematical tools.

The spin-$1/2$ operators  are given by the standard Pauli matrices
\be{02-Pauli}
\sigma^x=\begin{pmatrix} 0&1\\1&0 \end{pmatrix},\qquad
\sigma^y=\begin{pmatrix} 0&-i\\i&0 \end{pmatrix},\qquad
\sigma^z=\begin{pmatrix} 1&0\\0&-1 \end{pmatrix}.
\ee
However, the $\sigma$-matrices have an additional subscript in \eqref{02-QH-Ham-XXZ}. This means that these operators act in the tensor product of spaces
$V_1\otimes \dots \otimes V_N$, where $V_k \sim \mathbb {C}^2$, and $\sigma^{x,y,z}_k $ acts nontrivially in $V_k$ only. In other words,
\be{02-QH-sk}
\sigma^\alpha_k=\mathbf{1}\otimes\dots\otimes \sigma^\alpha \otimes\dots\otimes \mathbf{1},\qquad \alpha=x,y,z.
\ee
Thus, these operators are the matrices of the size $2^N\times 2^N$. Hence, the Hamiltonian also is the matrix  $2^N\times 2^N$.

The standard problem of the quantum mechanics is to find the spectrum of the Hamiltonian of a quantum system. Since in our case the Hamiltonian is a matrix
of a finite size, we are dealing with a linear algebra. Therefore, we could find the eigenvalues of the matrix $H$ by solving the characteristic equation
\be{02-equat}
\det(H-\lambda\mathbf{1})=0.
\ee
Let us, however, assume that the chain under consideration is a model of a one-dimensional macroscopic crystal. It means that
the number of sites of the chain $N$ is equal to the number of atoms in this macroscopic crystal, say, $N\sim 10^6$. Then the Hamiltonian is a matrix
the size of $2^{10^6}\times 2^{10^6}$, and equation \eqref{02-equat} is an algebraic equation of degree $2^{10^6}$. This number is so
huge that there is nothing to compare it with. Clearly, no computer can cope with such a task. We come to the conclusion that the standard
methods of linear algebra in this case are useless, and, therefore, we need to look for some alternative ways.

Such an alternative way was proposed by H.~Bethe in 1931 in the work \cite{Bet31}. Later on this method was called
Bethe ansatz. The reader can find the details of this approach in monographs \cite{Gaud83,KBIr}.

In the late 70's, early 80's of the last century in the works of the Leningrad School (see, for example, \cite{Skl80,FadST79,FadT79})
Quantum Inverse Scattering Method (QISM) was developed. One of the tools of this method is an
algebraic Bethe ansatz. In order to distinguish this new approach from the original method proposed by H.~Bethe, the latter is now called the coordinate
Bethe ansatz. The purpose of these lectures is to get acquainted with the algebraic
Bethe ansatz.

\section{Construction of integrable systems\label{02-Sec1}}

We mainly focus on the relatively recent results obtained in the field of the  algebraic Bethe ansatz. Therefore,
the main notions of the QISM are given fairly briefly, only to the extent that it is necessary to understand the subsequent stuff.
A more complete exposition of the main points of the QISM and the algebraic Bethe ansatz can be found in \cite{Takh85,FadLH96,KBIr}.
The $R$-matrix approach to classical exactly solvable systems is detailed in  \cite{FadT86}.

Let us describe the general procedure for constructing quantum integrable systems, which is applied within the framework of QISM.
Suppose we have a $2\times 2$  matrix $T(u)$. Its matrix elements $T^{ij}(u) $ are operators acting in some
Hilbert space $\mathcal{H}$ and depending on the complex variable $u$. Following the tradition, we will often denote
this matrix as
\be{02-T-ABCD}
T(u)=\begin{pmatrix}
A(u) & B(u)\\
C(u)& D(u)
\end{pmatrix}.
\ee
Let the commutation relations between the operators $ T^{ij}(u)$ be as follows:
\be{02-RTT}
R_{12}(u-v)T_1(u)T_2(v)=T_2(v)T_1(u)R_{12}(u-v),
\ee
where $R(u-v)$ is the solution of the Yang--Baxter equation \eqref{01-sol-YB}:
\be{02-Ruv}
R(u-v)=(u-v)\mathbf{I}+cP=\begin{pmatrix}
u-v+c & 0&0&0\\
0& u-v& c&0\\
0&c& u-v&0\\
0& 0& 0& u-v+c
\end{pmatrix}.
\ee
Equation \eqref{02-RTT} holds in the tensor product of three spaces $V_1\otimes V_2\otimes\mathcal{H}$, where
$V_k\sim \mathbb{C}^2$, $k=1,2$. The $R$-matrix  $R_{12}(u-v)$ acts nontrivially in $V_1\otimes V_2$ and as the identity operator in $\mathcal{H}$.
The matrices $T_k$ act nontrivially in $V_k\otimes\mathcal{H}$. Equation \eqref{02-RTT} sometimes is called an intertwining relation. We shall
call it the $RTT$-relation. Equation \eqref{02-RTT} also can be written in the form
\be{02-RTT-0}
R(u-v)\bigl(T(u)\otimes\mathbf{1}\bigr)\bigl(\mathbf{1}\otimes T(v)\bigr)
=\bigl(\mathbf{1}\otimes T(v)\bigr)\bigl(T(u)\otimes\mathbf{1}\bigr)R(u-v).
\ee
Here we omit the subscripts of the matrices $T$, showing explicitly  in which spaces each of them acts (nontrivially). Sometimes the form
\eqref{02-RTT-0} is more convenient.

In the case of the $R$-matrix \eqref{02-Ruv}, one can easily derive commutation relations of the matrix elements from the matrix equation \eqref{02-RTT}.
Indeed, let us write \eqref{02-RTT} in the form
\be{02-RTT-ind}
\bigl(R_{12}(u-v)\bigr)^{j\ell,\alpha\nu}\bigl(T_{1}(u)\bigr)^{\ell n,\nu\mu}\bigl(T_{2}(v)\bigr)^{nk,\mu\beta}=
\bigl(T_{2}(v)\bigr)^{j\ell,\alpha\nu}\bigl(T_{1}(u)\bigr)^{\ell n,\nu\mu}\bigl(R_{12}(u-v)\bigr)^{nk,\mu\beta},
\ee
and substitute here formulas for the matrix elements:
\be{02-mat-el}
\begin{aligned}
&\bigl(R_{12}(u-v)\bigr)^{jk,\alpha\beta} =(u-v)\delta^{jk}\delta^{\alpha\beta} +c \delta^{j\beta}\delta^{k\alpha},\\
&\bigl(T_{1}(u)\bigr)^{jk,\alpha\beta} = T^{jk}(u)\delta^{\alpha\beta},\\
&\bigl(T_{2}(v)\bigr)^{jk,\alpha\beta} = T^{\alpha\beta}(v)\delta^{jk}.
\end{aligned}
\ee
Then after simple calculations we obtain
\be{02-com-rel}
[T^{ij}(u), T^{kl}(v)]=\frac{c}{u-v}\bigl(T^{kj}(v)T^{il}(u) -T^{kj}(u)T^{il}(v)\bigr).
\ee

Let us turn back to equation \eqref{02-RTT}. Multiply it with $R^{-1}_{12}(u-v)$ from the right and take the trace with respect to the space $V_1\otimes V_2$. Then
\be{02-trace-1}
{\tr}_{12}T_1(u)T_2(v)=  {\tr}_{12}T_2(v)T_1(u),
\ee
where we used the possibility of cyclic permutation in the trace. Since the trace of the tensor product is equal
to the the product of traces, we obtain
\be{02-Tau-Tau}
\mathcal{T}(u)\mathcal{T}(v)=\mathcal{T}(v)\mathcal{T}(u),
\ee
where
\be{02-Tau}
\mathcal{T}(u)=\tr T(u)= A(u)+D(u).
\ee

Let us expand the operator $\mathcal{T}(u)$ into power series over $u$ centered in some point $u_0$
\be{02-expan-T}
\mathcal{T}(u)=\sum_k (u-u_0)^k I_k,
\ee
where the coefficients $I_k$ are some operators acting in the Hilbert space $\mathcal{H}$. In this formula, we do not specify the  summation area
of $k$. Generally, the sum in \eqref{02-expan-T} can be either finite or infinite. In any case, differentiating
\eqref{02-Tau-Tau} $k$ times over $u$ and $n$ times over $v$ at $u,v= u_0$, we find that all the operators $I_k$ commute
\be{02-com-IkIn1}
[I_k,I_n]=0, \qquad \forall k,n.
\ee
If we now set one of these $I_k$ to be the Hamiltonian of some quantum system, then we get a model with, generally speaking,
an infinite set of integrals of motion, that is, an integrable system. This is the general scheme for constructing integrable models in the framework of
QISM.

Let us stop here and discuss the results obtained. So far the role of the matrix $R (u-v)$ is rather mysterious. In particular, it is unclear why it
was taken  in the form of \eqref{02-Ruv}. Indeed, in the above derivation, we could replace this matrix by any other, and in doing so
nothing would have changed. Looking ahead, we say that this choice of the $R$-matrix guarantees the consistency of commutation relations between
the entries of the matrix $T(u)$. We will discuss this issue in more detail in the next lecture.

It should also be mentioned that the existence of an infinite set of integrals of motion does not yet guarantees that the model obtained describes any real
physical integrable system. The operators $I_k$ are not necessarily independent, generically they are not self-adjoint, are not local, and so on.
All these questions are certainly important, and they require a separate study. The reader can find the details in the already mentioned works
\cite{Takh85,FadLH96, KBIr}. We only note that in order to obtain the locality of the operators $I_k$, it is important to choose appropriately the decomposition point $u_0$. In addition, instead of expanding the operator $\mathcal{T}(u)$ into the series, one can expand a function
of this operator. Below, we demonstrate this with a concrete example.

Finally, it follows from the commutation relations \eqref{02-com-rel} that every operator $T^{ij}(u)$ commutes with itself
for arbitrary values of the argument
\be{02-TijTij}
[T^{ij}(u), T^{ij}(v)]=0.
\ee
Therefore, instead of taking the trace $\tr T(u)$, we could consider any matrix element of the matrix $T(u)$ and construct a set of commuting operators
via exactly the same scheme that we used above. In fact, sometimes this way is used. We have chosen  the trace $\tr T(u)$ as the generating function
of the integrals of motion, because we want to obtain a model with periodic boundary conditions. As we shall see later,
this boundary condition automatically arises due to the cyclicity  of the trace.

\section{Constructing the monodromy matrix\label{02-Sec2}}

In the framework of QISM, the matrix $T(u)$ is called
the {\it monodromy matrix}. This term is borrowed from the classical inverse scattering method, where an analog of
$T(u)$ consists of scattering data. The trace of the monodromy matrix $\mathcal{T}(u) =\tr T(u)$ is called the {\it transfer matrix}. In this
case the word `matrix' should be understood in the sense of an `operator' (similarly to the case of $S$-matrix).

One can ask, whether exist such $T(u)$ that satisfy equation \eqref{02-RTT}? It is clear that the identity matrix
satisfies this equation, but this example is not of much interest. In fact, we already have one non-trivial example:
this is the $R$-matrix. Indeed, due to the Yang--Baxter equation
\be{02-RRR}
R_{12}(u-v)R_{13}(u-w)R_{23}(v-w)=R_{23}(v-w)R_{13}(u-w)R_{12}(u-v),
\ee
we can say that $T(u)=R(u-w)$. We should only agreed upon how we understand this equation. We know that $R(u-w)$ is a
$4\times 4$ matrix. However, one can understand it as a $2 \times 2$  matrix, whose entries are operators acting in the Hilbert space
$\mathcal{H} = \mathbb{C}^2$. In other words, these operators are also  $2\times 2$ matrices. Let us set $w = c/2$ for convenience. Then
we have
\be{02-Ruc2}
R(u-\tfrac c2)=\begin{pmatrix}
u+\tfrac c2 & 0&0&0\\
0& u-\tfrac c2& c&0\\
0&c& u-\tfrac c2&0\\
0& 0& 0& u+\tfrac c2
\end{pmatrix}=
\begin{pmatrix}
u+\tfrac c2\sigma^z & c\sigma^-\\
c\sigma^+& u-\tfrac c2\sigma^z
\end{pmatrix},
\ee
where $\sigma^\pm=\tfrac12(\sigma^x\pm i\sigma^y)$. Thus,
\be{02-RRR-1}
R_{12}(u-v)R_{13}(u-\tfrac c2)R_{23}(v-\tfrac c2)=R_{23}(v-\tfrac c2)R_{13}(u-\tfrac c2)R_{12}(u-v),
\ee
and we see that we can set
\be{02-T-R}
T(u)=
\begin{pmatrix}
u+\tfrac c2\sigma^z & c\sigma^-\\
c\sigma^+& u-\tfrac c2\sigma^z
\end{pmatrix}.
\ee

We can verify by the direct calculation that the matrix elements of this monodromy matrix satisfy the commutation relations
\eqref{02-com-rel}. For example,
\be{02-primer-1}
[T^{12}(u), T^{21}(v)]=c^2[\sigma^-,\sigma^+]=-c^2\sigma^z.
\ee
On the other hand,
\begin{multline}\label{02-primer-2}
\frac{c}{u-v}\bigl(T^{22}(v)T^{11}(u) -T^{22}(u)T^{11}(v)\bigr)=\frac{c}{u-v}
\bigl((v-\tfrac c2\sigma^z)(u+\tfrac c2\sigma^z)-(u-\tfrac c2\sigma^z)(v+\tfrac c2\sigma^z)\bigr)\\
=\frac{c}{u-v}\bigl(\tfrac c2\sigma^z(v-u)-\tfrac c2\sigma^z(u-v)\bigr)=-c^2\sigma^z.
\end{multline}

Despite of we have managed to find a nontrivial example of the matrix $T(u)$,
the quantum model constructed on the base of this monodromy matrix is still rather poor, because
\be{02-examp-T}
\mathcal{T}(u)= (u+\tfrac c2\sigma^z)+ (u-\tfrac c2\sigma^z)=2u\mathbf{1}.
\ee

Nevertheless, this example is not useless, because it gives a hint  to construct a reasonable model. For this we consider
two $2\times 2$ matrices $X(u)$ and $Y(u)$ depending on the complex argument $u$.
Let the entries $X^{jk}(u)$ be operators acting in a space $\mathcal{H}_x$,
while the entries  $Y^{jk}(u)$ be operators acting in a space $\mathcal{H}_y$. These two spaces can
be finite-dimensional or even infinite-dimensional. It is crucial, however, that these spaces are different, hence,
$[X^{jk}(u_1),Y^{\ell m}(u_2)]=0$ for all $u_1$ and $u_2$.

Consider the spaces $V_1\otimes V_2\otimes \mathcal{H}_x$ and $V_1\otimes V_2\otimes \mathcal{H}_y$, where  $V_k\sim \mathbb{C}^2$,
$k=1,2$. As usual we can introduce the matrices $X_k$ and $Y_k$ for $k=1,2$. They act nontrivially only in
$V_k\otimes \mathcal{H}_x$ and $V_k\otimes \mathcal{H}_y$.

\begin{prop}\label{02-Predl2}
Let both matrices $X$ and $Y$ enjoy equation \eqref{02-RTT}
\begin{equation}\label{02-RTTXY}
\begin{aligned}
R_{12}(u_1-u_2)X_1(u_1)X_2(u_2)&=
X_{2}(u_2)X_{1}(u_1)R_{12}(u_1-u_2),\\
R_{12}(u_1-u_2)Y_1(u_1)Y_2(u_2)&=
Y_{2}(u_2)Y_{1}(u_1)R_{12}(u_1-u_2).
\end{aligned}
\end{equation}
Then their product $XY$ also enjoys equation \eqref{02-RTT}.
\end{prop}

{\sl Proof.}
Let us substitute the product $XY$ into the lhs of \eqref{02-RTT}
\begin{equation}\label{02-OM-LC}
R_{12}(u_1-u_2)X_1(u_1)Y_1(u_1)X_2(u_2)Y_2(u_2)=
R_{12}(u_1-u_2)X_1(u_1)X_2(u_2)Y_1(u_1)Y_2(u_2).
\end{equation}
Here we have used that $Y_1(u_1)$ acts in the space $V_1\otimes \mathcal{H}_y$, while
$X_2(u_2)$ acts in the space
$V_2\otimes \mathcal{H}_x$, and hence, these matrices commute.
Now, due to \eqref{02-RTTXY}
we can move the matrix $R_{12}(u_1-u_2)$  to the right
\begin{equation}\label{02-OM-LC1}
R_{12}(u_1-u_2)X_1(u_1)Y_1(u_1)X_2(u_2)Y_2(u_2)=
X_2(u_2)X_1(u_1)Y_2(u_2)Y_1(u_1)R_{12}(u_1-u_2).
\end{equation}
It remains to swap $X_1(u_1)$ and $Y_2(u_2)$, and we arrive at
\begin{equation}\label{02-OM-LC2}
R_{12}(u_1-u_2)X_1(u_1)Y_1(u_1)X_2(u_2)Y_2(u_2)=
X_2(u_2)Y_2(u_2)X_1(u_1)Y_1(u_1)R_{12}(u_1-u_2).
\end{equation}
\qed.

Consider now a space $\mathcal{H}$, that is the tensor product of $N$ spaces $\mathbb{C}^2$:
$\mathcal{H}=V_1\otimes\dots\otimes V_N$, where every $V_n\sim \mathbb{C}^2$. Let us build matrices
\be{02-R-Ln}
L_n(u)=
\begin{pmatrix}
u+\tfrac c2\sigma^z_n & c\sigma^-_n\\
c\sigma^+_n& u-\tfrac c2\sigma^z_n
\end{pmatrix},
\ee
where $\sigma_n^\alpha$ is defined by \eqref{02-QH-sk}. We call the matrices $L_n(u) $  $L$-operators. Each matrix element of every $L$-operator is in turn an operator acting in the space $\mathcal{H}$. Observe that  it acts nontrivially only in $V_n$. It is clear that the matrix elements of each $L$-operator
$L_n ^ {ij}(u)$ satisfy the commutation relations \eqref{02-com-rel}. Hence, every $L_n(u) $ satisfies the $RTT$-relation \eqref{02-RTT-0}
\be{02-RLL}
R(u-v)\bigl(L_n(u)\otimes\mathbf{1}\bigr)\bigl(\mathbf{1}\otimes L_n(v)\bigr)
=\bigl(\mathbf{1}\otimes L_n(v)\bigr)\bigl(L_n(u)\otimes\mathbf{1}\bigr)R(u-v).
\ee

We define the monodromy matrix as follows:
\be{02-T-LLL}
T(u)=L_N(u)\dots L_2(u)L_1(u).
\ee
Due to proposition~\ref{02-Predl2}  this matrix $T(u)$ enjoys equation \eqref{02-RTT-0}.

Consider an example when $T(u)$ is the product of two $L$-operators: $T(u)=L_2(u)L_1(u)$. It is easy to see
that in this case
\be{02-Tau-2}
\mathcal{T}(u)=(u+\tfrac c2\sigma^z_2)(u+\tfrac c2\sigma^z_1)+(u-\tfrac c2\sigma^z_2)(u-\tfrac c2\sigma^z_1)
+c^2\sigma^-_2\sigma^+_1+c^2\sigma^+_2\sigma^-_1.
\ee
Setting here $u=0$ we obtain
\be{02-Tau-20}
\tfrac{4}{c^2}\mathcal{T}(0)=2\sigma^z_2\sigma^z_1
+4\sigma^-_2\sigma^+_1+4\sigma^+_2\sigma^-_1,
\ee
or, using equations  $\sigma^\pm_k=\tfrac12(\sigma^x\pm i\sigma^y_k)$,
\be{02-Tau-21}
\tfrac{4}{c^2}\mathcal{T}(0)=2(\sigma^x_2\sigma^x_1+\sigma^y_2\sigma^y_1+\sigma^z_2\sigma^z_1).
\ee
We see that we have obtained the Hamiltonian of the $XXX$ spin chain, consisting of two sites.

\section{Hamiltonian of the $XXX$ spin chain \label{02-Sec3}}

We now show how one can use the monodromy matrix \eqref{02-T-LLL} for obtaining the Hamiltonian  of the $XXX$ chain in the general case.
It is easy to check that the Hamiltonian \eqref{02-QH-Ham-XXZ} can be written in the form
\be{02-QH-Ham-hkk}
H=2\sum_{k=1}^N P_{k\,k+1}-N,
\ee
where we omitted the identity operator at $N$ for brevity.  Indeed, the operator
\be{02-QH-Ham-XXZpm}
h_{k\,k+1}=\sigma_k^x\sigma_{k+1}^x+\sigma_k^y\sigma_{k+1}^y+\sigma_k^z\sigma_{k+1}^z
\ee
acts nontrivially $V_k\otimes V_{k+1}$, while it acts as the identity operator elsewhere.
Therefore, we can write it in the space $V_k\otimes V_{k+1}$ as a $4\times 4$ matrix
\be{02-QH-hkk}
h_{k\,k+1}=\begin{pmatrix}
1 &0&0&0\\ 0&-1 &2&0\\
0&2&-1 &0\\ 0&0&0&1
\end{pmatrix}=2P_{k\,k+1}-\mathbf{1}\otimes\mathbf{1}\,,
\ee
what implies \eqref{02-QH-Ham-hkk}.

We prove that the Hamiltonian  of the $XXX$ chain \eqref{02-QH-Ham-XXZ} can be obtained from the transfer matrix of the monodromy matrix \eqref{02-T-LLL}
by the following formula:
\be{02-QH-trident1}
H=2c\frac{d\mathcal{T}(u)}{du}\;\mathcal{T}^{-1}(u)\Bigr|_{u=\frac c2}-N\,.
\ee

First, let us agree upon notation. The entries of the monodromy matrix \eqref{02-T-LLL} act in the space $\mathcal{H}=V_1\otimes\dots\otimes V_N$,
where any $V_k\sim\mathbb{C}^2$. But the monodromy matrix itself is the $2 \times 2$ matrix  and acts in some space
$\mathbb{C}^2$. We denote this space by $V_0$. The space $V_0$ is called an auxiliary linear space. Thus,
the monodromy matrix $T(u)$ is an operator acting in the tensor product of $N + 1$ spaces $V_0\otimes V_1\otimes\dots\otimes V_N$.
In accordance with the accepted convention we can write $T(u)=T_{01\dots N}(u)$. Each $L$-operator \eqref {02-R-Ln} acts nontrivialy
in the space $V_0\otimes V_n$ and coincides with the $R$-matrix: $L_n(u)=R_{0n}(u-\frac c2)$.
Then, due to  \eqref{02-T-LLL}, we have
\be{02-QH-MonMat}
T(u)=L_N(u)\cdots L_1(u)= R_{0\,N}(u-\tfrac c2)\cdots R_{0\,1}(u-\tfrac c2)\,.
\ee
Thus,
\be{02-QH-TransfMat}
\mathcal{T}(u)={\tr}_0\bigl(R_{0\,N}(u-\tfrac c2)\cdots R_{0\,1}(u-\tfrac c2)\bigr)\,,
\ee
where the trace is taken over the auxiliary space. Since we can make
cyclic permutations inside the trace, we see that the transfer matrix \eqref{02-QH-TransfMat} automatically provides  periodic
boundary condition.

Let us calculate the derivative of  $\mathcal{T}(u)$ at
$u=\frac c2$.  We also take into account that
$R_{0 \, k}(0)$ coincides up to the factor $c$ with the permutation matrix  $P_{0\,k}$. Then
\be{02-QH-derT1}
\begin{aligned}
\frac{d\mathcal{T}(u)}{du}\Bigr|_{u=\frac c2}&=\frac{d}{du}
{\tr}_0\bigl(R_{0\,N}(u-\tfrac c2)\cdots R_{0\,1}(u-\tfrac c2)\bigr)\Bigr|_{u=\frac c2}\\
&=c^{N-1}\sum_{k=1}^N {\tr}_0\bigl(P_{0\,N}\cdots R'_{0\,k}(0)\cdots P_{0\,1}\bigr)\\
&=c^{N-1}\sum_{k=1}^N {\tr}_0\bigl(R'_{0\,k}(0)P_{0\,k-1}\cdots P_{0\,k+1}\bigr)\,.
\end{aligned}
\ee
Here we have used the ciclicity of the trace. Multiplying the
obtained product from the right with the identity matrix $P^2_{0\,k+1}$ and using again cyclic permutation we find
\be{02-QH-derT2}
\begin{aligned}
\frac{d\mathcal{T}(u)}{du}\Bigr|_{u=\frac c2}&
=c^{N-1}\sum_{k=1}^N {\tr}_0\bigl(R'_{0\,k}(0)P_{0\,k-1}\cdots P_{0\,k+1}P^2_{0\,k+1}\bigr)\\
&=c^{N-1}\sum_{k=1}^N {\tr}_0\bigl(P_{0\,k+1}\bigl[R'_{0\,k}(0)P_{0\,k-1}\cdots P_{0\,k+1}\bigr]P_{0\,k+1}\bigr)\,.
\end{aligned}
\ee
According to the general rule, we should replace the space $V_0$ by the space $V_ {k+1}$ in the matrix product between two permutation matrices
$P_{0\,k + 1}$,  that is
\be{02-QH-derT3}
\frac{d\mathcal{T}(u)}{du}\Bigr|_{u=\frac c2}
=c^{N-1}\sum_{k=1}^N {\tr}_0\bigl(R'_{k+1\,k}(0)P_{k+1\,k-1}\cdots P_{k+1\,k+2}P_{0\,k+1}\bigr)\,.
\ee
Now, there is only one matrix $P_{0\,k+1}$ acting in the space $V_0$. We know that its trace over the
space $V_0$ is equal to one. Hence we finally obtain
\be{02-QH-derT4}
\frac{d\mathcal{T}(u)}{du}\Bigr|_{u=\frac c2}
=c^{N-1}\sum_{k=1}^N R'_{k+1\,k}(0)P_{k+1\,k-1}\cdots P_{k+1\,k+2}
=c^{N-1}\sum_{k=1}^N P_{k+1\,k-1}\cdots P_{k+1\,k+2}\,,
\ee
because $R'_{k+1\,k}(0)=\mathbf{1}$.

Let us calculate now the transfer matrix at $u=\frac c2$. We have
\be{02-QH-T00}
\mathcal{T}(\tfrac c2)=c^{N}{\tr}_0\bigl( P_{0\,N}\cdots P_{0\,1}\bigr)\,.
\ee
To compute the trace, we fix an arbitrary $k$ and repeat all the calculations that we did
for the derivation  of the transfer matrix derivative. Then we obtain
\be{02-QH-T0}
\mathcal{T}(\tfrac c2)=c^{N} P_{k+1\,k}P_{k+1\,k-1}\cdots P_{k+1\,k+2}\,.
\ee
Then, using $P_{ab}^2=\mathbf{1}$, we find
\be{02-QH-T0-1}
\mathcal{T}^{-1}(\tfrac c2)=c^{-N} P_{k+1\,k+2}\cdots P_{k+1\,k-1}P_{k+1\,k}\,.
\ee
It is impportant to stress that in \eqref{02-QH-T0}, \eqref{02-QH-T0-1} the subscript $k$ is an arbitrary from the
set $\{1,\dots,N\}$.  Therefore, when we multiply the matrix $\mathcal{T}'(\tfrac c2)$ from the right by $\mathcal{T}^{-1}(\tfrac c2)$,
we can always choose for each term in the sum over $k$ in  \eqref{02-QH-derT4} the same subscript
$k$ in  \eqref{02-QH-T0-1}. Hence,
\be{02-QH-trident}
2 c\frac{d\mathcal{T}(u)}{du}\;\mathcal{T}^{-1}(u)\Bigr|_{u=\frac c2}=
2\sum_{k=1}^N P_{k+1\,k},
\ee
and we do reproduce \eqref{02-QH-Ham-hkk} up to a term proportional to the identity matrix.

Observe that in the example considered, we actually expanded the logarithm of  $\mathcal{T}(u)$ into a series, but not the transfer matrix itself.
It was also extremely important that the expansion was carried out at the point where the $L$-operator turns into the permutation matrix. It is these two
conditions have provided the locality of the resulting Hamiltonian. The latter  turned out to be
a sum of operators describing the interaction between the
nearest neighbors only.

%
%

\chapter{Algebraic Bethe ansatz\label{CHA-ABA}}

In this lecture we pass to a direct study of the algebraic Bethe ansatz method. We will show how, within the framework of
this method the eigenvectors of the transfer matrix can be found. We also obtain a system of Bethe equations that describes these vectors.
One of the main advantages of the algebraic Bethe ansatz is its universality: it is based only on the $RTT$-relation
and without any changes it can be applied to a wide class of models that can have a different physical interpretation. From the viewpoint of
this approach, one can say that different physical models are different representations of the same algebra. We will also return
to the Yang--Baxter equation and discuss its role in the algebraic Bethe  ansatz.

\section{Different $L$-operators\label{03-Sec1}}

In this section we give two more examples of the monodromy matrices satisfying the $RTT$-relation with the $R$-matrix \eqref{02-Ruv}.

\subsection{Inhomogeneous $XXX$ chain\label{03-Sec11}}

Let us use the fact that the $R$-matrix \eqref{02-Ruv}
depends on the difference $u-v$. This implies that the
$L$-operator $L_n(u-\xi_n)$ also satisfies the $RTT$-relation
\be{03-Lxi}
R(u,v)\bigl(L_n(u-\xi_n) \otimes I\bigr)\bigl(I \otimes L_n(v-\xi_n) \bigr)
= \bigl(I \otimes L_n(v-\xi_n) \bigr)\bigl(L_n(u-\xi_n) \otimes I\bigr)R(u,v),
\ee
for any $\xi_n$. Then a matrix
\be{03-TLxi}
T(u)=L_N(u-\xi_N)\dots L_1(u-\xi_1)
\ee
should satisfy the $RTT$-relation.

It is worth mentioning that  the variable $u$ is shifted by different $\xi_n$ at different sites of the chain. The parameters $\xi_n$
are called  inhomogeneities, and the model constructed via the monodromy matrix \eqref{03-TLxi} is called the inhomogeneous $XXX$ chain.
Following the standard procedure, one can construct a set of commuting operators by expanding the transfer matrix of this model in a series. However,
it is difficult to give a physical meaning to these operators, because there any spin operator interacts with all the other
operators of a chain, but not just with neighboring ones (nonlocality). Indeed, as we have already noted, in order to achieve locality, the decomposition of the transfer matrix was performed at the point $u = c/2$.  In this point {\it all} $L$-operators turn into the permutation matrix. In the case of the inhomogeneous chain, setting $u =\xi_n +c/2$, we can ensure that {\it only one} $L$-operator $L_n$ becomes the permutation matrix, which ultimately leads to the loss of locality.

Nevertheless, this model sometimes appears to be very useful in intermediate computations when working with the usual (homogeneous) $XXX$ chain. The latter
can be treated as a particular case of the inhomogeneous chain where all $\xi_n=0$.

\subsection{Quantum nonlinear Schr\"odinger equation \label{03-Sec12}}

Let us give one more example of the $L$-operator that satisfies the $RTT$-relation, but generates a completely different physical model.
Consider a set of operators $\psi_n$ and $\psi^\dagger_n$ ($n=1,\dots,N$) obeying the following commutation relations
\be{03-com-rel-psin}
[\psi_n, \psi^\dagger_m]=\frac{\delta_{nm}}\Delta,
\ee
where $\Delta$ is some number. Generically, this number can be complex, however, for physical applications it is real and positive.
One can consider the operators $\psi_n$ and $\psi^\dagger_n$ as a discretization of bosonic fields\footnote{Again,
following the tradition we use the notation $\psi$, although usually it is reserved for fermions. }. Indeed, let us consider
bosonic fields $\Psi(x)$ and $\Psi^\dagger(x)$ depending on a one-dimensional variable  $x$ and obeying canonical commutation relations
\be{03-com-rel-Psin}
[\Psi(x), \Psi^\dagger(y)]=\delta(x-y).
\ee
Consider also  an interval  $[0,L]$  and choose $N$ points  $x_1,\dots,x_N$ in this interval, such that  $x_n=\Delta n$, and
$x_N=L$. Then we can set
\be{03-lat-con}
\psi_n=\frac1\Delta\int_{x_{n-1}+0}^{x_n-0}\Psi(x)\,dx, \qquad \psi^\dagger_n=\frac1\Delta\int_{x_{n-1}+0}^{x_n-0}\Psi^\dagger(x)\,dx.
\ee
It is easy to check that the operators $\psi_n$ and $\psi^\dagger_n$ constructed in this way do satisfy the relations
\eqref{03-com-rel-psin}.

As usual, we allow the operators considered above to act in a Fock space with vacuum $|0\rangle$ in such a way that
\be{03-act-Fock}
\Psi(x)|0\rangle=0, \qquad \psi_n|0\rangle=0.
\ee

Consider an $L$-operator of the form
\begin{equation}\label{03-L-op22}
 L_n(u)=\begin{pmatrix}
1-\frac{iu\Delta}2+ \frac{\varkappa\Delta^2}2\psi^\dagger_n\psi_n&
-i\Delta\psi^\dagger_n \rho^+_n\\
i\Delta \rho^-_n \psi_n&
1+\frac{iu\Delta}2+\frac{\varkappa\Delta^2}2\psi^\dagger_n\psi_n
\end{pmatrix},
\end{equation}
where  $\varkappa$ is some parameter. Operators $\rho^\pm_n$ in \eqref{03-L-op22} satisfy two constraints.
First, they depend on the combination $\psi^\dagger_n\psi_n$ only, that is $\rho^\pm_n=\rho^\pm_n(\psi^\dagger_n\psi_n)$.
Second,
\be{03-rho}
\rho^+_n\rho^-_n=\varkappa+\frac{\varkappa^2\Delta^2}4\psi^\dagger_n\psi_n.
\ee
There are no other restrictions for the operators $\rho^\pm_n$. For instance, we can set one of them equal to the identity operator,
then the second operator is equal to the rhs of \eqref{03-rho}. We can also set them equal to the square root of the rhs of \eqref{03-rho} and so on.
Independently of the choice of  $\rho^\pm_n$, the $L$-operator \eqref{03-L-op22} satisfies the $RTT$-relation, if the parameter $\varkappa$
depends in a certain way on the  $R$-matrix constant $c$. We suggest the reader to check this statement.

Using the $L$-operator \eqref{03-L-op22} one can build the monodromy matrix \eqref{02-T-LLL}.
Then, the transfer matrix gives us a set of commuting operators. The corresponding model is known as a lattice
version of the quantum nonlinear Schr\"odinger equation (lattice QNLS model). In the continuous limit we obtain the following Hamiltonian
 \be{03-HamQ}
 H=\int_0^L\left(\partial_x\Psi^\dagger(x)\partial_x\Psi(x)+\varkappa\Psi^\dagger(x)\Psi^\dagger(x)\Psi(x)\Psi(x)\right)\,dx,
 \ee
where the parameter $\varkappa$ plays the role of the coupling constant.
The method to obtain the Hamiltonian \eqref{03-HamQ} from the transfer matrix is much more complicated than the derivation of the $XXX$ chain Hamiltonian.
The reader can find the details in Chapter VII of the book \cite{KBIr}. The Hamiltonian \eqref{03-HamQ} describes a one-dimensional Bose gas (also known as the Lieb-Liniger model) with point interaction \cite{LiebL63,Lieb63,YanY69,Gaud83,KBIr}.

\section{Hilbert space \label{03-Sec2}}

We have considered the examples of specific $L$-operators and the corresponding monodromy matrices. Recall that the elements of
the monodromy matrices act in some Hilbert space $\mathcal{H}$, which we have not yet discussed. In the framework of the algebraic Bethe ansatz,
very small requirements are imposed on this space. Namely, it is necessary that there exist a vacuum vector\footnote{One can also meet
with a terminology pseudovacuum or cyclic vector.} $|0\rangle$ such that
\be{03-actACD}
A(u)|0\rangle= a(u)|0\rangle, \qquad D(u)|0\rangle= d(u)|0\rangle, \qquad C(u)|0\rangle= 0.
\ee
Here $a(u)$ and $d(u)$ are some functions of $u$. Their explicit form depends on the specific model. In other words,
the vector $|0\rangle$ is the eigenvector for the operators $A(u)$ and $D(u)$, while the operator $C(u)$ annihilates it. The action of the operator $B(u)$
onto the vacuum is free. It is assumed that acting with this operator on $|0\rangle$ we generate all the space
$\mathcal{H}$ (taking into account the closure, if the space is infinite-dimensional).

In the examples considered above, the vacuum vector exists. In the  lattice version of the QNLS model
it coincides with the Fock vacuum \eqref{03-act-Fock}. In the case of the $XXX$ chains, this is a state with all spins up (ferromagnetic
state):
\be{03-vac-XXX}
|0\rangle = \left(\begin{smallmatrix}1\\0\end{smallmatrix}\right)_1\otimes\cdots\otimes \left(\begin{smallmatrix}1\\0\end{smallmatrix}\right)_N.
\ee
Let us check this. To this end, we prove the following statement.

\begin{prop}\label{03-PROP1}
Let the monodromy matrix $T(u)$ be given by \eqref{02-T-LLL},
and every $L$-operator have the form
\be{03-CR-Ln-abcd}
L_n(u)=\begin{pmatrix}\alpha_n(u)&\beta_n(u)\\ \gamma_n(u)&\delta_n(u)\end{pmatrix}, \qquad n=1,\dots,N.
\ee
Assume that there exists a vector $|0\rangle$ such that
\be{03-actacd}
\alpha_n(u)|0\rangle= \hat\alpha_n(u)|0\rangle, \qquad\delta_n(u)|0\rangle= \hat\delta_n(u)|0\rangle, \qquad \gamma_n(u)|0\rangle= 0, \qquad n=1,\dots,N,
\ee
where $\hat\alpha_n(u)$ and $\hat\delta_n(u)$ are some functions of $u$. Then the action of the monodromy matrix entries on $|0\rangle$
is given by formulas \eqref{03-actACD}, and
\be{03-mat-el-Tpl}
a(u)=\prod_{n=1}^N \hat\alpha_n(u),\qquad
d(u)=\prod_{n=1}^N \hat\delta_n(u).
\ee
\end{prop}

{\sl Proof. }
We use induction over $N$.  Proposition~\ref{03-PROP1} is obvious for  $N=1$. Let it be true for some
$N-1$. Then a matrix
\be{03-matTN-1}
\tilde T(u)=L_{N-1}(u)\dots L_{1}(u)=\begin{pmatrix}\tilde A(u)& \tilde B(u)\\ \tilde C(u)& \tilde D(u)\end{pmatrix}
\ee
has the following properties:
\be{03-actACDN-1}
\tilde A(u)|0\rangle= \tilde a(u)|0\rangle, \qquad \tilde D(u)|0\rangle= \tilde d(u)|0\rangle, \qquad \tilde C(u)|0\rangle= 0,
\ee
where
\be{03-mat-el-TplN-1}
\tilde a(u)=\prod_{n=1}^{N-1} \hat\alpha_n(u),\qquad
\tilde d(u)=\prod_{n=1}^{N-1} \hat\delta_n(u).
\ee

On the other hand, the matrix \eqref{02-T-LLL} is presented in the form
\be{03-matTN}
T(u)=L_{N}(u)\tilde T(u)=\begin{pmatrix}\alpha_{N}(u)&\beta_{N}(u)\\ \gamma_{N}(u)&\delta_{N}(u)\end{pmatrix}
\begin{pmatrix}\tilde A(u)& \tilde B(u)\\ \tilde C(u)& \tilde D(u)\end{pmatrix},
\ee
that is,
\be{03-matTN2}
T(u)=\begin{pmatrix} A(u)& B(u)\\ C(u)& D(u)\end{pmatrix}=\begin{pmatrix}\alpha_{N}(u)\tilde A(u)+\beta_{N}(u)\tilde C(u) \quad
&\alpha_{N}(u)\tilde B(u)+\beta_{N}(u)\tilde D(u) \\ \gamma_{N}(u)\tilde A(u)+\delta_{N}(u)\tilde C(u) \quad
&\gamma_{N}(u)\tilde B(u)+\delta_{N}(u)\tilde D(u) \end{pmatrix}.
\ee
From this representation, we immediately obtain
\be{03-actCind}
C(u)|0\rangle=\bigl(\gamma_{N}(u)\tilde A(u)+\delta_{N}(u)\tilde C(u)\bigr)|0\rangle=0,
\ee
because $\tilde C(u)|0\rangle=0$ and $\gamma_{N}(u)|0\rangle=0$. Similarly
\be{03-actADind}
\begin{aligned}
&A(u)|0\rangle=\bigl(\alpha_{N}(u)\tilde A(u)+\beta_{N}(u)\tilde C(u)\bigr)|0\rangle=\prod_{n=1}^{N} \hat\alpha_n(u)|0\rangle,\\
& D(u)|0\rangle=\bigl(\gamma_{N}(u)\tilde B(u)+\delta_{N}(u)\tilde D(u)\bigr)|0\rangle=\prod_{n=1}^{N} \hat\delta_n(u)|0\rangle,
\end{aligned}
\ee
where we used\footnote{Recall that $[L_{N}^{ij}(u),\tilde T^{kl}(v)]=0$ for any   $u$ and $v$ and for any $i,j,k,l$.}
$[\gamma_{N}(u),\tilde B(u)]=0$. Proposition~\ref{03-PROP1} is proved.

Note that the method described above also allows us to calculate the action of the operator $B(u)$ on the vector $|0\rangle$ in terms of
actions of local operators $\beta_n(u)$:
\be{03-actBN}
B(u)|0\rangle=\sum_{k=1}^N\left(\prod_{n=k+1}^N \hat\alpha_n(u)\right)\left(\prod_{n=1}^{k-1} \hat\delta_n(u)\right)\cdot \beta_k(u)|0\rangle.
\ee
This formula is easily proved by induction over $N$ using the representation \eqref{03-matTN2}. However, within the algebraic Bethe ansatz it is used
rarely, since the states of the form $B(u)|0\rangle$ correspond to single-particle excitations and are of little interest.

It is easy to see that in the case of the inhomogeneous $XXX$ chain the functions $\hat\alpha_n(u)$ and $\hat\delta_n(u)$ have the form
\be{03-aldel-1}
\hat\alpha_n(u)=u-\xi_n+\tfrac c2,\qquad  \hat\delta_n(u)=u-\xi_n-\tfrac c2.
\ee
Therefore,
\be{03-CR-ad-inh}
a(u)=\prod_{n=1}^N(u-\xi_n+\tfrac c2), \qquad
d(u)=\prod_{n=1}^N(u-\xi_n-\tfrac c2),
\ee
and the action of the $B(u)$ operator on $|0\rangle$  is given by
\begin{equation}\label{03-CR-Action-Binh}
B(u)|0\rangle
=c\sum_{k=1}^N \prod_{n=k+1}^N (u-\xi_n+\tfrac c2)
 \prod_{n=1}^{k-1}(u-\xi_n-\tfrac c2)\cdot \sigma^-_k|0\rangle.
\end{equation}
For the homogeneous chain one should simply set all $\xi_n=0$.

In the lattice QNLS model we have
\be{03-aldel-2}
\hat\alpha_n(u)=1-\frac{iu\Delta}2,\qquad  \hat\delta_n(u)=1+\frac{iu\Delta}2.
\ee
Hence,
\be{03-CR-ad-NS}
a(u)=\left(1-\frac{iu\Delta}2\right)^N, \qquad
d(u)=\left(1+\frac{iu\Delta}2\right)^N,
\ee
and the action of the $B(u)$ operator on $|0\rangle$ has the form (when choosing $\rho^+_n=\rho^-_n$)
\begin{equation}\label{03-CR-Action-BNS}
B(u)|0\rangle
=-i\Delta\sqrt{\varkappa}\sum_{k=1}^N \left(1-\frac{iu\Delta}2\right)^{N-k}
 \left(1+\frac{iu\Delta}2\right)^{k-1}\cdot \psi^\dagger_k|0\rangle.
\end{equation}
It is easy to verify that in the continuous limit $\Delta\to 0$,  $N\to\infty$, $N\Delta=L=const$ we obtain
\be{03-CR-ad-NS-2}
a(u)=e^{-iuL/2}, \qquad
d(u)=e^{iuL/2},
\ee
and
\begin{equation}\label{03-CR-Action-BNS-2}
B(u)|0\rangle
=-i\sqrt{\varkappa}e^{-iuL/2}\int_0^L e^{iux} \Psi^\dagger(x)\,dx|0\rangle.
\end{equation}

\section{Actions of the monodromy matrix entries onto vectors \label{03-Sec3}}

We have shown in the previous section that the space $\mathcal{H}$, in which the elements of the monodromy matrix act, can be either a finite-dimensional
($XXX$ chain) or an infinite-dimensional (QNLS model). In the framework of the algebraic Bethe ansatz this space is not specified. Also, the functions $a(u)$ and $d(u)$ remain free functional parameters. Nevertheless, these data are already sufficient
to formulate certain conditions on the eigenvectors of the transfer matrix $\mathcal{T}(u)=A(u)+D(u)$.
We recall that the transfer matrix $\mathcal{T}(u)$ is the generating function of the integrals of the motion of the model. By construction, the eigenfunctions (vectors) of the operator $\mathcal{T}(u)$ are simultaneously eigenfunctions for all integrals of motion, including the Hamiltonian.
Thus, instead of looking for the eigenfunctions of the Hamiltonian of a certain concrete model, we look for the eigenfunctions of the transfer matrix.

First of all let us agree upon a new notation. We denote sets of variables by bar:
\be{sets}
\bu=\{u_1,\dots,u_n\},\qquad \bv=\{v_1,\dots,v_m\}, \quad\text{and so on.}
\ee
If necessary, the cardinalities of the sets are given in special comments.

We introduce two new functions
\be{03-CR-fg}
f(u,v)=\frac{u-v+c}{u-v}\qquad\text{and}\qquad
g(u,v)=\frac{c}{u-v},
\ee
and write the  $R$-matrix in the form
\be{03-R}
R(u,v)=\mathbf{I}+\frac c{u-v}\;\mathbf{P}=
\begin{pmatrix}
f(u,v)&0&0&0\\
0&1&g(u,v)&0\\
0&g(u,v)&1&0\\
0&0&0&f(u,v)
\end{pmatrix}.
\ee
Pay attention that the $R$-matrix \eqref{03-R} differs from \eqref{02-Ruv}
by the factor  $(u-v)$. As already explained, multiplication of the $R$-matrix by an arbitrary scalar function is always possible.
This does not violate either the Yang--Baxter equation or the $RTT$-relation.

{\sl Remark.} Following the tradition we do not stress that the functions $f(u,v)$, $g(u,v)$, and the $R$-matrix \eqref{03-R} actually depend on the difference $u-v$. In some cases considered in the next lectures this appears to be useful.

We also use a shorthand notation for the products of the operators $T^{ij}(u)$, the functions $a(u)$ and $d(u)$, and also the functions $f(u,v)$ and $g(u,v)$ \eqref{03-CR-fg}.  Namely, if this operator or function depends on a set of variables, then this means that one should take a product of these operators
or functions over the corresponding set. For example,
\be{03-Conv}
B(\bu)=\prod_{u_j\in\bu}B(u_j),\qquad C(\bu)=\prod_{u_j\in\bu}C(u_j),\qquad a(\bv)=\prod_{v_i\in\bv} a(v_j),\quad\text{and so on.}
\ee
Recall that due to \eqref{02-TijTij} we have $[B(u),B(v)]=0$, $[C(u),C(v)]=0$, and hence, the operator products in \eqref{03-Conv} are well defined.

Similarly, we have for the functions  $g(u,v)$ and $f(u,v)$
\be{03-Snot-fg}
g(u,\bv)=\prod_{v_i\in\bv} g(u,v_i), \qquad f(\bu,\bv)=\prod_{u_j\in\bu}\prod_{v_i\in\bv} f(u_j,v_i).
\ee
In the second case we deal with the double product.

Below we often will deal with sets of variables, in which one of the elements is missing, for example, $\{u_1,\dots,u_n\}\setminus u_k$.
We denote such the sets by $\bu_k=\{u_1,\dots,u_n\}\setminus u_k$. Respectively, we have
\be{03-Conv-k}
B(\bu_k)=\prod_{\substack{u_j\in\bu\\ j\ne k}}B(u_j),\qquad
f(u_k,\bu_k)=\prod_{\substack{u_i\in\bu\\i\ne k}} f(u_k,u_i),\quad\text{and so on.}
\ee
Finally, we agree upon that a product over the empty set is equal to $1$ by definition. A double product is equal to $1$, if
at least one of the sets is empty.

\subsection{Commutation relations \label{03-Sec31}}

All the commutation relations between  $T^{jk}(u)$ follow from the  $RTT$-relation and are given by \eqref{02-com-rel}.
 Let us give them in details.
 \be{03-1TT}
 [T^{jk}(u),T^{jk}(v)]=0,\qquad j,k=1,2,
 \ee
\begin{align}
[A(u),D(v)]&=g(u,v)\bigl(C(v)B(u)-C(u)B(v)\bigr),\label{03-CR-AD}\\
[D(u),A(v)]&=g(u,v)\bigl(B(v)C(u)-B(u)C(v)\bigr),\label{03-CR-DA}\\
[C(u),B(v)]&=g(u,v)\bigl(A(v)D(u)-A(u)D(v)\bigr),\label{03-CR-CB}\\
[B(u),C(v)]&=g(u,v)\bigl(D(v)A(u)-D(u)A(v)\bigr),\label{03-CR-BC}
\end{align}
\begin{align}
A(v)B(u)&=f(u,v)B(u)A(v)+g(v,u)B(v)A(u),\label{03-CR-AB}\\
B(v)A(u)&=f(u,v)A(u)B(v)+g(v,u)A(v)B(u),\label{03-CR-BA}\\
D(v)C(u)&=f(u,v)C(u)D(v)+g(v,u)C(v)D(u),\label{03-CR-DC}\\
C(v)D(u)&=f(u,v)D(u)C(v)+g(v,u)D(v)C(u),\label{03-CR-CD}\\
A(u)C(v)&=f(u,v)C(v)A(u)+g(v,u)C(u)A(v),\label{03-CR-AC}\\
C(u)A(v)&=f(u,v)A(v)C(u)+g(v,u)A(u)C(v),\label{03-CR-CA}\\
B(u)D(v)&=f(u,v)D(v)B(u)+g(v,u)D(u)B(v),\label{03-CR-BD}\\
D(u)B(v)&=f(u,v)B(v)D(u)+g(v,u)B(u)D(v).\label{03-CR-DB}
\end{align}
Some of these relationships repeat each other, for instance, \eqref{03-CR-DA} follows from \eqref{03-CR-AD} and \eqref{03-CR-CB}.
One can also verify the consistency of certain relations, for instance,
applying \eqref{03-CR-BA} to the rhs of \eqref{03-CR-AB} we reproduce the lhs of \eqref{03-CR-AB}.

\subsection{Algebraic Bethe ansatz\label{03-Sec32}}

The main idea of the algebraic Bethe ansatz is that we look for the eigenvectors of the transfer matrix in the form $B(\bu)|0\rangle$.
In general, the set $\bu=\{u_1,\dots,u_n\}$ can contain  an arbitrary number of elements, and the parameters $u_k$ can take arbitrary complex values.

Recall that the space $\mathcal{H}$ consists of vectors of the form $B(\bu)|0\rangle$
and their linear combinations. The ansatz is that we exclude linear combinations from our consideration. In other words,  we assume
that eigenvectors of the transfer matrix can be represented as monomials in the operators $B$ applied to the vacuum.
We will call vectors of this kind Bethe vectors. We need, first, to check whether the Bethe vectors can indeed be eigenvectors
of the operator $\mathcal{T}(u)$. Secondly, under what conditions is it possible. Third, whether the resulting
vectors form a basis in the Hilbert space $\mathcal{H}$.

Strictly speaking, the last question goes beyond the framework of the algebraic Bethe ansatz. To answer it we need to know
much more about the space $\mathcal{H}$. At the same time, we do not even know whether it is finite-dimensional or not.
It is interesting to mention, however, that even for the finite-dimensional spaces $\mathcal{H}$  the question of the
completeness of the obtained system of eigenvectors
can be nontrivial. In particular, this problem takes place for the $XXX$ chain. In the general case of a chain of arbitrary length
an exhaustive proof of completeness was absent till very recently and was given in the works of \cite{MukTV09,MukTV13}. On the other hand, in the  QNLS model,
in which the Hilbert space is infinite-dimensional, the completeness of the eigenvectors system is proved at least
for positive coupling constant \cite{Gaud83}.

\subsection{Action of operators $A$ and $D$ on Bethe vectors\label{03-Sec33}}

Let us compute the action of the operator $A(v)$ onto the vector
$B(\bu)|0\rangle$
 \be{03-2ABBB}
 A(v)B(\bu)|0\rangle,
 \ee
assuming that the parameters $v,u_1,\dots,u_n$ are generic complex numbers
(that is, they are pair-wise distinct, they do not coincide with singular points of the function $a(u)$, and so on.).
Formally, we have all necessary tools for this. Using the commutation relations \eqref{03-CR-AB} we can move the operator
$A$ to the right until it arrives at the extreme right position. After this, it remains to act on the vacuum vector
via \eqref{03-actACD}. Thus, the action of $A(v)$ on the Bethe vector reduces to an $n$-fold application
of the formula \eqref{03-CR-AB}. It is clear, however, that at each step we will obtain two terms, so that in the end we will have
$2^n$ terms. For large $n$ such a result is difficult not only to obtain, but simply to write down. To solve this problem
we formulate the basic rules of the Bethe ansatz which must be taken into account when permuting a large number of operators.

\begin{itemize}

\item One should try to understand a general structure of the result. Namely,
how the result depends on the operators $T^{jk}(u)$ (or their eigenvalues $a(u)$ and $d(u)$).
At this stage we do not pay attention to the functions $f(u,v)$ and $g(u,v)$, which arise under the permutations of the operators.

\item It is necessary to fix a  term in the final result and try to obtain just this term, neglecting all other contributions.
In order to do this, it might be useful to reorder the operators in the original product.
\end{itemize}

Let us see how these rules work in practice. Applying the first rule to the formula \eqref{03-2ABBB},
we see that the number of linearly independent terms in the final
result anyway does not exceed $n+1$:
 \be{03-2ABBB-act}
A(v)B(\bu)|0\rangle=a(v)\Lambda(v|\bu)B(\bu)|0\rangle+
 \sum_{k=1}^n a(u_k)\Lambda_k(v|\bu)B(v)B(\bu_k)|0\rangle.
 \ee
Here $\Lambda(v|\bu)$ and $\Lambda_k(v|\bu)$ are some unknown coefficients to be determined.

Indeed, it follows from \eqref{03-CR-AB} that when we permute the operators $A$ and $B$,
they either preserve the original arguments, or exchange them. However, the `law of conservation of ar\-gu\-ments'
anyway takes place: if the operators in the lhs have arguments $u$ and $v$, then in the rhs they have
the same arguments. New arguments do not arise, and the original arguments do not disappear. In our case, in the lhs of the formula \eqref{03-2ABBB-act}
the arguments of the operators are the parameters $v,u_1,\dots,u_n$. Therefore, exactly the same parameters should appear in the rhs.
It is easy to see that \eqref{03-2ABBB-act} is the most general expression of this kind. It contains a term in which, after moving
$A(v)$ to the right all the operators have kept their original arguments. Then, the operator $A(v)$, acting on the vacuum, produces the function $a(v)$.
Besides, in \eqref{03-2ABBB-act} there are terms where, moving to the right, the operator $A(v)$ exchanged its argument one or several times with the operators
$B(u_i)$. When approaching  the extreme right position, the operator $A$ had the argument $u_k$ and, acting on the vacuum, eventually produced the function
$a(u_k)$. At the same time, in the final product of the operators $B(u_i)$ one of them disappeared, and $B(v)$ appeared instead.
When permutations of the operators occur, the functions $f(u,v)$ and $g(u,v)$ arise, but we did not pay attention to them, simply denoting
the coefficients in the final result by $\Lambda(v|\bu)$ and $\Lambda_k(v|\bu)$.
Later, we have to find them.

This concludes the first stage. We have determined the general structure of the final expression. Note that there are only
$n+1$ terms instead of the expected $2^n$ terms.

{\sl Remark. } Actually, the number of terms can be even less in specific models. The matter is that we considered
Bethe vectors with different sets of the arguments  as linearly independent. Clearly, generally speaking, this is not the case.
For example, in the $XXX$ model, the number of linearly independent
vectors can not exceed the dimension of a Hilbert space. The latter  is $2^N$, where $N$ is the number of sites of the chain. On the other hand, we can take
a vector $B(\bu)|0\rangle$, fix the parameters $u_2,\dots,u_n$, and move the variable $u_1$ within the whole complex plane. If with each new
value $u_1$ we get a vector that is linearly independent of all the previous ones, then the number of such vectors is obviously infinite.
Nevertheless, this observation does not invalidate the formula \eqref{03-2ABBB-act}. It still holds, however, if there are linearly dependent vectors among the terms in the rhs, then the number of terms in this formula can be reduced.

Let us now proceed to the second stage. We shall call the first term in the rhs of
\eqref{03-CR-AB} {\it the first commutation scheme} (when the operators keep their arguments). The term, in which the operators exchange their arguments,
will be called {\it the second commutation scheme}.

Let us compute the coefficient $\Lambda(v|\bu)$, that is, the coefficient of the vector $a(v)B(\bu)|0\rangle$. Obviously,
this coefficient arises in \eqref{03-2ABBB-act} if and only if we use only the first commutation scheme when permuting
the operators $A(v)$ and $B(u_j)$. Indeed, let the operator $A(v)$ exchange its argument with some operator $B(u_j)$ at least once. Then
we obtain a vector
\be{03-2sch-1}
B(u_1)\dots B(u_{j-1})B(v)A(u_j)B(u_{j+1})\dots B(u_{n})|0\rangle.
\ee
It is clear that when moving further to the right, the operator $A(u_j)$ can not accept the argument $v$, because it is to the left
from  $A$. Therefore, the function $a(v)$ cannot arise in the final result. In addition, the answer will contain the operator $B(v)$,
while this operator absences in the term of  our interest.

Thus, to obtain the term $a(v)B(\bu)|0\rangle$ we can use the first commutation scheme only.
Then, permuting $A(v)$ and $B(u_j)$ we obtain a coefficient $f(u_j,v)$. After all permutations we obtain a product of $f(u_j,v)$ over
$u_j$, while the operator $A(v)$, approaching the extreme right position and acting on the vacuum, gives the function $a(v)$. As the result we have
 \be{03-2Lam}
 \Lambda(v|\bu)=f(\bu,v).
 \ee

Let us find now  $\Lambda_k(v|\bu)$, that is, the coefficients of the vectors $a(u_k)B(v)B(\bu_k)|0\rangle$, $k=1,\dots,n$.
For this we fix some $k$. Using the fact that all the $B$ operators commute with each other,
we reorder the product $B(\bu)$ in such a way, that the operator $B(u_k)$ stands in the extreme left position:
 \be{03-2ABBB-act-ind}
 A(v)B(\bu)|0\rangle=
 A(v)B(u_k)B(\bu_k)|0\rangle.
 \ee
Now, in order to obtain a contribution proportional to $a(u_k)$, it is necessary to use the second commutation scheme when
permuting $A(v)$ and $B(u_k)$. Overwise we obtain a vector
\be{03-other}
B(u_k)A(v)B(\bu_k)|0\rangle,
\ee
and we see that when moving further to the right, the operator $A(v)$ can not accept the argument $u_k$, because it is to the left from $A$.
Besides, the answer will contain the operator $B(u_k)$,
while this operator absences in the term of  our interest. Thus, we find
 \be{03-2ABBB-act-ind-1}
   A(v)B(u_k)B(\bu_k)|0\rangle
  = g(v,u_k) B(v)A(u_k) B(\bu_k)|0\rangle+{\cal Z},
 \ee
where ${\cal Z}$ means terms, that anyway do not contribute to the coefficient under consideration. After this,
when moving further to the right,   $A(u_k)$ must keep its argument, therefore, we should use the first commutation scheme.
Indeed, if  there is an exchange at least once, then the argument $u_k$  remains on the left from $A$, and we can not get the function $a(u_k)$ in the
final result. After every permutation with the operators $B(u_j)$ we will have a factor $f(u_j,u_k)$, and as a result we obtain
 \be{03-2Lam-k}
 \Lambda_k(v|\bu)= g(v,u_k)f(\bu_k,u_k).
 \ee

Thus, we have found the action of the operator $A(v)$ on the Bethe vector $B(\bu)|0\rangle$. It is given by equation \eqref{03-2ABBB-act}, where
the coefficients $\Lambda(v|\bu)$ and $ \Lambda_k(v|\bu)$ respectively are given by \eqref{03-2Lam} and \eqref{03-2Lam-k}.

The action of the operator $D(v)$ on the Bethe vector $B(\bu)|0\rangle$ can be found exactly in the same way.
One can, however, do even easier. Observe that the commutation relations of the operators $D$ and $B$ \eqref{03-CR-DB}
can be obtained from those of $A$ and $B$ \eqref{03-CR-AB} under the formal replacement of the constant $c$ with $-c$. Indeed, it is easy to see that
\be{03-cc-fg}
f(u,v)\Bigr|_{c\to-c}=f(v,u),\qquad
g(u,v)|_{c\to-c}=g(v,u).
\ee
Thus, the action of $D(v)$ on the vector  $B(\bu)|0\rangle$ is given by \eqref{03-2ABBB-act}, where one should replace  $c$ with $-c$
and the function $a$ with the function $d$:
 \be{03-2DBBB-act}
 D(v)B(\bu)|0\rangle=d(v)\tilde\Lambda(v|\bu)B(\bu)|0\rangle+
 \sum_{k=1}^n d(u_k)\tilde\Lambda_k(v|\bu)B(v)B(\bu_k)|0\rangle,
 \ee
where
 \be{03-2barLam}
 \tilde\Lambda(v|\bu)=f(v,\bu),\qquad
 \tilde\Lambda_k(v|\bu)= g(u_k,v)f(u_k,\bu_k).
 \ee

The action of the operator $C(v)$ on the Bethe vector  $B(\bu)|0\rangle$ will be calculated in Lecture~\ref{CHA-SP}.
However, we recommend that the reader try to do it independently, since all necessary tools for such the calculation already are explained.

\subsection{Eigenvectors and Bethe equations}\label{03-Sec34}

Now we know the action of the transfer matrix on the vector $B(\bu)|0\rangle$
\begin{multline}\label{03-T-act}
\mathcal{T}(v)B(\bu)|0\rangle=\Bigl(a(v)\Lambda(v|\bu)+d(v)\tilde\Lambda(v|\bu)\Bigr)B(\bu)|0\rangle\\
+ \sum_{k=1}^n \Bigl(a(u_k)\Lambda_k(v|\bu)+d(u_k)\tilde\Lambda_k(v|\bu)\Bigr)B(v)B(\bu_k)|0\rangle.
 \end{multline}
In the first line of the rhs of this equation, we have the same vector $B(\bu)|0\rangle$ as in the lhs. In the second line,
we have a sum of vectors in which one of the operators $B$ depends on the variable $v$. There are no such vectors in the lhs.
However, if we impose conditions
\be{03-Cond}
a(u_k)\Lambda_k(v|\bu)+d(u_k)\tilde\Lambda_k(v|\bu)=0,\qquad k=1,\dots,n,
\ee
then the sum in the second line of \eqref{03-T-act} vanishes. Then the vector $B(\bu)|0\rangle$ becomes an eigenvector of the operator $\mathcal{T}(v)$
with an eigenvalue
\be{03-E-val}
\tau(v|\bu)=a(v)\Lambda(v|\bu)+d(v)\tilde\Lambda(v|\bu)=
a(v)f(\bu,v)+d(v)f(v,\bu).
\ee
In its turn, the condition \eqref{03-Cond} is usually written as follows:
\be{03-BE0}
\frac{a(u_k)}{d(u_k)}=\frac{f(u_k,\bu_k)}{f(\bu_k,u_k)}, \qquad k=1,\dots,n.
\ee
The system \eqref{03-BE0} is called the system of Bethe equations. The vector $B(\bu)|0\rangle$, where
the parameters  $\bu$ satisfy the system \eqref{03-BE0}, is often called on-shell Bethe vector. Thus,
on-shell Bethe vector is a synonym of the transfer matrix eigenvector.

It is worth mentioning that the obtained on-shell Bethe vectors are eigenvectors of the operator
$\mathcal{T}(v)$ for any $v$:
\be{03-eig}
\mathcal{T}(v)B(\bu)|0\rangle  =\tau(v|\bu)B(\bu)|0\rangle, \qquad \forall v\in \mathbb{C}.
\ee
Therefore, if the integrals of the motion $I_k$ of the model under consideration arise in the expansion
\be{03-decomp-T}
\mathcal{T}(v)=\sum_k (v-v_0)^k I_k,
\ee
then their eigenvalues $\hat I_k(\bu)$ arise in the similar expansion of the transfer matrix eigenvalue
\be{03-decomp-tau}
\tau(v|\bu)=\sum_k (v-v_0)^k \hat I_k(\bu).
\ee
Differentiating $k$ times equation \eqref{03-eig} over $v$ at $v=v_0$, we obtain
\be{03-spectr}
 I_k B(\bu)|0\rangle = \hat I_k(\bu)\; B(\bu)|0\rangle.
 \ee
Thus, if we solve the system of Bethe equations, we thereby find the eigenvalues of all the integrals of motion of the model under consideration and the corresponding eigenvectors.

Note that the derivation of Bethe equations is universal in the sense that it is model independent. We used only the commutation relations between the elements of the monodromy matrix (in fact, the $RTT$-relation) and the formulas of the action of the operators on the vacuum. If we consider a specific model, we should
only substitute the concrete values of the functions $a(u)$ and $d(u)$ into the system \eqref{03-BE0}. Thus, for the $XXX$ chain
the system of Bethe equations has the form
\be{03-BE-xxx}
\left(\frac{u_k+\frac c2}{u_k-\frac c2}\right)^N=\frac{f(u_k,\bu_k)}{f(\bu_k,u_k)}, \qquad k=1,\dots,n,
\ee
while for the QNLS model  (in the continuous limit) we have
\be{03-BE-NS}
e^{-iu_kL}=\frac{f(u_k,\bu_k)}{f(\bu_k,u_k)}, \qquad k=1,\dots,n.
\ee
Having solutions of these systems in each case, we can construct on-shell vectors $B(\bu)|0\rangle$.
The question of whether the obtained vectors form a basis, of course, still remains open for the time being.

\subsection{Example\label{03-Sec41}}

Consider the homogeneous $XXX$ chain and a Bethe vector $B(u)|0\rangle$, that depends on one variable  $u$ only. We have
an  explicit representation for this vector (see \eqref{03-CR-Action-Binh})
\begin{equation}\label{03-BV1}
B(u)|0\rangle
=c\sum_{k=1}^N (u+\tfrac c2)^{N-k}(u-\tfrac c2)^{k-1}\cdot \sigma^-_k|0\rangle.
\end{equation}
The operator $\sigma^-_k$, when acting on the state $|0\rangle$, flips the spin in the  $k$th site:
\be{03-sk-vac}
\sigma^-_k|0\rangle = \left(\begin{smallmatrix}1\\0\end{smallmatrix}\right)_1\otimes\cdots\otimes
 \left(\begin{smallmatrix}0\\1\end{smallmatrix}\right)_k\otimes
\cdots\otimes \left(\begin{smallmatrix}1\\0\end{smallmatrix}\right)_N.
\ee
Thus, the vector $B(u)|0\rangle$ is a linear combination of the states with one spin down and $N-1$ spins up.

Consider the subspace spanned by these vectors. Obviously, it is $N$-dimensional, and the vectors \eqref{03-BV1} belong
to this subspace. Let the Bethe vector \eqref{03-BV1} be an on-shell, that is, the parameter $u$ satisfies the system
Bethe equations \eqref{03-BE-xxx} for $n=1$:
\be{03-BE-xxx-1}
\left(\frac{u+\frac c2}{u-\frac c2}\right)^N=1.
\ee
Obviously, this equation admits an explicit solution. We propose to the reader to prove  that the system of on-shell Bethe vectors constructed from the solutions of  equation \eqref{03-BE-xxx} forms an orthogonal basis in a subspace with one spin down.

\section{Compatibility of commutation relations\label{04-Sec3}}

Let us write the action formula of $A(v)$ on the Bethe vector $B(\bu)|0\rangle$ \eqref{03-2ABBB-act} more detailed
 \be{04-2ABBB-act}
A(v)B(\bu)|0\rangle=a(v)f(\bu,v)B(\bu)|0\rangle+
 \sum_{k=1}^n a(u_k)g(v,u_k)f(\bu_k,u_k)B(v)B(\bu_k)|0\rangle.
 \ee
This formula is a consequence of the commutation relation \eqref{03-CR-AB}. There are two terms in the rhs of this commutation relations:
one of them is proportional to the function $f$, another is proportional to $g$. Acting with $A(v)$ onto $B(\bu)|0\rangle$ we applied
\eqref{03-CR-AB} several times. Therefore, different products of the functions $f$ appear in  \eqref{04-2ABBB-act}.
Surprisingly, however, this does not happen with the $g$ function. This function is included in the action \eqref{04-2ABBB-act} only
linearly. What is the reason for such an inequality between the functions $f$ and $g$?

Of course, we can say that $f=1+g$, hence, any expression containing these two functions can be converted to a form in which only one of them will appear. However, the real cause of the inequality is much deeper. For instance, in lecture~\ref{CHA-TRM},
we will deal with a trigonometric $R$-matrix, that can also be written in the form of \eqref{03-R}, but the functions $f$ and $g$ are already different and are independent. For models with such the $R$-matrix the commutation relation \eqref{03-CR-AB} is still valid, therefore, the action \eqref{04-2ABBB-act} also remains true. Thus, we are again faced with inequalities, but now we cannot refer to the dependence between the $f$ and $g$ functions.

To answer this question let us consider commutation relations of one operator $A(v)$ with a product
of two operators $B(u_1)$ and $B(u_2)$. Using \eqref{03-CR-AB} we obtain
\be{04-Eig-A2B1}
A(v)B(u_1)B(u_2)=\Bigl(f(u_1,v)B(u_1)A(v)+
g(v,u_1)B(v)A(u_1)  \Bigr)  B(u_2)\,.
\ee
Applying \eqref{03-CR-AB} once more we arrive at
\be{04-Eig-A2B2}
\begin{aligned}
A(v)B(u_1)B(u_2)&=
f(u_1,v)f(u_2,v)B(u_1)B(u_2)A(v)\\
&+g(v,u_1)f(u_2,u_1)B(v)B(u_2)A(u_1)\\
&+\Bigl(g(v,u_2)f(u_1,v)+
g(v,u_1)g(u_1,u_2)\Bigr)B(v)B(u_1)A(u_2)\,.
\end{aligned}
\ee

On the other hand, the operators $B(u_1)$ and $B(u_2)$ commute with each other, hence,
\be{04-Eig-tr}
A(v)B(u_1)B(u_2)=A(v)B(u_2)B(u_1)\,.
\ee
Thus, expression  \eqref{04-Eig-A2B2}
does not change, if we replace there $u_1$ with $u_2$:
\be{04-Eig-A2B3}
\begin{aligned}
A(v)B(u_1)B(u_2)&=
f(u_1,v)f(u_2,v)B(u_1)B(u_2)A(v)\\
&+g(v,u_2)f(u_1,u_2)B(v)B(u_1)A(u_2)\\
&+\Bigl(g(v,u_1)f(u_2,v)+
g(v,u_2)g(u_2,u_1)\Bigr)B(v)B(u_2)A(u_1)\,.
\end{aligned}
\ee
Comparing \eqref{04-Eig-A2B2} and \eqref{04-Eig-A2B3} we see that, generally speaking, these expressions are different.
The first terms in these formulas coincide, while the second and the third terms have different structure. In order to have the coincidence
we should require the following identity:
\be{04-Eig-identfg}
g(v,u_2)f(u_1,v)+
g(v,u_1)g(u_1,u_2)=g(v,u_2)f(u_1,u_2).
\ee
It is easy to check that if we substitute here $f(v,u)=\tfrac{v-u+ c}{v-u}$ and $g(v,u)=\tfrac{c}{v-u}$, then \eqref{04-Eig-identfg} does become the identity.
This equation shows the mechanism by which the expressions quadratic in $g$ (that is, the term $g(v,u_1)g(u_1,u_2)$) reduces to a linear expressions in $g$.

The example considered raises the question  of the commutation relations compatibility. Indeed, we have considered
the case when one operator $A(v)$ is reordered with two operators $B(u_1)B(u_2)$. But we could
take another triple of operators, for example, $C(v)$ and the product $D(u_1)D(u_2)$, and have the
same problem. Namely, we should check that the result of the permutation does not depend on the order of
the operators $D$: either $D(u_1)D(u_2)$ or $D(u_2)D(u_1)$.
It is clear that in doing so we obtain some identity on the $f$ and $g$ functions, which will have to be proved.

Instead of considering  all possible triples of operators, it is easier to work with one commutation relation for the monodromy matrix.
Indeed, all the scalar commutation relations are contained in a single matrix $RTT$-relation.

Consider a product of three monodromy matrices
$T_1(u_1)T_2(u_2)T_3(u_3)$. Here each matrix acts in its own auxiliary space.
Using the  $RTT$-relation we can reorder them to the opposite order
$T_3(u_3)T_2(u_2)T_1(u_1)$. For doing this we can use two different ways: either
 we move $T_1$  to the right and then we permute $T_2$ and $T_3$, or we can move
$T_3$ to the left and then permute $T_1$ and $T_2$:
\be{04-Eig-TTT1}
\begin{aligned}
&T_1(u_1)T_2(u_2)T_3(u_3)\to T_2(u_2)T_1(u_1)T_3(u_3)
\to  T_2(u_2)T_3(u_3)T_1(u_1)\to T_3(u_3)T_2(u_2)T_1(u_1),\\
&T_1(u_1)T_2(u_2)T_3(u_3)\to T_1(u_1)T_3(u_3)T_2(u_2)
 \to T_3(u_3) T_1(u_1)T_2(u_2) \to T_3(u_3)T_2(u_2)T_1(u_1).
\end{aligned}
\ee
Each permutation of two monodromy matrices can be done by means of the $R$-matrix
\begin{equation}\label{04-Eig-RTTR}
R_{jk}(u_j,u_k)  T_j(u_j)T_k(u_k)
R^{-1}_{jk}(u_j,u_k)=T_k(u_k)T_j(u_j),\qquad j,k=1,2,3\,.
\end{equation}
The first way of reordering the matrices $T_j$ leads us to
\begin{multline}\label{04-Eig-TTT2}
R_{23}(u_2,u_3)R_{13}(u_1,u_3)
R_{12}(u_1,u_2)  T_1(u_1)T_2(u_2)T_3(u_3)
R^{-1}_{12}(u_1,u_2)R^{-1}_{13}(u_1,u_3)R^{-1}_{23}(u_2,u_3)\\[2mm]
=T_3(u_3)T_2(u_2)T_1(u_1)\,.
\end{multline}
The second way gives
\begin{multline}\label{04-Eig-TTT3}
R_{12}(u_1,u_2)R_{13}(u_1,u_3)R_{23}(u_2,u_3)
  T_1(u_1)T_2(u_2)T_3(u_3)
R^{-1}_{23}(u_2,u_3)R^{-1}_{13}(u_1,u_3)R^{-1}_{12}(u_1,u_2)\\[2mm]
=T_3(u_3)T_2(u_2)T_1(u_1)\,.
\end{multline}
If we require that
\begin{equation}\label{04-Eig-YB}
R_{23}(u_2,u_3)R_{13}(u_1,u_3)R_{12}(u_1,u_2) =
R_{12}(u_1,u_2)R_{13}(u_1,u_3)R_{23}(u_2,u_3)\,,
\end{equation}
then the results of the both methods coincide. But  \eqref{04-Eig-YB}
is the Yang--Baxter equation. The $R$-matrix certainly satisfies it. Thus, the compatibility of the commutation relations is guaranteed by the Yang--Baxter equation. In turn, the identity \eqref{04-Eig-identfg} is just one of the scalar identities contained in the matrix identity \eqref{04-Eig-YB}.

Here it is appropriate to draw an analogy with Lie algebras. If the Lie bracket of  elements  $e_i$ and $e_j$ is determined by
the structure constants
 $c^k_{ij}$
\be{04-AlLie}
[e_i,e_j]=c^k_{ij}e_k,
\ee
then the properties of the bracket (antisymmetry, the Jacobi identity)  impose certain restrictions on the structure constants. One
can consider the $R$-matrix as an analog of the structure constants.
Then the $RTT$-relation \eqref{02-RTT-0}
imposes certain restriction on the $R$-matrix. The latter is just the Yang--Baxter equation.

It is worth mentioning that the Yang--Baxter equation is not necessary, but only a sufficient condition of the compatibility
of the commutation relations between the monodromy matrix entries.
However, an example of an $R$-matrix, that, on the one hand, provides the compatibility, but, on the other hand, does not satisfy the Yang--Baxter
equation, is not known.

It becomes now clear why we used the $R$-matrix \eqref{02-Ruv}
in section~\ref{02-Sec1}.
Of course, the commutativity of the transfer matrices follows from the $RTT$-relation for arbitrary $R$-matrix, however, the $RTT$-relation
imposes the restrictions on the $R$-matrix.


%
%

\chapter{Bethe equations\label{CHA-BE}}

In the previous lecture, we obtained the system of Bethe equations. The roots of this system determine the spectrum of the transfer matrix and,
thus, the spectrum of the Hamiltonian of the model under consideration. Of course, in the overwhelming majority of cases, it
is impossible to find these roots explicitly.
This is not surprising. Indeed, suppose that we have found solutions to the Bethe equations  for the $XXX$ model. This would mean,
that using such a method we were able to solve explicitly an algebraic equation of degree $2^N$. It would be naive to hope to be so lucky.

Nevertheless, even without an explicit solution, the Bethe equations give
a lot of useful information about the quantum system, that we have study.
In particular, it is possible to find the ground state of the system and to
describe excitations. In cases where the system consists of a large number of particles, one can describe its thermodynamic properties.
In addition, the system of Bethe equations can be solved
numerically, which ultimately makes it possible to compare the results of theoretical calculations with experimental data
obtained in real physical systems.

The main question concerning the system of Bethe equations is whether its solutions describe the complete set of the Hamiltonian eigenfunctions.
The answer to this question depends on the particular model. In some cases, it is relatively easy to prove the completeness (see e.g. \cite{Gaud83,Sla11}
for the QNLS model). However, in the model of  the $XXX$ Heisenberg chain, the proof is much more complicated,
and a lot of literature is devoted to this question (see \cite{Bax02,Tarv95b,MukTV09,MukTV13,NepW14}). In particular, the emerging difficulties
are associated with the presence of  `spurious' solutions. These are the roots that formally satisfy the Bethe equations, however, the vectors constructed from such solutions, are not on-shell vectors. In this lecture, we will consider several examples of such spurious solutions.

\section{Coinciding roots\label{04-Sec1}}

Formally, the system of Bethe equations
\be{04-BE0}
\frac{a(u_k)}{d(u_k)}=\frac{f(u_k,\bu_k)}{f(\bu_k,u_k)}
\equiv \prod_{\substack{j=1\\ j\ne k}}^n\frac{u_k-u_j+c}{u_k-u_j-c},  \qquad k=1,\dots,n,
\ee
might have coinciding roots. For instance, we can look for a solution of \eqref{04-BE0},
in which all the roots are equal to each other $u_j=u$, $j=1,\dots, n$. Then the system turns into one equation
\be{04-BE1}
\frac{a(u)}{d(u)}=(-1)^n.
\ee
For the models considered above ($XXX$ chain and QNLS) we obtain
\be{04-BE-BE-expl-all}
\begin{aligned}
\left(\frac{u+\frac{c}2}{u-\frac{c}2}\right)^N&=(-1)^n,\\
e^{-iuL}&=(-1)^n,
\end{aligned}
\ee
and we see that these equations have solutions. However, the vector $B(\bu)|0\rangle$, corresponding to these
solutions, generally speaking, is not an eigenvector. The point is that when deriving the Bethe equations, we actually assumed that
all the parameters $u_j$ were distinct. Let us return to the the action formula of the transfer matrix on the Bethe vector
\eqref{03-T-act}
\begin{equation}\label{04-T-act}
\mathcal{T}(v)B(\bu)|0\rangle=\tau(v|\bu)B(\bu)|0\rangle
+ \sum_{k=1}^n g(v,u_k)\Bigl(a(u_k)f(\bu_k,u_k)-d(u_k)f(u_k,\bu_k)\Bigr)B(v)B(\bu_k)|0\rangle.
 \end{equation}
For simplicity let us consider the case $n=2$. Then \eqref{04-T-act} takes the form
\begin{multline}\label{04-T-act2}
\mathcal{T}(v)B(\bu)|0\rangle=\tau(v|\bu)B(\bu)|0\rangle\\
\hspace{5mm}+ \frac{c}{(v-u_1)(u_2-u_1)}\Bigl(a(u_1)(u_2-u_1+c)-d(u_1)(u_2-u_1-c)\Bigr)B(v)B(u_2)|0\rangle\\
- \frac{c}{(v-u_2)(u_2-u_1)}\Bigl(a(u_2)(u_1-u_2+c)-d(u_2)(u_1-u_2-c)\Bigr)B(v)B(u_1)|0\rangle.
 \end{multline}
Obviously, a singularity arises at $u_1=u_2$. Taking the limit $u_1\to u_2=u$ we find
\begin{multline}\label{04-T-act3}
\mathcal{T}(v)B^2(u)|0\rangle=\tau(v|\bu)B^2(u)|0\rangle
+ \frac{c^2}{v-u}\Bigl(a(u)+d(u)\Bigr)B(v)B'(u)|0\rangle\\
+ \frac{c}{v-u}\Bigl(2\bigl(a(u)-d(u)\bigr)-c\bigl(a'(u)-d'(u)\bigr)-\frac{c}{v-u}\bigl(a(u)+d(u)\bigr)\Bigr)B(v)B(u)|0\rangle.
 \end{multline}
We see that besides the original vector $B^2(u)|0\rangle$ we obtain two new vectors. First, this is the vector $B(v)B(u)|0\rangle$. Second,
this is the vector $B(v)B'(u)|0\rangle$, that contains a derivative of the operator $B(u)$. One should set the coefficients of both vectors
equal to zero, otherwise the original vector is not on-shell. Hence, we obtain two conditions
\be{04-syst}
\begin{aligned}
&a(u)+d(u)=0,\\
&2\bigl(a(u)-d(u)\bigr)-c\bigl(a'(u)-d'(u)\bigr)=0.
\end{aligned}
\ee
Thus, although the number of variables has decreased (instead of $u_1$ and $u_2$ we have only one variable $u$), we still have two equations. The first of these equations formally coincides with the Bethe equations for $u_1=u_2=u$. But we also have an additional equation, and the system is overdetermined. The solvability of such the overdetermined system depends on the specific values of the parameters of the model under consideration. For example, in the QNLS case
the system \eqref{04-syst} has a solution under the condition $\varkappa L = -4$, where $\varkappa = -ic$ is the coupling constant.
Obviously, there are no solutions for a positive coupling constant.
However, even with a negative coupling constant, the solution exists only in the exceptional case.
Any deviation of the length of the interval $L$ from a given value leads to an unsolvable system \eqref{04-syst}.

It is now clear what happens in the case when the Bethe vector $B(\bu)|0\rangle$ depends on the $n$ variables $u_1,\dots,u_n$. If $u_j=u_k$, then the formula \eqref{04-T-act} has singularities due to the presence of the functions $f(u_j, u_k)$. After solving these singularities, we get vectors containing the derivative of the operator $B$. It is easy to verify that in the end we will always have $n$ vectors of different types. Their coefficients  should be set equal to zero. Therefore, we always obtain a system of $n$ equations. At the same time, the number of variables decreases. Therefore, the resulting system will be overdetermined. Generally, such an overdetermined system is unsolvable.

Let us stress once more that the system \eqref{04-BE0} does have coinciding roots. However, as we have seen, a Bethe vector corresponding to this
solution, generically is not on-shell.

\section{Bethe equations for the $XXX$ chain\label{04-Sec2}}

For the   $XXX$ Heisenberg chain, the system of Bethe equations has the form (at $c=i$)
\be{04-BE-BE-expl}
\left(\frac{u_j+\frac{i}2}{u_j-\frac{i}2}\right)^N=
\prod_{\substack{k=1\\k\ne j}}^n\frac{u_j-u_k+ i}{u_j-u_k- i}.
\ee
Here we have set $c=i$. Generically, the constant $c$ does not play an essential role in the $XXX$ model.
It is not for nothing that  the Hamiltonian does not depend on this constant.
We always can do a change of variables $u\to \alpha u$, what leads to the replacement $c\to c/\alpha$.
We have set $c=i$ for convenience, since in this case solutions of Bethe equations either consist of   real roots
or include complex conjugated pairs.

The analysis of the system \eqref{04-BE-BE-expl}  is quite a complex problem, and numerous literature is devoted to it (see, for example,
\cite{Tarv95b,MukTV09,MukTV13} and references therein).
In this section, we illustrate some subtleties hidden in the system \eqref{04-BE-BE-expl}.

\subsection{Infinite solutions\label{04-Sec21}}

If we send one parameter to infinity in the system \eqref{04-BE-BE-expl}  (for example, $u_n$), then one of equations
turns to the identity, and from the remaining equations this parameter disappears. The system simply turns exactly to the
same system, but for $n-1$ variable. The question arises whether such infinite roots should be considered.

An indirect answer to this question is given by the case $n=1$, when the system \eqref{04-BE-BE-expl} reduces to a single equation
\be{04-BE-xxx-1}
\left(\frac{u+\frac i2}{u-\frac i2}\right)^N=1.
\ee
From this we find
\be{04-BE-xxx-2}
\frac{u+\frac i2}{u-\frac i2}=e^{\frac{2\pi i k}N},\qquad k=0,\dots,N.
\ee
In particular, for $k=0$ we obtain that $u$ goes to infinity.
If we do not take this solution into account, then we have
only $N-1$ roots, and, as a consequence, we construct only $N-1$ eigenvectors. At the same time, the subspace of states
with one spin down (see the example from section~\ref{03-Sec41})
is $N$-dimensional, hence, the basis must consist of $N$ vectors.

Strictly speaking, the presence of infinite roots is a question of parametrization. Making a change of variables
\be{04-BE-Repar}
\frac{u_j+\frac{i}2}{u_j-\frac{i}2}=z_j,
\ee
we transform the system \eqref{04-BE-BE-expl} to
\be{04-BE-BE-expl-z}
z_j^N=(-1)^{n-1}\prod_{k=1}^n\frac{1-2 z_j+z_jz_k}{1-2 z_k+z_jz_k},\qquad j=1,\dots,n.
\ee
Clearly, in this case $u_n\to\infty$ corresponds to the root $z_n=1$.
However, even with such the parametrization, the system of Bethe equation  can have infinite solutions, which must be taken into account
when constructing the basis.

Another peculiarity of infinite solutions (in the initial parametrization) is that several roots can go to
different infinities. For example, we can specify $u_1\to\infty$ and $u_2\to\infty$, but the difference $u_1-u_2$ can be
either finite or infinite. However, if we pass  to the variables $z_j$ \eqref{04-BE-BE-expl-z}, then in this limit we get
coinciding roots $z_1=z_2=1$. The question of whether the Bethe vector constructed from such a solution, is an on-shell vector,
requires a separate analysis.

\subsection{Inadmissible solutions\label{04-Sec22}}

The writing of Bethe equations in the form of \eqref{04-BE0} is traditional, but not entirely correct. Indeed, it follows  from
\eqref{04-T-act} that the conditions of vanishing of the coefficients for unwanted vectors have the form
\be{04-BE-pol}
a(u_k)f(\bu_k,u_k)=d(u_k)f(u_k,\bu_k), \qquad k=1,\dots,n.
\ee
The system \eqref{04-BE-pol} is equivalent to \eqref{04-BE0}, if both sides of \eqref{04-BE-pol} do not vanish for all $k=1,\dots,n$.
However, one can find in the  $XXX$ model such solutions of \eqref{04-BE-pol}, that both sides are equal to zero. Let us consider again
the case $n=2$, and let $c=i$. Then the system \eqref{04-BE-pol} takes the form
\be{04-Eig-BE-expl}
\begin{aligned}
(u_1+\tfrac{i}2)^N(u_2-u_1+ i)=(u_1-\tfrac{i}2)^N(u_2-u_1- i),\\
(u_2+\tfrac{i}2)^N(u_1-u_2+ i)=(u_2-\tfrac{i}2)^N(u_1-u_2- i).
\end{aligned}
\ee
It is easy to see that the values $u_1=\tfrac{i}2$ and $u_2=-\tfrac{i}2$
are the roots of the system \eqref{04-Eig-BE-expl}. Apriori, these roots
are not worse and are not better than all other possible roots.
However, these roots show a very curious feature of the system \eqref{04-Eig-BE-expl}.
The matter is that the Bethe vector, constructed by the solution $u_1=\tfrac{i}2$ and $u_2=-\tfrac{i}2$,
turns out to be a null-vector. A complete proof of this fact will be given later, when we learn how to calculate scalar products of Bethe vectors.
In the meantime, we will consider a specific example of a chain consisting of four sites. This
example ($N=4$, $n=2$) is remarkable in that, on the one hand, it is non-trivial, and on the other hand, Bethe equations are
explicitly solvable.

So, we want to build a vector $B(-\tfrac i2)B(\tfrac i2)|0\rangle$. First, let us find a vector $|\phi\rangle=B(\tfrac i2)|0\rangle$. We
know the action of the $B(u)$ operator onto vacuum:
\begin{equation}\label{04-CR-Action-Binh}
B(u)|0\rangle
=i\sum_{k=1}^4  (u+\tfrac i2)^{4-k}
 (u-\tfrac i2)^{k-1}\cdot \sigma^-_k|0\rangle.
\end{equation}
Setting here $u=\tfrac i2$ we see that only the term  $k=1$ survives in the sum. Therefore,
\begin{equation}\label{04-phi1}
|\phi\rangle= \sigma^-_1|0\rangle,
\end{equation}
or equivalently
\be{04-phi2}
|\phi\rangle=
\left(\begin{smallmatrix}0\\1\end{smallmatrix}\right)_1\otimes \left(\begin{smallmatrix}1\\0\end{smallmatrix}\right)_2\otimes
\left(\begin{smallmatrix}1\\0\end{smallmatrix}\right)_3\otimes\left(\begin{smallmatrix}1\\0\end{smallmatrix}\right)_4.
\ee

Let us find now the operator $B(-\tfrac i2)$. The $L$-operator at $u=-\tfrac i2$  has the form
\be{04-Ln}
L_n(-\tfrac i2)=\begin{pmatrix}
\frac i2(1-\sigma_n^z) &i\;\sigma_n^-\\
i\;\sigma_n^+ &-\frac i2(1+\sigma_n^z)
\end{pmatrix}.
\ee
It is convenient to pass from the Pauli matrices to the elementary units $E^{jk}$:
\be{04-El-Ed}
E^{11}=\left(\begin{smallmatrix}1&0\\0&0\end{smallmatrix}\right),\qquad
E^{12}=\left(\begin{smallmatrix}0&1\\0&0\end{smallmatrix}\right),\qquad
E^{21}=\left(\begin{smallmatrix}0&0\\1&0\end{smallmatrix}\right),\qquad
E^{22}=\left(\begin{smallmatrix}0&0\\0&1\end{smallmatrix}\right).
\ee
In other words, the matrix $E^{jk}$ has a unit at the intersection of the $j$th row and the $k$th column, and all other entries are zero.
In terms of the matrices $E^{jk}$, the $L$-operator is written as
\be{04-Lop-i2}
L_n(-\tfrac i2)=i\begin{pmatrix} -E^{22}_n& E^{21}_n\\ E^{12}_n& -E^{11}_n
\end{pmatrix}.
\ee
Multiplying the $L$-operators over the four sites of the chain we find an explicit expression for $B(-\tfrac i2)$:
\be{04-B-i2}
B(-\tfrac i2)= -(E^{22}_4E^{22}_3 + E^{21}_4E^{12}_3)(E^{22}_2E^{21}_1 + E^{21}_2E^{11}_1)
-(E^{22}_4E^{21}_3 + E^{21}_4E^{11}_3)(E^{12}_2E^{21}_1 + E^{11}_2E^{11}_1).
\ee
It remains to act with this operator on the vector $|\phi\rangle$ \eqref{04-phi2}. Taking into account obvious formulas
\be{04-act-E}
\begin{aligned}
&E^{22}_k|\phi\rangle=0,\qquad E^{12}_k|\phi\rangle=0,\qquad k>2,\\
&E^{11}_1|\phi\rangle=0,\qquad  E^{21}_1|\phi\rangle=0,
\end{aligned}
\ee
we easily convince ourselves that $B(-\tfrac i2)|\phi\rangle=0$.

Thus, the Bethe vector corresponding to the solution $u_1=\tfrac i2$, $u_2=-\tfrac i2$ is the null-vector.
It seems that it is not difficult to fix the situation. Indeed, we can take the vector $B(u_1)B(u_2)|0\rangle$
and compute its norm $\mathcal{N}(u_1,u_2)$
in the case when the parameters $u_1$ and $u_2$ are generic complex numbers. Then the norm of the vector
\be{04-norm-vect}
\frac 1{\mathcal{N}(u_1,u_2)}\;B(u_1)B(u_2)|0\rangle
\ee
is equal to one for any $u_1$ and $u_2$. Hence, we never can obtain the null-vector. It remains only to take the limit
 $u_1\to\tfrac i2$, $u_2\to-\tfrac i2$ in \eqref{04-norm-vect}. The problem, however, is that this limit depends on how
 $u_1$ and $u_2$ tend to their limit values. The situation is quite similar to the case
\be{04-limit}
\lim_{x,y\to 0}\frac{x-y}{x+y}\;.
\ee
This limit can take any predefined value, depending on how $x$ and $y$ tend to zero.
In our case, we do not have any additional prescriptions on how exactly $u_1$ and $u_2$ go to $\tfrac i2$ and $-\tfrac i2$.
One of the possible ways of solving the problem is to introduce an additional parameter in the system of Bethe equations.
We will describe this method in the next lecture, where we continue to consider the example of the chain of $4$ sites.


%
%

\chapter{Trigonometric $R$-matrix\label{CHA-TRM}}

In this lecture we will study different modifications of already familiar objects. First, we consider a trigonometric $R$-matrix.
Secondly, we describe a one-parametric deformation of the monodromy matrix. The role of this additional parameter
will be shown using the example of the Bethe equations. Finally, thirdly, we construct the most general form of the monodromy matrix,
that possesses a vacuum vector and satisfies the $RTT$-relation with the $R$-matrix $R(u,v)=\mathbf{I}+g(u,v)\mathbf{P}$.

\section{Trigonometric $R$-matrix\label{05-Sec1}}

We have seen in section~\ref{04-Sec3}
that if the $R$-matrix satisfies the Yang--Baxter equation, then the commutation relations between the monodromy matrix entries are guaranteed to be compatible. However, the $R$-matrix used so far, is not the only solution of the Yang--Baxter equation.
Consider an $R$-matrix, which  has exactly the same form as before in terms of the functions $f$ and $g$
\be{05-R}
R(u,v)=
\begin{pmatrix}
f(u,v)&0&0&0\\
0&1&g(u,v)&0\\
0&g(u,v)&1&0\\
0&0&0&f(u,v)
\end{pmatrix},
\ee
however, this time we set
\be{05-CR-fg}
f(u,v)=\frac{\sinh(u-v+\eta)}{\sinh(u-v)}\qquad\text{and}\qquad
g(u,v)=\frac{\sinh\eta}{\sinh(u-v)},
\ee
where $\eta$ is some constant. This $R$-matrix also satisfies the Yang--Baxter equation and it is called
trigonometric $R$-matrix\footnote{This is a matter of taste to write  trigonometric  or hyperbolic sinus in the formulas \eqref{05-CR-fg}. Often the choice of a function is due to the desire to have the real roots of the Bethe equations.}. Pay attention that the previous  $R$-matrix actually depended on a
 one function $g$, while the trigonometric $R$-matrix depends on two functions, because $f\ne 1+g$. Nevertheless, identities between
$f$ and $g$, that we obtained before, still are valid. For example, we can check that, just as for the rational functions $f$ and $g$
(see \eqref{04-Eig-identfg}),
we have an identity
\be{05-Eig-identfg}
g(v,u_2)f(u_1,v)+ g(v,u_1)g(u_1,u_2)=g(v,u_2)f(u_1,u_2).
\ee
We recall that this identity is one of the scalar relations that arise if we write the matrix
Yang--Baxter equation in the components.

One can also see that in the limit $u=\epsilon u'$, $v=\epsilon v'$, $\eta=\epsilon c$, $\epsilon \to 0$ the trigonometric $R$-matrix turns into the
rational one. By the way, many authors prefer a different notation for the trigonometric $R$-matrix. Namely,
if we make a change of variables $e^{2u}=\lambda$,   $e^{2v}=\mu$, and $e^{\eta}=q$,
then the $f$ and $g$ functions  take the form
\be{05-CR-fg-q}
f(\lambda,\mu)=\frac{q\lambda-q^{-1}\mu}{\lambda-\mu}\qquad\text{and}\qquad
g(\lambda,\mu)=\frac{(q-q^{-1})\sqrt{\lambda\mu}}{\lambda-\mu}.
\ee
Using a special similarity transformation one can also get rid of the square root in the $g$ function, and then the trigonometric $R$-matrix
depends on rational functions. Therefore, in fact, the adequacy of the use of the terms `trigonometric' and `rational' depends
on the parametrization. Note also that, being written in the form \eqref{05-CR-fg-q}, the functions $f$ and $g$ no longer depend on
the difference of the arguments. It is for this reason that we write $f(u,v)$ and $g(u,v)$, without stressing the dependence on the difference,
even if it really is.

Now we can consider monodromy matrices satisfying the $RTT$-relation \eqref{02-RTT}
with the trigonometric $R$-matrix. It is obvious that all commutation relations between the operators $A$, $B$, $C$, and $D$,
remain the same if they are written in terms of the functions $f$ and $g$. The formulas for the action of operators on
the Bethe vectors also do not change, and therefore, the system of Bethe equations still
has the form
\be{05-BE0}
\frac{a(u_k)}{d(u_k)}=\frac{f(u_k,\bu_k)}{f(\bu_k,u_k)}, \qquad k=1,\dots,n.
\ee
In what follows we shall see that the formulas for scalar products also have this universality: they all remain
valid for the models with the rational $R$-matrix and for the models with the trigonometric $R$-matrix, if written in terms
of the functions $f$ and $g$. Therefore, we will continue to use this universal notation.

As for the physical models described by the trigonometric $R$-matrix, the most famous are the $XXZ$ Heisenberg chain and the Sin-Gordon model. In particular, the construction of the $XXZ$ chain Hamiltonian  is completely analogous to the case of the $XXX$ chain discussed above. The $R$-matrix \eqref{05-R} with the functions \eqref{05-CR-fg} generates an $L$-operator
\begin{equation}\label{05-CR-L-oper}
L_n(u)=\begin{pmatrix}
\sinh(u+\tfrac\eta2\sigma^z_n)&\sigma^-_n\sinh\eta\\
\sigma^+_n\sinh\eta&\sinh(u-\tfrac\eta2\sigma^z_n)
\end{pmatrix}.
\end{equation}
Recall that any matrix-valued function is understood as the corresponding Taylor series, that leads to
\begin{equation}\label{05-sz}
\sinh(u\pm\tfrac\eta2\sigma^z_n)=\sinh u\cosh\tfrac\eta2\pm \sigma^z_n\cosh u\sinh\tfrac\eta2.
\end{equation}
The monodromy matrix is equal to the product of the $L$-operators over all sites of the chain (see \eqref{02-T-LLL}),
and the Hamiltonian can be obtained from the transfer matrix by
\be{05-QH-trident1}
H=2\sinh\eta\frac{d\mathcal{T}(u)}{du}\;\mathcal{T}^{-1}(u)\Bigr|_{u=\frac \eta2}-N\Delta\,,
\ee
where $\Delta=\cosh\eta$.
The reader can convince himself that the formula \eqref{05-QH-trident1} leads to the Hamiltonian
\be{05-QH-Ham-XXZ}
H=\sum_{k=1}^N\Bigl(\sigma_k^x\sigma_{k+1}^x+\sigma_k^y\sigma_{k+1}^y+\Delta\sigma_k^z\sigma_{k+1}^z\Bigr).
\ee

Replacing in \eqref{02-T-LLL}
every $L$-operator  $L_n(u)$ with  $L_n(u-\xi_n)$, we obtain an inhomogeneous $XXZ$ chain. Similarly to the  $XXX$ case, this
inhomogeneous chain has no physical meaning, however, it appears to be useful in intermediate calculations.

Construction of the quantum Sin-Gordon model is described in chapter VII of the book \cite{KBIr}.

\section{Twisted monodromy matrix\label{05-Sec2}}

Let a $2\times 2$ matrix $\hat\kappa$ be such that its tensor square commutes with the $R$-matrix
\be{05-kR}
[\hat\kappa_1\hat\kappa_2,R_{12}(u,v)]=0.
\ee
Then, if a monodromy matrix $T(u)$ satisfies the $RTT$-relation, then the matrix $\hat\kappa T(u)$ also satisfies this relation
\be{05-RTT-twi}
R_{12}(u,v)\bigl(\hat\kappa T(u)\bigr)_1\bigl(\hat\kappa T(v)\bigr)_2=\bigl(\hat\kappa T(v)\bigr)_2\bigl(\hat\kappa T(u)\bigr)_1R_{12}(u,v).
\ee
To prove \eqref{05-RTT-twi} it is enough to multiply the original  $RTT$-relation by $\hat\kappa_1\hat\kappa_2$, say, from the left. Then we obtain in the
lhs of the $RTT$-relation
\be{05-RTT-kap}
\hat\kappa_1\hat\kappa_2 R_{12}(u,v) T_1(u)T_2(v)= R_{12}(u,v)\hat\kappa_1\hat\kappa_2 T_1(u)T_2(v)=
R_{12}(u,v)\bigl(\hat\kappa T(u)\bigr)_1\bigl(\hat\kappa T(v)\bigr)_2.
\ee
We have used the condition \eqref{05-kR} and the fact that the entries of the matrix $\hat\kappa$ commute with all matrix elements of the monodromy matrix,
what leads to $\hat\kappa_2 T_1(u)=T_1(u)\hat\kappa_2$. In the rhs of the
$RTT$-relation we have
\be{05-RTT-kap2}
\hat\kappa_1\hat\kappa_2  T_1(u)T_2(v)R_{12}(u,v)=\bigl(\hat\kappa T(v)\bigr)_2\bigl(\hat\kappa T(u)\bigr)_1R_{12}(u,v),
\ee
and we arrive at \eqref{05-RTT-twi}. It is easy to check that the initial monodromy matrix can be also multiplied by $\hat\kappa$ from the right,
and the new matrix $T(u)\hat\kappa $ also satisfies the $RTT$-relation.

It is obvious that the $R$-matrix $R(u,v)=\mathbf{I}+g(u,v)\mathbf{P}$ commute with $\hat\kappa_1\hat\kappa_2$ for an arbitrary
$2\times 2$ matrix $\hat\kappa$. For this reason, this $R$-matrix is called $GL(2)$-invariant. However, as a result of multiplying
$T(u)$ by $\hat\kappa$ we can break properties of vacuum. Indeed, if
\be{05-kappa}
\hat\kappa=\begin{pmatrix}
\kappa_{11} & \kappa_{12}\\
\kappa_{21}&\kappa_{22}
\end{pmatrix},
\ee
then
\be{05-T-twist}
\hat\kappa T(u)=\begin{pmatrix}
A_\kappa(u) & B_\kappa(u)\\
C_\kappa(u)& D_\kappa(u)
\end{pmatrix}=
\begin{pmatrix}
\kappa_{11}A(u) +\kappa_{12}C(u)& \kappa_{11}B(u)+\kappa_{12}D(u)\\
\kappa_{21}A(u)+\kappa_{22}C(u)& \kappa_{21}B(u)+\kappa_{22}D(u)
\end{pmatrix}.
\ee
We see that the vacuum vector $|0\rangle$ is no longer annihilated by the operator $C_\kappa(u)$ and is not an eigenvector for the operator\footnote{%
Note, however, that the violation of these properties of the vector $|0\rangle$ is not very crucial. Algebraic Bethe ansatz can be used
and in this case, although some techniques vary slightly (see e.g. \cite{WanYCS15}).} $D_\kappa(u)$.
If, however, we take the diagonal matrix $\hat\kappa =\diag(\kappa_1,\kappa_2)$, then the properties of the vacuum are not violated. In fact, without loss of generality we can take the matrix $\hat\kappa$ in the form $\hat\kappa =\diag(1,\kappa)$, because multiplication of the monodromy matrix by an arbitrary
constant is certainly not essential.

If the matrix $\hat\kappa$ is diagonal, then its tensor square commutes with the trigonometric $R$-matrix as well. Therefore, in this case
we can also consider similar deformation of the original monodromy matrix.

The new matrix $\hat\kappa T(u)$ obtained in this way is called a twisted monodromy matrix, and its trace $A(u)+\kappa D(u)$
is called a twisted transfer matrix. The parameter $\kappa$ is called the twist parameter. The properties of the twisted monodromy matrix are completely analogous to the properties of the original monodromy matrix. Only the value of the function $d(u)$ changes. Indeed, it is easy to see that
\be{05-actD-tw}
D_\kappa(u)|0\rangle=\kappa d(u)|0\rangle.
\ee
Therefore, in the action of the $D_\kappa(v)$ operator on the Bethe vectors  $B(\bu)$, the parameter $\kappa$ arises.  This ultimately leads to
the appearance of this parameter in the Bethe equations
\be{05-BE-tw}
\frac{a(u_k)}{d(u_k)}=\kappa\frac{f(u_k,\bu_k)}{f(\bu_k,u_k)}, \qquad k=1,\dots,n.
\ee
Eigenvectors of the twisted transfer matrix are called twisted on-shell vectors. They still have the form
$B(\bu)|0\rangle$, however, the parameters $\bu$ should satisfy twisted Bethe equations \eqref{05-BE-tw}.

The reader can ask the question: why is it necessary to introduce an additional parameter into the monodromy matrix? For
possible physical applications it makes sense to consider $\kappa=e^{i\varphi}$ where $\varphi$ is a real
number. Then we can study models not only with periodic, but also with quasiperiodic boundary conditions.
However, other values of the twist parameter are also useful, as we shall see in the next section.

\section{Inadmissible solutions for the $XXZ$ chain\label{05-Sec3}}

The appearance of the additional parameter $\kappa$ in the Bethe equations leads to completely unexpected effects. To illustrate
this, we  return to the example of a chain consisting of four sites: $N =4$. For the sake of generality, let it now be
the $XXZ$ chain. Let us construct the Bethe vector $B(u_1)B(u_2)|0\rangle$, dependent on two parameters $u_1$ and $u_2$.
This vector belongs to the subspace with two spins up and two spins
down. The dimension of this subspace is $\binom{4}{2}=6$.

We require the parameters  $u_1$ and $u_2$ to satisfy the twisted Bethe equations, that can be written in the form similar to
\eqref{04-Eig-BE-expl}:
\be{05-SBE0}
\begin{aligned}
&{\sinh}^4(u_1+\eta/2)\sinh(u_1-u_2-\eta)
=\kappa\;{\sinh}^4(u_1-\eta/2)\sinh(u_1-u_2+\eta),\\
&{\sinh}^4(u_2+\eta/2)\sinh(u_2-u_1-\eta)
=\kappa\;{\sinh}^4(u_2-\eta/2)\sinh(u_2-u_1+\eta).
\end{aligned}
\ee

It was shown in the section~\ref{04-Sec22}
that some solutions of the Bethe equations may cause problems in constructing the corresponding on-shell vector.
The same can happen with the twisted Bethe equations.
According to the results of the work \cite{Tarv95b}, the eigenvectors of the twisted transfer matrix correspond to such solutions of the system
\eqref{05-SBE0}, for which $u_1\ne u_2$ and both sides of the equations do not vanish. Thus, the solution $u_1=\eta/2$ and $u_2=-\eta/2$
is forbidden in the sense, that the corresponding vector is not a twisted on-shell vector. Note that this solution exists for any $\kappa$, and it
is an analog of the solution $u_1=\tfrac{i}2$ and $u_2=-\tfrac{i}2$ to the system  \eqref{04-Eig-BE-expl}.
We have seen that a vector constructed from such the solution is the null-vector.

If both sides of equations \eqref{05-SBE0} do not vanish, then we can rewrite them in the form
\be{05-SBE}
\begin{aligned}
&\left(\frac{\sinh(u_1+\eta/2)}
{\sinh(u_1-\eta/2)}\right)^4
\frac{\sinh(u_1-u_2-\eta)}
{\sinh(u_1-u_2+\eta)}
=\kappa,\\
&\left(\frac{\sinh(u_2+\eta/2)}
{\sinh(u_2-\eta/2)}\right)^4
\frac{\sinh(u_2-u_1-\eta)}
{\sinh(u_2-u_1+\eta)}
=\kappa.
\end{aligned}
\ee
Setting here $\kappa=\theta^{-2}$ and
\be{05-setting}
\frac{\sinh(u_j-\eta/2)}
{\sinh(u_j+\eta/2)}=w_j,
\ee
we obtain a new system for $w_1$ and $w_2$:
\be{05-SBEw}
\begin{aligned}
& w_1^4\cdot\frac{1+w_1w_2-2\Delta w_2}
{1+w_1w_2-2\Delta w_1}=-\theta^2,\\[2mm]
& w_2^4\cdot\frac{1+w_1w_2-2\Delta w_1}
{1+w_1w_2-2\Delta w_2}=-\theta^2,
\end{aligned}
\ee
where $\Delta=\cosh\eta$. Observe that in such the parametrization, the system of Bethe equations is equally
suitable for both  $XXZ$  and $XXX$ chains.
In the latter case, we simply put $\Delta = 1$.

Taking the product of both equations we obtain $(w_1w_2)^4=\theta^4$. Hence, there are four possibilities:
$w_1w_2=\pm i\theta$ and $w_1w_2=\pm \theta$.

Let $w_2=\pm i\theta/w_1$. Substituting this into the first equation \eqref{05-SBEw} we obtain
\be{05-w_i_theta} [w_1^2\mp i\theta][(1\pm i\theta)(w_1^2\pm i\theta)
\mp 2i\Delta\theta w_1]=0.
\ee
If $w_1^2=\pm i\theta$, then $w_1=w_2$, and this solution does not correspond to an eigenvector of the twisted transfer matrix.
Thus, it remains to solve the quadratic equation
\be{05-q_eq_1}
(1\pm i\theta)(w_1^2\pm i\theta) \mp 2i\Delta\theta
w_1=0.
\ee
If $w_1w_2=i\theta$, then we obtain a solution
\be{05-solut_1}
w_{1}=\frac{i\Delta\theta}{1+i\theta}
+\sqrt{-i\theta-\frac{\Delta^2\theta^2}
{(1+i\theta)^2}},\qquad w_{2}=\frac{i\Delta\theta}{1+i\theta}
-\sqrt{-i\theta-\frac{\Delta^2\theta^2}
{(1+i\theta)^2}}.
\ee
Pay attention that we are speaking about solutions of the system \eqref{05-SBEw}, but not about solutions of the equation \eqref{05-q_eq_1}. The equation \eqref{05-q_eq_1} has two roots $w_{1}^+$ and  $w_{1}^-$ at $w_1w_2=i\theta$. They are different by the sign of the square root.
We have set $w_{1}=w_{1}^+$ in the solution \eqref{05-solut_1}. It is easy to see that if we put
$w_{1}=w_{1}^-$, then it would be equivalent to the replacement    $w_1$ with  $w_2$ in the solution of the system \eqref{05-SBEw}.
But we do not distinguish between solutions of the Bethe equations, which differ from each other only by permutation, since they correspond to the same eigenvector.

Similarly, for the case $w_1w_2=-i\theta$, we find
\be{05-solut_2}
w_{1}=\frac{-i\Delta\theta}{1-i\theta}
+\sqrt{i\theta-\frac{\Delta^2\theta^2}
{(1-i\theta)^2}},\qquad w_{2}=\frac{-i\Delta\theta}{1-i\theta}
-\sqrt{i\theta-\frac{\Delta^2\theta^2}
{(1-i\theta)^2}}.
\ee

Let now $w_1w_2=\pm\theta$. Then we obtain the following equation for the parameter $w_1$:
\be{05-w_theta}
(1\pm\theta)(w_1^4+\theta^2)\mp2\Delta\theta w_1(w_1^2\pm\theta)=0.
\ee
Dividing this equation by $w_1^2$ and setting
\be{05-setting2}
x=w_1\pm\theta/w_1,
\ee
we obtain a quadratic equation for $x$:
\be{05-q_eq_x}
(1\pm\theta)(x^2\mp2\theta)\mp2\Delta\theta x=0.
\ee
Thus, in the case $w_1w_2=\theta$, we obtain two solutions
\be{05-sol3}
w_{1}=\frac x2+\sqrt{\frac{x^2}4-\theta},\qquad w_{2}=\frac x2 -\sqrt{\frac{x^2}4-\theta},
\ee
where $x$ can take two values
\be{05-x1}
x=\frac{\Delta\theta}{1+\theta}
\pm\sqrt{\frac{\Delta^2\theta^2}{(1+\theta)^2}+2\theta}.
\ee
Similarly, in the case $w_1w_2=-\theta$ we have
\be{05-sol4}
w_{1}=\frac x2+\sqrt{\frac{x^2}4+\theta},\qquad w_{2}=\frac x2-\sqrt{\frac{x^2}4+\theta},
\ee
where
\be{05-x2}
x=-\frac{\Delta\theta}{1-\theta}
\pm\sqrt{\frac{\Delta^2\theta^2}{(1-\theta)^2}-2\theta}.
\ee

Thus, we have obtained $6$ solutions to the twisted Bethe equations, for which we can
construct $6$ eigenvectors of the twisted transfer matrix. Of course, it is also necessary to verify linear independence
of these vectors, but at least their number coincides with the dimension of the subspace with two spins up and two spins down.
Therefore, there is a chance that this system of vectors forms a basis.

Let us now see what happens if from the very beginning we consider the usual system of Bethe equations, and not the twisted one. To do this,
we simply put $\theta = 1$ in the equations \eqref{05-SBEw}. Then for $w_1w_2= -1$ equation \eqref{05-w_theta} takes the form
\be{05-w_theta-eq1}
w_1(w_1^2-1)=0,
\ee
leading to only one solution of Bethe equations:  $w_1=1$ and $w_2=-1$ (or vice versa).
The solution $w_1 = 0$ does not seem to suit us, as in this case we can not satisfy  the condition $w_1w_2 = -1$.
However, looking at the formula \eqref{05-sol4} for $\theta \ne 1$, we
immediately see that this solution can not be discarded. Indeed, for $w_1w_2 = -1$ and $\theta \to 1$, we obtain either $x = 0$ or $x = \infty$.
In the first case we reproduce the solution $w_1 = 1$ and $w_2 = -1$, and in the second we obtain $w_1 = 0$ and $w_2 =\infty$, and we need to provide that the product of zero and infinity equals to $-1$.
Clearly, we could not have found such a `strange solution' if we had worked with the usual Bethe equations system from the very beginning. As a result
we would get only five solutions, and accordingly five on-shell vectors, which obviously can not form a basis in the $6$-dimensional space.

Thus, we are dealing with a very curious situation. We have $6$ admissible solutions of twisted Bethe equations for $\kappa \ne 1$.
However, in the limit $\kappa \to 1$, one of these admissible solutions tends to an inadmissible solution $u_1=\eta/2$, $u_2=-\eta/2$.
Recall once more that the latter exists for any $\kappa$. If from the very beginning we considered the usual Bethe equations (with $\kappa = 1$), then
we could not distinguish between an admissible and an inadmissible solution, and then we would construct only $5$ solutions of equations, which is certainly not enough for the construction of a basis.

We have already mentioned that if we normalize the original vector $B(u_1)B(u_2)|0\rangle$ and then consider the limit $u_1\to\eta/2$, $u_2\to-\eta/2$,
then the result depends on how the parameters $u_1$ and $u_2$ tend to their limit values. The example considered above gives a hint to solve
the problem. We should consider the twisted Bethe equations and take such a solution $u_1(\kappa)$ and $u_2(\kappa)$, that goes to the inadmissible solution
in the limit $\kappa\to 1$. Then the limit for the normalized Bethe vector must be uniquely determined, since
this time we have fixed the way how $u_1$ and $u_2$ go to the limit. We will consider this question later.

\section{Classification of $L$-operators\label{05-Sec4}}

It is clear that multiplying the monodromy matrix by an arbitrary scalar function $T(u)\to \phi(u)T(u)$
leads only to a change of the functions $a(u)\to \phi(u)a(u)$ and $d(u)\to \phi(u)d(u)$.
Thus, essentially different monodromy matrices are parameterized by different ratios
$a(u)/d(u)$. It is not accidental that just this ratio is included in the Bethe equations. In this section, we
consider models with $R$-matrix of the form $R(u,v)=\mathbf{I}+g(u,v)\mathbf{P}$
and show that we can construct a monodromy matrix for which the function $a(u)/d(u)$ is the ratio of two arbitrary polynomials of the same degree.

We know already the $L$-operator of the lattice QNLS model
\begin{equation}\label{05-L-op22}
 L_n(u)=\begin{pmatrix}
1-\frac{iu\Delta_n}2+ \frac{\varkappa\Delta_n^2}2\psi^\dagger_n\psi_n&
-i\Delta_n\psi^\dagger_n \rho^+_n\\
i\Delta_n \rho^-_n \psi_n&
1+\frac{iu\Delta_n}2+\frac{\varkappa\Delta_n^2}2\psi^\dagger_n\psi_n
\end{pmatrix}.
\end{equation}
Recall that here $\rho^\pm_n=\rho^\pm_n(\psi^\dagger_n\psi_n)$ and
\be{05-rho}
\rho^+_n\rho^-_n=\varkappa+\frac{\varkappa^2\Delta_n^2}4\psi^\dagger_n\psi_n.
\ee
The parameter $\varkappa$ is related to the $R$-matrix constant $c$ by $c=-i\varkappa$.
The constant $\Delta_n$ is an arbitrary complex number\footnote{Do not confuse this constant with the anisotropy parameter in the $XXZ$ Heisenberg
chain \eqref{05-QH-Ham-XXZ}, which is traditionally denoted by the same symbol.}
therefore, we added a sibscript $n$, thus emphasizing that this parameter can take different values at different lattice sites. We can add three more arbitrary constants to the $L$-operator \eqref{05-L-op22}. First, one can shift the variable $u$ to some number $\xi_n$, because the $R$-matrix depends on the difference of the arguments. Secondly, we can multiply the $L$-operator by the matrix  $\hat\kappa_n=\diag(\kappa_{n1},\kappa_{n2})$.
As a result, we obtain
\begin{equation}\label{05-L-op22-t}
 \tilde L_n(u)=\begin{pmatrix}
\kappa_{n1}\left(1-\frac{i(u-\xi_n)\Delta_n}2+ \frac{\varkappa\Delta_n^2}2\psi^\dagger_n\psi_n\right)&
-i\kappa_{n1}\Delta_n\psi^\dagger_n \rho^+_n\\
i\kappa_{n2}\Delta_n \rho^-_n \psi_n&
\kappa_{n2}\left(1+\frac{i(u-\xi_n)\Delta_n}2+\frac{\varkappa\Delta_n^2}2\psi^\dagger_n\psi_n\right)
\end{pmatrix}.
\end{equation}
{\sl Remark}. Actually, we can also consider the case where one of the parameters $\kappa_{nj}$ is zero, but
then we will get a too poor model. Therefore, we assume that the parameters $\kappa_{n1}$ and $\kappa_{n2}$ are nonvanishing.

The action of diagonal matrix elements on the vacuum $|0\rangle$ is now given by
\be{05-eig-n}
\tilde L_n^{11}(u)|0\rangle=(a_n^{(1)}u+a_n^{(0)})|0\rangle,\qquad
\tilde L_n^{22}(u)|0\rangle=(d_n^{(1)}u+d_n^{(0)})|0\rangle,
\ee
where
 \be{05-ad}
 \begin{aligned}
 a_n^{(1)}&=-\frac{i\kappa_{n1}\Delta_n}2,\qquad a_n^{(0)}&=\kappa_{n1}+\frac{i\kappa_{n1}\xi_n\Delta_n}2,\\
 d_n^{(1)}&=\frac{i\kappa_{n2}\Delta_n}2, \qquad d_n^{(0)}&=\kappa_{n2}-\frac{i\kappa_{n2}\xi_n\Delta_n}2.
 \end{aligned}
 \ee
Since $\Delta_n$, $\kappa_{n1}$, $\kappa_{n2}$, and $\xi_n$, are arbitrary complex, the values of
$a_n^{(0)}$ and $d_n^{(0)}$ also are arbitrary, while $a_n^{(1)}$, $d_n^{(1)}$ are arbitrary nonvanishing (see the remark above).

Now we can build the monodromy matrix via the usual method
\be{05-Mon-mat}
\tilde T(u)= \tilde L_{N}(u)\cdots \tilde L_{1}(u).
\ee
Obviously, for this monodromy matrix, the ratio of the vacuum eigenvalues of the operators $A(u)$ and $D(u)$ is
\be{05-aovd}
\frac{a(u)}{d(u)}=\prod_{n=1}^N\frac{a_n^{(1)}u+a_n^{(0)}}{d_n^{(1)}u+d_n^{(0)}},
\ee
where $N$ is an arbitrary natural number. Thus, we have constructed a monodromy matrix for which the ratio $a(u)/d(u)$ is an arbitrary rational function bounded at infinity by a nonzero constant.
Furthermore, in the limit $N\to\infty$ we can obtain functions that have essential singu\-la\-ri\-ties, cuts, etc. Thus, if we do not
want to consider too exotic functions
$a(u)/d(u)$, then the matrix $T(u)$ \eqref{05-Mon-mat}
is the monodromy matrix of the most general form, satisfying the $RTT$-relation and possessing vacuum.

In particular, we should reproduce the monodromy matrix of the $XXX$ chain. Indeed, this is possible. Let
\begin{equation}\label{05-L-op22-sig}
 \tilde L_n(u)=\frac{2i}\Delta\; \sigma^z L_n(u)=\begin{pmatrix}
u+\frac{2i}\Delta+i\varkappa\Delta\psi^\dagger_n\psi_n&
2\Delta\psi^\dagger_n \rho^+_n\\
2\Delta \rho^-_n \psi_n&
u-\frac{2i}\Delta-i\varkappa\Delta\psi^\dagger_n\psi_n
\end{pmatrix},
\end{equation}
where we have set $\Delta_n=\Delta$ for all $n$. It is easy to see that the $L$-operator  \eqref{05-L-op22-sig} can be presented in the form
\be{05-Lt}
\tilde L_n(u)=u-i\varkappa\sum_{\alpha=x,y,z}\sigma^\alpha t_{n}^\alpha,
\ee
where
\be{05-tj}
\begin{aligned}
t_{n}^x&=\frac i\varkappa(\psi^\dagger_n \rho^+_n+\rho^-_n \psi_n),\\
t_{n}^y&=\frac 1\varkappa(\rho^-_n \psi_n-\psi^\dagger_n \rho^+_n),\\
t_{n}^z&=-\frac 2{\varkappa\Delta}-\Delta\psi^\dagger_n\psi_n.
\end{aligned}
\ee
A direct calculation shows that the operators $t_{n}^\alpha$  obey commutation relations
\be{05-comm-t}
[t_{n}^\alpha,t_{n}^\beta]=i\varepsilon_{\alpha\beta\gamma}t_{n}^\gamma.
\ee
Here  $\varepsilon_{\alpha\beta\gamma}$ is the antisymmetric tensor. In addition, it is easy to check that
\be{05-t-squared}
(t^{n})^2\equiv (t_{n}^x)^2 +(t_{n}^y)^2 +(t_{n}^z)^2 =s(s+1), \qquad\text{where}\qquad s=-\frac 2{\varkappa\Delta},
\ee
and hence, we obtained a representation of the $su(2)$ algebra of arbitrary spin. In fact, we have reproduced a well-known
Holstein--Primakov transformation from the creation and annihilation operators to the spin operators \cite{HolP40}.
Generally speaking, the constructed representation is infinite-dimensional, but in the particular case $s=1/2$ we can put
$t_{n}^\alpha=\frac{\sigma_n^\alpha}2$,
$\alpha=x,y,z$. Then, taking into account that $-i\varkappa=c$, we do reproduce the $L$-operator of the $XXX$ chain.


%
%
\chapter{Six-vertex model\label{CHA-6V}}

This lecture is devoted to a six-vertex model. At first glance, this is a step away from the description of the  Bethe ansatz method.
Actually, it is quite the opposite. The main task of this lecture is to show that the six-vertex model is not only
closely related to the $XXZ$ and $XXX$ Heisenberg chains, but  in essence this is the same thing. The reader can learn more about the six-vertex model
and other exactly solvable models of two-dimensional statistical physics from the book \cite{Bax82}.

\section{Definition of the six-vertex model\label{08-Sec1}}

Consider an $N\times N$ squared lattice. Let a two-values classical variable  be located at  the edges of this lattice.
Assume that these values are $1$ and $2$.
Let us match these values to the arrows. If the variable has the value $1$, then it corresponds to the
arrows $\rightarrow$ or $\uparrow$, depending on whether on horizontal or vertical
edge is this variable. Similarly, let the value $2$ correspond to the arrows $\leftarrow$
and $\downarrow$. We get a lattice  with the arrows on the edges.


\begin{figure}[ht!]
\begin{picture}(400,160)
\put(137,0){%
\begin{picture}(200,170)
\multiput(30,0)(30,0){5}{\line(0,1){160}}
\multiput(10,20)(0,30){5}{\line(1,0){160}}
\put(30,0){\vector(0,1){10}}
\put(60,0){\vector(0,1){10}}
\put(90,15){\vector(0,-1){10}}
\put(120,0){\vector(0,1){10}}
\put(150,15){\vector(0,-1){10}}
\put(30,27){\vector(0,1){10}}
\put(60,42){\vector(0,-1){10}}
\put(90,42){\vector(0,-1){10}}
\put(120,42){\vector(0,-1){10}}
\put(150,42){\vector(0,-1){10}}
\put(30,57){\vector(0,1){10}}
\put(60,57){\vector(0,1){10}}
\put(90,72){\vector(0,-1){10}}
\put(120,72){\vector(0,-1){10}}
\put(150,57){\vector(0,1){10}}
\put(30,102){\vector(0,-1){10}}
\put(60,87){\vector(0,1){10}}
\put(90,87){\vector(0,1){10}}
\put(120,87){\vector(0,1){10}}
\put(150,87){\vector(0,1){10}}
\put(30,129){\vector(0,-1){8}}
\put(60,129){\vector(0,-1){8}}
\put(90,129){\vector(0,-1){8}}
\put(120,117){\vector(0,1){10}}
\put(150,129){\vector(0,-1){8}}
\put(30,145){\vector(0,1){10}}
\put(60,145){\vector(0,1){10}}
\put(90,159){\vector(0,-1){10}}
\put(120,159){\vector(0,-1){10}}
\put(150,145){\vector(0,1){10}}
\put(10,20){\vector(1,0){10}}
\put(10,50){\vector(1,0){10}}
\put(25,80){\vector(-1,0){10}}
\put(25,110){\vector(-1,0){10}}
\put(25,140){\vector(-1,0){10}}
\put(37,20){\vector(1,0){10}}
\put(52,50){\vector(-1,0){10}}
\put(52,80){\vector(-1,0){10}}
\put(37,110){\vector(1,0){10}}
\put(37,140){\vector(1,0){10}}
\put(82,20){\vector(-1,0){10}}
\put(67,50){\vector(1,0){10}}
\put(82,80){\vector(-1,0){10}}
\put(82,110){\vector(-1,0){10}}
\put(67,140){\vector(1,0){10}}
\put(97,20){\vector(1,0){10}}
\put(97,50){\vector(1,0){10}}
\put(97,80){\vector(1,0){10}}
\put(97,110){\vector(1,0){10}}
\put(115,140){\vector(-1,0){10}}
\put(127,20){\vector(1,0){10}}
\put(142,50){\vector(-1,0){10}}
\put(142,80){\vector(-1,0){10}}
\put(127,110){\vector(1,0){10}}
\put(127,140){\vector(1,0){10}}
%
\put(157,20){\vector(1,0){10}}
\put(157,50){\vector(1,0){10}}
\put(157,80){\vector(1,0){10}}
\put(157,110){\vector(1,0){10}}
\put(170,140){\vector(-1,0){8}}
\end{picture}}
\end{picture}
\caption{\label{08-6V-16-vertex} $16$-vertex model}
\end{figure}
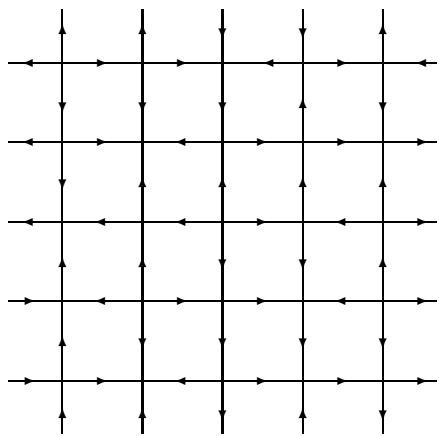

Every vertex can be associated with  four edges.  The arrow can have
one of two directions on each of these edges. Hence, there are $16$ types of different vertices. Such a model is
called $16$-vertex\footnote[1]{It would be more correct to call it `$16$-type-vertex'
model, but this name is too bulky.} by the number of different types of vertices (see Fig.~\ref{08-6V-16-vertex}).

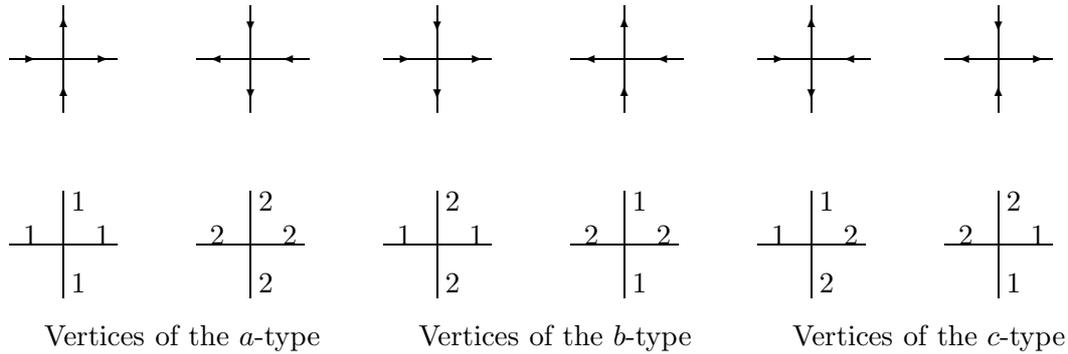
\begin{figure}[t!]
\begin{picture}(440,120)
\put(0,90){%
\begin{picture}(400,70)
\multiput(30,0)(70,0){6}{\line(0,1){40}}
\multiput(10,20)(70,0){6}{\line(1,0){40}}
\put(30,0){\vector(0,1){10}}
\put(100,14.5){\vector(0,-1){10}}
\put(170,14.5){\vector(0,-1){10}}
\put(240,0){\vector(0,1){10}}
\put(380,0){\vector(0,1){10}}
\put(310,14.5){\vector(0,-1){10}}
%
\put(30,26){\vector(0,1){10}}
\put(100,40){\vector(0,-1){10}}
\put(170,40){\vector(0,-1){10}}
\put(240,26){\vector(0,1){10}}
\put(380,40){\vector(0,-1){10}}
\put(310,26){\vector(0,1){10}}
\put(10,20){\vector(1,0){10}}
\put(95,20){\vector(-1,0){10}}
\put(150,20){\vector(1,0){10}}
\put(235,20){\vector(-1,0){10}}
\put(290,20){\vector(1,0){10}}
\put(375,20){\vector(-1,0){10}}
\put(37,20){\vector(1,0){10}}
\put(122,20){\vector(-1,0){10}}
\put(177,20){\vector(1,0){10}}
\put(262,20){\vector(-1,0){10}}
\put(332,20){\vector(-1,0){10}}
\put(387,20){\vector(1,0){10}}
\end{picture}}
\put(0,20){%
\begin{picture}(490,70)
\multiput(30,0)(70,0){6}{\line(0,1){40}}
\multiput(10,20)(70,0){6}{\line(1,0){40}}
\put(33,2){$1$}
\put(103,2){$2$}
\put(173,2){$2$}
\put(243,2){$1$}
\put(313,2){$2$}
\put(383,2){$1$}
%
\put(33,33){$1$}
\put(103,33){$2$}
\put(173,33){$2$}
\put(243,33){$1$}
\put(313,33){$1$}
\put(383,33){$2$}
\put(15,20){$1$}
\put(85,20){$2$}
\put(155,20){$1$}
\put(225,20){$2$}
\put(295,20){$1$}
\put(365,20){$2$}
\put(42,20){$1$}
\put(112,20){$2$}
\put(182,20){$1$}
\put(252,20){$2$}
\put(322,20){$2$}
\put(392,20){$1$}
\put(23,-18){Vertices of the $a$-type }
\put(163,-18){Vertices of the $b$-type}
\put(303,-18){Vertices of the $c$-type}
\end{picture}}
%
\end{picture}
\caption{\label{08-6V-types-2} Six types of vertices.}
\end{figure}

Let us impose an additional condition on the vertices. Let at each vertex the number of incoming
arrows be equal to the number of outgoing (that is, both those and the other exactly two). If we understand the arrows as
graphic representation of a certain flow, then we can say that the flow through each vertex is zero.
It is easy to see that there are six types of vertices possessing this property. Indeed,
two edges with incoming arrows can be chosen by $\binom{4}{2} = 6$ ways. On the remaining two edges
the arrows should automatically be outgoing. It is natural to call such a model $6$-vertex. Figure~\ref{08-6V-types-2} shows all six
types of vertices. The upper figure shows the arrows. The lower figure  shows
the corresponding values of the variable.

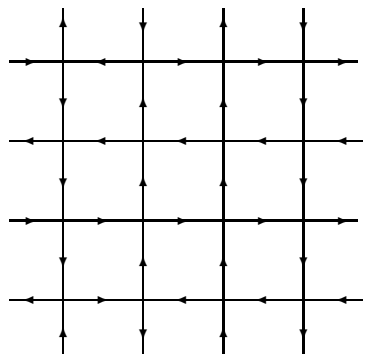
\begin{figure}[ht!]
\begin{picture}(440,140)
\put(150,0){%
\begin{picture}(200,200)
\multiput(30,0)(30,0){4}{\line(0,1){130}}
\multiput(10,20)(0,30){4}{\line(1,0){130}}
\put(30,0){\vector(0,1){10}}
\put(90,0){\vector(0,1){10}}
\put(60,15){\vector(0,-1){10}}
\put(120,15){\vector(0,-1){10}}
\put(90,27){\vector(0,1){10}}
\put(60,27){\vector(0,1){10}}
\put(30,42){\vector(0,-1){10}}
\put(120,42){\vector(0,-1){10}}
\put(90,57){\vector(0,1){10}}
\put(60,57){\vector(0,1){10}}
\put(30,72){\vector(0,-1){10}}
\put(120,72){\vector(0,-1){10}}
\put(30,102){\vector(0,-1){10}}
\put(60,87){\vector(0,1){10}}
\put(120,102){\vector(0,-1){10}}
\put(90,87){\vector(0,1){10}}
\put(30,117){\vector(0,1){10}}
\put(60,129){\vector(0,-1){8}}
\put(120,129){\vector(0,-1){8}}
\put(90,117){\vector(0,1){10}}
\put(10,110){\vector(1,0){10}}
\put(10,50){\vector(1,0){10}}
\put(25,80){\vector(-1,0){10}}
\put(25,20){\vector(-1,0){10}}
\put(37,20){\vector(1,0){10}}
\put(37,50){\vector(1,0){10}}
\put(52,80){\vector(-1,0){10}}
\put(52,110){\vector(-1,0){10}}
\put(82,20){\vector(-1,0){10}}
\put(67,50){\vector(1,0){10}}
\put(82,80){\vector(-1,0){10}}
\put(67,110){\vector(1,0){10}}
\put(112,20){\vector(-1,0){10}}
\put(97,50){\vector(1,0){10}}
\put(112,80){\vector(-1,0){10}}
\put(97,110){\vector(1,0){10}}
\put(127,110){\vector(1,0){10}}
\put(127,50){\vector(1,0){10}}
\put(142,80){\vector(-1,0){10}}
\put(142,20){\vector(-1,0){10}}
\end{picture}}
%
\end{picture}
\caption{\label{08-6V-6-vertex} $6$-vertex model}
\end{figure}

The general form of the lattice corresponding to the $6$-vertex model is shown in Fig.~\ref{08-6V-6-vertex}.


\section{Partition function of the $6$-vertex model \label{08-Raz-15-08-SS}}

In this section we consider the question of calculating the partition function of the $6$-vertex model. Our main task is to show,
that in this model, already familiar objects naturally arise: the $R$-matrix, the monodromy matrix, and the transfer matrix.

\subsection{General settings\label{08-Sec21}}

Traditional objects of statistical mechanics are systems consisting of a large number of particles. Let some system have a set of states $s$, so that each state corresponds to the energy $E(s)$. To find the thermodynamic properties of this system, it suffices to calculate its partition function, defined by the expression
\be{08-Stat-sum-gen}
Z=\sum_se^{\left(-E(s)/T\right)},
\ee
where $T$ is the temperature measured in units of energy, and the summation is taken over all possible states of the system.

According to the Gibbs principle, the probability to find the system in the state $s$ is
\be{08-prob}
p(s)=Z^{-1}e^{\left(-E(s)/T\right)}=Z^{-1}W(s),
\ee
where $W(s)=e^{\left(-E(s)/T\right)}$ is called a statistical weight of the state $s$.
Let $X$ be some measurable characteristic of the system (energy, magnetization, etc.), and let
it have the value $X(s)$ in the state $s$. Then the average thermodynamic value of $X$ is
\be{08-Srednee}
\langle X\rangle=Z^{-1}\sum_s X(s)e^{\left(-E(s)/T\right)}.
\ee

\subsection{Periodic boundary condition\label{08-Sec22}}

In order to formulate the problem of calculating the partition function of the $6$-vertex model, we should
assign to each vertex type some statistical weight. We make an additional assumption that the values $1$ and $2$ are equally probable.
Then the statistical weight of a vertex should not change with a simultaneous change in the direction of all the arrows. In this case, there are three
statistical weights, which we denote $a$, $b$, and $c$ (following  the tradition introduced by R.~Baxter).
The corresponding vertices will be called vertices of types $a$, $b$, or $c$ (see Fig.~\ref{08-6V-types-2}).
In the language of the arrows, vertices of the $c$-type are characterized by the fact that an arrow, passing through this vertex,
changes its direction. In vertices of the $a$-type and $b$-type, the arrows preserve their direction when passing through the vertex.
In the language of indices, vertices of the  $a$-type differ from all the others in that they have the same  indices on each edge.

To calculate the partition function, we first have to fix some configuration of the arrows on the entire lattice. The energy
of such a configuration is equal to the sum of the energies of all the vertices entering in this configuration. Hence, its statistical weight is equal to the product of statistical weights of all vertices. After this, we need to take the sum over all possible configurations
of arrows on the entire lattice. In this way, the partition function $Z_N$ has the form
\be{08-6V-StatSum-def}
Z_N=\sum_{\text{configurations}}\quad\prod_{\text{vertices}} W_{ij}.
\ee
Here $W_{ij}$ is the statistical weight of the vertex located at the intersection of the $i$th horizontal and $j$th vertical lines. Depending on the
vertex type, it can take three values: either $a$, or $b$, or $c$.

{\bf Remark.} Consider the $6$-vertex model in which all statistical weights are multiplied by a common constant $\gamma$.
It is clear that the partition function of this new model is related to the partition function of the initial model by a simple relation
\be{08-Nov-Star}
Z_N(\gamma a,\gamma b,\gamma c)=\gamma^{N^2} Z_N(a,b,c).
\ee
Therefore, two $6$-vertex models are essentially different if they have  different ratios,
for example, $a/c$ and $b/c$. Thus, essentially different models are
parameterized by two parameters, not three.

We can sum up over all possible configurations, including different orientations of the arrows on the boundary edges. In this case, we deal with free
boundary conditions. Another possibility is to fix the directions of the arrows on the boundary edges and to take the sum of all possible configurations
inside the lattice. We can also assume that the lattice is placed on the torus, and then in fact it has no boundary (periodic boundary conditions).
The latter option is considered most often, since it is in this case that translational invariance of
physical quantities is ensured from the very beginning.  In this section we consider the case of periodic boundary conditions.

To begin with, we slightly  change the definition of the
$6$-vertex model. Suppose that all $16$ types of vertices are allowed, but for those vertices,
in which the number of incoming arrows is not equal to the number of outgoing ones, the statistical weights are identically equal to
zero. Vanishing of the statistical weight means that the probability to find such a vertex is equal to  zero.
It is therefore clear that the new definition is equivalent to the old one. Now, however, we formally have
$16$ statistical weights (ten of them are identically zero) that we can write
in the form of a $4\times 4$ matrix.
\begin{figure}[ht!]
\begin{picture}(440,90)
\put(50,10){%
\begin{picture}(400,70)
\put(40,0){\line(0,1){50}}
\put(15,25){\line(1,0){50}}
\put(18,18){$\alpha$}
\put(58,28){$\beta$}
\put(43,5){$j$}
\put(33,40){$k$}
\put(240,0){\line(0,1){50}}
\put(280,0){\line(0,1){50}}
\put(215,25){\line(1,0){95}}
\put(218,18){$\alpha$}
\put(258,28){$\alpha'$}
\put(243,5){$j_2$}
\put(230,40){$k_2$}
\put(302,16){$\beta$}
\put(283,5){$j_1$}
\put(270,40){$k_1$}
\put(40,-20){(a)}
\put(260,-20){(b)}
\end{picture}}
\end{picture}
\caption{\label{08-6V-vertex1-2} One vertex and two neighbouring vertices.}
\end{figure}
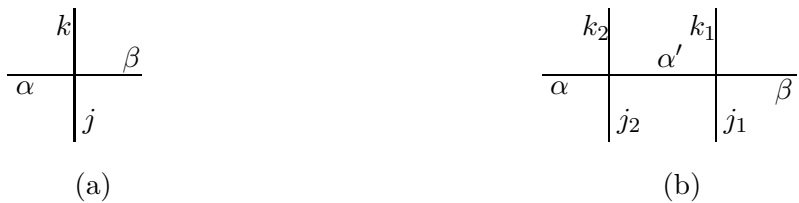
Consider a vertex on Fig.~\ref{08-6V-vertex1-2} (a). It has indices
$\alpha$ and $\beta$ on the horizontal line. They can take  values either $1$ or $2$. Similarly, the vertex has indices
$j$ and $k$ on the vertical line,  that also may take values either $1$ or $2$. Let us assign a statistical weight $R^{\alpha\beta,jk}$ to  this
vertex, where $R$ is a matrix. It should be understood as a $2\times2$
block-matrix, where each block in its turn is a $2\times2$ matrix. The first pair of indices gives the number of the block, the second pair
gives the number of the element in the block. Non-vanishing matrix elements are
\begin{equation}\label{08-6V-Mat-el}
\begin{aligned}
R^{11,11}&=R^{22,22}=a,\\
R^{11,22}&=R^{22,11}=b,\\
R^{12,21}&=R^{21,12}=c.
\end{aligned}
\end{equation}
All other entries are equal to zero. It is quite natural to call $R$ the matrix of statistical weights. It has the form
\begin{equation}\label{08-6V-Stat-ves-R}
R=\begin{pmatrix}
a&0&0&0\\
0&b&c&0\\
0&c&b&0\\
0&0&0&a
\end{pmatrix}.
\end{equation}
We see that, by construction, the matrix of statistical weights $R$ acts in the tensor product of two spaces
$\mathbb{C}^2\otimes \mathbb{C}^2$. The first space $\mathbb{C}^2$ is called
a horizontal space (since the indices $\alpha$ and $\beta$ are on the horizontal edges of the vertex),
while the second space $\mathbb{C}^2$ is a vertical space (since the indices $j$ and $k$ are on the vertical edges of the vertex).

One can see with the naked eye that the structure of the matrix of statistical weights coincides with the structure of the $R$-matrix. Indeed,
it is enough to make a change of variables
\be{08-reparametr}
a=\gamma f(u,\xi)=\gamma \frac{\sinh(u-\xi+\eta)}{\sinh(u-\xi)}, \qquad
b=\gamma, \qquad
c=\gamma g(u,\xi)=\frac{\gamma \sinh\eta}{\sinh(u-\xi)},
\ee
and the matrix of statistical weights takes a familiar form
\begin{equation}\label{08-6V-Stat-ves-Ruv}
R=R(u,\xi)=\gamma \begin{pmatrix}
f(u,\xi)&0&0&0\\
0&1&g(u,\xi)&0\\
0&g(u,\xi)&1&0\\
0&0&0&f(u,\xi)
\end{pmatrix}.
\end{equation}
Note that it is the trigonometric $R$-matrix that corresponds to the case when all three statistical weights $a$, $b$, and $c$ are
independent. The rational $R$-matrix describes the $6$-vertex model with the additional constraint $(c +1)/a = 1$.

Let us now do an exercise. Namely, we want to calculate the partition function on a lattice consisting of two vertices
(see Fig.~\ref{08-6V-vertex1-2} (b)). To calculate the partition function
we must compute the product of the statistical weights of both vertices for each configuration, and then sum up the result
over all possible configurations. We assume that the indices on the external edges are fixed. Then there are only two configurations: in the first the intermediate horizontal edge has the index $\alpha'=1$, in the second configuration it has the index $\alpha'=2$.
In the first case, the product of the statistical weights is equal to $R^{\alpha1,j_2k_2}R^{1\beta,j_1k_1}$, in the second case we obtain
$R^{\alpha2,j_2k_2}R^{2\beta,j_1k_1}$. Thus, the partition function $Z$ is equal to
\begin{equation}\label{08-6V-Stat-exer}
Z=R^{\alpha\alpha',j_2k_2}R^{\alpha'\beta,j_1k_1},
\end{equation}
where, as usual, the summation  over the repeated index $\alpha'$ is assumed. Thus, the procedure of calculating the partition function
reduces to the multiplying the matrices acting in the tensor product of spaces. Let the horizontal line in Fig.~\ref{08-6V-vertex1-2} (b)
correspond to a space $V_0$, and two vertical lines correspond to spaces $V_2$ and $V_1$. Then
\begin{equation}\label{08-6V-Stat-exer1}
Z=\bigl(R_{02}R_{01}\bigr)^{\alpha\beta,j_2k_2,j_1k_1}.
\end{equation}
Thus, the partition function of two vertices is equal to some matrix element of the product $R_{02}R_{01}$.

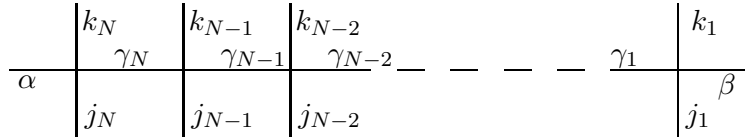
\begin{figure}[h!]
\begin{picture}(440,70)
\put(50,0){%
\begin{picture}(400,70)
\put(40,0){\line(0,1){50}}
\put(80,0){\line(0,1){50}}
\put(120,0){\line(0,1){50}}
\put(265,0){\line(0,1){50}}
\put(15,25){\line(1,0){135}}
\multiput(160,25)(20,0){4}{\line(1,0){10}}
\put(240,25){\line(1,0){50}}
\put(18,18){$\alpha$}
\put(54,28){$\gamma_N$}
\put(94,28){$\gamma_{N-1}$}
\put(134,28){$\gamma_{N-2}$}
\put(43,5){$j_N$}
\put(42,40){$k_N$}
\put(280,16){$\beta$}
\put(83,5){$j_{N-1}$}
\put(82,40){$k_{N-1}$}
\put(123,5){$j_{N-2}$}
\put(122,40){$k_{N-2}$}
\put(268,5){$j_1$}
\put(270,40){$k_1$}
\put(240,28){$\gamma_{1}$}
\end{picture}}
\end{picture}
\caption{\label{08-6V-TM}Monodromy matrix and transfer matrix}
\end{figure}

The result obtained is obviously generalized to the calculation of the partition function of the horizontal line consisting
of $N$ vertices (see Fig.~\ref{08-6V-TM}).
We introduce a space $V_0\otimes V_1\otimes \cdots \otimes V_N$, where $V_0$ is associated with the horizontal
line, and the remaining $V_k$ correspond to the vertical lines. After this it suffices to take the tensor product of the $R$-matrices over all
vertices. The matrix element $\bigl(R_{0N}\dots R_{01})^{\alpha\beta,j_Nk_N,\dots ,j_1k_1}$  is the required partition function.

We see that for fixed external indices $\alpha$ and $\beta$, the partition function is equal to the matrix element $T^{\alpha\beta}$ of the monodromy matrix
of the $XXZ$ chain. In the case of periodic boundary conditions, we must put $\alpha = \beta$ and sum up the result over the values $\alpha = 1,2$.
In other words, we take the trace over the space $V_0$, that is, we arrive at the transfer matrix
\begin{equation}\label{08-6V-TM-def}
\mathcal{T}={\tr}_0R_{0N}\dots R_{01}.
\end{equation}
Obviously, $\mathcal{T}$ is a $2^N\times 2^N$ matrix acting in the space $V_1\otimes \cdots \otimes V_N$.
From the point of view of the $6$-vertex model, it describes the transition from vertical edges adjacent to the bottom of one horizontal line to the edges adjacent to the same horizontal line from above. Therefore, it is natural to call $\mathcal{T}$ the transfer matrix of the horizontal line.

\begin{figure}[ht!]
\begin{picture}(440,100)
\put(70,0){%
\begin{picture}(400,100)
\put(40,0){\line(0,1){80}}
\put(80,0){\line(0,1){80}}
\put(120,0){\line(0,1){80}}
\put(265,0){\line(0,1){80}}
\put(15,25){\line(1,0){135}}
\multiput(160,25)(20,0){4}{\line(1,0){10}}
\put(240,25){\line(1,0){50}}
\put(15,60){\line(1,0){135}}
\multiput(160,60)(20,0){4}{\line(1,0){10}}
\put(240,60){\line(1,0){50}}
\put(18,18){$\alpha_1$}
\put(18,63){$\alpha_2$}
\put(43,5){$j_N$}
\put(43,40){$\ell_N$}
\put(43,70){$k_N$}
\put(285,16){$\alpha_1$}
\put(285,63){$\alpha_2$}
\put(83,5){$j_{N-1}$}
\put(83,40){$\ell_{N-1}$}
\put(83,70){$k_{N-1}$}
\put(123,5){$j_{N-2}$}
\put(123,40){$\ell_{N-2}$}
\put(123,70){$k_{N-2}$}
\put(268,5){$j_1$}
\put(268,40){$\ell_1$}
\put(268,70){$k_1$}
\end{picture}}
\end{picture}
\caption{\label{08-6V-TM-2}Two transfer matrices}
\end{figure}
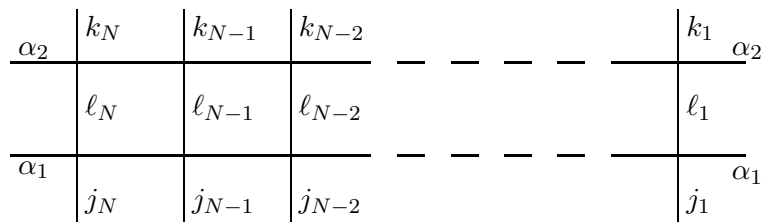

In order to calculate the partition function of two adjacent horizontal lines, it is
enough to multiply the corresponding matrices $\mathcal{T}$ (Fig.~\ref{08-6V-TM-2}).
Indeed, in this case different configurations correspond to different values of the variables $\ell_k$ on
the intermediate vertical edges, that is,
\begin{equation}\label{08-6V-ZT2}
Z=\mathcal{T}^{j_N\ell_N,\dots,j_1\ell_1}\mathcal{T}^{\ell_Nk_N,\dots,\ell_1,k_1}=
\bigl(\mathcal{T}\mathcal{T}\bigr)^{j_Nk_N,\dots,j_1k_1}.
\end{equation}
Continuing this process and taking into account periodic boundary conditions  we obtain
\begin{equation}\label{08-6V-ZTN}
Z_N=\tr\bigl(\mathcal{T}^N\bigr),
\end{equation}
where the trace is taken with respect to the space $V_1\otimes \cdots \otimes V_N$.

If we diagonalize the transfer matrix $\mathcal{T}$, then
\be{08-trT}
Z_N=\tr(\mathcal{T})^N=\sum_{k=1}^{2^N}\Lambda^N_k,
\ee
where $\Lambda_k$ are the  transfer matrix eigenvalues.
Thus, the problem of calculating the partition function reduces to the finding the eigenvalues of
the $XXZ$ chain   transfer matrix. In its turn, the latter problem can be solved by the algebraic Bethe ansatz.

Thus, we showed that the $6$-vertex model is very closely related to the $XXZ$ spin chain. In both models
the same objects appear: the $R$-matrix, the monodromy matrix, and the transfer matrix. Moreover, the problem of calculating the partition function
of the $6$-vertex model with periodic boundary conditions coincides  with the problem of studying the spectrum of the $XXZ$ Hamiltonian.
We refer the interested reader to the book \cite{Bax82}, where these questions are considered in more detail.
There one can also learn about how the $6$-vertex model was studied  before the QISM was created.

\subsection{Inhomogeneous model\label{08-Sec23}}

Let us assume that some field acts in parallel to the plane of the lattice. Then the energy of each vertex (and, accordingly,
its statistical weight) can depend not only on the type of the vertex, but also on its position on the lattice, that is, $W =W_ {ij}$,
where $i$ and $j$ are the indices of the horizontal and vertical lines. Such a model is called inhomogeneous. Generally
the partition function of the inhomogeneous model cannot be  calculated exactly. However, this can be done, if  statistical weights
depend on the position of the vertex in a special way.

Let  a parameter $u_i$  match to the $i$th horizontal line  and a parameter $\xi_j$ match to the $j$th vertical line, as shown on
Fig.~\ref{08-6V-6-vertex-inh}. Let the matrix of statistical weights at each vertex have the form
\begin{equation}\label{08-6V-Stat-ves-Ruvj}
R(u_i,\xi_j)=\gamma \begin{pmatrix}
f(u_i,\xi_j)&0&0&0\\
0&1&g(u_i,\xi_j)&0\\
0&g(u_i,\xi_j)&1&0\\
0&0&0&f(u_i,\xi_j)
\end{pmatrix}.
\end{equation}
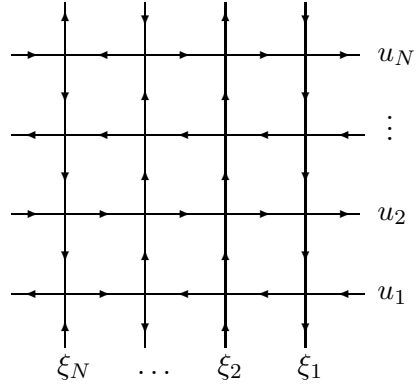
\begin{figure}[h!]
\begin{picture}(440,160)
\put(0,10){%
\begin{picture}(420,150)
\put(150,0){%
\begin{picture}(200,200)
\multiput(30,0)(30,0){4}{\line(0,1){130}}
\multiput(10,20)(0,30){4}{\line(1,0){130}}
\put(30,0){\vector(0,1){10}}
\put(90,0){\vector(0,1){10}}
\put(60,15){\vector(0,-1){10}}
\put(120,15){\vector(0,-1){10}}
\put(90,27){\vector(0,1){10}}
\put(60,27){\vector(0,1){10}}
\put(30,42){\vector(0,-1){10}}
\put(120,42){\vector(0,-1){10}}
\put(90,57){\vector(0,1){10}}
\put(60,57){\vector(0,1){10}}
\put(30,72){\vector(0,-1){10}}
\put(120,72){\vector(0,-1){10}}
\put(30,102){\vector(0,-1){10}}
\put(60,87){\vector(0,1){10}}
\put(120,102){\vector(0,-1){10}}
\put(90,87){\vector(0,1){10}}
\put(30,117){\vector(0,1){10}}
\put(60,129){\vector(0,-1){8}}
\put(120,129){\vector(0,-1){8}}
\put(90,117){\vector(0,1){10}}
\put(10,110){\vector(1,0){10}}
\put(10,50){\vector(1,0){10}}
\put(25,80){\vector(-1,0){10}}
\put(25,20){\vector(-1,0){10}}
\put(37,20){\vector(1,0){10}}
\put(37,50){\vector(1,0){10}}
\put(52,80){\vector(-1,0){10}}
\put(52,110){\vector(-1,0){10}}
\put(82,20){\vector(-1,0){10}}
\put(67,50){\vector(1,0){10}}
\put(82,80){\vector(-1,0){10}}
\put(67,110){\vector(1,0){10}}
\put(112,20){\vector(-1,0){10}}
\put(97,50){\vector(1,0){10}}
\put(112,80){\vector(-1,0){10}}
\put(97,110){\vector(1,0){10}}
\put(127,110){\vector(1,0){10}}
\put(127,50){\vector(1,0){10}}
\put(142,80){\vector(-1,0){10}}
\put(142,20){\vector(-1,0){10}}
\end{picture}}
\put(297,108){$u_N$}
\put(300,78){$\vdots$}
\put(297,48){$u_2$}
\put(297,18){$u_1$}
\put(177,-10){$\xi_N$}
\put(207,-10){$\dots$}
\put(237,-10){$\xi_2$}
\put(267,-10){$\xi_1$}
\end{picture}}
\end{picture}
\caption{\label{08-6V-6-vertex-inh} Inhomogeneous $6$-vertex model}
\end{figure}
that is,
\be{08-reparametrij}
a_{ij}=\gamma f(u_i,\xi_j), \qquad
b_{ij}=\gamma,\qquad
c_{ij}=\gamma g(u_i,\xi_j).
\ee
Then the partition function of each horizontal line is the transfer matrix of the inhomogeneous $XXZ$ chain
\be{08-TM-inh}
\mathcal{T}(u_i)=\mathcal{T}(u_i|\bar\xi)={\tr}_0\bigl(R_{0N}(u_i,\xi_N)\dots R_{02}(u_i,\xi_2) R_{01}(u_i,\xi_1)\bigr),
\ee
and whole the partition function is given by
\begin{equation}\label{08-6V-ZTN-inh}
Z_N(\bu|\bxi)=\tr\bigl(\mathcal{T}(u_N)\dots \mathcal{T}(u_2)\mathcal{T}(u_1)\bigr),
\end{equation}
where the trace is taken over the space $V_1\otimes \cdots \otimes V_N$. We know that the transfer matrices of the inhomogeneous $XXZ$ chain
commute with each other for arbitrary values of the arguments $u_k$. Therefore, they are all simultaneously diagonalizable.
Thus, the problem is again reduced to the finding of the transfer matrix eigenvalues.

\section{Domain wall boundary conditions\label{08-Sec3}}

In this section, we consider one more special case of the $6$-vertex model introduced in \cite{Kor82} and called
domain wall boundary conditions.
The meaning of these boundary conditions is very simple: on the left and right
faces all arrows enter the lattice, while on the upper and lower faces arrows exit the lattice (see Fig.~\ref{08-6V-6-vertex-dw} (a)).
The terminology `domain wall' has appeared due to one more possible graphic presentation
of a variable located on the edges of the lattice. We can match the two values to arrows, orthogonal
to the plane of the lattice. For example, we assign the value $1$ to the up arrow (black circle in Fig.~\ref{08-6V-vertexP}),
and the value $2$ is the down arrow (white circle in Fig.~\ref{08-6V-vertexP}). If we now look at the lattice under some
angle, then on the bottom and right faces all the arrows will be directed downwards, and on the left and the upper faces all
the arrows will point up (see Fig.~\ref{08-6V-Dom-wall}).


\begin{figure}[ht!]
\begin{picture}(450,140)
\put(70,-10){(a)}
\put(320,-10){(b)}
\put(0,0){%
\begin{picture}(200,200)
\multiput(30,0)(30,0){4}{\line(0,1){130}}
\multiput(10,20)(0,30){4}{\line(1,0){130}}
\put(30,15){\vector(0,-1){10}}
\put(90,15){\vector(0,-1){10}}
\put(60,15){\vector(0,-1){10}}
\put(120,15){\vector(0,-1){10}}
\put(30,42){\vector(0,-1){10}}
\put(60,42){\vector(0,-1){10}}
\put(90,27){\vector(0,1){10}}
\put(120,42){\vector(0,-1){10}}
\put(30,72){\vector(0,-1){10}}
\put(60,72){\vector(0,-1){10}}
\put(90,57){\vector(0,1){10}}
\put(120,57){\vector(0,1){10}}
\put(30,102){\vector(0,-1){10}}
\put(60,87){\vector(0,1){10}}
\put(90,87){\vector(0,1){10}}
\put(120,87){\vector(0,1){10}}
\put(30,117){\vector(0,1){10}}
\put(60,117){\vector(0,1){10}}
\put(120,117){\vector(0,1){10}}
\put(90,117){\vector(0,1){10}}
\put(10,110){\vector(1,0){10}}
\put(10,50){\vector(1,0){10}}
\put(10,80){\vector(1,0){10}}
\put(10,20){\vector(1,0){10}}
\put(37,20){\vector(1,0){10}}
\put(37,50){\vector(1,0){10}}
\put(37,80){\vector(1,0){10}}
\put(52,110){\vector(-1,0){10}}
\put(67,20){\vector(1,0){10}}
\put(67,50){\vector(1,0){10}}
\put(82,80){\vector(-1,0){10}}
\put(82,110){\vector(-1,0){10}}
\put(112,20){\vector(-1,0){10}}
\put(97,50){\vector(1,0){10}}
\put(112,80){\vector(-1,0){10}}
\put(112,110){\vector(-1,0){10}}
\put(142,110){\vector(-1,0){10}}
\put(142,50){\vector(-1,0){10}}
\put(142,80){\vector(-1,0){10}}
\put(142,20){\vector(-1,0){10}}
\end{picture}}
\put(250,0){%
\begin{picture}(200,200)
\multiput(30,0)(30,0){4}{\line(0,1){130}}
\multiput(10,20)(0,30){4}{\line(1,0){130}}
\put(27.5,6){$\circ$}
\put(87.5,6){$\circ$}
\put(57.5,6){$\circ$}
\put(117.5,6){$\circ$}
\put(27.5,32){$\circ$}
\put(57.5,32){$\circ$}
\put(87.5,32){$\bullet$}
\put(117.5,32){$\circ$}
\put(27.5,62){$\circ$}
\put(57.5,62){$\circ$}
\put(87.5,62){$\bullet$}
\put(117.5,62){$\bullet$}
\put(27.5,92){$\circ$}
\put(57.5,92){$\bullet$}
\put(87.5,92){$\bullet$}
\put(117.5,92){$\bullet$}
\put(27.5,120){$\bullet$}
\put(57.5,120){$\bullet$}
\put(117.5,120){$\bullet$}
\put(87.5,120){$\bullet$}
\put(12,107){$\bullet$}
\put(12,47){$\bullet$}
\put(12,77){$\bullet$}
\put(15,17){$\bullet$}
\put(41,17){$\bullet$}
\put(41,47){$\bullet$}
\put(41,77){$\bullet$}
\put(41,107){$\circ$}
\put(71,17){$\bullet$}
\put(71,47){$\bullet$}
\put(71,77){$\circ$}
\put(71,107){$\circ$}
\put(101,17){$\circ$}
\put(101,47){$\bullet$}
\put(101,77){$\circ$}
\put(101,107){$\circ$}
\put(131,107){$\circ$}
\put(131,47){$\circ$}
\put(131,77){$\circ$}
\put(131,17){$\circ$}
\end{picture}}
\end{picture}
\caption{\label{08-6V-6-vertex-dw} $6$-vertex model with domain wall boundary condition. On Fig. (a)
the edge variable  is shown by the standard arrows. On Fig. (b) this variable
is shown by arrows orthogonal to the plane of the lattice: $\bullet$ is up arrow (that is, the variable has the
value $1$); $\circ$ is down arrow (that is, the variable has the value $2$). }
\end{figure}
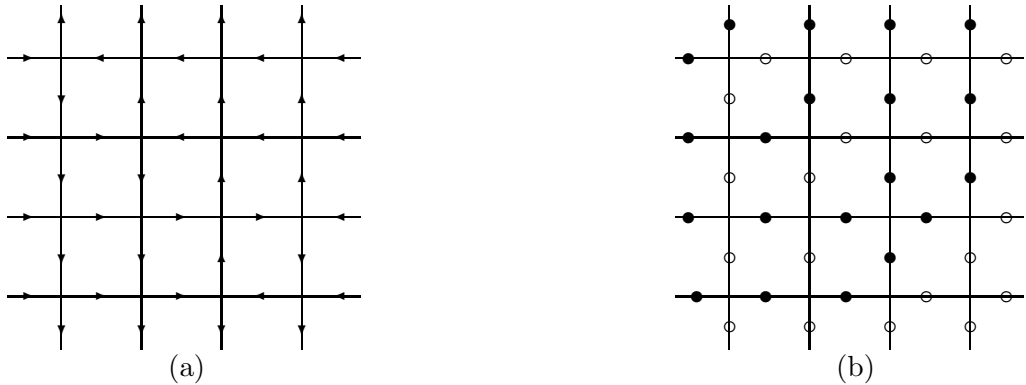
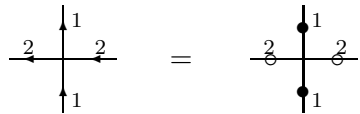
\begin{figure}[ht!]
\begin{picture}(140,50)
\put(150,0){%
\begin{picture}(50,50)
\put(25,20){\vector(-1,0){10}}
\put(50,20){\vector(-1,0){10}}
\put(30,0){\vector(0,1){10}}
\put(30,25){\vector(0,1){10}}
%
\put(30,0){\line(0,1){40}}
\put(10,20){\line(1,0){40}}
\put(15,22){$\scriptstyle 2$}
\put(42,22){$\scriptstyle 2$}
\put(33,32){$\scriptstyle 1$}
\put(33,2){$\scriptstyle 1$}
\end{picture}}
\put(220,17){$=$}
\put(240,0){%
\begin{picture}(50,50)
\put(27,5){$\bullet$}
\put(27,29){$\bullet$}
\put(15,17){$\circ$}
\put(40,17){$\circ$}
%
\put(30,0){\line(0,1){40}}
\put(10,20){\line(1,0){40}}
\put(15,22){$\scriptstyle 2$}
\put(42,22){$\scriptstyle 2$}
\put(33,32){$\scriptstyle 1$}
\put(33,2){$\scriptstyle 1$}
\end{picture}}
\end{picture}
\caption{\label{08-6V-vertexP} Identification of two vertices in the standard and new parameterizations. }
\end{figure}
Thus, the lattice separates two domains:
one of them consists of all `spins' down, another consists of all `spins' up.
\begin{figure}[h!]
\begin{picture}(300,100)
\put(100,20){%
\begin{picture}(200,150)
\multiput(18,0)(5,0){24}{\line(3,1){130}}
\multiput(15,2)(5,1.7){24}{\line(1,0){130}}
\multiput(15,2)(5,1.7){24}{\vector(0,1){20}}
\multiput(144,1)(5,1.7){24}{\vector(0,-1){20}}
\multiput(18,-1)(5,0){24}{\vector(0,-1){20}}
\multiput(147,43)(5,0){24}{\vector(0,1){20}}
\end{picture}}
\end{picture}
\caption{\label{08-6V-Dom-wall} Domain wall. }
\end{figure}
The lattice itself plays the role of the domain wall.

In what follows, in order to avoid too long terminology, we will use the abbreviation of the DWPF (the domain wall partition function). This partition function plays extremely important role in the calculation of the scalar products of Bethe vectors. As we shall see below, the commutation relations
of the monodromy matrix entries can be also written with the help of the DWPF.

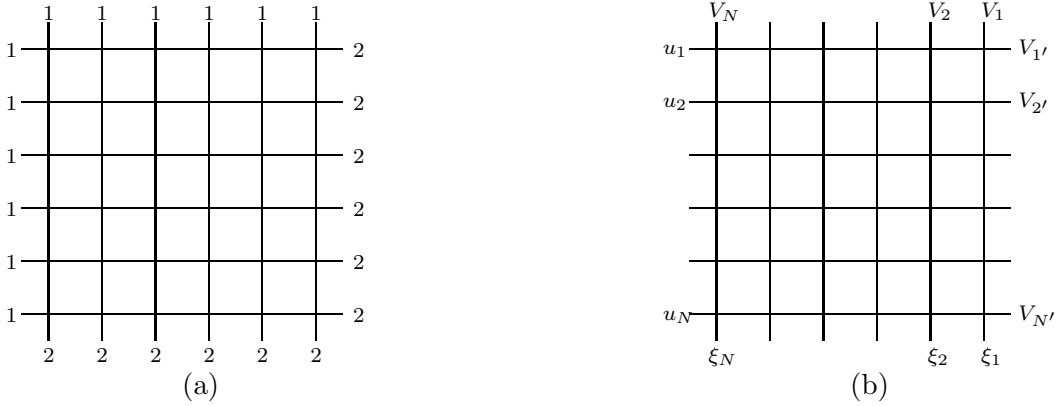
\begin{figure}[h!]
\begin{picture}(450,140)
\put(70,-10){(a)}
\put(320,-10){(b)}
\put(10,10){%
\begin{picture}(200,200)
\multiput(10,0)(20,0){6}{\line(0,1){120}}
\multiput(0,10)(0,20){6}{\line(1,0){120}}
\multiput(8,-8)(20,0){6}{$\scriptstyle 2$}
\multiput(8,121)(20,0){6}{$\scriptstyle 1$}
\multiput(-6,7)(0,20){6}{$\scriptstyle 1$}
\multiput(124,7)(0,20){6}{$\scriptstyle 2$}

\end{picture}}
\put(260,10){%
\begin{picture}(200,200)
\multiput(10,0)(20,0){6}{\line(0,1){120}}
\multiput(0,10)(0,20){6}{\line(1,0){120}}
\put(7,122){$\scriptstyle V_{N}$}
\put(109,122){$\scriptstyle V_{1}$}
\put(89,122){$\scriptstyle V_{2}$}
\put(7,-8){$\scriptstyle \xi_{N}$}
\put(109,-8){$\scriptstyle \xi_{1}$}
\put(89,-8){$\scriptstyle \xi_{2}$}
\put(-10,8){$\scriptstyle u_{N}$}
\put(-10,88){$\scriptstyle u_{2}$}
\put(-10,108){$\scriptstyle u_{1}$}
\put(123,8){$\scriptstyle V_{N'}$}
\put(123,88){$\scriptstyle V_{2'}$}
\put(123,108){$\scriptstyle V_{1'}$}
\end{picture}}
\end{picture}
\caption{\label{08-6V-6-vertex-DW} $6$-vertex model: (a) domain wall boundary condition; (b) inhomogeneities and the spaces,
in which the $R$-matrices act. }
\end{figure}

Consider the inhomogeneous model (see Fig.~\ref{08-6V-6-vertex-DW} (b)). We assume that the matrix of the statistical weights of the vertex located at the intersection $i$th and $j$th lines, is given by the \eqref{08-6V-Stat-ves-Ruvj} and acts in the tensor product of the spaces $V_{i'}\otimes V_j$
(see Fig.~\ref{08-6V-6-vertex-DW} (b)). We set $\gamma = 1$ in the formula \eqref{08-6V-Stat-ves-Ruvj} for simplicity (that is, the weight $b = 1$). The partition function of such a model is denoted by the symbol $K_N(\bu|\bxi)$.
%
%

As we have seen in the previous section, the product of the  $R$-matrices along a horizontal line
$i$ gives the monodromy matrix of the inhomogeneous  $XXZ$ chain:
\begin{equation}\label{08-6V-Mon-mat-inh}
T(u_i|\bxi)=R_{i'N}(u_i,\xi_N)\cdots R_{i'2}(u_i,\xi_2)R_{i'1}(u_i,\xi_1).
\end{equation}
In the case of periodic boundary conditions, we had to take the trace of the monodromy matrix, that is,
$A(u_i)+D(u_i) $. Now, due to the domain wall boundary condition (see Fig.~\ref{08-6V-6-vertex-DW}(a)), we should
take the matrix element $T^{12}(u_i)$, that is, $B(u_i)$. To calculate the partition function
it is necessary to multiply the contributions from each horizontal line, that is, to take the product
$B(u_N)\cdots B(u_1)$. Due to the boundary conditions on the upper and lower faces of the lattice,
the DWPF is equal to the following matrix element of the resulting product
\be{08-6V-ParFun}
K_N(\bu|\bxi)=\bigl(B(u_N)\cdots B(u_1)\bigr)^{21,\dots,21}.
\ee
It follows from the representation \eqref{08-6V-ParFun} that DWPF $K_N$
is a symmetric function over $\bu$, because  $[B(u_j),B(u_k)]=0$.
One can also think that $K_N$ is  a symmetric function of the parameters $\bxi$ as well. This is really so,
and this is easy to verify. For this, we compute $K_N$ by taking the product of the $R$-matrices in a different order.
The Yang--Baxter equation guarantees that the result does not change.

Define a `vertical' monodromy matrix as a product of the $R$-matrices along a vertical line
\begin{equation}\label{08-6V-Ver-Mon-mat}
\tilde T(\xi_j|\bu)=R_{N'j}(u_{N'},\xi_j)\cdots R_{2'j}(u_{2'},\xi_j)R_{1'j}(u_{1'},\xi_j)=
\begin{pmatrix} \tilde A(\xi_j)&\tilde B(\xi_j)\\ \tilde C(\xi_j)&\tilde D(\xi_j)
\end{pmatrix}.
\end{equation}
Due to the boundary conditions on the upper and lower faces, the contribution from one vertical line to the partition function
is equal to $\tilde T^{21}(\xi_j)=\tilde C(\xi_j)$. Therefore, in complete analogy to the formula \eqref{08-6V-ParFun}, we obtain
another presentation for the DWPF
\be{08-6V-ParFun2}
K_N(\bu|\bxi)=\bigl(\tilde C(\xi_N)\cdots \tilde C(\xi_1)\bigr)^{12,\dots,12}.
\ee
We suggest the reader to check that the vertical monodromy matrix satisfies the $RTT$-relation\footnote{In proving this statement,
the key fact is  that  the vertical monodromy matrix is the product of the $R$-matrices that satisfy the Yang--Baxter equation.}
\be{08-6V-RTT-TTR}
R_{12}(\xi_1,\xi_2)\tilde T_{1}(\xi_1)\tilde T_{2}(\xi_2) =\tilde T_{2}(\xi_2)\tilde T_{1}(\xi_1)R_{12}(\xi_1,\xi_2).
\ee
We know already that equation \eqref{08-6V-RTT-TTR} implies $[\tilde C(\xi_1),\tilde C(\xi_2)]=0$. Hence, the partition function $K_N$ is symmetric over
the parameters $\bxi$.

\section{Properties of the partition function\label{08-S-6V-Prop-DW}}

In this section, we prove several properties of DWPF, that allow one
to compute $K_N$ recursively.

\subsection{Decreasing property\label{08-Sec41}}

Consider a reparameterization $q=e^\eta$, $x_j=e^{2u_j}$, and $y_j=e^{2\xi_j}$, $j=1,\dots,N$. In order to avoid new notation, we agree,
that the notation $f(x_j,y_k)$ and $g(x_j,y_k)$ means that one should substitute $u_j$, $\xi_k$, and $\eta$ expressed in terms of $x_j$, $y_k$, and $q$,
into the functions $f(u_j,\xi_k)$ and $g(u_j,\xi_k)$. Then with the new variables we have
\be{08-6V-fg-new}
\begin{aligned}
f(x_j,y_k)&=\frac{x_jq-y_kq^{-1}}{x_j-y_k},\\
g(x_j,y_k)&=\frac{(q-q^{-1})\sqrt{x_jy_k}}{x_j-y_k}.
\end{aligned}
\ee
Respectively, the partition function becomes a function of variables $\{x\}$ and $\{y\}$: $K_N=K_N(\bar x|\bar y)$.

\begin{prop}\label{08-6V-Prop1}
DWPF can be presented in the form
\be{08-6V-K-tK}
K_N(\bar x|\bar y)=\prod_{j=1}^N\sqrt{x_jy_j}\cdot\widetilde{K}_N(\bar x|\bar y),
\ee
where $\widetilde{K}_N$ is a rational function of variables $\{x\}$ and $\{y\}$. The function $\widetilde{K}_N$ goes to zero if one of its arguments goes to infinity at other arguments fixed.
\end{prop}

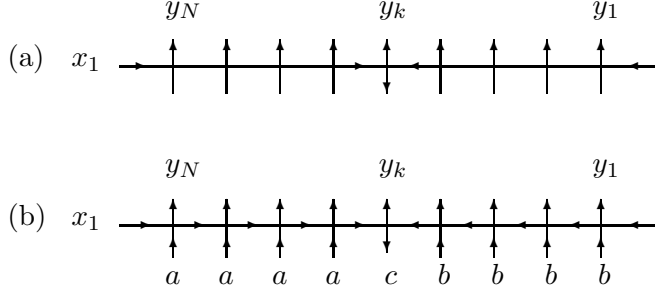
\begin{figure}[ht!]
\begin{picture}(450,120)
\put(0,60){%
\begin{picture}(450,50)
\put(100,-10){%
\begin{picture}(200,40)
\put(20,30){\line(1,0){200}}
\multiput(40,20)(20,0){9}{\vector(0,1){20}}
\put(20,30){\vector(1,0){10}}
\put(220,30){\vector(-1,0){10}}
\put(102,30){\vector(1,0){10}}
\put(138,30){\vector(-1,0){10}}
\put(120,30){\vector(0,-1){10}}
\put(-22,30){(a)}
\put(2,30){$x_1$}
\put(37,50){$y_N$}
\put(197,50){$y_{1}$}
\put(117,50){$y_{k}$}
\end{picture}}
\end{picture}}
\put(0,0){%
\begin{picture}(450,50)
\put(100,-10){%
\begin{picture}(200,40)
\put(20,30){\line(1,0){200}}
\multiput(40,20)(20,0){9}{\vector(0,1){20}}
\multiput(40,18)(20,0){4}{\vector(0,1){8}}
\multiput(140,18)(20,0){4}{\vector(0,1){8}}
\multiput(22,30)(20,0){4}{\vector(1,0){10}}
\put(220,30){\vector(-1,0){10}}
\put(102,30){\vector(1,0){10}}
\multiput(138,30)(20,0){4}{\vector(-1,0){10}}
\put(120,30){\vector(0,-1){10}}
\put(-22,30){(b)}
\put(2,30){$x_1$}
\put(37,50){$y_N$}
\put(197,50){$y_{1}$}
\put(117,50){$y_{k}$}
\multiput(37,8)(20,0){4}{$a$}
\multiput(139,8)(20,0){4}{$b$}
\put(119,8){$c$}
\end{picture}}
\end{picture}}
\end{picture}
\caption{Upper face of the lattice. If we  specify a vertex where the arrow changes direction (Fig. (a)),
then all other vertices are fixed uniquely (Fig. (b))).\label{08-6V-Upper-line}}
\end{figure}

{\sl Proof.} Consider $K_N(\bar x|\bar y)$ as a function of $x_1$.  The parameter $x_1$ enters only the statistical weights of the vertices
in the upper horizontal line (see Fig.~\ref{08-6V-Upper-line}).
Due to the boundary conditions at the right and left ends, the arrow, passing through the vertices,
must  change the direction at least once. It is easy to see that such a change in direction should
happen exactly once, because the boundary condition on the upper face forbids vertices with two incoming arrows on both vertical
lines. The change in the direction of the arrow is possible only at the $c$-type vertex. Its statistical weight
is equal to $g(x_1,y_k)$. We come to
the conclusion that there is exactly one vertex of the $c$-type on the upper line, and all the others are either vertices of the $a$-type or
the $b$-type. It is easy to see that all the vertices to the left of the $c$-type vertex  are of $a$-type, and all vertices to the right are of $b$-type.
Thus, with a fixed number $k$ of the $c$-vertex  the statistical weight of the upper horizontal
line has the form
\be{08-6V-1line}
Z_1=\prod_{\ell=1}^{k-1} f(x_1,y_\ell)\cdot\frac{(q-q^{-1})\sqrt{x_1y_k}}{x_1-y_k},
\qquad k=1,\dots,N.
\ee
Hence, the partition function can be presented as follows:
\be{08-6V-1line1}
K_N=\sqrt{x_1}\sum_{k=1}^N\frac{(q-q^{-1})}{x_1-y_k}\prod_{\ell=1}^{k-1} f(x_1,y_\ell)\cdot W_k,
\ee
where $W_k$  does not depend on $x_1$. We see that $K_N$ as a function of $x_1$, is presented as a product of
$\sqrt{x_1}$ and some rational function of $x_1$. The latter goes to zero at $x_1\to\infty$. Due to the symmetry
of $K_N$ over $\bar x$, the same representation exists for any other $x_j$. This implies
\be{08-6V-1line2}
K_N=\prod_{j=1}^N\sqrt{x_j}\cdot \widetilde{\widetilde{K}}_N,
\ee
where $\widetilde{\widetilde{K}}_N$ is a rational function of any $x_j$ that vanishes at $x_j\to\infty$,
if all other variables are fixed.

The dependence of the partition function of the parameters  $\bar y$ can be considered in the similar way, and we arrive at the proof of the statement.

\subsection{Poles of the partition function\label{08-Sec42}}

Thus, we have extracted explicitly the dependence $K_N(\bar x|\bar y)$ of the square roots $\sqrt{x_j}$ and $\sqrt{y_j}$. The remaining part
is a rational function of all variables. Obviously, it has poles at the points
$x_j=y_k$, $j,k=1,\dots,N$. Let us study the residues in these poles. Due to the symmetry it is enough to consider the residue at
$x_N=y_N$.

\begin{prop}\label{08-6V-Prop2}
At $x_N\to y_N$ the singular part of $K_N$ reduces to  $K_{N-1}$
\be{08-6V-K-Kn-1}
K_N(\bar x|\bar y)\Bigr|_{x_N\to y_N}= g(x_N,y_N) f(\bar x_N,x_N) f(y_N,\bar y_N)
K_{N-1}(\bar x_N|\bar y_N)+\text{\rm Reg},
\ee
where $\text{\rm Reg}$ means regular terms. Recall also that
$\bar x_N=\bar x\setminus x_N$ and $\bar y_N=\bar y\setminus y_N$.
\end{prop}

{\sl Proof.}

The pole at $x_N \to y_N$ occurs if and only if the extreme south-western vertex
is of $c$-type. Then it corresponds to the weight $g(x_N,y_N)$. But as soon as the type of the extreme south-western vertex is fixed,
immediately, the types of vertices on the southern and western faces are uniquely determined. Indeed, the neighboring vertex on the right
%
%
\begin{figure}[h!]
\begin{picture}(450,140)
\put(160,10){%
\begin{picture}(200,200)
\multiput(10,0)(20,0){6}{\line(0,1){120}}
\multiput(0,10)(0,20){6}{\line(1,0){120}}
\multiput(10,10)(20,0){6}{\vector(0,-1){10}}
\multiput(10,111)(20,0){6}{\vector(0,1){10}}
\multiput(-2,10)(0,20){6}{\vector(1,0){6}}
\multiput(121,10)(0,20){6}{\vector(-1,0){6}}
\multiput(24,10)(20,0){5}{\vector(-1,0){6}}
\multiput(30,21)(20,0){5}{\vector(0,-1){6}}
\multiput(10,15)(0,20){5}{\vector(0,1){6}}
\multiput(15,30)(0,20){5}{\vector(1,0){6}}
\put(7,-8){$\scriptstyle y_{N}$}
\put(27,-8){$\scriptstyle y_{N-1}$}
\put(107,-8){$\scriptstyle y_{1}$}
\put(87,-8){$\scriptstyle y_{2}$}
\put(57,-8){$\cdots$}
\put(-16,4){$\scriptstyle x_{N}$}
\put(-16,24){$\scriptstyle x_{N-1}$}
\put(-16,104){$\scriptstyle x_{1}$}
\put(-16,84){$\scriptstyle x_{2}$}
\put(-12,44){$\cdot$}
\put(-12,54){$\cdot$}
\put(-12,64){$\cdot$}
\end{picture}}
\end{picture}
\caption{\label{08-6V-6-vertex-DW-red} Reduction $K_N$ to $K_{N-1}$.}
\end{figure}
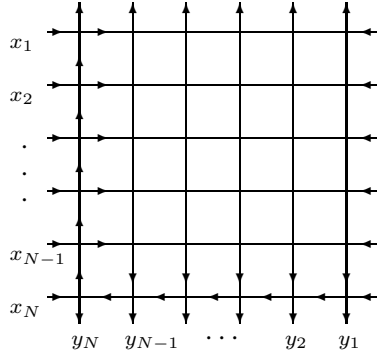
has two outgoing arrows. Hence, on the remaining edges there must be incoming arrows, and this vertex has
$a$-type. It is easy to see that, by the same principle, all other vertices on the southern face are of $a$-type
(see Fig.~\ref{08-6V-6-vertex-DW-red}), and their
total contribution to the partition function is
\be{08-6V-Vklad-1}
 f(x_N,\bar y_N)=  f(y_N,\bar y_N)\qquad \text{at}\qquad x_N= y_N.
\ee
Similarly, all the vertices on the western face have  $a$-type (see Fig.~\ref{08-6V-6-vertex-DW-red}), and their
total contri\-bu\-tion is
\be{08-6V-Vklad-2}
f(y_N,\bar x_N)=f(x_N,\bar x_N)\qquad \text{at}\qquad x_N= y_N.
\ee
The remaining sublattice still has the domain wall boundary condition, so that the contribution
of this sublattice is simply $K_{N-1}(\bar x_N|\bar y_N)$. Multiplying the statistical weights $g(x_N,y_N)$, \eqref{08-6V-Vklad-2},
and \eqref{08-6V-Vklad-1}, we arrive at \eqref{08-6V-K-Kn-1}.

\subsection{Explicit expression for DWPF\label{08-Sec43}}

The properties listed above together with initial condition
\be{08-6V-init}
K_1(x|y)= g(x,y)
\ee
allow us to construct  $K_N(\bar x|\bar y)$ recursively. Suppose that we know an explicit expression
for $K_n(\bar x|\bar y)$ for some $n$ (for $n = 1$ this is true).
Consider $K_{n+1}(\bar x|\bar y)$
as a function of $x_{n+1}$. Up to the factor $\sqrt{x_{n+1}}$
this is a rational function decreasing as $x_{n+1}\to\infty$.
Therefore, it can be presented in the form of the ratio of two polynomials in
$x_{n+1}$, and the degree of the polynomial in the numerator must be less than the degree of the polynomial
in the denominator. The latter is known, because its zeros coincide with the poles of $K_{n + 1}$. Thus, we have
\be{08-6V-Kn1x1}
K_{n+1}(x_{n+1})=\sqrt{x_{n+1}}\frac{P_{n}(x_{n+1})}{\prod_{k=1}^{n+1}(x_{n+1}-y_k)},
\ee
where the degree of the polynomial $P_{n}(x_{n+1})$ does not exceed $n$. Due to the formula \eqref{08-6V-K-Kn-1} we know
the values of this polynomial in $n + 1$ points $y_1,\dots,y_{n+1}$. Hence, we can uniquely
recover this polynomial by the Lagrange interpolation formula.

\begin{prop}\label{08-PROP1}
An explicit expression for DWPF reads (see \cite{Ize87})
\be{08-K-def-fg}
K_{n}(\bar x|\bar y)=\left(\prod_{n\ge \ell>m\ge 1} g(x_\ell,x_m)g(y_m,y_\ell)\right) \;
\frac{f(\bar x,\bar y)}{g(\bar x,\bar y)}
\det_n\left(\frac{g^2(x_j,y_k)}{f(x_j,y_k)}\right).
\ee
\end{prop}

Let us also give \eqref{08-K-def-fg} in a more detailed form, first in variables $\bar x$ and $\bar y$
\be{08-K-defxy}
K_{N}(\bar x|\bar y)=\prod_{\ell=1}^N\sqrt{x_\ell\, y_\ell}\;\frac{\prod_{\ell,m=1}^N (qx_\ell-y_mq^{-1})}{\prod_{N\ge \ell>m\ge 1} (x_\ell-x_m)(y_m-y_\ell)}
\det_N\left(\frac{q-q^{-1}}{(x_j-y_k)(qx_j-y_kq^{-1})}\right),
\ee
and then in variables $\bu$ and $\bxi$
\be{08-K-defuv}
K_{N}(\bar u|\bxi)=\frac{\prod_{\ell,m=1}^N \sinh(u_\ell-\xi_m+\eta)}{\prod_{N\ge \ell>m\ge 1} \sinh(u_\ell-u_m)\sinh(\xi_m-\xi_\ell)}
\det_N\left(\frac{\sinh\eta}{\sinh(u_j-\xi_k)\sinh(u_j-\xi_k+\eta)}\right).
\ee

{\sl Proof.} In fact, we need to check that the function given by \eqref{08-K-defxy} satisfies all
the properties of DWPF (symmetry, decreasing at infinity, residues at the poles, initial condition). Obviously, up to the product of
the square roots $\sqrt{x_\ell\, y_\ell}$ the function \eqref{08-K-defxy} is a rational function that is symmetric in $\bar x$
and is symmetric in $\bar y$. For $x_N\to\infty$, the prefactor at the determinant behaves as $x_N$. At the same time, all elements of the last row
of the determinant behave as $x_N^{-2}$, and hence the entire determinant behaves as $x_N^{-2}$. As a result, the rational part of the
function \eqref{08-K-defxy} behaves as $x_N^{-1}$.

Let us check the property of residues. Let $x_N\to y_N$. Then the matrix element $(q-q^{-1})/(x_N-y_N)(qx_N-y_Nq^{-1})$
becomes singular, and the determinant reduces to the product of this matrix element by the corresponding minor
 \be{08-sing-min}
\det_N\left(\frac{q-q^{-1}}{(x_j-y_k)(qx_j-y_kq^{-1})}\right)\Bigr|_{x_N\to y_N}=
\frac1{x_N(x_N-y_N)}\;\det_{N-1}\left(\frac{q-q^{-1}}{(x_j-y_k)(qx_j-y_kq^{-1})}\right)+\text{Reg}.
\ee
Setting $x_N=y_N$ in the prefactor we obtain
\begin{multline}\label{08-sing-pref}
\frac{\prod_{\ell=1}^N\sqrt{x_\ell\, y_\ell}\;\prod_{\ell,m=1}^N (qx_\ell-y_mq^{-1})}
{\prod_{N\ge \ell>m\ge 1} (x_\ell-x_m)(y_m-y_\ell)}\Bigr|_{x_N\to y_N}
=\prod_{\ell=1}^{N-1}\frac{(qx_N-y_\ell q^{-1})(qx_\ell-y_N q^{-1})}{(x_N-x_\ell)(y_\ell-y_N)}\Bigr|_{x_N\to y_N}\\[2mm]
\times \sqrt{x_N\;y_N}(qx_N-y_Nq^{-1})\Bigr|_{x_N\to y_N} \frac{\prod_{\ell=1}^{N-1}\sqrt{x_\ell\, y_\ell}\;\prod_{\ell,m=1}^{N-1}
(qx_\ell-y_mq^{-1})}{\prod_{N-1\ge \ell>m\ge 1} (x_\ell-x_m)(y_m-y_\ell)}\\[2mm]
= x_N (q-q^{-1})\prod_{\ell=1}^{N-1}f(x_\ell,x_N)f(y_N,y_\ell)\; \frac{\prod_{\ell=1}^{N-1}\sqrt{x_\ell\, y_\ell}\;\prod_{\ell,m=1}^{N-1}
(qx_\ell-y_mq^{-1})}{\prod_{N-1\ge \ell>m\ge 1} (x_\ell-x_m)(y_m-y_\ell)}.
\end{multline}
Combining \eqref{08-sing-min} and \eqref{08-sing-pref} we immediately arrive at \eqref{08-6V-K-Kn-1}.

It remains to check that \eqref{08-K-defxy} gives $g(x,y)$ at  $N=1$, and thus, the representation \eqref{08-K-defxy} possesses all
necessary properties  of DWPF.  As we have already seen, these conditions fix the DWPF unambiguously.


%
%


\chapter{Alternating sign matrices \label{CHA-ASM}}

In this lecture, we really make a step away from the main stream. We will speak about {\it alternating sign matrices}.
These matrices arise in various problems on the partitions of the plane and in the generalization of the concept of
determinant. More information on this topic can be found in the monograph \cite{Bressoud}. At first sight,
the subject of alternating sign matrices is completely unrelated
with the previous stuff. We will see, however, that this is not at all the case. This example perfectly
illustrates the fact that the methods discussed above can be applied in the very
different problems, far from statistical physics and quantum mechanics. In addition, some formulas obtained
in this lecture, will be used in the calculation of the correlation functions in Lecture~15. 

\section{Definition of the alternating sign matrices \label{09-S-ASM-DEF}}

Alternating sign matrices are square matrices whose entries take three values: $0$, or $1$, or $-1$. Besides, the arrangement of
the matrix elements must satisfy the following conditions:

\begin{itemize}
\item when moving along a row (or along a column), the signs of non-zero elements alternate;

\item the sum of all elements in each row and each column is equal to $1$.
\end{itemize}

We see that the formulated rules are the same for rows and for columns, so in what follows
we will often speak about rows. It is easy to understand how an arbitrary row of an alternating sign matrix  of size
$N \times N$ looks like. First, we write a sequence of the form $ + 1, -1, + 1, -1, \dots, + 1 $. The first and last element
of the sequences should be equal to $+1 $, since only in this case the sum of all elements is $1$.
The number of elements in this sequence should be less than or equal to $N$. After that
we insert in arbitrary way zeros between  $+ 1 $ and $ -1 $, so that the total number of all elements becomes $N$:
\be{ASM-row-ex}
\underbrace{0,0,+1, -1,0,+1,0,0,0,0,-1,0,0,0,\dots,+1}_N
\ee
This is how an arbitrary row of an  alternating sign  matrix looks like. Each column has exactly the same form.
As an illustration, we give a concrete example of the $4\times 4$ matrix:
\be{ASM-example}
A=\begin{pmatrix}
0&0&1&0\\
1&0&-1&1\\
0&0&1&0\\
0&1&0&0
\end{pmatrix}.
\ee

It is easy to establish several properties of the alternating sign matrices. For example, each line
and each column must contain at least one non-zero element. This property is
obvious. Less obvious is another property:  there is exactly one nonzero element equal to $1$ in the
first row of the alternating sign matrix. Indeed, suppose the opposite. Then in the first line, there is at least one non-zero element,
equal to $-1$. But then there is a column that begins with $-1$. This implies that the sum
of all elements in this column is either $0$ (if the number of non-zero elements is even) or
is equal to $-1$ (if the number of non-zero elements is odd). In any case, it can not be
equal to $1$, what contradicts the second condition.

It is easy to see that the last line, as well as the first and last columns, has the same property.

\vspace{3mm}

{\sl Exercise.}  Prove that  there is at most one $-1$ in the second row of the  alternating sign matrix.

\vspace{3mm}

One can put a question: what is the number of different  alternating sign matrices  of a given size $ N \times N$?
Let us  denote this number by $\mathcal{A}_N$. Obviously, there exists only one such matrix of size $ 1\times 1$,
that is, $\mathcal{A}_1 = 1$. It is also easy to find all alternating sign  matrices  of size $2\times 2$
\be{ASM-2t2}
\left(\begin{smallmatrix}
1&0\\
0&1
\end{smallmatrix}\right)\quad
\left(\begin{smallmatrix}
0&1\\
1&0
\end{smallmatrix}\right)\;.
\ee
There are only two such matrices, thus $\mathcal{A}_2 = 2$. There are no elements equal to $-1 $ in these matrices. For $N =3$ the alternating sign
matrices have the form
\be{ASM-3t3}
\left(\begin{smallmatrix}
1&0&0\\
0&1&0\\
0&0&1
\end{smallmatrix}\right)\quad
\left(\begin{smallmatrix}
1&0&0\\
0&0&1\\
0&1&0
\end{smallmatrix}\right)\quad
\left(\begin{smallmatrix}
0&1&0\\
1&0&0\\
0&0&1
\end{smallmatrix}\right)\quad
\left(\begin{smallmatrix}
0&1&0\\
0&0&1\\
1&0&0
\end{smallmatrix}\right)\quad
\left(\begin{smallmatrix}
0&0&1\\
1&0&0\\
0&1&0
\end{smallmatrix}\right)\quad
\left(\begin{smallmatrix}
0&0&1\\
0&1&0\\
1&0&0
\end{smallmatrix}\right)\quad
\left(\begin{smallmatrix}
0&1&0\\
1&-1&1\\
0&1&0
\end{smallmatrix}\right)\;.
\ee
In this case, $\mathcal{A}_3=7$, and there is one matrix with $-1$ element. After this, the number of alternating sign matrices rapidly increases
with growth $N$: $\mathcal{A}_4=42$, $\mathcal{A}_5=429$, $\mathcal{A}_6=7436$ and so on. As a hypothesis, the explicit formula for the numbers
$\mathcal{A}_N$ was first given in the paper \cite{MilMR82}
\be{AMS-EXpl-form}
\mathcal{A}_N=\prod_{k=0}^{N-1}\frac{(3k+1)!}{(N+k)!}.
\ee
The first and very long proof of the formula \eqref{AMS-EXpl-form} was given in \cite{Zei96}. We will obtain
formula \eqref {AMS-EXpl-form}, using the method proposed in \cite{Kup96}. This method
is based on the following statement.

\begin{prop}\label{09-PROP1}
There is a one-to-one correspondence between the  alternating sign matrices and the configurations of the arrows
in the $6$-vertex model with the domain wall boundary condition.
\end{prop}

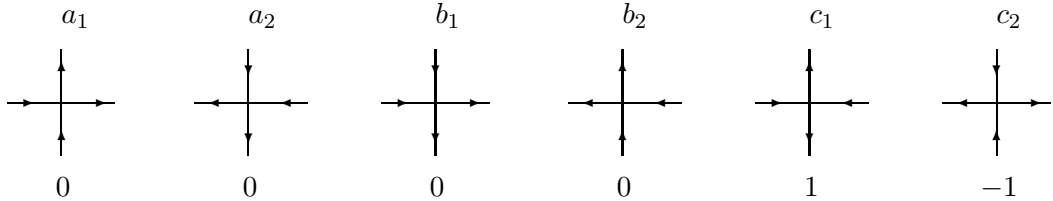
\begin{figure}[h!]
\begin{picture}(450,80)
\put(0,15){%
\begin{picture}(450,70)
\put(30,50){$a_1$}
\put(100,50){$a_2$}
\put(170,50){$b_1$}
\put(240,50){$b_2$}
\put(310,50){$c_1$}
\put(380,50){$c_2$}
\multiput(30,0)(70,0){6}{\line(0,1){40}}
\multiput(10,20)(70,0){6}{\line(1,0){40}}
\put(30,0){\vector(0,1){10}}
\put(100,14.5){\vector(0,-1){10}}
\put(170,14.5){\vector(0,-1){10}}
\put(240,0){\vector(0,1){10}}
\put(380,0){\vector(0,1){10}}
\put(310,14.5){\vector(0,-1){10}}
%
\put(30,26){\vector(0,1){10}}
\put(100,40){\vector(0,-1){10}}
\put(170,40){\vector(0,-1){10}}
\put(240,26){\vector(0,1){10}}
\put(380,40){\vector(0,-1){10}}
\put(310,26){\vector(0,1){10}}
\put(10,20){\vector(1,0){10}}
\put(95,20){\vector(-1,0){10}}
\put(150,20){\vector(1,0){10}}
\put(235,20){\vector(-1,0){10}}
\put(290,20){\vector(1,0){10}}
\put(375,20){\vector(-1,0){10}}
\put(37,20){\vector(1,0){10}}
\put(122,20){\vector(-1,0){10}}
\put(177,20){\vector(1,0){10}}
\put(262,20){\vector(-1,0){10}}
\put(332,20){\vector(-1,0){10}}
\put(387,20){\vector(1,0){10}}
\end{picture}}
\put(28,0){$0$}
\put(98,0){$0$}
\put(168,0){$0$}
\put(238,0){$0$}
\put(308,0){$1$}
\put(374,0){$-1$}
\end{picture}
\caption{\label{ASM-types} Six types of vertices and their labels.}
\end{figure}

{\sl Proof.}
Consider the $6$-vertex model and assign every type of vertices its own `label', as shown in Fig.~\ref{ASM-types}.
We have already agreed upon to distinguish vertices of three types: $a$, $b$, and $c$. Actually, there are two vertices corresponding to each type.
It is convenient to distinguish between these two vertices, for example, a vertices of type
$a_1$ and $a_2$ (see Fig.~\ref{ASM-types}). At the same time, both vertices have the same statistical weight $a$.

Let us associate the label $0$ with vertices of the type $a_k$ and $b_k$. At the same time, we put  the $c_1$-type vertices  in correspondence
with the label $1$, and the $c_2$-type vertices with the label $-1$ (see Fig.~\ref{ASM-types}). Pay attention that the numbers
introduced have nothing to do with the statistical weights. Just every vertex is now
supplied with an additional label.

It is important to stress a significant difference between the vertices of types $a_k$ and $b_k$ ($ k = 1,2 $) on the one hand,
and vertices of the type $c_1$ and $c_2$ on the other.
Passing through the $a_k$-types or $b_k$-types vertices, the arrow does not change direction. That is why all these vertices have the same
zero label. In contrast, in vertices of the type $c_1$ and $c_2$, the arrow changes its direction. As we shall see below, this difference plays a fundamental role.

\begin{figure}[h!]
\begin{picture}(450,70)
\put(0,25){%
\begin{picture}(200,45)
\multiput(100,0)(30,0){10}{\line(0,1){30}}
\put(70,15){\line(1,0){330}}
\multiput(70,15)(30,0){3}{\vector(1,0){15}}
\multiput(190,15)(30,0){4}{\vector(-1,0){15}}
\multiput(280,15)(30,0){3}{\vector(1,0){15}}
\put(400,15){\vector(-1,0){15}}
\put(160,15){\vector(0,-1){15}}
\put(280,25){\vector(0,-1){5}}
\put(370,15){\vector(0,-1){15}}
\put(160,15){\vector(0,1){15}}
\put(280,5){\vector(0,1){5}}
\put(370,15){\vector(0,1){15}}
\end{picture}}
\multiput(98,10)(30,0){2}{$0$}
\multiput(188,10)(30,0){3}{$0$}
\multiput(308,10)(30,0){2}{$0$}
\put(158,10){$1$}
\put(271,10){$-1$}
\put(368,10){$1$}
\end{picture}
\caption{\label{AMS-6V-ASM} A horizontal line of the $6$-vertex model and row in an alternating sign matrix. }
\end{figure}
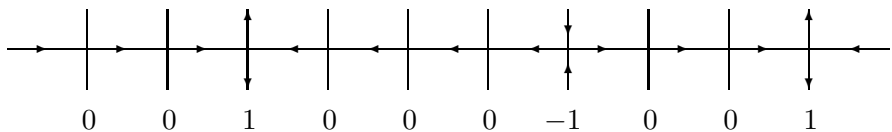

It is clear that if we have some configuration of the arrows of the $6$-vertex model with the domain wall boundary condition
and instead of each vertex we write its label, then we get a square matrix whose matrix elements are $0$, or
$1$, or $-1$. Let us check that the rules for arranging these matrix elements are fulfilled. Let us consider an arbitrary horizontal
line and start moving along it from left to right. Due to the boundary condition at the left end of this line, we have an incoming
arrow. Let the arrow go through several vertices without changing the direction (in Fig.~\ref{AMS-6V-ASM}  two vertices are passed).
These vertices have the type $a_1$ or $b_1$. Therefore, they correspond to the label $0$, no matter how the arrows are directed on the
vertical edges.
At some point we reach a vertex, where the arrow changes direction (the third vertex in Fig.~\ref{AMS-6V-ASM}).
At least one such vertex must exist because of the boundary condition on the right side of the lattice. This vertex
is of the $c_1$-type, and it corresponds to the label $1$. After this, the arrow again does not change direction several times, and these vertices
again correspond to zeros. It may happen that the arrow  reaches the right side without changing the direction, but if we meet a vertex where the direction of the arrow has changed again, then this vertex is of the $c_2$-type. It is clear that after this
vertices of types $c_1$ and $c_2$ will alternate, and between them there might be an arbitrary number of vertices
of the types $a_k$ and $b_k$. Due to the boundary condition, the last vertex in which the direction of the arrow changes,
is of the $c_1$-type. Thus, in the obtained row of the matrix, the non-zero elements alternate the sign. In addition, the first and last
of them are equal to $1$. This means that the sum of all the elements in the line is $1$. Hence, the conditions for the arrangement of non-zero
elements in an arbitrary row of the alternating sign matrix  are satisfied. Similarly, it can be checked that these conditions are also satisfied for the columns. Thus, any configuration of the arrows in the $6$-vertex model with the domain wall boundary condition
corresponds to a certain alternating sign  matrix.

We now prove the converse statement, namely, starting from the given alternating sign  matrix we construct a configuration of arrows
of the $6$-vertex model with the domain wall boundary condition. At first glance it seems that this can be done
in different ways, since the label $0$  corresponds to four different vertices. However, if we begin with the upper face of the lattice,
then the arrows can be restored unambiguously.
\begin{figure}[h!]
\begin{picture}(450,70)
\put(160,5){%
\begin{picture}(200,45)
\multiput(100,0)(30,0){5}{\vector(0,1){60}}
\put(70,45){\line(1,0){180}}
\put(70,15){\line(1,0){180}}
\multiput(73,45)(30,0){4}{\vector(1,0){15}}
\multiput(218,45)(30,0){2}{\vector(-1,0){15}}
\multiput(100,18)(30,0){3}{\vector(0,1){15}}
\put(220,18){\vector(0,1){15}}
\put(190,40){\vector(0,-1){15}}
\put(73,15){\vector(1,0){15}}
\end{picture}}
\multiput(60,50)(20,0){3}{$0$}
\put(120,50){$1$}
\put(140,50){$0$}
\put(60,20){$0$}
\put(80,20){$1$}
\put(100,20){$0$}
\put(112,20){$-1$}
\put(140,20){$1$}
\end{picture}
\caption{\label{AMS-ASM-6V} Correspondence of the upper horizontal line of the $6$-vertex model and the first row of the alternating sign matrix.}
\end{figure}
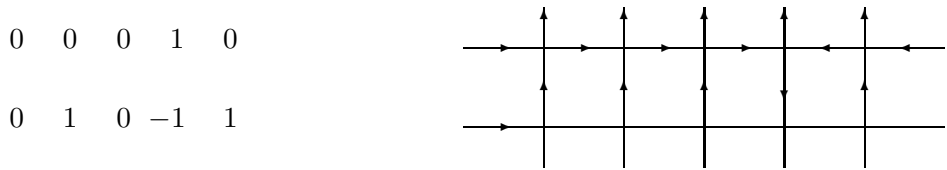
Fig.~\ref{AMS-ASM-6V} shows the first two rows of an alternating sign  matrix of size $5 \times 5$. Let us draw
a $5\times 5$ square lattice  and fix the boundary conditions on the left and upper faces. Assume that
few elements in the first row of the matrix are zeros (in Fig.~\ref{AMS-ASM-6V} there are three zeros). Hence, they correspond
to vertices in which the arrow does not change direction. But at the extreme north-western vertex the direction of the arrows
on two edges are already fixed by the boundary conditions.
Then the directions of the arrows on the two remaining edges
are uniquely determined (this is a vertex of the type $a_1$). Passing to the second vertex, we see that again the directions of the arrows
on two edges are already fixed, and hence this vertex also has the type $a_1$. It is clear that  this process continues:
every time, passing to a new vertex, we will have fixed directions of arrows on its two edges. In this way
we will obtain the  $a_1$-type vertices  until we reach the vertex corresponding to the element $1$.
There is no longer any arbitrariness here, as the number $1$ corresponds to the  $c_1$-type vertex. After that,
we again have zeros (since the first row of the alternating sign  matrix has only one non-zero element), and it is clear that all remaining vertices
have the type $b_2$. In doing so, we automatically obtain the correct boundary condition on the right boundary.

When the arrows on the upper horizontal line are determined, we can proceed to the construction of the arrows
on the second line. It is clear that the process described above is completely repeated, since all vertices have already a fixed
direction of the arrows on the upper edges, as well as the extreme left vertex has a fixed arrow on the left edge.
We suggest the reader to make check that in the second line the first vertex is of the type $a_1$,
the second and fifth are the $c_1$-type, the third is the $b_2$-type, and the fourth is the $c_2$-type.
Since the last non-zero element is equal to $1$ in any row of the alternating sign matrix,
we come to the conclusion that the last vertex in which the arrow changes direction is of the type $c_1$. After this, the arrow reaches
the right side of the lattice, without changing the direction, and hence the boundary condition on the right face is automatically
correct.

In this way, we unambiguously recover all the arrows on the edges of the lattice according to a given  alternating sign matrix.
It remains to check only that the boundary condition on the lower face turned out to be correct. This is almost obvious.
Indeed, at each vertex of the lattice the number of incoming and outgoing arrows are equal. Hence, the same is true for the entire lattice.
We already have $2N$ incoming arrows on the left and right faces.
We also have $N$ outgoing arrows on the upper face. Hence, on the lower face all the arrows must be outgoing.

The proved one-to-one correspondence allows us to make the following statement.

\begin{cor}
The number of the alternating sign matrices of size $N\times N$ is equal to the number of configurations in the $6$-vertex model
on the $N \times N$ lattice  with the domain wall boundary condition.
\end{cor}

Equivalently, this statement can be formulated as follows.

\begin{cor}
The number of the alternating sign matrices of size $N\times N$ is equal to the partition function of
the $6$-vertex model  on the $N \times N$ lattice  with the domain wall boundary condition, provided all statistical weights
equal to $1$.
\end{cor}

Indeed, if the statistical weights of all vertices are $1$, then the partition function is  equal to the number of all possible states of the system.

\section{Calculating the partition function\label{09-Sec2}}

We know already the answer for DWPF of the inhomogeneous model. Explicitly it reads
\be{ASM-Izer1}
K_N(\bu|\bxi)=
\frac{\prod_{j,k=1}^N\sinh(u_j-\xi_k+\eta)}{\prod_{j<k}^N \sinh(u_j-u_k)\sinh(\xi_k-\xi_j)}\cdot
\det_N\left(\frac{\sinh\eta}{\sinh(u_j-\xi_k)\sinh(u_j-\xi_k+\eta)}\right).
\ee
Here a vertex lying on the intersection of $j$th horizontal line and
$k$th vertical line has the statistical weights
\be{ASM-statvesa}
a=\frac{\sinh(u_j-\xi_k+\eta)}{\sinh(u_j-\xi_k)},\qquad
b=1,\qquad
c=\frac{\sinh\eta}{\sinh(u_j-\xi_k)}.
\ee
Can we make all statistical weights equal to $1$? It is obvious that
first, we need to put $u_j = u$ for all $j=1,\dots,N$,
and $\xi_k = \xi$ for all $k=1,\dots,N$. After this, putting $a = c = 1$ in the equations \eqref{ASM-statvesa},
we easily find $u-\xi=\pm i\pi/3$ and $\eta=\pm i\pi/3$.
Without loss of generality we can choose any specific solution, for example, with a plus sign.

Unfortunately, we cannot simply put,  for example, $u_j=i\pi/3$, $\xi_k=0$, and $\eta=i\pi/3$, in the formula \eqref{ASM-Izer1},
because of singularities. Indeed, with this choice, all $\sinh(u_j-u_k)$ and
$\sinh(\xi_k-\xi_j)$ in the denominator go to zero. At the same time,  all rows and columns in the determinant coincide. Therefore,
we set $\eta=i\pi/3$, $u_j=v_j+i\pi/3$ in \eqref{ASM-Izer1} and consider the limit
$v_j\to0$, $\xi_j\to0$  for all $j=1,\dots,N$. As we already know, in this limit the partition function $K_N$ gives the number of the
alternating signs matrices of size $N \times N$, and we obtain
\be{ASM-Izer2}
\mathcal{A}_N=\frac{(-1)^{\frac{N^2-N}2}3^{\frac{N^2+N}2}}{2^{N^2-N}}\lim_{\substack{v_1,\dots,v_N\to0\\
\xi_1,\dots,\xi_N\to0}}
\prod_{j<k}^N \Bigl(\sinh(v_j-v_k)\sinh(\xi_k-\xi_j)\Bigr)^{-1}\cdot
\det_N\left(\frac{\sinh(v_j-\xi_k)}{\sinh3(v_j-\xi_k)}\right).
\ee
Here we have used $\sinh\frac{i\pi}3=\frac{i\sqrt{3}}2$ and $4\sinh x\sinh( x-\frac{i\pi}3)
\sinh( x+\frac{i\pi}3)=\sinh3x$.

To compute the limit we could use the following statement.

\begin{prop}\label{ASM-Pred-limdet}
For any sufficiently smooth function of two variables $\Phi(v,\xi)$
\be{ASM-lim-det}
\lim_{\substack{v_1,\dots,v_N\to v\\
\xi_1,\dots,\xi_N\to\xi}}
\frac{\det_N\Phi(v_j,\xi_k)}{\prod_{j<k}^N \sinh(v_j-v_k)\sinh(\xi_j-\xi_k)}
=\prod_{k=1}^{N-1}\left(\frac1{k!}\right)^2\cdot\det_N\left(\frac{\partial^{j+k-2}\Phi(v,\xi)}{\partial v^{j-1}\partial\xi^{k-1}}\right).
\ee
In this case the words `sufficiently smooth' mean the existence of all partial derivatives in the rhs of
\eqref{ASM-lim-det}.
\end{prop}

Thus, computing the derivatives of the function $\sinh(v-\xi)/\sinh3(v-\xi)$ at the point $v=0$, $\xi=0$,
we would get a determinant of a matrix consisting of  numbers, and could try to calculate this
numerical determinant. We, however, use a different method. Therefore, we suggest the readers
to prove the proposition~\ref{ASM-Pred-limdet} or at least to check its validity for small $N$ (such
checking will suggest the idea of proof).

We will only use the fact that the limit in \eqref{ASM-lim-det} does not depend on how exactly
the parameters $v_j$ and $\xi_k$ tend to their limit values. Therefore, one can choose
quite a specific way of taking the limit, the result does not depend on it. In particular, we can put
in \eqref{ASM-Izer2} $v_j =j\epsilon$, $\xi_k=(1-k)\epsilon$ and consider the limit $\epsilon \to 0 $. We obtain
\be{ASM-Izer3}
\mathcal{A}_N=\frac{(-1)^{\frac{N^2-N}2}3^{\frac{N^2+N}2}}{2^{N^2-N}}\lim_{\epsilon\to 0}
\prod_{j<k}^N \Bigl(\sinh\epsilon(j-k)\Bigr)^{-2}\cdot
\det_N\left(\frac{\sinh\epsilon(j+k-1)}{\sinh3\epsilon(j+k-1)}\right).
\ee

It turns out that the determinant in the rhs of equation \eqref{ASM-Izer3} can be explicitly calculated.
However, for this we have to prove an auxiliary lemma.

\begin{lemma}\label{ASM-Cauchy}(Cauchy determinant.) For arbitrary $x_1,\dots,x_N$ and $y_1,\dots,y_N$
\be{ASM-detCauchy}
\det_N\left(\frac1{x_j-y_k}\right)=\frac{\prod_{j<k}^N(x_j-x_k)(y_k-y_j)}
{\prod_{j,k=1}^N(x_j-y_k)}.
\ee
\end{lemma}

{\sl Proof. } There are several relatively simple ways to prove this formula. One can, for example, use induction on $N$.
One can also prove \eqref{ASM-detCauchy} almost without calculations. The lhs
is a rational function of the variables $x_1,\dots,x_N$ and $y_1,\dots,y_N$,
which has poles at the points $x_j=y_k$, $j,k=1,\dots,N$. If $x_j = x_k$ (or $y_j = y_k$), then there are two
coinciding rows in the determinant
(or two columns), so at these points the determinant has zeros. If some $x_\ell$ tends to
infinity, while the remaining variables are fixed, then the whole determinant goes to zero, because
the $\ell $th row in the matrix tends to zero. A similar property holds if some $y_\ell\to\infty$.
The listed properties are sufficient to write down the determinant in the form
\be{ASM-detCauchy1}
\det_N\left(\frac1{x_j-y_k}\right)=C\frac{\prod_{j<k}^N(x_j-x_k)(y_k-y_j)}
{\prod_{j,k=1}^N(x_j-y_k)},
\ee
where a constant $C$ does not depend on either $x_1,\dots,x_N$, or $y_1,\dots,y_N$.
Indeed, the poles of the determinant uniquely fix the denominator in the rhs of \eqref{ASM-detCauchy1}. The zeros at $x_j=x_k$
and $y_j=y_k$ give the numerator. It is easy to show that the determinant  cannot have other zeros. Indeed, suppose that the determinant has
an additional zero for $x_\ell=x_0$, where $x_\ell $ is an arbitrary element of the set $x_1,\dots,x_N$, while   $x_0$
does not belong to this set. Then there would be an additional factor $(x_\ell-x_0)$ in the rhs of \eqref {ASM-detCauchy1}. In this case
the rhs does not tend to zero as $x_\ell\to\infty$. Similar considerations lead us to
the conclusion that the determinant has no additional zeros for $y_\ell = y_0$. Hence, the rational functions
in both sides of the formula \eqref{ASM-detCauchy1} can differ from each other only by a constant that does not depend
on either  $x_1,\dots,x_N$, or $y_1,\dots,y_N$. In order to find this constant, it is sufficient to consider
a special case $x_j = y_j + \varepsilon$, $j=1,\dots,N$, and to calculate the coefficients of the maximal degree
$1/\varepsilon^N$. In the lhs, this coefficient is $1$ (it occurs when taking the product of diagonal elements).
It is easy to check that the coefficient of $1/\varepsilon^N$ in the rhs is equal to $C$. Hence, $C = 1$.

\begin{cor}\label{ASM-Cauchy-S} For arbitrary $x_1,\dots,x_N$ and $y_1,\dots,y_N$
\be{ASM-detCauchy-S}
\det_N\left(\frac1{\sinh(x_j-y_k)}\right)=\frac{\prod_{j<k}^N\sinh(x_j-x_k)\sinh(y_k-y_j)}
{\prod_{j,k=1}^N\sinh(x_j-y_k)}.
\ee
\end{cor}

{\sl Proof.} It suffices to pass to the variables $u_j=e^{2x_j}$ and $v_j=e^{2y_j}$.
Then in the new variables, the determinant in \eqref{ASM-detCauchy-S} reduces to the Cauchy determinant.
It remains to use \eqref {ASM-detCauchy} and return to the original variables.

Let us turn now to the calculating the determinant in \eqref{ASM-Izer3}.

\begin{lemma}\label{ASM-Lem-Kuper}
For any $\alpha$ and $\beta$
\be{ASM-det-Kuper}
\det_N\left(\frac{\sinh\alpha(j+k-1)}{\sinh\beta(j+k-1)}\right)=
2^{N^2-N}\prod_{j>k}^N{\sinh}^2\beta(j-k)\prod_{j,k=1}^N
\frac{\sinh(\alpha+\beta(j-k))}{\sinh\beta(j+k-1)}.
\ee
\end{lemma}

{\sl Proof.} We transform the determinant in the lhs of \eqref{ASM-det-Kuper} as follows:
\begin{multline}\label{ASM-doc-Kuper}
\det_N\left(\frac{\sinh\alpha(j+k-1)}{\sinh\beta(j+k-1)}\right)=
\det_N\left(\frac{e^{\alpha(j+k-1)}-e^{-\alpha(j+k-1)}}{2\sinh\beta(j+k-1)}\right)\num
=\prod_{j=1}^Ne^{-\alpha(j-1)}\prod_{k=1}^Ne^{-\alpha k}\cdot
\det_N\left(\frac{e^{2\alpha(j+k-1)}-1}{2\sinh\beta(j+k-1)}\right)
=e^{-\alpha N^2}\cdot
\det_N\left(\frac{e^{2\alpha(j+k-1)}-1}{2\sinh\beta(j+k-1)}\right).
\end{multline}
The determinant in the rhs of \eqref{ASM-doc-Kuper} is a polynomial in $e^{2\alpha}$. In order to find the degree of this polynomial,
one should neglect the term $-1$ in the combination $e^{2\alpha(j+k-1)}-1$. After this we immediately find that the
degree of the polynomial is $N^2$. Thus,
\be{ASM-doc-Kuper2}
\det_N\left(\frac{\sinh\alpha(j+k-1)}{\sinh\beta(j+k-1)}\right)
=e^{-\alpha N^2}\; P_{N^2}\left(e^{2\alpha}\right).
\ee
In order to find a polynomial, it is sufficient to find its zeros and the coefficient of the leading term $e^{2\alpha N^2}$.
The zeros can be easily found. First, for $\alpha = 0$, all matrix elements of the matrix in the lhs of \eqref{ASM-doc-Kuper}
vanish. We are dealing with a matrix of zero rank. Therefore, for $\alpha = 0$, the determinant has zero of order $N$.

If $\alpha=\beta$, then all the matrix elements are equal to $1$, and we obtain a matrix of the first rank. Therefore, for $\alpha=\beta$
the determinant has zero of order $N-1$. Similarly, the determinant has zero of order $N-1$ for $\alpha=-\beta$.

It is easy to see that for $\alpha=\pm 2\beta$
we obtain a matrix of the second rank (the sum of two matrices of the first rank), therefore, at these points the determinant
has zeros of order $N-2$. And so on up to $\alpha=\pm (N-1)\beta$, where the determinant has first order zeros.
If we now sum up the orders of all the found zeros, then we obtain $N^2$, hence,
there are no other zeros.

It remains to find the coefficient of the leading term $e^{2\alpha N^2}$.
To do this, we send $\alpha\to\infty$ in
the rhs of \eqref{ASM-doc-Kuper} and calculate the coefficient of $e^{2\alpha N^2}$.
It is easy to see that this coefficient is
\be{ASM-doc-Kuper3}
2^{-N}\det_N\left(\frac{1}{\sinh\beta(j+k-1)}\right).
\ee
This is a special case of the Cauchy determinant calculated in Corollary~\ref{ASM-Cauchy-S}. Indeed,
setting in \eqref{ASM-detCauchy-S} $x_j=j\beta$ and $y_k=(1-k)\beta$, we arrive at \eqref{ASM-doc-Kuper3}.
Therefore,
\be{ASM-doc-Kuper4}
2^{-N}\det_N\left(\frac{1}{\sinh\beta(j+k-1)}\right)=
2^{-N}\frac{\prod_{j>k}^N\sinh^2\beta(j-k)}{\prod_{j,k=1}^N\sinh\beta(j+k-1)}.
\ee

Thus, we arrive at expression
\begin{multline}\label{ASM-doc-Kuper5}
\det_N\left(\frac{\sinh\alpha(j+k-1)}{\sinh\beta(j+k-1)}\right)
=e^{-\alpha N^2}\; 2^{-N}\frac{\prod_{j>k}^N\sinh^2\beta(j-k)}{\prod_{j,k=1}^N\sinh\beta(j+k-1)}\num
\times\left(e^{2\alpha}-1\right)^N \prod_{m=1}^{N-1}
\left(e^{2\alpha}-e^{2m\beta}\right)^{N-k}\left(e^{2\alpha}-e^{-2m\beta}\right)^{N-k}.
\end{multline}
Comparing \eqref{ASM-doc-Kuper5} with  \eqref{ASM-det-Kuper}, we see that we should prove that
\be{ASM-doc-Kuper6}
2^{N^2}\prod_{j,k=1}^N\sinh(\alpha+\beta(j-k))= e^{-\alpha N^2}\;
\left(e^{2\alpha}-1\right)^N \prod_{m=1}^{N-1}
\left(e^{2\alpha}-e^{2m\beta}\right)^{N-k}\left(e^{2\alpha}-e^{-2m\beta}\right)^{N-k}.
\ee
Of course, it is not difficult to prove this equality by direct calculation. There exists, however, a more simple way.
We already know that the rhs of \eqref{ASM-doc-Kuper6} is the product of $e^{-\alpha N^2}$ and a polynomial in $e^{2\alpha}$
of degree $N^2$. It is easy to see that the lhs is also representable in the same form.
The coefficients of the leading terms of the polynomials in both sides are equal to one. It remains to check that zeros
of both polynomials coincide. The zeros of the polynomial in the rhs are already known, so we investigate now the product
in the lhs of \eqref{ASM-doc-Kuper6}.

In this product, there are $N$ factors for which $j = k$. Obviously,
all these factors turn to zero for $\alpha = 0$. Thus, the point $\alpha = 0$ is  zero of order $N$
for the lhs of \eqref{ASM-doc-Kuper6}. If $j-k + 1 =0$, then the product under consideration vanishes
for $\alpha = \beta$. This is zero of order $N-1$, because there are $N-1$ factors for which $j = k-1$.
Similarly,  zero of order $N-1$ occurs when $\alpha = -\beta$ (it is given by the factors for which $j-k-1 = 0$).
In general, if $\alpha =\pm m \beta$, where $m=0,1, \dots, N-1$, then we get zero of order $N-m$, since
there are exactly $N-m$ factors for which $j-k \mp m = 0$. Thus, the coincidence  of two polynomials is
proved.

We can now apply Lemma~\ref{ASM-Lem-Kuper} to the calculating the determinant \eqref{ASM-Izer3}. For this it is enough to set
$\alpha=\epsilon$ and $\beta=3\epsilon$. We obtain
\be{ASM-Izer-Kuper}
\mathcal{A}_N=(-1)^{\frac{N^2-N}2}3^{\frac{N^2+N}2}\lim_{\epsilon\to 0}
\prod_{j<k}^N \frac{\sinh^{2}3\epsilon(j-k)}{\sinh^{2}\epsilon(j-k)}\prod_{j,k=1}^N
\frac{\sinh\epsilon(1+3(j-k))}{\sinh3\epsilon(j+k-1)}.
\ee
The limit $\epsilon\to 0$ becomes trivial, and we arrive at
\be{ASM-pochti-ans}
\mathcal{A}_N=(-3)^{\frac{N^2-N}2}
\prod_{j,k=1}^N\frac{3(j-k)+1}{j+k-1}.
\ee
We provide the reader with the opportunity to complete the calculation  and make sure that the obtained
double product is reduced to the formula \eqref{AMS-EXpl-form}. Starting with the representation  \eqref{ASM-pochti-ans}
one can also obtain another formula
\be{AMS-EXpl-form2}
\mathcal{A}_N=\frac{(\sqrt{3})^{3N^2}}{4^{N^2}}\prod_{k=1}^{N}\frac{\Gamma(k+\frac13)\Gamma(k-\frac13)}
{\Gamma(k+\frac12)\Gamma(k-\frac12)},
\ee
which is equivalent to \eqref{AMS-EXpl-form}.


%
%

\chapter{Multiple commutation relations\label{CHA-MultCR}}

The $RTT$-relation defines the commutation relations between  the monodromy matrix elements $T^{ij}(u)$ and $T^{kl}(v)$.
In this section we proceed to the study of multiple commutation relations, when we permute a product of several
operators $T^{ij}(\bu)$ with a product of several operators $T^{kl}(\bv)$. In the particular case, such multiple commutation relations allow one to
calculate scalar products of Bethe vectors.

\section{Summation over partitions\label{06-Sec2}}

Starting from this section we will permanently deal with sums over partitions of sets of variables into disjoint
subsets. To denote subsets, we  mainly use Roman numerals (all other notation will be specified separately). For example, notation
$\bv\mapsto\{\bv_{\so},\bv_{\st}\}$ means that the set $\bv$ is divided into subsets $\bv_{\so}$
and $\bv_{\st}$, such that $\bv_{\so}\cap\bv_{\st}=\emptyset$ and $\bv_{\so}\cup\bv_{\st}=\bv$.
Summation over partitions of some set of parameters means taking a sum over all possible subsets
of this set (including empty set). Let us give an example. Let $\bv=\{v_1,v_2,v_3\}$. Then
\begin{multline}\label{06-MCR-Sum-part}
\sum_{\bv\mapsto\{\bv_{\so},\bv_{\st}\}}f(\bv_{\so},\bv_{\st})=
1+\bigl[f(v_{1},v_{2})f(v_{1},v_{3})+f(v_{2},v_{1})f(v_{2},v_{3})+
f(v_{3},v_{1})f(v_{3},v_{2})\bigr]\\
+\bigl[f(v_{2},v_{1})f(v_{3},v_{1})+f(v_{1},v_{2})f(v_{3},v_{2})+
f(v_{1},v_{3})f(v_{2},v_{3})\bigr]+1.
\end{multline}
Recall that $f(\bv_{\so},\bv_{\st})$  is the shorthand notation for the product
\be{06-short}
f(\bv_{\so},\bv_{\st})=\prod_{v_i\in \bv_{\so}}\prod_{v_k\in\bv_{\st}} f(v_i,v_k),
\ee
and if at least one of the subsets is empty, then the product is equal to $1$. Then
the first term in \eqref{06-MCR-Sum-part} corresponds to the partition $\bv_{\so}=\emptyset$, $\bv_{\st}=\{v_1,v_2,v_3\}$.
Then a group of three terms follows that corresponds to three different partitions in which $\#\bv_{\so}=1$
and $\#\bv_{\st}=2$. The following three terms correspond to three partitions where $\#\bv_{\so}=2$
and $\#\bv_{\st}=1$. Finally, the last term arises from the partition
$\bv_{\so}=\{v_1,v_2,v_3\}$, $\bv_{\st}=\emptyset$. Pay attention to the fact that the order
of the elements in each subset is not essential. So, we consider the subsets  $\{v_1,v_2\}$
and $\{v_2,v_1\}$ as equivalent, and we take them into account only once when summing up.

Sometimes we will deal with sums over partitions with restrictions on the cardinalities of the subsets.
For example, if we impose a restriction on the sum  \eqref{06-MCR-Sum-part} $\#\bv_{\so}=1$
and $\#\bv_{\st}=2$, then only three terms in square brackets survive in the first line of the sum
\eqref{06-MCR-Sum-part}. To facilitate the notation, we will indicate restrictions on subsets only
in the comments to the formulas.

\section{Identity for the partition function\label{06-Sec3}}

For definiteness and simplicity we consider models described by the rational $R$-matrix $R(u,v)=\mathbf{I}+g(u,v)\mathbf{P}$.
However, all of the following entirely applies to the models with trigonometric $R$-matrix (unless otherwise specified).

In the previous sections we got acquainted with
the partition function of the six-vertex model with the domain wall boundary condition  (DWPF). Now we will see the role
of this object in the commutation relations between the  monodromy matrix entries.
Recall that we denote DWPF by the symbol $K_{n}(\bar u|\bar v)$.
This is a function that depends on two sets of variables $\bu$ and $\bv$ with the same number of elements. The subscript $n$ shows
the cardinalities of the sets: $\#\bu=\#\bv=n$.
In terms of the functions $f$ and $g$, the DWPF is given by the formula
(6.4.11). 
In the case of the rational $R$-matrix this formula takes the form:
\be{06-K-def}
K_{n}(\bar u|\bar v)=\frac{\prod_{\ell,m=1}^n (u_\ell-v_m+c)}{\prod_{n\ge \ell>m\ge 1} (u_\ell-u_m)(v_m-v_\ell)}
\det_n\left(\frac{c}{(u_j-v_k)(u_j-v_k+c)}\right).
\ee

Some properties of DWPF were described in section~\ref{08-S-6V-Prop-DW}.
One of the most important properties is a reduction of DWPF in the poles \eqref{08-6V-K-Kn-1}.
One more reduction has the following form:
 \begin{equation}\label{06-K-red}
K_{n+1}(\{\bu, z-c\}|\{\bv,z\}) =K_{n+1}(\{\bu, z\}|\{\bv,z+c\})=-K_{n}(\bu|\bv).
\end{equation}
The proof of this formula is very similar to the proof of the properties of the DWPF residues at the poles.
Indeed, if among the elements of the first and second sets there are those that differ by a shift by $c$, then the double product
in \eqref{06-K-def} vanishes. On the other hand, one of the matrix elements in the determinant becomes singular. It is therefore clear,
that the determinant reduces to the product of this element by the corresponding minor, which ultimately leads to the formula
\eqref {06-K-red}.

We will also need the properties of DWPF under the replacement of the subsets    $\bu\leftrightarrow\bv$.
First of all, this replacement is equivalent to the change $c$ by $-c$. This immediately follows from the analogous
properties of the  $f$ and $g$ functions and representation \eqref{08-K-def-fg}.
The replacement of the sets $\bu$ and  $\bv$ also can be achieved due to
 \begin{equation}\label{06-K-invers}
  K_{n}( \bu-c|\bv)=K_{n}( \bu|\bv+c) = (-1)^{n}  \frac{K_{n}(\bv|\bu)}{f(\bv,\bu)}\,.
\end{equation}
This formula can be easily derived from the explicit representation \eqref{06-K-def}. The notation $\bu-c$ means that all the elements of the set
$\bu$ are shifted by the constant  $c$ (and similarly for $\bv+c$).

Let us now pass to more sophisticated properties of DWPF, that allow one to compute certain sums over partitions.

\begin{lemma}\label{06-MCR-main-ident}
Let $\bar\xi$, $\bu$, and $\bv$ be sets of complex numbers such that  $\#\bu=m$,
$\#\bv=n$, and  $\#\bar\xi=m+n$. Then
\begin{equation}\label{06-MCR-Part-old1}
  \sum_{\bar\xi \mapsto\{\bar\xi_{\so},\bar\xi_{\st}\}}
 K_{m}(\bar\xi_{\so}|\bu)K_{n}(\bv|\bar\xi_{\st})f(\bar\xi_{\st},\bar\xi_{\so})
 = (-1)^{m}f(\bar\xi,\bu) K_{m+n}(\{\bu-c,\bv\}|\bar\xi).
 \end{equation}
The sum is taken over partitions of the set $\bar\xi$ into two subsets
$\bar\xi_{\so}$ and $\bar\xi_{\st}$, such that  $\#\bar\xi_{\so}=m$ and $\#\bar\xi_{\st}=n$.
\end{lemma}

{\sl Remark. } We have explicitly indicated that we are dealing with summation over partitions with restrictions on the cardinalities
of the subsets. However, we could avoid this, because the restrictions automatically follow from the
form of the formula. Indeed, by definition each of the DWPF should have an equal number of
arguments to the left and to the right from the vertical line. Then $\#\bu=m$ yields $\#\bar\xi_{\so}=m$,
and $\#\bv=n$  yields $\#\bar\xi_{\st}=n$. Only in this case the formula \eqref{06-MCR-Part-old1} makes sense.

{\sl Proof. } First of all let us find the properties of both sides of \eqref{06-MCR-Part-old1}
in the poles. Consider, for example, the lhs as a function of $u_1$ at all other variables fixed. Let
\begin{equation}\label{06-MCR-Fal}
 F_{m,n}^{(l)}(\bu;\bxi)= \sum_{\bar\xi \mapsto\{\bar\xi_{\so},\bar\xi_{\st}\}}
 K_{m}(\bar\xi_{\so}|\bu)K_{n}(\bv|\bar\xi_{\st})f(\bar\xi_{\st},\bar\xi_{\so}).
 \end{equation}
Obviously, the function $F_{m,n}^{(l)}(\bu;\bxi)$ is analytic function of  $u_1$ in the whole complex plane
except the points $\xi_k$, where it has simple poles. Here
$\xi_k$ is an arbitrary element of the set $\bxi$. Indeed, the parameter
$u_1$ is included into DWPF $K_{m}(\bar\xi_{\so}|\bu)$ only. The latter has simple poles at the points where $u_1=\xi_k$.
Let us fix $\xi_k$. Then the singularity in the rhs of
\eqref{06-MCR-Fal} arises if and only if $\xi_k\in\bxi_{\so}$.
Let us put $\bar\xi_{\so'}= \bar\xi_{\so}\setminus\xi_k$, \ $\bar\xi_k=\bar\xi\setminus\xi_k$, and $\bu_1=\bu\setminus u_1$.
Then, sending  $u_1\to\xi_k$ and using \eqref{08-6V-K-Kn-1},
we obtain
\begin{multline}\label{06-MCR-Fa-res}
 \Bigl.F_{m,n}^{(l)}(\bu;\bxi)\Bigr|_{u_1\to\xi_k}= \sum_{\bar\xi_k \mapsto\{\bar\xi_{\so'},\bar\xi_{\st}\}}
  g(\xi_k,u_1)f(u_1,\bu_1)
 f(\bar\xi_{\so'},\xi_k)\\
 \times K_{m-1}(\bar\xi_{\so'}|\bu_1)K_{n}(\bv|\bar\xi_{\st})
 f(\bar\xi_{\st},\bar\xi_{\so'})f(\bar\xi_{\st},\xi_k)+\text{Reg},
 \end{multline}
and we recall that $\text{Reg}$ means the terms that remains regular at $u_1=\xi_k$.
We also have used an equality
\be{06-MCR-rav}
f(\bar\xi_{\st},\bar\xi_{\so})=f(\bar\xi_{\st},\bar\xi_{\so'})f(\bar\xi_{\st},\xi_k).
\ee
Note that now the sum is taken over the partitions of the set $\bar\xi_k$ into subsets
$\bar\xi_{\so'}$ and $\bar\xi_{\st}$. Then the factors $g(\xi_k,u_1)f(u_1,\bu_1)$ do not depend on the partitions, hence,
they can be taken outside the sum. Similarly, it is easy to see that the product
$f(\bar\xi_{\so'},\xi_k)f(\bar\xi_{\st},\xi_k)=f(\bar\xi_k,\xi_k)$ does not depend on the partitions and it thus also can be taken out
the sum. We obtain
\begin{equation}\label{06-MCR-Fa-res1}
 \Bigl.F_{m,n}^{(l)}(\bu;\bxi)\Bigr|_{u_1\to\xi_k}= g(\xi_k,u_1)f(u_1,\bu_1)
 f(\bar\xi_k,\xi_k)\sum_{\bar\xi_k \mapsto\{\bar\xi_{\so'},\bar\xi_{\st}\}}
  K_{m-1}(\bar\xi_{\so'}|\bu_1)K_{n}(\bv|\bar\xi_{\st})
 f(\bar\xi_{\st},\bar\xi_{\so'})+\text{Reg}.
 \end{equation}
Obviously, the remaining sum over partitions is nothing but $F_{m-1,n}^{(l)}(\bu_1;\bxi_k)$. Hence,
\begin{equation}\label{06-MCR-Fa-res2}
 \Bigl.F_{m,n}^{(l)}(\bu;\bxi)\Bigr|_{u_1\to\xi_k}= g(\xi_k,u_1)\;f(u_1,\bu_1)
 f(\bar\xi_k,\xi_k)F_{m-1,n}^{(l)}(\bu_1;\bxi_k)+\text{Reg}.
 \end{equation}
Thus, the residue of the function $F_{m,n}^{(l)}(\bu;\bxi)$   at $u_1\to\xi_k$
reduces to the function $F_{m-1,n}^{(l)}(\bu_1;\bxi_k)$.

Consider now the behaviour of the rhs of \eqref{06-MCR-Part-old1} at $u_1\to\xi_k$. Let
\begin{equation}\label{06-MCR-RHS}
  F_{m,n}^{(r)}(\bu;\bxi)
 = (-1)^{m}f(\bar\xi,\bu) K_{m+n}(\{\bu-c,\bv\}|\bar\xi).
 \end{equation}
As in the case considered above, the function $F_{m,n}^{(r)}(\bu;\bxi)$ is an analytic function of the variable $u_1$ in the whole complex plane
with the exception of the points $\xi_k$, where it has simple poles. Although
the DWPF $K_{m+n}(\{\bu-c,\bv\}|\bar\xi)$ has poles at $u_1-c = \xi_k$, but the product $f(\bar\xi,\bu)$ vanishes at these points.
Therefore, the function $F_{m,n}^{(r)}(\bu;\bxi)$ turns out to be holomorphic at $u_1-c = \xi_k$.
The only singularities arise in the product $f(\bar\xi,\bu)$
at $u_1=\xi_k$. It is obvious that
\be{06-MCR-pref}
 f(\bar\xi,\bu)\Bigr|_{u_1\to\xi_k}=
  g(\xi_k,u_1)f(u_1,\bu_1)f(\bar\xi_k,\xi_k)
 f(\bar\xi_k,\bu_1)+\text{Reg}.
 \ee
Due to \eqref{06-K-red} we also have
\be{06-MCR-Kred}
K_{m+n}(\{\bu-c,\bv\}|\bar\xi)\Bigr|_{u_1=\xi_k}=-K_{m+n-1}(\{\bu_1-c,\bv\}|\bar\xi_k).
\ee
Substituting this into \eqref{06-MCR-RHS} we find
\be{06-MCR-RHS-res}
 F_{m,n}^{(r)}(\bu;\bxi)\Bigr|_{u_1\to\xi_k}=
g(\xi_k,u_1)\;f(u_1,\bu_1)f(\bar\xi_k,\xi_k)
F_{m-1,n}^{(r)}(\bu_1;\bxi_k)+\text{Reg}.
 \ee
Hence, taking into account \eqref{06-MCR-Fa-res2}, we obtain
\begin{multline}\label{06-MCR-LRHS-res}
  \Bigl(F_{m,n}^{(l)}(\bu;\bxi) - F_{m,n}^{(r)}(\bu;\bxi)\Bigr)\Bigr|_{u_1\to\xi_k}=
  g(\xi_k,u_1)\;f(u_1,\bu_1)f(\bar\xi_k,\xi_k) \\
  \times\Bigl(F_{m-1,n}^{(l)}(\bu_1;\bxi_k)-F_{m-1,n}^{(r)}(\bu_1;\bxi_k)\Bigr)+\text{Reg}.
 \end{multline}

Equation \eqref{06-MCR-LRHS-res} allows us to prove Lemma~\ref{06-MCR-main-ident} by induction. Let $m=0$.
Then $\bu=\emptyset$, and we have only one term in the sum over partitions:
$\bxi_{\so}=\emptyset$, $\bxi_{\st}=\bxi$.
The statement of the lemma becomes trivial. Suppose that the lemma holds
for some $m-1$ and consider the difference $F_{m,n}^{(l)}(\bu;\bxi) - F_{m,n}^{(r)}(\bu;\bxi)$
as a function of $u_1$ for all other variables fixed. Then by virtue of \eqref{06-MCR-LRHS-res}
the function $F_{m,n}^{(l)}(\bu;\bxi) - F_{m,n}^{(r)}(\bu;\bxi)$
has no poles at the points $u_1 = \xi_k$. In other words, the function $F_{m,n}^{(l)}(\bu;\bxi) - F_{m,n}^{(r)}(\bu;\bxi)$
turns out to be holomorphic in the whole complex plane. Hence, due to the Liouville theorem, it is identically
constant. Since $K_m$ and $K_{m + n}$ tend to zero as $u_1\to\infty$,
this constant is zero, that is, $F_{m,n}^{(l)}(\bu;\bxi) - F_{m,n}^{(r)}(\bu;\bxi)= 0$.

\begin{cor}\label{06-MCR-main-ident2}
Under the conditions of lemma~\ref{06-MCR-main-ident}
\begin{equation}\label{06-MCR-Part-old2}
  \sum
 K_{m}(\bar\xi_{\so}|\bu)K_{n}(\bv|\bar\xi_{\st})f(\bar\xi_{\st},\bar\xi_{\so})
 = (-1)^{n}f(\bv,\bar\xi) K_{m+n}(\bar\xi|\{\bu,\bv+c\}).
 \end{equation}
\end{cor}

Identity \eqref{06-MCR-Part-old2} follows from \eqref{06-MCR-Part-old1} due to \eqref{06-K-invers}.

\section{Multiple commutation relations\label{06-Sec4}}

In this section, we consider an example of the multiple commutation relations between the monodromy matrix entries.
Let us recall once more the commutation relations between the operators $A$ and $B$:
\begin{equation}
A(v)B(u)=f(u,v)B(u)A(v)+g(v,u)B(v)A(u).\label{06-MCR-AB}
\end{equation}
We have already considered the question of how to carry the operator
$A(v)$ through the product of several operators\footnote {The fact that the product $B(\bu)$ acted on the vacuum vector $|0\rangle$,
was not crucial. This simply led to the replacement of the operator $A$ by the function $a$.} $B(\bu)$. Now we  consider more
complex case of the permutation of the product of operators $A(\bv)$ with the product of the operators $B(\bu)$. It is clear that even if we use
equation
 \be{06-ABBB-act}
A(v)B(\bu)=f(\bu,v)B(\bu)A(v)+
 \sum_{k=1}^n g(v,u_k)f(\bu_k, u_k) B(v)B(\bu_k)A(u_k),
 \ee
and apply it several times, then the result will be too cumbersome. Meanwhile, there exist relatively compact formulas
for the multiple commutation relations of $A(\bv)$ and $B(\bu)$. We give them below.

\begin{prop}\label{06-MCR-Lem1}
Let $\#\bv=m$ and $\#\bu=n$. Then
\begin{equation}\label{06-MCR-two}
A(\bv)B(\bu)=(-1)^{n}\sum_{\{\bv,\bu\}\mapsto\{\bar w_{\so},\bar w_{\st}\}} K_{n}(\bu|\bar w_{\st}+c)f(\bar w_{\st},\bar w_{\so})
B(\bar w_{\st})A(\bar w_{\so}),
\end{equation}
and
\begin{equation}\label{06-MCR-two-t}
A(\bv)B(\bu)=(-1)^{m}\sum_{\{\bv,\bu\}\mapsto\{\bar w_{\so},\bar w_{\st}\}} K_{m}(\bar  w_{\so}|\bv+c)f(\bar w_{\st},\bar w_{\so})
B(\bar w_{\st})A(\bar w_{\so}).
\end{equation}
In these formulas, the sets $\bu$ and $\bv$ are combined into one set $\bar w$, that is $\{\bu,\bv\}=\bar w$. The
sum is taken over partitions $\bar w\mapsto\{\bar w_{\so},\bar w_{\st}\}$, such that $\#\bar w_{\so}=m$ and $\#\bar w_{\st}=n$.
\end{prop}

{\sl Remark.} Despite of both formulas look rather compact, the sums \eqref{06-MCR-two} and \eqref{06-MCR-two-t} contain $\tbinom{n+m}n$ terms each.

First, it is worth mentioning that equations \eqref{06-MCR-two} and \eqref{06-MCR-two-t} are equivalent.
They differ from each other only by different DWPF. But
using the property \eqref{06-K-red}, it is easy to prove that these DWPF are equal to each other
\begin{multline}\label{06-MCR-Ident-K}
(-1)^{n} K_{n}(\bu|\bar w_{\st}+c)=(-1)^{n+m} K_{n+m}(\{\bu,\bv\}|\{\bar w_{\st}+c,\bv+c\})\\
=(-1)^{n+m} K_{n+m}(\{\bar w_{\so},\bar w_{\st}\}|\{\bar w_{\st}+c,\bv+c\})
=(-1)^{m} K_{m}(\bar w_{\so}|\bv+c).
\end{multline}
Therefore, if we prove one of these formulas, then the other is proved automatically.

{\sl Proof.} We use induction over $n$ and $m$. For $n=m=1$, the set  $\bar w$ consists of two elements: $v$ and $u$. Thus, only
two partitions are possible:
either $\bar w_{\so}=v$, $\bar w_{\st}=u$, or  $\bar w_{\so}=u$, $\bar w_{\st}=v$. Then we obtain
\begin{equation}\label{06-MCR-A1B1}
A(v)B(u)=- K_{1}(u|u+c)f(u,v)
B(u)A(v)- K_{1}(u|v+c)f(v,u)
B(v)A(u).
\end{equation}
A direct calculation shows that
\begin{equation}\label{06-MCR-K,K}
K_{1}(u|u+c)=-1,\qquad K_{1}(u|v+c)f(v,u)=-g(v,u),
\end{equation}
and substituting these expressions into \eqref{06-MCR-A1B1} we arrive at \eqref{06-MCR-AB}.

Assume now that equations \eqref{06-MCR-two}, \eqref{06-MCR-two-t} are valid for
$m=\#\bv=1$ and some  $n-1=\#\bu$, where $n\ge 1$. In spite of the set $\bv$ consists of only one element, we still write a bar
over $v$. We thus emphasise that we treat $\bv$ as a set of variables.

Consider the case $\#\bu=n$ and let  $\bu_n=\bu\setminus u_n$.
Then due to the induction assumption
\begin{equation}\label{06-MCR-A1Bn-1}
A(\bv)B(\bu)=(-1)^{n-1}\sum_{\{\bv,\bu_n\}\mapsto\{\bar w_{\so},\bar w_{\st}\}} K_{n-1}(\bu_n|\bar w_{\st}+c)
f(\bar w_{\st},\bar w_{\so})B(\bar w_{\st})A(\bar w_{\so})B(u_n).
\end{equation}
Here we used equation \eqref{06-MCR-two}. Then we permute the operators $A(\bar w_{\so})B(u_n)$ via
\eqref{06-MCR-two-t}:
\begin{multline}\label{06-MCR-A1Bn-2}
A(\bv)B(\bu)=(-1)^{n}\sum_{\substack{\{\bv,\bu_n\}\mapsto\{\bar w_{\so},\bar w_{\st}\}\\
\{\bar w_{\so},u_n\}\mapsto\{\bar w_{\sth},\bar w_{\stf}\}}} K_{n-1}(\bu_n|\bar w_{\st}+c)
f(\bar w_{\st},\bar w_{\so})B(\bar w_{\st})\\
\times K_1(u_n|\bar w_{\stf}+c)f(\bar w_{\stf},\bar w_{\sth})B(\bar w_{\stf}) A(\bar w_{\sth}).
\end{multline}
In fact, we have a double sum over partitions in  \eqref{06-MCR-A1Bn-2}. First, we divide the set
$\{\bv,\bu_n\}$ into subsets $\bar w_{\so}$ and $\bar w_{\st}$. Then the set
$\{\bar w_{\so},u_n\}$  is divided into subsets $\bar w_{\sth}$ and $\bar w_{\stf}$.

The goal of further transformations is to calculate partly the sum  over partitions. For this, we have only one tool, namely,
lemma~\ref{06-MCR-main-ident}. Therefore, our task is to
transform  the expression \eqref{06-MCR-A1Bn-2} to such a form that application of lemma~\ref{06-MCR-main-ident} becomes possible.

Pay attention that in the formula \eqref{06-MCR-A1Bn-2},
we deal with the division of the complete set $\{\bv,\bu\}$ into three
subsets $\bar w_{\st}$, $\bar w_{\sth}$, and $\bar w_{\stf}$.
Although, as we have already mentioned above, this division was done in two stages.
In the first stage, we also had the subset $\bar w_ {\so}$. Let us get rid of this intermediate subset.

The subset $\bar w_{\so}$ is contained only in the product $f(\bar w_{\st},\bar w_{\so})$. Let us multiply and divide it by $f(\bar w_{\st},u_n)$:
\be{06-MCR-fff-0}
f(\bar w_{\st},\bar w_{\so})=\frac{f(\bar w_{\st},\bar w_{\so})f(\bar w_{\st},u_n)}{f(\bar w_{\st},u_n)}.
\ee
Then, using that $\{\bar w_{\so},u_n\}=\{\bar w_{\sth},\bar w_{\stf}\}$, we recast it in the form
\be{06-MCR-fff}
f(\bar w_{\st},\bar w_{\so})=\frac{f(\bar w_{\st},\bar w_{\so})f(\bar w_{\st},u_n)}{f(\bar w_{\st},u_n)}
=\frac{f(\bar w_{\st},\bar w_{\sth})f(\bar w_{\st},\bar w_{\stf})}{f(\bar w_{\st},u_n)}.
\ee
Substituting this into \eqref{06-MCR-A1Bn-2} we obtain
\begin{multline}\label{06-MCR-A1Bn-3}
A(\bv)B(\bu)=(-1)^{n}\sum_{\substack{\{\bv,\bu_n\}\mapsto\{\bar w_{\st},\bar w_{\sth}, \bar w_{\stf}\}\\
u_n\notin\bar w_{\st}}}
K_1(u_n|\bar w_{\stf}+c)K_{n-1}(\bu_n|\bar w_{\st}+c)\\
\times \frac{f(\bar w_{\st},\bar w_{\sth})f(\bar w_{\st},\bar w_{\stf})f(\bar w_{\stf},\bar w_{\sth})}
{f(\bar w_{\st},u_n)}\;B(\bar w_{\st})B(\bar w_{\stf}) A(\bar w_{\sth}).
\end{multline}
Now we do not have the subset $\bar w_{\so}$ in the formula \eqref{06-MCR-A1Bn-3}, and the complete set
$\{\bv,\bu\}$ indeed is divided into subsets
$\bar w_{\st}$, $\bar w_{\sth}$, and $\bar w_{\stf}$. There is, however, one restriction (besides natural restrictions on
the cardinalities of the subsets). The parameter $u_n$ does not belong to the subset $\bar w_{\st}$.
Indeed, $\bar w_{\st}$ is the subset of the set $\{\bv,\bu_n\}$. The latter does not contain the parameter $u_n$.
However, we have a factor $f^{-1}(\bar w_{\st},u_n)$ in  \eqref{06-MCR-A1Bn-3}, that vanishes as soon as
$u_n\in\bar w_{\st}$. Thus, the condition $u_n\notin\bar w_{\st}$ holds automatically, and we can say that the sum is taken
over partitions of the set $\{\bv,\bu\}$
into subsets $\bar w_{\st}$, $\bar w_{\sth}$, and $\bar w_{\stf}$ without any restrictions.

One can also say that the product in the rhs of \eqref{06-MCR-fff} is a projection of the product  $f(\bar w_{\st},\bar w_{\so})$
on such the partitions, for which $u_n\notin\bar w_{\st}$. Indeed, if $u_n\in\bar w_{\st}$, then the rhs of \eqref{06-MCR-fff}
vanishes. If, however, $u_n\notin\bar w_{\st}$, then $u_n\in \{\bar w_{\sth},\bar w_{\stf}\}$, and we can say that
$\{\bar w_{\sth}, \bar w_{\stf}\}=\{\bar w_{\so}, u_n\}$, what give us the lhs of \eqref{06-MCR-fff}.

Thus, we obtain
\begin{multline}\label{06-MCR-A1Bn-4}
A(\bv)B(\bu)=(-1)^{n}\sum_{\{\bv,\bu\}\mapsto\{\bar w_{\st},\bar w_{\sth},\bar w_{\stf}\}}
K_1(u_n|\bar w_{\stf}+c)K_{n-1}(\bu_n|\bar w_{\st}+c)\\
\times \frac{f(\bar w_{\st},\bar w_{\sth})f(\bar w_{\st},\bar w_{\stf})f(\bar w_{\stf},\bar w_{\sth})}
{f(\bar w_{\st},u_n)}\;B(\bar w_{\st})B(\bar w_{\stf}) A(\bar w_{\sth}).
\end{multline}

Now we need to understand which partitions anyway cannot be summed up. It is easy to see that in this case this is any partition,
that affects the subset $\bar w_{\sth}$. Indeed, the formula \eqref{06-MCR-A1Bn-4} has the product $A(\bar w_{\sth})$.
Hence, the terms of the sum corresponding to different subsets of $\bar w_{\sth}$ contain different products of the operators $A$. On the contrary, summation over partitions into subsets of $\bar w_{\st}$ and $\bar w_{\stf}$ can be attempted. Indeed,
let us introduce a new subset $\{\bar w_{\st},\bar w_{\stf}\}=\bar w_{0}$.
We see that the product of the operators $B$ in the formula \eqref{06-MCR-A1Bn-4}
is taken exactly over this subset. Therefore, for any partition $\bar w_{0}\mapsto\{\bar w_{\st},\bar w_{\stf}\}$
the product of the operators $B$ remains the same, but only the coefficient of this product changes. This coefficient is a rational function of the parameters
$\bar w$, and we can try to calculate the sum of rational functions.

We rewrite the formula \eqref{06-MCR-A1Bn-4} with the new subset $\bar w_{0}$:
\begin{multline}\label{06-MCR-A1Bn-40}
A(\bv)B(\bu)=(-1)^{n}\sum_{\{\bv,\bu\}\mapsto\{\bar w_{0},\bar w_{\sth}\}}\;B(\bar w_{0}) A(\bar w_{\sth})f(\bar w_{0},\bar w_{\sth})\\
\times \sum_{\bar w_{0}\mapsto\{\bar w_{\st},\bar w_{\stf}\}}
K_1(u_n|\bar w_{\stf}+c)K_{n-1}(\bu_n|\bar w_{\st}+c)
 \frac{f(\bar w_{\st},\bar w_{\stf})}{f(\bar w_{\st},u_n)}.
\end{multline}
Now the summation is organized as follows. First, the set $\{\bv,\bu\}$ is divided into two subsets
$\bar w_{0}$ and $\bar w_{\sth}$. Then, the subsets $\bar w_{0}$  is divided again into two `sub-subsets'
$\bar w_{\st}$ and $\bar w_{\stf}$.
We should transform the sum over partitions in the second line of  \eqref{06-MCR-A1Bn-40} to the form \eqref{06-MCR-Part-old1}. For this
we transform the function $K_1(u_n|\bar w_{\stf}+c)$ via \eqref{06-K-invers}
\be{06-MCR-K,K2}
K_1(u_n|\bar w_{\stf}+c)=-f^{-1}(\bar w_{\stf},u_n)\;K_1(\bar w_{\stf}|u_n),
\ee
Equation \eqref{06-MCR-A1Bn-40} takes the form
\begin{multline}\label{06-MCR-A1Bn-5}
A(\bv)B(\bu)=(-1)^{n-1}\sum_{\{\bv,\bu\}\mapsto\{\bar w_{0},\bar w_{\sth}\}}\;B(\bar w_{0}) A(\bar w_{\sth})
\frac{f(\bar w_{0},\bar w_{\sth})}{f(\bar w_{0},u_n)}\\
\times \sum_{\bar w_{0}\mapsto\{\bar w_{\st},\bar w_{\stf}\}}
K_1(\bar w_{\stf}|u_n)K_{n-1}(\bu_1-c|\bar w_{\st})f(\bar w_{\st},\bar w_{\stf}),
\end{multline}
where we have used an evident relation $K_{n-1}(\bu_n|\bar w_{\st}+c)=K_{n-1}(\bu_n-c|\bar w_{\st})$.
Now the sum over partitions
$\bar w_{0}\mapsto\{\bar w_{\st},\bar w_{\stf}\}$ can be calculated via lemma~\ref{06-MCR-main-ident}
\be{06-MCR-w0}
\sum_{\bar w_{0}\mapsto\{\bar w_{\st},\bar w_{\stf}\}}
K_1(\bar w_{\stf}|u_n)K_{n-1}(\bu_1-c|\bar w_{\st})f(\bar w_{\st},\bar w_{\stf})=
-f(\bar w_{0},u_n)K_n(\bu-c|\bar w_{0}).
\ee
Substituting this into \eqref{06-MCR-A1Bn-5} we obtain
\begin{equation}\label{06-MCR-A1Bn-6}
A(\bv)B(\bu)=(-1)^{n}\sum_{\{\bv,\bu\}\mapsto\{\bar w_{0},\bar w_{\sth}\}}
K_n(\bu-c|\bar w_{0}) f(\bar w_{0},\bar w_{\sth})B(\bar w_{0}) A(\bar w_{\sth}).
\end{equation}
It remains to redenote
$\bar w_{0}\to \bar w_{\st}$ and $\bar w_{\sth}\to \bar w_{\so}$, and we arrive at
\eqref{06-MCR-two} with $\#\bv=1$ and $\#\bu=n$.

Thus, we proved equations \eqref{06-MCR-two}, \eqref{06-MCR-two-t} for $m=1$ and arbitrary $n$. Now we should use
induction over $m$. This can be done in a completely analogous way to induction over $n$. Let us give, nevertheless, the key
formulas. Let \eqref{06-MCR-two}, \eqref{06-MCR-two-t} be true for some $m-1$, and let $\bv_m=\bv\setminus v_m$.
Then, due to \eqref{06-MCR-two-t}
\begin{equation}\label{06-MCR-AmBn-1}
A(\bv)B(\bu)=(-1)^{m-1}\sum_{\{\bv_m,\bu\}\mapsto\{\bar w_{\so},\bar w_{\st}\}} K_{m-1}(\bar w_{\so}|\bv_m+c)
f(\bar w_{\st},\bar w_{\so})A(v_m)B(\bar w_{\st})A(\bar w_{\so}).
\end{equation}
Then we permute the operators $A(v_m)B(\bar w_{\st})$ via
\eqref{06-MCR-two}:
\begin{multline}\label{06-MCR-AmBn-2}
A(\bv)B(\bu)=(-1)^{m}\sum_{\substack{\{\bv_m,\bu\}\mapsto\{\bar w_{\so},\bar w_{\st}\}\\
\{\bar w_{\st},v_n\}\mapsto\{\bar w_{\sth},\bar w_{\stf}\}}} K_{m-1}(\bar w_{\so}|\bv_m+c)
f(\bar w_{\st},\bar w_{\so})B(\bar w_{\stf})\\
\times K_1(\bar w_{\sth}|v_m+c)f(\bar w_{\stf},\bar w_{\sth})A(\bar w_{\sth}) A(\bar w_{\so}).
\end{multline}
We again deal with a double sum over partitions: first, the set
$\{\bv_m,\bu\}$ is divided into subsets $\bar w_{\so}$ and $\bar w_{\st}$, then the set
$\{\bar w_{\st},v_m\}$  is divided into subsets  $\bar w_{\sth}$ and $\bar w_{\stf}$.
Similarly to equation \eqref{06-MCR-fff}, we have
\be{06-MCR-fff-2}
f(\bar w_{\st},\bar w_{\so})=\frac{f(\bar w_{\st},\bar w_{\so})f(v_m,\bar w_{\so})}{f(v_m,\bar w_{\so})}
=\frac{f(\bar w_{\sth},\bar w_{\so})f(\bar w_{\stf},\bar w_{\so})}{f(v_m,\bar w_{\so})},
\ee
therefore,
\begin{multline}\label{06-MCR-AmBn-3}
A(\bv)B(\bu)=(-1)^{m}\sum_{\{\bv,\bu\}\mapsto\{\bar w_{\so},\bar w_{\sth},\bar w_{\stf}\}}
K_{m-1}(\bar w_{\so}|\bv_m+c)K_1(\bar w_{\sth}|v_m+c)
\\
\times
\frac{f(\bar w_{\sth},\bar w_{\so})f(\bar w_{\stf},\bar w_{\so})f(\bar w_{\stf},\bar w_{\sth})}{f(v_m,\bar w_{\so})}
B(\bar w_{\stf})A(\bar w_{\sth}) A(\bar w_{\so}).
\end{multline}
Here the sum is taken over partitions of the complete set $\{\bv,\bu\}$ into three subsets $\bar w_{\so}$, $\bar w_{\sth}$, and $\bar w_{\stf}$.
The initial restriction  $v_m\notin\bar w_{\so}$ holds automatically due to the product
$f^{-1}(v_m,\bar w_{\so})$.

Let us introduce a new set $\bar w_{0}=\{\bar w_{\so},\bar w_{\sth}\}$ and transform $K_1(\bar w_{\sth}|v_m+c)$
in the spirit of \eqref{06-MCR-K,K2}. Then we obtain
\begin{multline}\label{06-MCR-AmBn-4}
A(\bv)B(\bu)=(-1)^{m-1}\sum_{\substack{\{\bv,\bu\}\mapsto\{\bar w_{0},\bar w_{\stf}\}\\
\bar w_{0}\mapsto\{\bar w_{\so},\bar w_{\sth}\}  }}
K_{m-1}(\bar w_{\so}|\bv_m+c)K_1(v_m|\bar w_{\sth})f(\bar w_{\sth},\bar w_{\so})
\\
\times
\frac{f(\bar w_{\sth},\bar w_{0})}{f(v_m,\bar w_{0})}
B(\bar w_{\stf})A(\bar w_{0}).
\end{multline}
In the formula \eqref{06-MCR-AmBn-4}, the set $\{\bv,\bu\}$ is firstly divided into subsets $\bar w_{0}$ and $\bar w_{\stf}$,
and after this the subset $\bar w_{0}$  is divided into $\bar w_{\so}$ and $\bar w_{\sth}$. The sum over partitions
$\bar w_{0}\mapsto\{\bar w_{\so},\bar w_{\sth}\}$ can be computed via \eqref{06-MCR-Part-old2}
\begin{equation}\label{06-MCR-w0-w1w3}
\sum_{\bar w_{0}\mapsto\{\bar w_{\so},\bar w_{\sth}\}  }
K_{m-1}(\bar w_{\so}|\bv_m+c)K_1(v_m|\bar w_{\sth})f(\bar w_{\sth},\bar w_{\so})=-
f(v_m,\bar w_{0})\;K_{m}(\bar w_{0}|\bv+c).
\end{equation}
Taking into account \eqref{06-MCR-w0-w1w3} we eventually obtain
\begin{equation}\label{06-MCR-AmBn-5}
A(\bv)B(\bu)=(-1)^{m}\sum_{\{\bv,\bu\}\mapsto\{\bar w_{0},\bar w_{\stf}\}}
K_{m}(\bar w_{0}|\bv+c)f(\bar w_{\sth},\bar w_{0})
B(\bar w_{\stf})A(\bar w_{0}),
\end{equation}
what coincides with \eqref{06-MCR-two-t} up to the notation.

{\sl Example}. Consider a particular case of the formula \eqref{06-MCR-two-t} for  $m=1$, that is, $\bv=v$. Then  $\bar  w=\{v,\bu\}$, and the subset
$\bar  w_{\so}$ consists of one element. There exist two possibilities: either $\bar  w_{\so}=v$, and then $\bar  w_{\st}=\bu$; or $\bar  w_{\so}=u_k$,
and then $\bar  w_{\st}=\{v,\bu_k\}$, where $k=1,\dots,n$. In the first case we have
\be{06-1case}
K_{1}(\bar  w_{\so}|\bv+c)f(\bar w_{\st},\bar w_{\so})=K_{1}(\bv|\bv+c)f(\bu,v)=-f(\bu,v).
\ee
In the second case
\be{06-2case}
K_{1}(\bar  w_{\so}|\bv+c)f(\bar w_{\st},\bar w_{\so})=K_{1}(u_k|\bv+c)f(v,u_k)f(\bu_k,u_k)=-g(v,u_k)f(\bu_k,u_k).
\ee
Obviously, all together these contributions give equation \eqref{06-ABBB-act}.

\section{Other commutation relations\label{06-Sec5}}

Let us give some other multiple commutation relations

\begin{prop}\label{06-MCR-LemCD}
Let $\#\bv=m$ and $\#\bu=n$. Then
\begin{enumerate}
\item
\begin{equation}\label{06-MCR-two0}
C(\bv)D(\bu)=(-1)^{n}\sum_{\{\bv,\bu\}\mapsto\{\bar w_{\so},\bar w_{\st}\}} K_{n}(\bu|\bar w_{\st}+c)f(\bar w_{\st},\bar w_{\so})
D(\bar w_{\st})C(\bar w_{\so}),
\end{equation}
and
\begin{equation}\label{06-MCR-two-t0}
C(\bv)D(\bu)=(-1)^{m}\sum_{\{\bv,\bu\}\mapsto\{\bar w_{\so},\bar w_{\st}\}} K_{m}(\bar  w_{\so}|\bv+c)f(\bar w_{\st},\bar w_{\so})
D(\bar w_{\st})C(\bar w_{\so}).
\end{equation}
\item
\be{06-MCR-DB}
D(\bv)B(\bu)=(-1)^{n}\sum_{\{\bv,\bu\}\mapsto\{\bar w_{\so},\bar w_{\st}\}} K_{n}(\bar w_{\st}|\bu+c)f(\bar w_{\so},\bar w_{\st})
B(\bar w_{\st})D(\bar w_{\so}),
\ee
and
\be{06-MCR-DB-t}
D(\bv)B(\bu)=(-1)^{m}\sum_{\{\bv,\bu\}\mapsto\{\bar w_{\so},\bar w_{\st}\}} K_{m}(\bv|\bar w_{\so}+c)f(\bar w_{\so},\bar w_{\st})
B(\bar w_{\st})D(\bar w_{\so}).
\ee
\item
\be{06-MCR-CA}
C(\bv)A(\bu)=(-1)^{n}\sum_{\{\bv,\bu\}\mapsto\{\bar w_{\so},\bar w_{\st}\}} K_{n}(\bar w_{\st}|\bu+c)f(\bar w_{\so},\bar w_{\st})
A(\bar w_{\st})C(\bar w_{\so}),
\ee
and
\be{06-MCR-CA-t}
C(\bv)A(\bu)=(-1)^{m}\sum_{\{\bv,\bu\}\mapsto\{\bar w_{\so},\bar w_{\st}\}} K_{m}(\bv|\bar w_{\so}+c)f(\bar w_{\so},\bar w_{\st})
A(\bar w_{\st})C(\bar w_{\so}).
\ee
\end{enumerate}
In all these formulas, the sets  $\bu$ and $\bv$ are combined into one set $\bar w$, that is $\{\bu,\bv\}=\bar w$. The sums are taken
over partitions $\bar w\mapsto\{\bar w_{\so},\bar w_{\st}\}$, such that  $\#\bar w_{\so}=m$ and $\#\bar w_{\st}=n$.
\end{prop}

Actually, we proved already the relations \eqref{06-MCR-two0}, \eqref{06-MCR-two-t0} because the commutation relations for the operators
$C$ and $D$ are the same as for  $A$ and $B$. The remaining commutation relations can be proved in the same way as we did above. One can, however,
use a more simple method by remembering the possibility of replacing $c$ with $-c$. For example, the commutation relations for
$D$ and $B$ are obtained from the ones
for $C$ and $D$ when $c$ is replaced by $-c$. Consider, for example, the formula \eqref{06-MCR-two-t0} and replace the operator $C$ by $D$, the operator
$D$ by $B$, and the constant $c$ on $-c$. Then, taking into account
\be{06-uchet}
\begin{aligned}
&f(\bar w_{\st},\bar w_{\so})\Bigr|_{c\to -c}=f(\bar w_{\so},\bar w_{\st}),\\
&K_{m}(\bar  w_{\so}|\bv+c)\Bigr|_{c\to -c}= K_{m}(\bv-c|\bar  w_{\so})=K_{m}(\bv|\bar  w_{\so}+c),
\end{aligned}
\ee
we immediately arrive at the commutation relation \eqref{06-MCR-DB-t}.

\subsection{One more identity for DWPF\label{06-Sec6}}

The following statement takes place for the models with the rational $R$-matrix only. It is not valid for the models with the
trigonometric $R$-matrix.

\begin{prop}\label{06-PROP-fun}
Let  $\#\bu=n$ and $\#\bv=m$. Then
\begin{equation}\label{06-MCR-two-g-rat2}
\sum_{\{\bv,\bu\}\mapsto\{\bar w_{\so},\bar w_{\st}\}} K_{n}(\bu|\bar w_{\st}+c)f(\bar w_{\st},\bar w_{\so})
=(-1)^{n}.
\end{equation}
Here the sets  $\bu$ and $\bv$ are combined into one set $\bar w$, that is $\{\bu,\bv\}=\bar w$. The sum is taken over partitions
$\bar w\mapsto\{\bar w_{\so},\bar w_{\st}\}$, such that  $\#\bar w_{\so}=m$ and $\#\bar w_{\st}=n$.
\end{prop}

This identity can be proved, for example, by induction, but in fact it has already been implicitly proven. Did the reader notice this?
As a hint, we once again draw the attention of the reader to the fact that  \eqref{06-MCR-two-g-rat2} is valid only
in the case when the DWPF is a rational function \eqref{06-K-def}.


%
%


\chapter{Scalar products of Bethe vectors\label{CHA-SP}}

This lecture is devoted to the scalar products of Bethe vectors.
It is to these objects the calculation of all correlation functions eventually reduces.
Therefore, it is so important to obtain convenient and compact representations
for the scalar products. In the case when both vectors are generic Bethe vectors, such representations
for today are not found. However, if one of the vectors is a (twisted) on-shell vector, then
a representation  for the scalar product is known in the form of a determinant of some matrix \cite{Sla89}.
We will obtain this formula in this lecture.

The determinant representations are quite convenient both for analytical and numerical calculations.
The matter is that with increasing matrix size, the time required to calculate its determinant grows polynomially\footnote{More precisely, the determinant of
an $n\times n$ matrix  can be calculated in time $O(n^{2.376})$.}. This is the main advantage of determinant representations in comparison, say,
with representations in the form of sums over partitions of a certain set of elements, where  the number of terms in the sum
grows exponentially, as the number of elements in the set increases.

For definiteness, in this lecture we are still considering models with the rational $R$-matrix. However, all the presentation can be
easily generalized to the case of the trigonometric $R$-matrix. In particular, for this reason, most of the formulas are given in the universal
form (that is, in terms of the functions $f$, $g$ and their combinations). We recommend the reader to check independently (where required)
the validity of different statements for the case of the trigonometric $R$-matrix.

\section{Dual Bethe vectors\label{07-Sec1}}

The elements of the monodromy matrix act in a Hilbert space $\mathcal{H}$ with the vacuum vector $|0\rangle$.
Assume that they also act in a dual space $\mathcal{H}^*$ with a dual vacuum vector $\langle0|$, that has properties
\be{07-dvac}
\langle0|A(u)=a(u)\langle0|,\qquad \langle0|D(u)=d(u)\langle0|,\qquad\langle0|B(u)=0,
\ee
where functions $a(u)$ and $d(u)$ are the same as in the case of the vacuum vector $|0\rangle$. We also assume that
\be{07-norm}
\langle 0|= |0\rangle^\dagger,\qquad  \langle 0|0\rangle=1.
\ee

In the dual space, the operator $C$ plays the role of the creation operator. Using this operator
one can  construct the dual Bethe vectors $\langle0|C(\bu)$. Properties of these vectors are completely analogous to the properties
of the ordinary Bethe vectors. In particular, it is easy to show that the dual vector becomes an eigenvector of the transfer matrix (an on-shell dual
Bethe vector), if the parameters $\bu $ satisfy the system of Bethe equations.

Are the dual vectors Hermitian conjugated to the usual Bethe vectors? The answer to this question depends on the specific model and on the values of its parameters. For example, in the case of the $XXX$ chain for $c = i$, we have $D(u^*)=A^\dagger(u)$ and  $C(u^*)=-B^\dagger(u)$.
These properties can easily be proved by induction on the number of sites in the chain. For one $L$-operator
\be{07-Ln}
L_n(u)=\begin{pmatrix}
u+\frac i2\;\sigma_n^z &i\;\sigma_n^-\\
i\;\sigma_n^+ &u-\frac i2\;\sigma_n^z
\end{pmatrix}
\ee
they are correct. Next we assume that they are true for $T(u)$, which is the product of $(N-1)$ $L$-operators, and multiply
this monodromy matrix with the $L$-operator in the  $N$th site. By a direct calculation, we see that the relations between the elements
of the monodromy matrices are still valid.

Hence, if $\bu^*=\bu$ (that is, the set $\bu=\{u_1,\dots,u_n\}$ consists of real or complex conjugated numbers), then
\be{07-conj}
\langle 0|C(\bu)= (-1)^n\bigl(B(\bu)|0\rangle\bigr)^\dagger.
\ee
The solutions of the Bethe equations do have the property $\bu^*=\bu$, therefore, in the case of on-shell vectors the property \eqref{07-conj} is satisfied\footnote{%
The factor $(-1)^n$ does not play any role, because eigenvectors of operators anyway are determined up to a phase factor.}.
However, if we consider the inhomogeneous $XXX$ chain or we deal with twisted Bethe equations, then all of the above properties
of the operators and vectors can be violated. Generally speaking, for such `non-physical' models, the  basis of the on-shell vectors $B(\bu)|0\rangle$
is not orthogonal. Therefore, in this case, we should find in the dual space a dual basis, which does not have to be Hermitian conjugated to the original basis. As we will see later, the dual Bethe vectors $\langle 0|C(\bu)$ do form the desired dual basis.

\section{Action of $C$ operators on Bethe vectors\label{07-Sec2}}

Multiple commutation relations for the products of the operators $C$ and $B$ have a much more complex structure than the formulas considered in the previous lecture. However, the problem is somewhat simplified if the product of the operators $B(\bu)$
acts on the vacuum $|0\rangle$.
This particular case of the multiple commutation relations is very important, because it leads to formulas for the scalar products of Bethe
vectors\footnote{Here and below, for brevity, we do not distinguish between Bethe vectors and dual Bethe vectors,
if this does not lead to misunderstanding.}.

\begin{prop}\label{07-SP-actCBBB}
Let $\#\bv=n$, $\#\bu=m$, and $m\ge n$. Then
\begin{multline}\label{07-SP-actC}
C(\bv)B(\bu)|0\rangle=\sum_{\{\bv,\bu\}\mapsto\{\bar \xi_{\so},\bar \xi_{\st},\bar \xi_{\sth}\}}
 d(\bar \xi_{\so})a(\bar \xi_{\st})K_n(\bv|\bar \xi_{\so}+c) K_n(\bar \xi_{\st}|\bv+c) \\
\times f(\bar \xi_{\so},\bar \xi_{\st})
 f(\bar \xi_{\so},\bar \xi_{\sth})f(\bar \xi_{\sth},\bar \xi_{\st})
\;B(\bar \xi_{\sth})|0\rangle\,.
\end{multline}
Here $K_n$ are DWPF. The sum is taken over partitions of the set
$\{\bv,\bu\}$ into three subsets $\bar \xi_{\so}$, $\bar \xi_{\st}$, and $\bar \xi_{\sth}$, such that
$\#\bar \xi_{\so}=\#\bar \xi_{\st}=n$.
\end{prop}

One can prove this proposition similarly to the proof of the multiple commutation relations for the products of the operators
$A(\bv)$ and $B(\bu)$, that is, using double induction over $m$ and $n$. However, we will use another way. We first obtain
equation \eqref{07-SP-actC} for $n=1$ and arbitrary $m$ using the standard considerations of the algebraic Bethe ansatz,
and only then we use induction over $n$.

\subsection{Action of one operator $C$ on Bethe vector\label{07-Sec21}}

If you have already done this exercise, then this section can be skipped. However, you can compare your own calculations with those,
that are given below.

Recall that first we need to understand a general form of the result. Namely, we should describe the types
of the contributions that appear in the result of  the operator $C(v)$  action  on the vector $B(\bu)|0\rangle$.
Then we choose some specific term and try to obtain it, neglecting all other contributions. In doing so, we can use
the fact that the operators $B$ commute with each other, therefore, they can be reordered.

Permuting the operators $C$ and $B$ via the relation  \eqref{03-CR-BC}

\be{07-CB-AD}
[C(v),B(u)]=g(v,u)\bigl(D(v)A(u)-D(u)A(v)\bigr)
\ee
we obtain the operators $A$ and $D$.
The latter, in turn, when moving to the right will be permuted with the operators $B (u_j)$. As a result of these permutations
the operators $A$ and $D$ can either keep their arguments, or exchange them with the arguments of the  $B$ operators.
Our goal is  to  reorder the $B$ operators in such a way, that for obtaining the desired contribution, the operators $A$ and $D$ can move to the right in the only possible way. Let us now consider this scheme in more detail.

When we act with the operator $C(v)$ on the vector $B(\bu)|0\rangle$, we should obtain at some stage
the commutator $[C(v),B(u_{\ell})]$, because $C(v)|0\rangle=0$:
\be{07-SP-sum-comm}
C(v)B(\bu)|0\rangle=\sum_{\ell=1}^m B(u_1)\cdots B(u_{\ell-1})
[C(v),B(u_{\ell})] B(u_{\ell+1})\cdots B(u_m)|0\rangle\,.
\ee
Due to relation \eqref{07-CB-AD} we obtain
\be{07-SP-sum-comm1}
C(v)B(\bu)|0\rangle=\sum_{\ell=1}^m g(v,u_{\ell}) B(u_1)\cdots B(u_{\ell-1})
\bigl(D(v)A(u_{\ell})-D(u_{\ell})A(v)\bigr) B(u_{\ell+1})\cdots B(u_m)|0\rangle\,.
\ee
Then we should move the operators $A$ and $D$ to the right until they act on $|0\rangle$ and respectively give the functions
$a$ and $d$. Then we find the action of the operator $C(v)$ on the vector $B(\bu)|0\rangle$.

Note that there exists three possibilities after the operators $A$ and $D$ have approached the extreme right position: the operator $A$
depends on $v$, while the operator $D$ depends on some $u_k$; the operator $A$
depends on some $u_k$, while the operator $D$ depends on $v$; the operator $A$
depends on some $u_j$, while the operator $D$ depends on some $u_k$. As a result of the action of these operators on the
vacuum vector $|0\rangle$  we obtain contributions proportional to either
$a(v)d(u_k)$, or  $d(v)a(u_k)$, or  $a(u_j)d(u_k)$. It is also easy to see that in the first and second cases we loose the operator $B(u_k)$ in the product
of the $B$ operators. In the third case, we loose the operators  $B(u_k)$ and $B(u_j)$, but instead we obtain a new operator $B(v)$. Therefore, the result
can be written in the following form:
\begin{multline}\label{07-SP-ans-vid}
C(v)B(\bu)|0\rangle=\sum_{k=1}^m \bigl(d(v)a(u_k) X_k + a(v)d(u_k)\tilde X_k \bigr)
B(\bu_k)|0\rangle\\
+ \sum_{j<k}^m \bigl(a(u_j)d(u_k)Y_{jk}+a(u_k)d(u_j)Y_{kj}\bigr)
B(v)B(\bu_{jk})|0\rangle\,.
\end{multline}
Here $X_k$, $\tilde X_k$, and $Y_{jk}$ are some numerical coefficients to be determined. The symbol $\bu_{jk}$ means $\bu_{jk}=\bu\setminus\{u_j,u_k\}$.
Thus, we have found a schematic form of the result, and the first stage of the derivation is completed.

At the second stage we fix some $k$ and compute the coefficient $X_k$. All other contributions should be neglected\footnote[1]{%
For convenience, one can say that $d(u_s)=0$ for all $s=1,\dots,m$, as well as  $a(v)=0$ and $a(u_s)=0$
for all $s\ne k$. Then only one term survives in \eqref{07-SP-ans-vid}, and this contribution is proportional to
$X_k$.}. Using commutativity of the $B$ operators we can move $B(u_k)$ to the extreme left position with respect to all other
$B$ operators:
\be{07-SP-k-perv}
C(v)B(\bu)|0\rangle=C(v)B(u_k)B(\bu_k)|0\rangle\,.
\ee
Now it becomes obvious, that moving the operator $C(v)$ to the right, we should obtain the commutator
$[C(v),B(u_k)]$ at the first step. Otherwise we obtain $B(u_k)C(v)B(\bu_k)|0\rangle$, and hence,
the operator $B(u_k)$ is  to the left from the operator $C(v)$. Therefore, when moving $C(v)$ further to the right
we do not touch  $B(u_k)$, and this operator remains in the final result. At the same time, the coefficient
$d(v)a(u_k)X_k$ is multiplied by a vector that does not depend on $B(u_k)$. Hence, we can write
\be{07-SP-first-com}
C(v)B(\bu)|0\rangle=g(v,u_k)\bigl(D(v)A(u_k)-D(u_k)A(v)\bigr) B(\bu_k)|0\rangle
+\mathcal{Z}\,,
\ee
where we have denoted by $\mathcal{Z}$ the terms that do not contribute to the coefficient $X_k$.

It is easy to see that $D(u_k)A(v)$ also does not contribute to the desired result. Indeed, when the operator
$A(v)$ moves to the right, it can keep its argument $v$, or exchange it with an argument of one of the  $B$ operators. In the first case, after the action on the vacuum it gives $a(v)$, in the second case it gives $a(u_j)$ with $u_j\ne u_k$. No one of these functions can appear with the coefficient $X_k$. Thus,
\be{07-SP-first-com1}
C(v)B(\bu)|0\rangle=g(v,u_k)D(v)A(u_k)B(\bu_k)|0\rangle
+\mathcal{Z}\,.
\ee
Now, when moving  $A(v)$ to the right, we should keep its argument $u_k$. If we allow the operator
$A$ to exchange its argument with the one of the operator $B$ at least once, then we obtain a product
$B(u_k)A(u_j)$. Hence, the operator $A$ has lost its argument $u_k$, and it cannot get it again when moving further to the right. Thus, we cannot obtain
the function $a(u_k)$ in the final result. This means that
\be{07-SP-AB}
A(u_k)B(u_j)=f(u_j,u_k)B(u_j)A(u_k)+\mathcal{Z}\,,
\ee
where $\mathcal{Z}$ still means the terms that do not contribute to the result. Then
\be{07-SP-vkladA}
C(v)B(\bu)|0\rangle=g(v,u_k)a(u_k)f(\bu_{k},u_k)\;D(v)B(\bu_k)|0\rangle
+\mathcal{Z}\,.
\ee

Similarly, we conclude that the operator $D(v)$ should keep its argument when moving to the right. It means that we
can write the commutation relation for the operators $D$ and $B$ in the form
\be{07-SP-DB}
D(v)B(u_j)=f(v,u_j)B(u_j)D(v)+\mathcal{Z}\,.
\ee
Thus, we obtain
\be{07-SP-vkladAD}
C(v)B(\bu)|0\rangle=g(v,u_k)a(u_k)d(v)f(\bu_{k},u_k)f(v,\bu_{k})\;B(\bu_k)|0\rangle
+\mathcal{Z}\,,
\ee
what implies
\be{07-SP-ansLk}
X_k=g(v,u_k)f(\bu_{k},u_k)f(v,\bu_{k})\,.
\ee

The coefficient  $\tilde X_k$ can be obtained in the similar manner. The same considerations lead us to the formula
\eqref{07-SP-first-com}. In this case the term $D(v)A(u_k)$ does not contribute to the desired result, therefore,
instead of \eqref{07-SP-first-com1} we have
\be{07-SP-first-com2}
C(v)B(\bu)|0\rangle=g(u_k,v)D(u_k)A(v)B(\bu_k)|0\rangle
+\mathcal{Z}\,.
\ee
Then the operators $D(u_k)$ and $A(v)$ should move to the right keeping their arguments, what eventually leads us to the following result:
\be{07-SP-anstLk}
\tilde X_k=g(u_k,v)f(u_k,\bu_{k})f(\bu_{k},v)\,.
\ee

The coefficient $Y_{jk}$ can be found approximately in the same spirit.
For example, let us calculate a contribution proportional to $a(u_j)d(u_k)Y_{jk}$.
It is clear that first of all it is necessary to reorder the operators  $B$ as follows:
\be{07-SP-kj-perv}
C(v)B(\bu)|0\rangle=C(v)B(u_k)B(u_j)B(\bu_{jk})|0\rangle\,.
\ee
Again, at the first step, only the commutator $[C(v),B(u_k)]$ gives a contribution to the desired result
\be{07-SP-com-mk}
C(v)B(\bu)|0\rangle=g(v,u_k)\bigl(D(v)A(u_k)-D(u_k)A(v)\bigr) B(u_j)B(\bu_{jk})|0\rangle
+\mathcal{Z}\,,
\ee
After this, however, the situation slightly changes. Consider the successive action of the operators $D(v)A(u_k)$.
Clearly, when permuting the operators $A(u_k) $ and $B(u_j) $ we must exchange their arguments; otherwise, with further permutations, the operator $A(u_k)$
cannot get the argument $u_j$ and cannot turn into the function $a(u_j)$. Similarly,
the operator $D(v)$ must exchange its argument with the operator $B(u_k)$ in order to acquire the required argument $u_k$. Hence,
\be{07-SP-effCR1}
D(v)A(u_k)B(u_j)= g(u_k,u_j)g(u_k,v)\;B(v)D(u_k)A(u_j)+\mathcal{Z}\,.
\ee

Now we consider the action of the operators $D(u_k)A(v)$.
Here,  the operator $A(v)$ also must exchange its
argument with the one of the operator $B(u_j)$ to get the required argument $u_j$. However, the operator $D(u_k)$
when permuted with the operator $B(v)$, should keep its argument, that is
\be{07-SP-effCR2}
D(u_k)A(v)B(u_j)= g(v,u_j)f(u_k,v)\;B(v)D(u_k)A(u_j)+\mathcal{Z}\,.
\ee
Then we eventually obtain
\be{07-SP-effCR3}
\bigl(D(v)A(u_k)-D(u_k)A(v)\bigr)B(u_j)= \bigl(g(u_k,u_j)g(u_k,v)-g(v,u_j)f(u_k,v)\bigr)\;B(v)D(u_k)A(u_j)+\mathcal{Z}\,.
\ee
It is not difficult to check that
\be{07-SP-legko}
g(u_k,u_j)g(u_k,v)-g(v,u_j)f(u_k,v)=f(u_k,u_j)g(u_j,v)\,,
\ee
and substituting  \eqref{07-SP-effCR3}, \eqref{07-SP-legko} into \eqref{07-SP-com-mk} we arrive at
\be{07-SP-com-mkjeff}
C(v)B(\bu)|0\rangle=g(v,u_k)g(u_j,v)f(u_k,u_j)\;B(v)D(u_k)A(u_j)B(\bu_{jk})|0\rangle
+\mathcal{Z}\,.
\ee
After this, when moving to the right, the operators $D(u_k)$ and $A(u_j$) must keep their arguments, what gives us
\be{07-SP-Mjk-act}
C(v)B(\bu)|0\rangle=g(v,u_k)g(u_j,v)f(u_k,u_j)f(u_k,\bu_{jk})
f(\bu_{jk},u_j)d(u_k)a(u_j)
\;B(v)B(\bu_{jk})|0\rangle
+\mathcal{Z}\,.
\ee
This implies
\be{07-SP-Mjk}
Y_{jk}=g(v,u_k)g(u_j,v)f(u_k,u_j)f(u_k,\bu_{jk})
f(\bu_{jk},u_j)\,.
\ee

Thus, all unknown coefficients in  \eqref{07-SP-ans-vid} are found, and now we can compare it with equation
\eqref{07-SP-actC}  for $n=1$. There are three essentially different partitions in   \eqref{07-SP-actC}:

\begin{itemize}

\item $\bar\xi_{\so}=v$;\ \  $\bar\xi_{\st}=u_k$, $k=1,\dots,m$;\ \  $\bar\xi_{\sth}=\bu_k$;

\item $\bar\xi_{\so}=u_k$, $k=1,\dots,m$;\ \  $\bar\xi_{\st}=v$;\ \  $\bar\xi_{\sth}=\bu_k$;

\item $\bar\xi_{\so}=u_k$, $k=1,\dots,m$;\ \  $\bar\xi_{\st}=u_j$, $j=1,\dots,m$, $j\ne k$;\ \
$\bar\xi_{\sth}=\{v,\bu_{jk}\}$;
\end{itemize}

We give the reader an opportunity to check that the first partition gives the coefficient
$X_k$, the second gives  $\tilde X_k$, and the third gives $Y_{jk}$.

\subsection{Action  of the $C$ operators product on Bethe vector\label{07-Sec22}}

Equation \eqref{07-SP-actC}  is proved for $n=1$, and now we can use induction over $n$. Assume that \eqref{07-SP-actC} is valid for
some $n-1$. Let us act with the product $C(\bv)$ onto vector $B(\bu)|0\rangle$. We first act with the product
$C(\bv_n)$ via \eqref{07-SP-actC}, where one should replace $n$ by $n-1$:
\begin{multline}\label{07-SP-actCn-1}
C(\bv)B(\bu)|0\rangle=\sum_{\{\bv_n,\bu\}\mapsto\{\bar \xi_{\so},\bar \xi_{\st},\bar \xi_{\sth}\}}
 d(\bar \xi_{\so})a(\bar \xi_{\st})K_{n-1}(\bv_n|\bar \xi_{\so}+c) K_{n-1}(\bar \xi_{\st}|\bv_n+c) \\
\times f(\bar \xi_{\so},\bar \xi_{\st})
 f(\bar \xi_{\so},\bar \xi_{\sth})f(\bar \xi_{\sth},\bar \xi_{\st})\;
C(v_n)B(\bar \xi_{\sth})|0\rangle\,.
\end{multline}
Then we act with the remaining  operator $C(v_n)$ on $B(\bar \xi_{\sth})|0\rangle$. We again can use equation
\eqref{07-SP-actC}, setting there $n=1$:
\begin{multline}\label{07-SP-actCn}
C(\bv)B(\bu)|0\rangle=\sum_{\{\bv_n,\bu\}\mapsto\{\bar \xi_{\so},\bar \xi_{\st},\bar \xi_{\sth}\}}
 d(\bar \xi_{\so})a(\bar \xi_{\st})K_{n-1}(\bv_n|\bar \xi_{\so}+c) K_{n-1}(\bar \xi_{\st}|\bv_n+c) \\
\times f(\bar \xi_{\so},\bar \xi_{\st})
 f(\bar \xi_{\so},\bar \xi_{\sth})f(\bar \xi_{\sth},\bar \xi_{\st})
 \sum_{\{v_n,\bar \xi_{\sth}\}\mapsto\{\bar \xi_{\rm i},\bar \xi_{\rm ii},\bar \xi_{\rm iii}\}}
 d(\bar \xi_{\rm i})a(\bar \xi_{\rm ii})K_{1}(v_n|\bar \xi_{\rm i}+c) K_{1}(\bar \xi_{\rm ii}|v_n+c) \\
 \times f(\bar \xi_{\rm i},\bar \xi_{\rm ii})
 f(\bar \xi_{\rm i},\bar \xi_{\rm iii})f(\bar \xi_{\rm iii},\bar \xi_{\rm ii})\;
B(\bar \xi_{\rm iii})|0\rangle\,.
\end{multline}
In equation \eqref{07-SP-actCn}, the set $\{\bv_n,\bu\}$ first is divided into three subsets $\bar \xi_{\so}$, $\bar \xi_{\st}$,
$\bar \xi_{\sth}$. After this, the remaining parameter $v_n$ is combined with the subset $\bar \xi_{\sth}$ into a new set
$\{v_n,\bar \xi_{\sth}\}$. The latter, in its turn is divided into subsets  $\bar \xi_{\rm i}$, $\bar \xi_{\rm ii}$,
and $\bar \xi_{\rm iii}$. In addition, we require that $\#\bar \xi_{\so}=\#\bar \xi_{\st}=n-1$ and
$\#\bar \xi_{\rm i}=\#\bar \xi_{\rm ii}=1$. The sum is taken over all such partitions. We can say that actually we take the sum
over partitions of the set $\{\bv,\bu\}$ into five subsets $\bar \xi_{\so}$, $\bar \xi_{\st}$,
$\bar \xi_{\rm i}$, $\bar \xi_{\rm ii}$, and $\bar \xi_{\rm iii}$, where $v_n\notin\bar \xi_{\so}$ and
$v_n\notin\bar \xi_{\st}$.

Let us briefly recall a general strategy of further transformations. The goal is to take partly the sum over partitions via
lemma~\ref{06-MCR-main-ident}.
%
%
For this, we first of all should rewrite \eqref{07-SP-actCn} in such a way, that the restrictions $v_n\notin\bar \xi_{\so}$ and
$v_n\notin\bar \xi_{\st}$ would hold automatically.
Second, it is necessary to determine which sums over partitions anyway cannot be calculated.
It is already clear from the formula \eqref{07-SP-actCn} that the subset $\bar \xi_{\rm iii}$
is fixed by the product of the $B$ operators. Similarly, the product of the $d$ functions  fixes the union of the subsets
$\{\bar \xi_{\rm i},\bar \xi_{\so}\}$, while the product of the $a$ functions  fixes the union $\{\bar \xi_{\rm ii},\bar \xi_{\st}\}$.
Thus, for any transformations of the expression
\eqref{07-SP-actCn} these three sets remain unchanged. However,  the partitions inside the subsets $\{\bar \xi_{\rm i},\bar \xi_{\so}\}$
and $\{\bar \xi_{\rm ii},\bar \xi_{\st}\}$ only lead to a change in the rational coefficient. Therefore, one can try to compute the
sums over these partitions. To do this, we need to transform some of DWPF so as to reduce these sums to the identity~(8.2.4). 

Since $\bar \xi_{\sth}=\{\bar \xi_{\rm i},\bar \xi_{\rm ii},\bar \xi_{\rm iii}\}\setminus v_n$, we have
\be{07-SP-x1x3}
\begin{aligned}
f(\bar \xi_{\so},\bar \xi_{\sth})&=\frac{f(\bar \xi_{\so},\bar \xi_{\rm i})f(\bar \xi_{\so},\bar \xi_{\rm ii})
f(\bar \xi_{\so},\bar \xi_{\rm iii})}{f(\bar \xi_{\so},v_n)}\,,\\
f(\bar \xi_{\sth},\bar \xi_{\st})&=\frac{f(\bar \xi_{\rm i},\bar \xi_{\st})f(\bar \xi_{\rm ii},\bar \xi_{\st})
f(\bar \xi_{\rm iii},\bar \xi_{\st})}{f(v_n,\bar \xi_{\st})}\,.
\end{aligned}
\ee
The expressions in the rhs of \eqref{07-SP-x1x3} vanish as soon as $v_n\in\bar \xi_{\so}$ or
$v_n\in\bar \xi_{\st}$. Substituting them into \eqref{07-SP-actCn} instead of
$f(\bar \xi_{\so},\bar \xi_{\sth})$ and $f(\bar \xi_{\sth},\bar \xi_{\st})$, we obtain a sum over partitions into five
subsets, but without any restrictions:
\begin{multline}\label{07-SP-actCn-5sets}
C(\bv)B(\bu)|0\rangle=\sum_{\{\bv,\bu\}\mapsto\{\bar \xi_{\so},\bar \xi_{\st},
\bar \xi_{\rm i},\bar \xi_{\rm ii},\bar \xi_{\rm iii}\}}
 d(\bar \xi_{\so}) d(\bar \xi_{\rm i})a(\bar \xi_{\st})a(\bar \xi_{\rm ii})\num
 \times
 K_{n-1}(\bv_n|\bar \xi_{\so}+c)K_{1}(v_n|\bar \xi_{\rm i}+c)\cdot
  K_{n-1}(\bar \xi_{\st}|\bv_n+c) K_{1}(\bar \xi_{\rm ii}|v_n+c)  \num
\times f(\bar \xi_{\so},\bar \xi_{\st})f(\bar \xi_{\so},\bar \xi_{\rm i})f(\bar \xi_{\so},\bar \xi_{\rm ii})
f(\bar \xi_{\so},\bar \xi_{\rm iii})f(\bar \xi_{\rm i},\bar \xi_{\st})f(\bar \xi_{\rm ii},\bar \xi_{\st})
f(\bar \xi_{\rm iii},\bar \xi_{\st})f(\bar \xi_{\rm i},\bar \xi_{\rm ii})
 f(\bar \xi_{\rm i},\bar \xi_{\rm iii})f(\bar \xi_{\rm iii},\bar \xi_{\rm ii})\num
 \times  f^{-1}(\bar \xi_{\so},v_n)f^{-1}(v_n,\bar \xi_{\st})\;
B(\bar \xi_{\rm iii})|0\rangle\,.
\end{multline}

Let us introduce new sets
$\{\bar \xi_{\so},\bar \xi_{\rm i}\}=\bar \xi_{0}$ and $\{\bar \xi_{\st},\bar \xi_{\rm ii}\}=\bar \xi_{0'}$. One can easily convince oneself
that equation \eqref{07-SP-actCn-5sets} then takes the following form:
\begin{multline}\label{07-SP-actCn-00}
C(\bv)B(\bu)|0\rangle=\sum_{\{\bv,\bu\}\mapsto\{\bar \xi_{0},\bar \xi_{0'},
\bar \xi_{\rm iii}\}}
\; d(\bar \xi_{0}) a(\bar \xi_{0'})f(\bar \xi_{0},\bar \xi_{0'})f(\bar \xi_{0},\bar \xi_{\rm iii})f(\bar \xi_{\rm iii},\bar \xi_{0'})\;B(\bar \xi_{\rm iii})|0\rangle\num
 \times \sum_{\bar \xi_{0}\mapsto\{\bar \xi_{\so},
\bar \xi_{\rm i}\}}
 K_{n-1}(\bv_n|\bar \xi_{\so}+c)K_{1}(v_n|\bar \xi_{\rm i}+c)
  \frac{ f(\bar \xi_{\so},\bar \xi_{\rm i})}
{f(\bar \xi_{\so},v_n)}  \num
\times \sum_{\bar \xi_{0'}\mapsto\{\bar \xi_{\st},\bar \xi_{\rm ii}\}}
K_{n-1}(\bar \xi_{\st}|\bv_n+c) K_{1}(\bar \xi_{\rm ii}|v_n+c)\frac{ f(\bar \xi_{\rm ii},\bar \xi_{\st})}
{f(v_n,\bar \xi_{\st})}\,.
\end{multline}
Now the sum over partitions is organized as follows. First, the set
$\{\bv,\bu\}$ is divided into three subsets $\bar \xi_{0}$, $\bar \xi_{0'}$, and
$\bar \xi_{\rm iii}$. After this, the subset
$\bar \xi_{0}$ is divided once more into $\bar \xi_{\so}$ and $\bar \xi_{\rm i}$, and the subset
$\bar \xi_{0'}$  is divided into $\bar \xi_{\st}$ and $\bar \xi_{\rm ii}$. Now we use the following property of the DWPF:
\be{07-SP-Ktransf}
K_{1}(v_n|\bar \xi_{\rm i}+c)=-f^{-1}(\bar \xi_{\rm i},v_n)K_{1}(\bar \xi_{\rm i}|v_n)\,,\qquad
K_{1}(\bar \xi_{\rm ii}|v_n+c)=-f^{-1}(v_n,\bar \xi_{\rm ii})K_{1}(v_n|\bar \xi_{\rm ii})\,.
\ee
Substituting these expressions into \eqref{07-SP-actCn-00} we obtain
\begin{multline}\label{07-SP-actCn-001}
C(\bv)B(\bu)|0\rangle=\sum_{\{\bv,\bu\}\mapsto\{\bar \xi_{0},\bar \xi_{0'},
\bar \xi_{\rm iii}\}}d(\bar \xi_{0}) a(\bar \xi_{0'})
\frac{f(\bar \xi_{0},\bar \xi_{0'})f(\bar \xi_{0},\bar \xi_{\rm iii})f(\bar \xi_{\rm iii},\bar \xi_{0'})}
{f(\bar \xi_{0},v_n)f(v_n,\bar \xi_{0'})}\;B(\bar \xi_{\rm iii})|0\rangle\num
\times \sum_{\bar \xi_{0}\mapsto\{\bar \xi_{\so},\bar \xi_{\rm i}\}}
K_{n-1}(\bv_n-c|\bar \xi_{\so})K_{1}(\bar \xi_{\rm i}|v_n)f(\bar \xi_{\so},\bar \xi_{\rm i})\\
\times \sum_{\bar \xi_{0'}\mapsto\{\bar \xi_{\st},\bar \xi_{\rm ii}\}}
   K_{n-1}(\bar \xi_{\st}|\bv_n+c) K_{1}(v_n|\bar \xi_{\rm ii})
f(\bar \xi_{\rm ii},\bar \xi_{\st})\,.
\end{multline}
The sums in the second and third lines of the formula \eqref{07-SP-actCn-001} can be computed via lemma~\ref{06-MCR-main-ident}:
\be{07-SP-1stsum}
\sum_{\bar \xi_{0}\mapsto\{\bar \xi_{\so},\bar \xi_{\rm i}\}}
K_{n-1}(\bv_n-c|\bar \xi_{\so})K_{1}(\bar \xi_{\rm i}|v_n)f(\bar \xi_{\so},\bar \xi_{\rm i})=
-f(\bar \xi_{0},v_n)K_{n}(\bv-c|\bar \xi_{0})\,,
\ee
and
\be{07-SP-2ndsum}
\sum_{\bar \xi_{0'}\mapsto\{\bar \xi_{\st},\bar \xi_{\rm ii}\}}
   K_{n-1}(\bar \xi_{\st}|\bv_n+c) K_{1}(v_n|\bar \xi_{\rm ii})f(\bar \xi_{\rm ii},\bar \xi_{\st})  =
   -f(v_n,\bar \xi_{0'})K_{n}(\bar \xi_{0'}|\bv+c)\,.
   \ee
Substituting these expressions into \eqref{07-SP-actCn-001} we finally obtain
\begin{multline}\label{07-SP-actCn-00fin}
C(\bv)B(\bu)|0\rangle=\sum_{\{\bv,\bu\}\mapsto\{\bar \xi_{0},\bar \xi_{0'},
\bar \xi_{\rm iii}\}}d(\bar \xi_{0}) a(\bar \xi_{0'})
f(\bar \xi_{0},\bar \xi_{0'})f(\bar \xi_{0},\bar \xi_{\rm iii})f(\bar \xi_{\rm iii},\bar \xi_{0'})
\\
\times K_{n}(\bv|\bar \xi_{0}+c)K_{n}(\bar \xi_{0'}|\bv+c)\;B(\bar \xi_{\rm iii})|0\rangle\,.
\end{multline}
It remains to redenote $\bar \xi_{0}\to\bar \xi_{\so}$, $\bar \xi_{0'}\to\bar \xi_{\st}$,
and $\bar \xi_{\rm iii}\to\bar \xi_{\sth}$, and we arrive at \eqref{07-SP-actC}  for $\#\bv=n$.

\section{Scalar product\label{07-Sec3}}

It was assumed in \eqref{07-SP-actC}  that  $n\le m$. Indeed, since
$\#\bar \xi_{\so}=\#\bar \xi_{\st}=n$, then $\#\bar \xi_{\sth}=m-n$, and hence, for  $n>m$ this formula becomes senseless.
On the other hand, it is quite obvious, that if the number of the  $C$ operators is more than the number of the creation operators
$B$, then the action of the product $C(\bv)$ on the vector $B(\bu)|0\rangle$
gives zero. Therefore, the case  $n>m$ is not interesting. Much more interesting and important is the special case
$n=m$, in which we immediately obtain a formula for the scalar product of the dual vector $\langle0|C(\bv)$
and the vector $B(\bu)|0\rangle$. For $n=m$, we get $\#\bar \xi_{\sth}=0$,
that is, $\bar \xi_{\sth}=\emptyset$, and the formula \eqref{07-SP-actC} is simplified
\begin{equation}\label{07-SP-actCB}
C(\bv)B(\bu)|0\rangle=\sum_{\{\bv,\bu\}\mapsto\{\bar \xi_{\so},\bar \xi_{\st}\}}
 d(\bar \xi_{\so})a(\bar \xi_{\st})K_n(\bv|\bar \xi_{\so}+c) K_n(\bar \xi_{\st}|\bv+c)
 f(\bar \xi_{\so},\bar \xi_{\st}) |0\rangle\,.
\end{equation}
Taking into account that $\langle0|0\rangle=1$, we immediately obtain
\begin{equation}\label{07-SP-SP}
\langle0|C(\bv)B(\bu)|0\rangle=\sum_{\{\bv,\bu\}\mapsto\{\bar \xi_{\so},\bar \xi_{\st}\}}
 d(\bar \xi_{\so})a(\bar \xi_{\st})K_n(\bv|\bar \xi_{\so}+c) K_n(\bar \xi_{\st}|\bv+c)
 f(\bar \xi_{\so},\bar \xi_{\st})\,.
\end{equation}

The formula \eqref{07-SP-SP} is perhaps the most compact form of the scalar product,
if we do not  impose any additional conditions  for the parameters $\bv$ and $\bu$. It turns out, however, that if one of the
vectors in the formula \eqref {07-SP-SP} is an eigenvector of the transfer matrix, then the sum over the partitions
reduces to a certain determinant. In order to see this, one needs to write \eqref{07-SP-SP} in more detail,
namely, to pass from partitions of the combined set $\{\bv, \bu \}$ to independent partitions of the sets
$\bv$ and $\bu$. Let
\be{07-SP-parlm}
\begin{array}{l}
\bar \xi_{\so}=\{\bv_{\st},\bu_{\so}\},\\
\bar \xi_{\st}=\{\bv_{\so},\bu_{\st}\},
\end{array}
\qquad\qquad
\begin{array}{l}
\#\bv_{\st}=\#\bu_{\st}=k,\\
\#\bv_{\so}=\#\bu_{\so}=n-k,
\end{array}
\qquad k=0,1,\dots,n\,.
\ee
It is easy to see that
\be{07-SP-ad}
d(\bar \xi_{\so})a(\bar \xi_{\st})=d(\bv_{\st})d(\bu_{\so})a(\bv_{\so})a(\bu_{\st})\,,
\ee
and
\be{07-SP-ff}
f(\bar \xi_{\so},\bar \xi_{\st})=f(\bv_{\st},\bv_{\so})f(\bu_{\so},\bu_{\st})
f(\bv_{\st},\bu_{\st})f(\bu_{\so},\bv_{\so})\,.
\ee
The DWPF $K_n$ are transformed via the reduction formulas, for example,
\begin{equation}\label{07-SP-K1}
\begin{aligned}
K_n(\bv|\bar \xi_{\so}+c)&=K_n(\{\bv_{\st},\bv_{\so}\}|\{\bv_{\st}+c, \bu_{\so}+c\})\\
&=(-1)^kK_{n-k}(\bv_{\so}| \bu_{\so}+c)=
(-1)^n f^{-1}( \bu_{\so},\bv_{\so})K_{n-k}( \bu_{\so}|\bv_{\so})\,.
\end{aligned}
\end{equation}
Similarly,
\begin{equation}\label{07-SP-K2}
\begin{aligned}
K_n(\bar \xi_{\st}|\bv+c)&=K_n(\{\bv_{\so},\bu_{\st}\}|\{\bv_{\st}+c, \bv_{\so}+c\})\\
&=(-1)^{n-k}K_{k}(\bu_{\st}| \bv_{\st}+c)=
(-1)^n f^{-1}(\bv_{\st}, \bu_{\st})K_{k}( \bv_{\st}|\bu_{\st})\,.
\end{aligned}
\end{equation}
Substituting \eqref{07-SP-ad}--\eqref{07-SP-K2} into \eqref{07-SP-SP} we obtain
\begin{equation}\label{07-SP-SPlm}
\langle0|C(\bv)B(\bu)|0\rangle=\sum_{\substack{\bv\mapsto\{\bv_{\so},\bv_{\st}\}\\
\bu\mapsto\{\bu_{\so},\bu_{\st}\}}}
d(\bv_{\st})a(\bu_{\st})d(\bu_{\so})a(\bv_{\so})\,
K_{k}( \bv_{\st}|\bu_{\st})K_{n-k}( \bu_{\so}|\bv_{\so})\,f(\bv_{\st},\bv_{\so})f(\bu_{\so},\bu_{\st})\,.
\end{equation}

{\sl Remark.} It is clear from representation \eqref{07-SP-SPlm} that
\begin{equation}\label{07-SP-symm}
\langle0|C(\bv)B(\bu)|0\rangle=\langle0|C(\bu)B(\bv)|0\rangle\,.
\end{equation}
Indeed, if we replace $\bu \leftrightarrow \bv$ in  \eqref{07-SP-SPlm} and change everywhere the summation indices $\so$ to $\st$,
then this formula will pass into itself.

\begin{prop}\label{07-Prop-zero} Let parameters $u_1$ and $u_2$ be such that $a(u_1)=0$, $d(u_2)=0$, and  $u_2=u_1+c$. If $\{u_1,u_2\}\subset\bu$, then
a vector $B(\bu)|0\rangle$  is the null-vector.
\end{prop}

{\sl Proof.} Consider the scalar product of the vector $B(\bu)|0\rangle$
with an arbitrary dual vector $\langle0|C(\bv)$.
It follows from the conditions $a(u_1) = 0$ and $d(u_2) = 0$ that the formula \eqref{07-SP-SPlm} contains only those partitions of the set $\bu$,
for which $u_1\in\bu_{\so}$ and $u_2\in\bu_{\st}$.
But then the product $f(\bu_{\so},\bu_{\st})$ contains the function $f(u_1, u_2)$,
that vanishes due to the condition $u_2 =u_1 + c$. Thus, the scalar product of the vector $B(\bu)|0\rangle$
with any dual vector vanishes, hence $B(\bu)|0\rangle$
is the null-vector.

The situation described in proposition~\ref{07-Prop-zero} has already been considered  earlier for the special case of the $XXX$ chain
consisting of four sites.
We recall that this was connected with the study of solutions of the Bethe equations. The matter is that in the $XXX$ model there are formal solutions
$u_1 = -c/2$ and $u_2=c/2$ to these
equations\footnote{In the models with trigonometric $R$-matrix these are the roots $u_1 = -\eta/2$ and $u_2 =\eta/2$.}. However, as we now see, the vector corresponding to this solution is the null-vector. Hence,
it by definition cannot be an eigenvector of the transfer matrix (and of any other operator).

Let us turn back to the study of the scalar products. We introduce a new notation
\be{07-SP-SPdef}
S_n(\bv|\bu)=\frac{\langle0|C(\bv)B(\bu)|0\rangle}{d(\bv)d(\bu)}\,,
\ee
and a new function $r(u)=a(u)/d(u)$. Then  \eqref{07-SP-SPlm} is written in the form
\begin{equation}\label{07-SP-SPlm1}
S_n(\bv|\bu)=\sum_{\substack{\bv\mapsto\{\bv_{\so},\bv_{\st}\}\\
\bu\mapsto\{\bu_{\so},\bu_{\st}\}}}
r(\bu_{\st})r(\bv_{\so})\,
K_{k}( \bv_{\st}|\bu_{\st})K_{n-k}( \bu_{\so}|\bv_{\so})\,f(\bv_{\st},\bv_{\so})f(\bu_{\so},\bu_{\st})\,,
\end{equation}
where we used the shorthand notation for the products of the functions $r$ over subsets $\bu_{\st}$ and $\bv_{\so}$.

{\sl Question. }  Let $r(u)$ be identically $1$. Then equation \eqref{07-SP-SPlm1} takes the form
\begin{equation}\label{07-SP-r1}
S_n(\bv|\bu)=\sum_{\substack{\bv\mapsto\{\bv_{\so},\bv_{\st}\}\\
\bu\mapsto\{\bu_{\so},\bu_{\st}\}}}
K_{k}( \bv_{\st}|\bu_{\st})K_{n-k}( \bu_{\so}|\bv_{\so})\,f(\bv_{\st},\bv_{\so})f(\bu_{\so},\bu_{\st})\,.
\end{equation}
We claim that $S_n(\bv|\bu)=0$  for  $n>0$. This statement can be proved at least in two ways. One can turn back to the formula
\eqref{07-SP-SP} and use lemma~\ref{06-MCR-main-ident}.
But there exists another way, that does not require any calculations. Try to find this way.

\vspace{5mm}

Assume now that the vector $B(\bu)|0\rangle$  is an eigenvector of the twisted transfer matrix\footnote{%
In particular, the twist parameter $\kappa $ can be equal to $1$. In the case under consideration, there is no fundamental difference between
usual and twisted on-shell vectors, and we consider the latter solely for the sake of generality.}.
This means that the parameters $\bu$ satisfy twisted Bethe equations
\be{07-SP-BE}
r(u_j)=\kappa\;\frac{f(u_j,\bu_j)}{f(\bu_j,u_j)},\qquad j=1,\dots,n\,.
\ee
It follows from this that
\be{07-SP-BEmany}
r(\bu_{\st})f(\bu_{\so},\bu_{\st})=\kappa^k\;f(\bu_{\st},\bu_{\so})\,,
\ee
where $k=\#\bu_{\st}$. Substituting \eqref{07-SP-BEmany}  into \eqref{07-SP-SPlm1} we obtain
\begin{equation}\label{07-SP-SPlm2}
S_n(\bv|\bu)=\sum_{\substack{\bv\mapsto\{\bv_{\st},\bv_{\so}\}\\
\bu\mapsto\{\bu_{\st},\bu_{\so}\}}}
\kappa^k\; r(\bv_{\so})\,
K_{k}( \bv_{\st}|\bu_{\st})K_{n-k}( \bu_{\so}|\bv_{\so})\,f(\bv_{\st},\bv_{\so})f(\bu_{\st},\bu_{\so})\,,
\end{equation}
and we see that the sum over partitions of the set $\bu$ can be computed via lemma~\ref{06-MCR-main-ident}:
\begin{equation}\label{07-SP-sumlam}
\sum_{\bu\mapsto\{\bu_{\so},\bu_{\st}\}}
K_{k}( \bv_{\st}|\bu_{\st})K_{n-k}( \bu_{\so}|\bv_{\so})\,f(\bu_{\st},\bu_{\so})=
(-1)^{n-k}f(\bu,\bv_{\so})K_{n}( \{\bv_{\so}-c, \bv_{\st}\}|\bu)\,.
\end{equation}
Thus, we arrive at the following formula:
\begin{equation}\label{07-SP-SPlm3}
S_n(\bv|\bu)=\sum_{\bv\mapsto\{\bv_{\so},\bv_{\st}\}}
(-1)^{n-k}\kappa^k\;r(\bv_{\so})\,
K_{n}( \{\bv_{\so}-c, \bv_{\st}\}|\bu)\,f(\bv_{\st},\bv_{\so})f(\bu,\bv_{\so})\,.
\end{equation}

The sum in the rhs of equation \eqref{07-SP-SPlm3} directly reduces to the determinant of an $n\times n$ matrix.
However, it is not easy to see this, therefore, we make a stop here and in the next section we obtain
some identities for determinants of general form.

\section{Identities for determinants\label{07-Sec4}}

So far we have dealt with the sums of the partitions of functions that were
symmetric in their arguments. Therefore, the order of the elements in each
subset was not essential. When we are dealing with determinants, we have functions that are antisymmetric in their arguments.
Therefore, we must agree upon the order of the elements in each subset. We assume that if a set $\bv $
is divided into two (or more) subsets  $\bv^{\so}$ and $\bv^{\st}$,
then in each of them the elements are ordered in increasing order of indices.
For example: $\bv=\{v_1,\dots,v_5\}$,  $\bv^{\so}=\{v_1,v_4\}$ and $\bv^{\st}=\{v_2,v_3,v_5\}$.
Such the ordering will be called {\it natural order}.

We also introduce the notion of {\it parity of a partition}. By the parity of the partition $P(\bv^{\so},\bv^{\st})$
we mean the parity of the permutation that maps the union of two subsets $\{\bv^{\so},\bv^{\st}\}$
(in each of them the elements should be in the natural order)
to the naturally ordered set $\bv$. Thus, in the above example, the union of two subsets has the form
$\{\bv^{\so},\bv^{\st}\}=\{v_1,v_4, v_2,v_3,v_5\}$.
It is easy to see that the permutation that maps this sequence to the naturally ordered set $\bv=\{v_1, v_2,v_3,v_4,v_5\}$
is even, therefore, this partition also is even.

Finally, let us introduce a new notation for special products
\be{07-SP-defD}
\Delta_n(\bv)=\prod_{1\le j<k\le n}g(v_k,v_j),\qquad
\Delta'_n(\bv)=\prod_{1\le j<k\le n}g(v_j,v_k)=(-1)^{n(n-1)/2}\Delta_n(\bv)\,.
\ee

\begin{prop}\label{07-Vand}
For any partition of a set $\bv$ into two subsets $\bv^{\so}$ and $\bv^{\st}$, such that  $\#\bv^{\so}=m$,
$\#\bv^{\st}=n-m$, where $m$ is any from the set $0,1,\dots,n$, the following identity holds:
\be{07-SP-Del-Del}
\Delta_n(\bv)=(-1)^{P(\bv^{\so},\bv^{\st})}\Delta_m(\bv^{\so})\Delta_{n-m}(\bv^{\st})\;
g(\bv^{\st},\bv^{\so})\,.
\ee
\end{prop}

{\sl Proof}.
Obviously, up to a constant factor, the product $\Delta^{-1}_n(\bv)$ is Vandermonde determinant of the parameters $\bv$:
\be{07-SP-Del-Wand1}
\Delta^{-1}_n(\bv)=c^{-n(n-1)/2}\det_n\left(v_k^{j-1}\right)\,.
\ee
For any fixed partition of the set $\bv$ into two subsets $\bv^{\so}$ and $\bv^{\st}$,
we reorder the columns of the determinant so that all $v_k\in\bv^{\so}$ would be
on the left, and all $v_k\in\bv^{\st}$ would be on right. In each subset $\bv^{\so}$ and $\bv^{\st}$,
the columns are ordered in the natural order. Then
\be{07-SP-Del-Wand2}
\Delta^{-1}_n(\bv)=(-1)^{P(\bv^{\so},\bv^{\st})}c^{-n(n-1)/2}\det_n\Bigl((v_k^{\so})^{j-1}\;\Bigr|(v_k^{\st})^{j-1}\Bigr)\,.
\ee
Obviously, the determinant in the rhs of \eqref{07-SP-Del-Wand2} is equal to
\begin{equation}\label{07-SP-Del-Wand3}
\det_n\Bigl((v_k^{\so})^{j-1}\;\Bigr|(v_k^{\st})^{j-1}\Bigr)=\prod_{\substack{j<k\\v_j,v_k\in\bv^{\so}}}
(v_k-v_j)  \prod_{\substack{j<k\\v_j,v_k\in\bv^{\st}}}
(v_k-v_j)  \prod_{v_j\in\bv^{\so}}\prod_{v_k\in\bv^{\st}}(v_k-v_j)\,,
\end{equation}
and, taking inverse of this equality, we arrive at the assertion of the proposition~\ref{07-Vand}.

We now turn to the expansion of determinants into sums over partitions.
Suppose that we now have two sets $\bu=\{u_1,\dots,u_n\}$ and $\bv=\{v_1,\dots,v_n\}$.
Let us define on these sets two functions of two variables $\Phi_\ell(u_j,v_k)$,  $\ell=1,2$,
and form a matrix
\be{07-SP-MatPhi}
M_{jk}=\Phi_1(u_j,v_k)+ \Phi_2(u_j,v_k)\,.
\ee

\begin{prop}\label{07-SP-exp-det}
\begin{equation}\label{07-SP-ident-det}
\det_n(M_{jk})
=\sum_{\bv\mapsto\{\bv^{\so},\bv^{\st}\}} (-1)^{P(\bv^{\so},\bv^{\st})}
\det_n\Bigl[\Phi_1(u_j,v^{\so}_k)\;\Bigr|\;\Phi_2(u_j,v^{\st}_k)\Bigr]\,.
\end{equation}
The summation is over  partitions of the set $\bv$ into two naturally ordered subsets $\bv^{\so}$ and $\bv^{\st}$,
such that $\#\bv^{\so}=m$, $\#\bv^{\st}=n-m$ and $m=0,1,\dots,n$.
The matrix whose determinant enters the rhs of \eqref{07-SP-ident-det} consists of two parts. The first $m$
columns correspond to the subset $\bv^{\so}$, and there the matrix elements are equal to the function
$\Phi_1(u_j,v^{\so}_k)$. The remaining $n-m$ columns correspond to the subset $\bv^{\st}$, and there the matrix elements are
equal to the function $\Phi_2(u_j,v^{\st}_k)$.
\end{prop}

We first of all illustrate the idea of the proof by a simple example. Let $n=2$.
We use the property of linearity of any determinant with respect to the elements of the first column. Then
\begin{equation}\label{07-SP-expan1}
\det_2\begin{pmatrix}
        M_{11} &  M_{12} \\
        M_{21} & M_{22}
      \end{pmatrix} =
      \det_2\begin{pmatrix}
       \Phi_1(u_1,v_1)& M_{12}\\
       \Phi_1(u_2,v_1)& M_{22}
      \end{pmatrix}+
      \det_2\begin{pmatrix}
       \Phi_2(u_1,v_1)& M_{12}\\
       \Phi_2(u_2,v_1)& M_{22}
      \end{pmatrix}.
\end{equation}
Using the linearity of the obtained determinants with respect to the elements of the second columns we find
\begin{multline}\label{07-SP-expan2}
\det_2\begin{pmatrix}
        M_{11} &  M_{12} \\
        M_{21} & M_{22}
      \end{pmatrix} =
      \det_2\begin{pmatrix}
       \Phi_1(u_1,v_1)& \Phi_1(u_1,v_2)\\
       \Phi_1(u_2,v_1)& \Phi_1(u_2,v_2)
      \end{pmatrix}+ \det_2\begin{pmatrix}
       \Phi_1(u_1,v_1)& \Phi_2(u_1,v_2)\\
       \Phi_1(u_2,v_1)& \Phi_2(u_2,v_2)
      \end{pmatrix}\\
      + \det_2\begin{pmatrix}
       \Phi_2(u_1,v_1)& \Phi_1(u_1,v_2)\\
       \Phi_2(u_2,v_1)& \Phi_1(u_2,v_2)
      \end{pmatrix}+ \det_2\begin{pmatrix}
       \Phi_2(u_1,v_1)& \Phi_2(u_1,v_2)\\
       \Phi_2(u_2,v_1)& \Phi_2(u_2,v_2)
      \end{pmatrix}.
\end{multline}
Now we reorder the columns of the obtained determinants in such a way that all the columns containing $\Phi_1$ would be on the left
and all the columns containing $\Phi_2$ would be on the right. In fact, only the columns of the third determinant in the rhs of
\eqref{07-SP-expan2} should be permuted, because in all other determinants the columns are already ordered in the required way. We obtain
\begin{multline}\label{07-SP-expan3}
\det_2\begin{pmatrix}
        M_{11} &  M_{12} \\
        M_{21} & M_{22}
      \end{pmatrix} =
      \det_2\begin{pmatrix}
       \Phi_1(u_1,v_1)& \Phi_1(u_1,v_2)\\
       \Phi_1(u_2,v_1)& \Phi_1(u_2,v_2)
      \end{pmatrix}+ \det_2\begin{pmatrix}
       \Phi_1(u_1,v_1)& \Phi_2(u_1,v_2)\\
       \Phi_1(u_2,v_1)& \Phi_2(u_2,v_2)
      \end{pmatrix}\\
      - \det_2\begin{pmatrix}
        \Phi_1(u_1,v_2)&\Phi_2(u_1,v_1)\\
       \Phi_1(u_2,v_2)& \Phi_2(u_2,v_1)
      \end{pmatrix}+ \det_2\begin{pmatrix}
       \Phi_2(u_1,v_1)& \Phi_2(u_1,v_2)\\
       \Phi_2(u_2,v_1)& \Phi_2(u_2,v_2)
      \end{pmatrix}.
\end{multline}
It is easy to see that the four terms in the rhs of \eqref{07-SP-expan3} correspond to four possible partitions of the set $\bv$ into
two subsets:
\be{07-SP-part2}
\begin{aligned}
&\bv^{\so}=\{v_1,v_2\}, &\quad& \bv^{\st}=\emptyset;\\
&\bv^{\so}=v_1, &\quad& \bv^{\st}=v_2;\\
&\bv^{\so}=v_2, &\quad& \bv^{\st}=v_1;\\
&\bv^{\so}=\emptyset , &\quad& \bv^{\st}=\{v_1,v_2\}.
\end{aligned}
\ee
It is also clear that the parity of the third partition is odd. Therefore, the third determinant in the rhs of \eqref{07-SP-expan3} has the sign minus.
The parities of all other partitions are even, and thus, the corresponding determinants enter \eqref{07-SP-expan3} with the sign plus.

{\sl Proof of proposition~\ref{07-SP-exp-det}.}
The proof of proposition~\ref{07-SP-exp-det} in the general case now becomes obvious. Using successively the linearity of the determinant with respect to
the elements of any column we arrive at
\begin{equation}\label{07-SP-ident-det01}
\det_n(M_{jk})
=\sum_{\bv\mapsto\{\bv^{\so},\bv^{\st}\}}
\det_n\bigl(\Phi(u_j,v_k)\bigr)\,,
\end{equation}
where
\be{07-SP-Phi12}
\Phi(u_j,v_k)=\left\{\begin{array}{l}
\Phi_1(u_j,v_k),\qquad\text{if}\qquad v_k\in\bv^{\so},\\
\Phi_2(u_j,v_k),\qquad\text{if}\qquad v_k\in\bv^{\st}.
\end{array}\right.
\ee
Moving all the columns with $\Phi_1$ to the left we obtain the sign $(-1)^{P(\bv^{\so},\bv^{\st})}$ and reproduce \eqref{07-SP-ident-det}.

\begin{cor}\label{07-SP-exp-detD}
\begin{equation}\label{07-SP-ident-detD}
\Delta_n(\bv)\det_n(M_{jk})
=\sum_{\bv\mapsto\{\bv^{\so},\bv^{\st}\}} \Delta_m(\bv^{\so})\Delta_{n-m}(\bv^{\st})g(\bv^{\st},\bv^{\so})
\det_n\Bigl[\Phi_1(u_j,v^{\so}_k)\;\Bigr|\;\Phi_2(u_j,v^{\st}_k)\Bigr]\,.
\end{equation}
Here the sum over partitions and the notation is the same as in proposition~\ref{07-SP-exp-det}
\end{cor}
{\sl Proof.} This formula immediately follows from propositions~\ref{07-Vand} and~\ref{07-SP-exp-det}.

\begin{prop}\label{07-SP-exp-det3}
Let $\Omega(\bv|\bu)$ be a function of two sets of variables $\bv=\{v_1,\dots,v_n\}$ and $\bu=\{u_1,\dots,u_n\}$. Let this
function have the following representation
\begin{equation}\label{07-SP-repOm}
\Omega(\bv|\bu)=\Delta_n(\bv)\det_n\bigl(\Phi(u_j,v_k)\bigr),
\end{equation}
where $\Phi(u_j,v_k)$ is a function of two variables. Then
\begin{equation}\label{07-SP-ident-detD3}
\sum_{\bv\mapsto\{\bv^{\so},\bv^{\st}\}} f(\bv^{\st},\bv^{\so})\Omega(\{\bv^{\so}-c,\bv^{\st}\}|\bu)=
\Delta_n(\bv)\det_n\Bigl(\Phi(u_j,v_k-c)+\Phi(u_j,v_k)\Bigr).
\end{equation}
Here the sum is taken over all partitions of the set $\bv$ into two subsets $\bv^{\so}$ and $\bv^{\st}$,
such that $\#\bv^{\so}=m$, $\#\bv^{\st}=n-m$ and $m=0,1,\dots,n$.
\end{prop}
{\sl Proof.} Let us fix a partition $\bv\mapsto\{\bv^{\so},\bv^{\st}\}$. Since $\Omega(\bv|\bu)$ is symmetric over $\bv$, we can reorder the columns
of $\det_n\bigl(\Phi(u_j,v_k)\bigr)$ in such a way that all the columns depending on $v_k\in\bv^{\so}$ would be on the left in the natural order. Of course, the same reordering
of the parameters $\bv$ should be done in the product $\Delta_n(\bv)$. Then we obtain
\begin{multline}\label{07-SP-Om-part}
\Omega(\{\bv^{\so}-c,\bv^{\st}\}|\bu)=
\Delta_n(\{\bv^{\so}-c,\bv^{\st}\})\det_n\Bigl[\Phi(u_j,v^{\so}_k-c)\;\Bigr|\;\Phi(u_j,v^{\st}_k)\Bigr]\\
=\Delta_m(\bv^{\so})\Delta_{n-m}(\bv^{\st})g(\bv^{\st},\bv^{\so}-c)\det_n\Bigl[\Phi(u_j,v^{\so}_k-c)\;\Bigr|\;\Phi(u_j,v^{\st}_k)\Bigr]\,.
\end{multline}
Using
\begin{equation}\label{07-SP-ghf}
g(x,y-c)=\frac c{x-y+c}=\frac{g(x,y)}{f(x,y)},
\end{equation}
we find
\begin{equation}\label{07-SP-Om-part1}
\Omega(\{\bv^{\so}-c,\bv^{\st}\}|\bu)f(\bv^{\st},\bv^{\so})=
\Delta_m(\bv^{\so})\Delta_{n-m}(\bv^{\st})g(\bv^{\st},\bv^{\so})\det_n\Bigl[\Phi(u_j,v^{\so}_k-c)\;\Bigr|\;\Phi(u_j,v^{\st}_k)\Bigr]\,,
\end{equation}
and applying corollary~\ref{07-SP-exp-detD} to the sum \eqref{07-SP-ident-detD3} we complete the proof.

\begin{cor}\label{07-SP-exp-det3C}
Let $\alpha_1(z)$ and $\alpha_2(z)$ be two functions  of complex variable $z$.  Then under the conditions
of proposition~\ref{07-SP-exp-det3}
\begin{multline}\label{07-SP-ident-detD3C}
\sum_{\bv\mapsto\{\bv^{\so},\bv^{\st}\}} \alpha_1(\bv^{\so}) \alpha_2(\bv^{\st}) f(\bv^{\st},\bv^{\so})\Omega(\{\bv^{\so}-c,\bv^{\st}\}|\bu)\\
=\Delta_n(\bv)\det_n\Bigl(\alpha_1(v_k)\Phi(u_j,v_k-c)+\alpha_2(v_k)\Phi(u_j,v_k)\Bigr).
\end{multline}
Here we have extended the convention on the shorthand notation to the products of the functions $\alpha_\ell(z)$ ($\ell=1,2$).
\end{cor}

{\sl Proof.} Multiplying \eqref{07-SP-Om-part1} by $\alpha_1(\bv^{\so}) \alpha_2(\bv^{\st})$ we obtain
\begin{multline}\label{07-SP-ident-detD3C1}
\alpha_1(\bv^{\so}) \alpha_2(\bv^{\st}) f(\bv^{\st},\bv^{\so})\Omega(\{\bv^{\so}-c,\bv^{\st}\}|\bu)\\
=\alpha_1(\bv^{\so}) \alpha_2(\bv^{\st})\Delta_m(\bv^{\so})\Delta_{n-m}(\bv^{\st})g(\bv^{\st},\bv^{\so})\det_n\Bigl[\Phi(u_j,v^{\so}_k-c)\;\Bigr|\;\Phi(u_j,v^{\st}_k)\Bigr]\\
=\Delta_m(\bv^{\so})\Delta_{n-m}(\bv^{\st})g(\bv^{\st},\bv^{\so})\det_n\Bigl[\alpha_1(\bv^{\so}_k)\Phi(u_j,v^{\so}_k-c)\;\Bigr|\;\alpha_2(\bv^{\st}_k)\Phi(u_j,v^{\st}_k)\Bigr].
\end{multline}
Then the use of corollary~\ref{07-SP-exp-detD} completes the proof.

\section{Determinant formula for the scalar product\label{07-Sec5}}

Before calculating the sum over partitions in \eqref{07-SP-SPlm3}, we introduce two new functions
\be{07-ht}
h(u,v)=\frac{f(u,v)}{g(u,v)}=\frac{u-v+c}{c},\qquad
t(u,v)=\frac{g(u,v)}{h(u,v)}=\frac{c^2}{(u-v)(u-v+c)}.
\ee
In what follows, we will often deal with these combinations, therefore, it makes sense to introduce a special notation for them.
We will also extend to these functions the convention on the shorthand notation for the products.

Using these functions we can write down
the explicit expression for the DWPF in the form
\be{07-SP-K-expl}
K_n(\bv|\bu)=\Delta'_n(\bu) \Delta_n(\bv)\;\det_n\Bigl( t(v_k,u_j)h(v_k,\bu)\Bigr)\,.
\ee

Now one can easily see that the sum in \eqref{07-SP-SPlm3} can be calculated via corollary~\ref{07-SP-exp-det3C}.
Indeed, setting $k=n-m$ in \eqref{07-SP-SPlm3} we obtain
\begin{equation}\label{07-SP-SPlm3new}
S_n(\bv|\bu)=\sum_{\bv\mapsto\{\bv_{\so},\bv_{\st}\}}
(-1)^{m}\kappa^{n-m}\;r(\bv_{\so})\,
K_{n}( \{\bv_{\so}-c, \bv_{\st}\}|\bu)\,f(\bv_{\st},\bv_{\so})f(\bu,\bv_{\so})\,.
\end{equation}
If we set
\begin{equation}\label{07-SP-settings}
\Phi(u_j,v_k)=t(v_k,u_j)h(v_k,\bu),\qquad \alpha_1(z)=-r(z)f(\bu,z), \qquad \alpha_2(z)=\kappa,
\end{equation}
then equation \eqref{07-SP-ident-detD3C} turns into\footnote{%
The fact that $\Phi(u_j,v_k)$ and $\alpha_1(z)$ have additional dependence on  the set $\bu$
does not play any role.} \eqref{07-SP-SPlm3new}. Thus, due to corollary~\ref{07-SP-exp-det3C}
we find
\begin{equation}\label{07-SP-SPlm3new1}
S_n(\bv|\bu)=\Delta'_n(\bu) \Delta_n(\bv)\;\det_n\Bigl( -r(v_k)f(\bu,v_k)t(v_k-c,u_j)h(v_k-c,\bu)
+\kappa t(v_k,u_j)h(v_k,\bu)\Bigr)\,.
\end{equation}
Using obvious identities
\be{07-evid1}
t(v-c,u)=t(u,v), \qquad  f(u,v)h(v-c,u)=\frac{(u-v+c)(v-u)}{c(u-v)}=-h(u,v),
\ee
we arrive at
\begin{equation}\label{07-SP-SPlmexpl}
S_n(\bv|\bu)=\Delta'_n(\bu)
\Delta_n(\bv)\;\det_n\bigl[(-1)^{n+1}r(v_k)t(u_j,v_k)h(\bu,v_k)+  \kappa\,t(v_k,u_j)h(v_k,\bu)\bigr].
\end{equation}
After simple algebra this formula also can be written in the form
\begin{equation}\label{07-SP-det-pres1}
S_n(\bv|\bu)=\Delta'_n(\bu)\Delta_n(\bv)\; h(\bv,\bu)
\;\det_n\mathcal{M}_{jk}\,,
\end{equation}
where
\begin{equation}\label{07-SP-det-pres0}
\mathcal{M}_{jk}
=t(v_k,u_j)
\left(\kappa-r(v_k)\frac{f(\bu_j,v_k)}{f(v_k,\bu_j)}\right)\,.
\end{equation}

{\sl Remark}. We have considered the case of the scalar product in which the vector $B(\bu)|0\rangle$ is
the twisted on-shell  vector, and the dual vector $\langle0|C(\bv)$ is arbitrary. One can consider
the case when the  dual vector $\langle0|C(\bv)$ is an eigenvector of the twisted transfer matrix, while the vector $B(\bu)|0\rangle$ is arbitrary.
It is enough to use the equation \eqref{07-SP-symm}, and the problem reduces to the case considered above.
The result is still given by the formulas \eqref{07-SP-det-pres1}, \eqref{07-SP-det-pres0}, where one needs to make the replacement
$\bv\leftrightarrow\bu$. The reader can easily verify this independently.

{\sl Remark}. All the considerations above remain true  for the models with trigonometric $R$-matrix.

To conclude this section we note that one can rewrite the formula for the scalar product \eqref{07-SP-SPlmexpl} in terms
of Jacobian of the transfer matrix eigenvalue
\be{07-E-val}
\tau(v|\bu)=a(v)f(\bu,v)+d(v)f(v,\bu).
\ee
Indeed, it is not difficult to check that representation \eqref{07-SP-SPlmexpl} is equivalent to the formula
\begin{equation}\label{07-SP-Jac-pres0}
S_n(\bv|\bu)=\frac{\Delta'_n(\bu)\Delta_n(\bv)}{g(\bv,\bu)}
\;\det_n\left(\frac{c}{d(v_k)}\frac{\partial\tau(v_k|\bu)}{\partial u_j}
\right)\,.
\end{equation}

\section{Orthogonality of the eigenvectors\label{07-Sec6}}

The formula obtained above allows us to consider a particular case of the scalar product when both vectors are
twisted  on-shell Bethe vectors. We will see that if the two vectors are different, then their scalar product vanishes
in complete agreement with the general theory.

We start with the case when $v_j\ne u_k$ for all $j,k=1,\dots,n$. If the vector $\langle0|C(\bv)$ is an eigenvector
of the twisted transfer matrix, then the parameters $\bv$ satisfy twisted Bethe equations
\be{07-SP-BEmu}
r(v_k)=\kappa\,\frac{f(v_k,\bv_k)}{f(\bv_k,v_k)}\,.
\ee
Substituting the function $r(v_k)$  \eqref{07-SP-BEmu} into \eqref{07-SP-det-pres0} we obtain
after simple algebra
\begin{equation}\label{07-SP-cM}
\mathcal{M}_{jk}=t(v_k,u_j)+t(u_j,v_k)\;Y_k^{(n)}=\frac{\kappa\, c^2}{(v_k-u_j)(v_k-u_j+c)}+
\frac{ \kappa\,c^2}{(v_k-u_j)(v_k-u_j-c)}\;Y_k^{(n)}\,,
\end{equation}
where
\begin{equation}\label{07-SP-Yk}
Y_{k}^{(n)}=\frac{h(\bu,v_k)h(v_k,\bv)}{h(v_k,\bu)h(\bv,v_k)}=\prod_{m=1}^n\frac{(v_k-u_m-c)}{(v_k-u_m+c)}
\frac{(v_k-v_m+c)}{(v_k-v_m-c)}\,.
\end{equation}
In these formulas we give expressions for $\mathcal{M}_{jk}$ and $Y_{k}^{(n)}$ in two forms: first, in terms of the functions $h$ and $t$;
then explicitly. The first form is universal, because it is valid both for the models with rational and trigonometric $R$-matrices.

Let us prove that $\det_n\mathcal{M}_{jk}=0$. For this we prove that the rows of the matrix $\mathcal{M}_{jk}$ are linearly dependent.

\begin{prop}\label{07-SP-prop-ortho}
Let
\be{07-SP-nuk}
\nu_j^{(n)}=\frac{g(u_j,\bu_j)}{g(u_j,\bv)}=\frac1{c}\;\prod_{m=1}^n(u_j-v_m)\prod_{\substack{m=1\\m\ne j}}^n(u_j-u_m)^{-1}\,.
\ee
Then
\be{07-SP-lin-dep}
\frac1{\kappa}\sum_{j=1}^m\nu_j^{(n)}\mathcal{M}_{jk}=0\,.
\ee
\end{prop}

{\sl Proof}.
The linear combination \eqref{07-SP-lin-dep} can be presented in the form
\be{07-SP-2sum}
\frac1{\kappa}\sum_{j=1}^n\nu_j^{(n)}\mathcal{M}_{jk}=U_k^+ +U_k^-\;Y_k^{(n)}\,,
\ee
where
\be{07-SP-Upm}
U_k^\pm=\sum_{j=1}^n \frac{\nu_j^{(n)} c^2}{(v_k-u_j)(v_k-u_j\pm c)}\,.
\ee
To compute $U_k^\pm$  we consider an auxiliary contour integral over a circle of a big radius
\be{07-SP-Int-def}
I_\pm=\frac1{2\pi i}\oint_{|z|=R\to\infty} \frac{ c\,dz}{(v_k-z)(v_k-z\pm c)}
\prod_{m=1}^n\frac{z-v_m}{z-u_m}\,.
\ee
Since the integrand behaves as $z^{-2}$ at $z\to\infty$,
we conclude that $I_\pm=0$. On the other hand, the integral \eqref{07-SP-Int-def} is equal to the sum of residues in the poles
within the integration contour. First, these are poles at $z=u_\ell$, $\ell=1,\dots,n$.
It is easy to see that the sum of the residues in these poles gives exactly $U_k^\pm$. The pole in the point $z=v_k$ actually
is spurious, because it is compensated by the factor $(z-v_k)$. There remains one more pole in the point
$z=v_k\pm c$. Obviously,
\be{07-SP-res}
\Res \frac{ c}{(v_k-z)(v_k-z\pm c)}
\prod_{m=1}^n\frac{z-v_m}{z-u_m}\Bigr|_{z=v_k\pm c}=
\pm \prod_{m=1}^n\frac{v_k-v_m\pm c}{v_k-u_m\pm c}\,.
\ee
We arrive at an identity
\be{07-SP-int-sum}
I_\pm=0=U_k^\pm\pm \prod_{m=1}^n\frac{v_k-v_m\pm c}{v_k-u_m\pm c}\,,
\ee
what implies
\be{07-SP-Upm-res}
U_k^\pm=\mp \prod_{m=1}^n\frac{v_k-v_m\pm c}{v_k-u_m\pm c}\,.
\ee
Substituting these expressions for $U_k^\pm$ into \eqref{07-SP-2sum} we immediately arrive at \eqref{07-SP-lin-dep}.

{\sl Question }. In the trigonometric case, the rows of the matrix $\mathcal{M}_{jk}$ also are linearly dependent. The coefficients of the linear combination
still are given by \eqref{07-SP-nuk}, if we use the trigonometric analog of the function $g$. However, the auxiliary integral
\eqref{07-SP-Int-def} is different. What kind of changes occur in this integral?

The proof of orthogonality slightly changes if some variables from the set $\bv$
coincide with the variables from the set $\bu$. In this case, some matrix elements of the matrix \eqref{07-SP-det-pres1}
have indeterminate form. Without loss of generality we can assume that $v_j=u_j$ for $j=\ell+1,\dots,n$, where $\ell<n$,
while  $v_j\ne u_j$ for $j=1,\dots,\ell$. Consider the diagonal elements of the matrix $\mathcal{M}_{jk}$
\eqref{07-SP-det-pres0} at $j=\ell+1,\dots,n$
\begin{equation}\label{07-SP-det-pres2}
\mathcal{M}_{jj}
=\frac{ c^2}{(v_j-u_j)(v_j-u_j+c)}
\left(\kappa-r(v_j)\prod_{\substack{m=1\\m\ne j}}^n\frac{v_j-u_m-c}{v_j-u_m+c}\right)\,.
\end{equation}
We see that if we set   $v_j=u_j$,
then we obtain zero in the denominator, but at the same time the expression in the parenthesis vanishes due to the Bethe equations.
Therefore, first of all, we should solve this indeterminateness, and only after this we can express
the function $r(v)$ in terms of solutions of  Bethe equations \eqref{07-SP-BEmu}. We will consider this issue in more detail in the following
section, but for now we will study the matrix elements of $\mathcal{M}_{jk}$
for $j=1,\dots,\ell$.  There are no singularities in these matrix elements, therefore, as before,
we can simply express the function $r(v_k)$ in terms of solutions of the Bethe equations and obtain
\begin{equation}\label{07-SP-cM1}
\mathcal{M}_{jk}=\frac{ \kappa\,c^2}{(v_k-u_j)(v_k-u_j+c)}+
\frac{ \kappa\,c^2}{(v_k-u_j)(v_k-u_j-c)}\;Y_k^{(\ell)}\,,
\end{equation}
where
\begin{equation}\label{07-SP-Ykl}
Y_{k}^{(\ell)}=Y_{k}^{(n)}\Bigr|_{\substack{v_{s}=u_{s}\\s>\ell}}
=\prod_{m=1}^\ell\frac{(v_k-u_m-c)}{(v_k-u_m+c)}
\frac{(v_k-v_m+c)}{(v_k-v_m-c)}\,.
\end{equation}
Note that the function $Y_{k}^{(\ell)}$ differs from the previous function $Y_{k}^{(n)}$ \eqref{07-SP-Yk}
only in that the  upper limit in the product over $m$ is now bounded by $\ell$.
We can thus say that the matrix elements $\mathcal{M}_{jk}$ with $j=1,\dots,\ell$
depend  on variables $v_1,\dots,v_\ell$ and $u_1,\dots,u_\ell$ only.

A similar effect occurs with the coefficients $\nu_j^{(n)}$  $j=1,\dots,\ell$:
\be{07-SP-nuj}
\nu_j^{(n)}\Bigr|_{\substack{v_{s}=u_{s}\\s>\ell}}=\nu_j^{(\ell)}=
\prod_{m=1}^\ell(u_j-v_m)\prod_{\substack{m=1\\m\ne j}}^\ell(u_j-u_m)^{-1},
\qquad j=1,\dots,\ell\,.
\ee
Therefore, replacing  $n$ by $\ell$ in the proof considered above, we immediately obtain
\be{07-SP-lin-dep1}
\sum_{j=1}^\ell\nu_j^{(\ell)}\mathcal{M}_{jk}=0\,.
\ee
The remaining part of the linear combination is definitely  zero, since $\nu_j^{(n)}=0$ for $j=\ell+1,\dots,n$.
Therefore, regardless of the explicit form of the matrix elements in the rows with  $j=\ell+1,\dots,n$
this part of the linear combination does not make any contribution. Hence, the equation \eqref{07-SP-lin-dep} is still
valid. Thus, the orthogonality of eigenvectors is fulfilled also in the case when some parameters of the set $\bv $ coincide
with parameters from the set $\bu $.

\section{Norm of the twisted on-shell vector\label{07-Sec7}}

The only case when the proof does not work is the case $\bv =\bu$. Then we are dealing with a scalar product
of an eigenvector with itself, that is, the square of the norm\footnote{Such a scalar product is usually called the square of the norm,
even if the dual vector is not Hermitian conjugated to the initial vector.}. It is natural to expect that the square of the norm of
the eigenvector is not equal to zero. Indeed, if $\bv =\bu$, then all the coefficients $\nu_j^{(n)}$ vanish, therefore,
we cannot prove the linear dependence of the rows of the matrix $\mathcal{M}_{jk}$.

As we already noted, in the case when the parameters $\bv$ and $\bu$ coincide, the diagonal elements of the matrix
$\mathcal{M}_{jj}$ have indeterminant form (see \eqref{07-SP-det-pres2}), that should be resolved.
It is important to emphasize that it would be wrong
first,  to express the function $r(v)$ in terms of solutions of the Bethe equations in  \eqref{07-SP-det-pres2}, and then
to take the limit $v_j\to u_j$. Indeed, one cannot send  two different solutions of any system of equations to each other.
Let us illustrate this with a simple example.

Consider the scalar product $S_1(v|u)$, in which the vectors  $\langle0|C(v)$ and $B(u)|0\rangle$
depend on one variable each. This scalar product is elementary calculated either by means of commutation
relations \eqref{07-CB-AD}, or using the formula \eqref{07-SP-SPlm1}:
\be{07-SP-n1}
S_1(v|u)=-c\;\frac{ r(v)-r(u)}{v-u}\,.
\ee
This formula holds for any complex $v$ and $u$. In particular, for
$v=u$ we obtain
\be{07-SP-n1ll}
S_1(u|u)=-c \;r'(u)\,.
\ee

If, on the other hand, both vectors are eigenvectors of the twisted transfer matrix, then $v$ and $u$ satisfy the system of twisted Bethe equations.
The latter  in this case has the form
\be{07-SP-BEn1r}
r(v)=\kappa,\qquad r(u)=\kappa\,.
\ee
If we substitute this into \eqref{07-SP-n1}, we immediately obtain $S_1(v|u)=0$, and taking
the limit $v=u $ we still have zero. Of course, the described calculation is incorrect. Indeed, consider for simplicity
the QNLS model, where $ r(u)=e^{-iuL}$. Then the equations \eqref{07-SP-BEn1r} take the form
\be{07-SP-BEn1}
e^{-ivL}=\kappa,\qquad
e^{-iuL}=\kappa\,,
\ee
what implies
\be{07-SP-root}
v=\log\kappa+\frac{2\pi is}L, \qquad u=\log\kappa+\frac{2\pi ir}L,
\ee
where $s$ and $r$ are integers. If  we assume from the very beginning that
$v\ne u$, then we obtain that $s\ne r$, and  we cannot consider the limit of two different
integers going to each other. If, on the other hand, we say that $v=u$, then we must return to the formula \eqref{07-SP-n1}, calculate
the limit \eqref{07-SP-n1ll}, and only after this we can substitute an explicit expression for $u$ \eqref{07-SP-root} into the obtained result
(if necessary).

One can also require that  one of the variables, for example $u$, is a root
of twisted Bethe equation, while $v$ is generic. Then the formula \eqref{07-SP-n1}
turns into
\be{07-SP-n12}
S_1(v|u)=-c\;\frac{ r(v)-\kappa}{v-u}\,.
\ee
Here the limiting procedure $ v\to u$ is possible, since $v$ remains an arbitrary complex number,
and we do can consider the limit when this number goes to the root of the twisted Bethe equation. The only  wrong method is
to take the limit in which two different roots go to each other.

Let us turn back to the calculating the norm of the twisted on-shell Bethe vector in the general case.
Setting in \eqref{07-SP-det-pres2} $v_j=u_j+\varepsilon$, we obtain
\begin{equation}\label{07-SP-Matel-diag}
\mathcal{M}_{jj}=\lim_{\varepsilon\to 0}
\frac{ c}{\varepsilon}
\left(\kappa-r(u_j+\varepsilon)\prod_{\substack{m=1\\m\ne j}}^n\frac{(u_j-u_m-c+\varepsilon)}{(u_j-u_m+c+\varepsilon)}\right)\,.
\end{equation}
Taking into account that the parameters $\bu$ satisfy twisted Bethe equations we find after simple algebra
\begin{equation}\label{07-SP-Matel-diag1}
\mathcal{M}_{jj}=
-c\kappa\Bigl(\log'r(u_j)+\sum_{\substack{m=1\\m\ne j}}^n\mathcal{K}(u_j-u_m)\Bigr)\,,
\end{equation}
where
\be{07-SP-calK}
\mathcal{K}(u)=\frac{2 c}
{(u+c)(u-c)}\,.
\ee
Observe that in spite of the function $r(u_j)$ can be expressed from (twisted) Bethe equations in terms of the product of the $f$-functions,
its logarithmic derivative $\log'r(u_j)$ cannot be found from these equations. Bethe equations
do not imply any conditions on the derivatives of the function $r(u)$.

In order to find off-diagonal matrix elements, we do not need to calculate the limit. Instead,
we can simply put $v_j=u_j$
\begin{equation}\label{07-SP-Matel-offdiag}
\mathcal{M}_{jk}=
\frac{c^2}{(u_k-u_j)(u_k-u_j+c)}
\left(\kappa-r(u_k)\prod_{\substack{m=1\\m\ne j}}^n\frac{(u_k-u_m-c)}
{(u_k-u_m+c)}\right),
\qquad  j\ne k\,.
\end{equation}
Substituting here $r(u_k)$ in terms of twisted Bethe equations we obtain after some algebra
\begin{equation}\label{07-SP-Matel-offdiag1}
\mathcal{M}_{jk}=c\kappa\,\mathcal{K}(u_j-u_k), \qquad  j\ne k\,,
\end{equation}
where $\mathcal{K}(u)$ is defined by \eqref{07-SP-calK}. Combining \eqref{07-SP-Matel-diag1} and \eqref{07-SP-Matel-offdiag1}
we find
\begin{equation}\label{07-SP-Mtilde}
\mathcal{M}_{jk}\Bigr|_{\bv=\bu}= c\kappa\, \widetilde{\mathcal{M}}_{jk},
\end{equation}
where
\begin{equation}\label{07-SP-Matel-all}
\widetilde{\mathcal{M}}_{jk}=
-\delta_{jk}\Bigl(\log'r(u_j)+\sum_{m=1}^n\mathcal{K}(u_j-u_m)\Bigr)
+\mathcal{K}(u_j-u_k)\,.
\end{equation}
Thus, we finally obtain for the square of the norm of the twisted on-shell Bethe vector the following result:
\be{07-SP-Norm}
S_n(\bu|\bu)=(c\kappa)^n\prod_{\substack{j,k=1\\j\ne k}}^n
f(u_j,u_k)\; \det_n\widetilde{\mathcal{M}}_{jk}\,.
\ee

It is interesting to note that, as in the case of the scalar product, the determinant in \eqref{07-SP-Norm} has the meaning of Jacobian.
Let us introduce a function
\be{07-SP-Psi}
\Psi_j\equiv \Psi_j(\bu)=\log r(u_j)+\sum_{\substack{k=1\\k\ne j}}^n\log\left(\frac{u_k-u_j+c}
{u_k-u_j-c}\right)\,.
\ee
Then twisted Bethe equations \eqref{07-SP-BE} for the parameters $\bu$  take a very simple form
\be{07-SP-BE-log}
\Psi_j=2\pi in_j+\log\kappa\,,
\ee
where $n_j$ are integers. In fact, $\Psi_j(\bu)$ is the logarithm of the system of twisted Bethe equations. It is not difficult to check that
\be{07-SP-Jak-norm}
\widetilde{\mathcal{M}}_{jk}=-\frac{\partial\Psi_j}{\partial u_k}\,,
\ee
and thus, the square of the norm of the twisted transfer matrix eigenvector is proportional to the Jacobian of the system
of twisted Bethe equations in the logarithmic  form.

To conclude this section we recall once more that all the formulas obtained above remain valid in the particular case
$\kappa=1$.

{\sl Question}.  We suggest the reader to derive the formula for the square of the norm of the twisted on-shell vector in the case of the trigonometric $R$-matrix.

\section{Answers to some questions\label{07-Sec8}}

In this section we answer two questions formulated in this and the previous lectures.

\subsection{Proof of proposition~8.4.2.\label{07-Sec81}}

We actually proved in section~\ref{06-Sec4}
that if commutation relations of some operators $X(u)$ and $Y(u)$ are given by
\begin{equation}
X(v)Y(u)=f(u,v)Y(u)X(v)+g(v,u)Y(v)X(u),\label{07-MCR-AB}
\end{equation}
then their multiple commutation relations have the form
\begin{equation}\label{07-MCR-two}
X(\bv)Y(\bu)=(-1)^{n}\sum_{\{\bv,\bu\}\mapsto\{\bar w_{\so},\bar w_{\st}\}} K_{n}(\bu|\bar w_{\st}+c)f(\bar w_{\st},\bar w_{\so})
Y(\bar w_{\st})X(\bar w_{\so}).
\end{equation}
We considered the case  $X(u)=A(u)$ and $Y(u)=B(u)$. However, we could put  $X(u)=A(u)$ and $Y(u)=A(u)$.  Indeed, since $[A(u),A(v)]=0$,
the equality
\begin{equation}
A(v)A(u)=f(u,v)A(u)A(v)+g(v,u)A(v)A(u)\label{07-MCR-AY}
\end{equation}
is correct, because $f=1+g$ for the rational
$R$-matrix. But then \eqref{07-MCR-AY} implies
\begin{equation}\label{07-MCR-twoAY}
A(\bv)A(\bu)=(-1)^{n}\sum_{\{\bv,\bu\}\mapsto\{\bar w_{\so},\bar w_{\st}\}} K_{n}(\bu|\bar w_{\st}+c)f(\bar w_{\st},\bar w_{\so})
A(\bar w_{\st})A(\bar w_{\so}).
\end{equation}
Here we have the same operator products in the both sides, therefore, we obtain
\begin{equation}\label{07-MCR-twoAY1}
A(\bv)A(\bu)\left(1-(-1)^{n}\sum_{\{\bv,\bu\}\mapsto\{\bar w_{\so},\bar w_{\st}\}} K_{n}(\bu|\bar w_{\st}+c)f(\bar w_{\st},\bar w_{\so})\right)
=0.
\end{equation}
This is possible only if the condition~\eqref{06-MCR-two-g-rat2}
is fulfilled.

\subsection{Calculating the sum \eqref{07-SP-r1}\label{07-Sec82}}

The formula for the scalar product \eqref{07-SP-SPlm1} is valid for {\it any} model whose monodromy matrix satisfies
the $RTT$-relation and has a vacuum. A particular case of such a monodromy matrix is the identity matrix. First, it certainly satisfies
the $RTT$-relation. Secondly, we can consider its matrix elements as operators acting in a one-dimensional
Hilbert space, and as a vacuum vector we can choose any nonzero vector of this space. Then the scalar
product of two Bethe vectors in this model is given by the formula \eqref{07-SP-SPlm1}. It is also obvious that in this model
$r(u) =1$, and we arrive at the formula \eqref {07-SP-r1}. But on the other hand, in this monodromy matrix the operators $B(u)$ and $C(u)$ are
identically equal zero. Hence, all Bethe vectors (except the vacuum) are null-vectors, and hence their scalar products vanish.


%
%
\chapter{Quantum inverse problem\label{CHA-QIP}}

In $XXX$ and $XXZ$ Heisenberg chains, the monodromy matrix is the product of $L$-operators (2.3.14). 
Therefore, the operators $A$, $B$, $C$, and $D$ depend on the local spin operators $\sigma_k^\alpha$, $\alpha=x,y,z$.
However, for the chains of a long  length, one can hardly  hope to obtain an explicit expression for the monodromy matrix
elements in terms of $\sigma$-matrices.  The inverse problem --- to express the local spin operators $\sigma_k^\alpha$
in terms of the monodromy matrix entries seems to be even more hopeless. This problem all the more seems unsolvable, because the operators
$A$, $B$, $C$, and $D$  are global in the sense that they depend on the spin operators at all sites of the chain
$\sigma_1^\alpha,\dots,\sigma_N^\alpha$.
And if we can still hope to express some global object in terms of local objects, then the idea of expressing local objects in terms of global ones looks absurd. However, the example of $XXX$ and $XXZ$ chains refutes all these `obvious' arguments. In these models, it is possible to obtain simple and, we will not be afraid of this word, very nice formulas for the operators $\sigma_k^\alpha$
in terms of  the monodromy matrix elements \cite{KitMT99}. These formulas are called {\it Quantum inverse problem}, which
is considered below.

\section{Cyclic permutations in the monodromy matrix\label{10-Sec1}}

We consider an inhomogeneous  $XXZ$ chain, in which the monodromy matrix depends on inhomogeneities $\bxi=\{\xi_1,\dots,\xi_N\}$
\begin{multline}\label{10-LISP-Def-T}
T_0(u)=T_{0;1\dots N}(u|\{\xi\})=R_{0N}(u,\xi_N)\cdots R_{01}(u,\xi_1)\\
=\begin{pmatrix}
A_{1\dots N}(u|\{\xi\})& B_{1\dots N}(u|\{\xi\})\\
C_{1\dots N}(u|\{\xi\})&D_{1\dots N}(u|\{\xi\}).
\end{pmatrix}_0
\end{multline}
This matrix acts in the tensor product of spaces $V_0\otimes V_1\otimes\cdots\otimes V_N$,
where each space $V_k\sim \mathbb{C}^2$. The space $V_0$ is auxiliary. In this space, the monodromy matrix can be
written as a $2\times 2$ matrix, whose matrix elements are operators acting in the space
$V_1\otimes\cdots\otimes V_N$. All this is stressed by the notation in \eqref{10-LISP-Def-T}. The short
notation $T_0(u)$ means that being a $2\times 2$ matrix, the monodromy matrix acts in the
space $ V_0 $. The more detailed notation $T_{0;1\dots N}(u|\{\xi\})$ shows that
the matrix elements of $T_0(u)$ act in the space $V_1\otimes\cdots\otimes V_N$,
and  each space $V_k$ is associated with the inhomogeneity $\xi_k$ . The subscript $0$ is separated from the rest of the indices by a semicolon,
to emphasize that $V_0$ is the auxiliary space. This is also shown by the $0$ subscript of the matrix
in the second line of \eqref{10-LISP-Def-T}.

Let the $R$-matrix be normalized so as it turns into the permutation matrix if their arguments coincide: $R(u,u)=P$. Consider the
transfer matrix $\mathcal{T}(u)$ at a point of inhomogeneity $u=\xi_k$:
\begin{equation}\label{10-LISP-Def-Tau}
\mathcal{T}(\xi_k)={\tr}_0 T_0(\xi_k)=A_{1\dots N}(\xi_k|\{\xi\})+D_{1\dots N}(\xi_k|\{\xi\}).
\end{equation}
The trace is taken over the space $V_0$, what is stressed by the subscript $0$ of the symbol $\tr$. Substituting
the product of the $R$-matrices from \eqref{10-LISP-Def-T} into \eqref{10-LISP-Def-Tau} and taking into account the cyclicity of the trace, we find
\begin{multline}\label{10-LISP-res-tau1}
\mathcal{T}(\xi_k)={\tr}_0 T_{0;1\dots N}(\xi_k|\{\xi\})={\tr}_0 \Bigl(R_{0N}(\xi_k,\xi_N)\cdots R_{01}(\xi_k,\xi_1)\Bigr)\\
{\tr}_0\Bigl(P_{0k}R_{0k-1}(\xi_k,\xi_{k-1})\cdots R_{01}(\xi_k,\xi_1)R_{0N}(\xi_k,\xi_N) \cdots R_{0k+1}(\xi_k,\xi_{k+1})\Bigr).
\end{multline}
Here we also used $R_{0k}(\xi_k,\xi_k)=P_{0k}$. Let us add the identity operator in the form $\mathbf{1}=P_{0k}^2$
on the right of the product of all the matrices:
\begin{equation}\label{10-LISP-res-tau2}
\mathcal{T}(\xi_k)={\tr}_0\Bigl(P_{0k}R_{0k-1}(\xi_k,\xi_{k-1})\cdots R_{01}(\xi_k,\xi_1)R_{0N}(\xi_k,\xi_N) \cdots R_{0k+1}(\xi_k,\xi_{k+1})P_{0k}P_{0k}\Bigr).
\end{equation}
Due to the properties of the permutation matrix, we should replace the space $V_0$  by $V_k$ in the product
$R_{0k-1}(\xi_k,\xi_{k-1})\cdots R_{0k+1}(\xi_k,\xi_{k+1})$
between two $P_{0k}$. We obtain
\begin{equation}\label{10-LISP-res-tau3}
\mathcal{T}(\xi_k)={\tr}_0\Bigl(R_{kk-1}(\xi_k,\xi_{k-1})\cdots R_{k1}(\xi_k,\xi_1)R_{kN}(\xi_k,\xi_N) \cdots R_{kk+1}(\xi_k,\xi_{k+1})P_{0k}\Bigr).
\end{equation}
We arrive at the expression of the form
\begin{equation}\label{10-LISP-Exerc}
{\tr}_0\Bigl(Z_{1\dots N}P_{0k}\Bigr),
\end{equation}
where the matrix $Z_{1\dots N}$ acts in the space $V_0$ as the identity operator.
That is, in the product $Z_{1\dots N}P_{0k}$ only the permutation matrix $P_{0k}$ acts nontrivially in the space $V_0$, therefore
the matrix $Z_{1\dots N}$ can be taken out from the trace symbol ${\tr}_0$. Since ${\tr}_0P_{0k}=\mathbf{1}$,
we find
\begin{equation}\label{10-LISP-Exerc1}
{\tr}_0\Bigl(Z_{1\dots N}P_{0k}\Bigr)=Z_{1\dots N}\cdot{\tr}_0 P_{0k}=Z_{1\dots N},
\end{equation}
and hence,
\begin{equation}\label{10-LISP-res-tau4}
\mathcal{T}(\xi_k)=R_{kk-1}(\xi_k,\xi_{k-1})\cdots R_{k1}(\xi_k,\xi_1)R_{kN}(\xi_k,\xi_N) \cdots R_{kk+1}(\xi_k,\xi_{k+1}).
\end{equation}

\vspace{5mm}

It turns out that the products of transfer matrices $\mathcal{T}(\xi_k)$ are the operators of cyclic permutations.
More precisely,
\begin{multline}\label{10-LISP-1-cycl-n}
\prod_{k=1}^m\mathcal{T}(\xi_k)\cdot T_{0;1\dots N}(u|\{\xi\})\cdot\prod_{k=1}^m\mathcal{T}^{-1}(\xi_k)\\
=R_{0m}(u,\xi_m)\cdots R_{01}(u,\xi_1)R_{0N}(u,\xi_N)\cdots R_{0m+1}(u,\xi_{m+1}),
\end{multline}
that is,  the operators $\prod_{k=1}^m\mathcal{T}(\xi_k)$ make a cyclic permutation of the $R$-matrices when acting on the
monodromy matrix $T_{0;1\dots N}(u|\{\xi\})$. Let us check this.

We begin with the case $m=1$. First of all, we recall the Yang--Baxter equation in the tensor product of the spaces
$V_a\otimes V_b\otimes V_c$:
\begin{equation}\label{10-LISP-YB-gen}
R_{ab}(\mu_a,\mu_b)R_{ac}(\mu_a,\mu_c)R_{bc}(\mu_b,\mu_c)=R_{bc}(\mu_b,\mu_c)R_{ac}(\mu_a,\mu_c)R_{ab}(\mu_a,\mu_b).
\end{equation}

Let us consider the product $\mathcal{T}(\xi_1) T_{0}(u)$:
\begin{equation}\label{10-LISP-1-cycl}
\mathcal{T}(\xi_1) T_{0;1\dots N}(u|\{\xi\})=
\mathcal{T}(\xi_1)R_{0N}(u,\xi_N)\cdots R_{02}(u,\xi_2)R_{01}(u,\xi_1).
\end{equation}
Substituting here $\mathcal{T}(\xi_1)$ from \eqref{10-LISP-res-tau4}, we find
\begin{equation}\label{10-LISP-1-cycl-1}
\mathcal{T}(\xi_1) T_{0;1\dots N}(u|\{\xi\})=R_{1N}(\xi_1,\xi_{N})\cdots \underline{R_{12}(\xi_1,\xi_2)}
\; R_{0N}(u,\xi_N)\cdots \underline{R_{02}(u,\xi_2)R_{01}(u,\xi_1)}.
\end{equation}
Observe that the combination of underlined terms coincides with the rhs of the equation \eqref{10-LISP-YB-gen}, if we put in the latter $a=0$, $b=1$, and $c=2$.  In addition, the $R$-matrix $R_{12}(\xi_1,\xi_2)$ acts trivially in the space $V_0\otimes V_3\otimes\cdots\otimes V_N$,
and therefore, it commutes with all $R$-matrices $R_{0k}(u,\xi_k)$ for $k=3,\dots,N$. Therefore, we can carry
$R_{12}(\xi_1,\xi_2)$ to the right through the product $R_{0N}(u,\xi_N)\cdots R_{03}(u,\xi_3)$.
After this we get exactly the rhs of the equation \eqref{10-LISP-YB-gen}. Replacing it with the lhs
we arrive at the expression
\begin{multline}\label{10-LISP-1-cycl-2}
\mathcal{T}(\xi_1) T_{0;1\dots N}(u|\{\xi\})=R_{1N}(\xi_1,\xi_{N})\cdots \underline{R_{13}(\xi_1,\xi_3)}\\
\times R_{0N}(u,\xi_N)\cdots \underline{R_{03}(u,\xi_3)R_{01}(u,\xi_1)} R_{02}(u,\xi_2)R_{12}(\xi_1,\xi_2).
\end{multline}
We see that we again can apply the Yang--Baxter equation to the new underlined terms in \eqref{10-LISP-1-cycl-2},
what gives us
\begin{multline}\label{10-LISP-1-cycl-3}
\mathcal{T}(\xi_1) T_{0;1\dots N}(u|\{\xi\})
=R_{1N}(\xi_1,\xi_{N})\cdots \underline{R_{14}(\xi_1,\xi_4)}\\
\times
R_{0N}(u,\xi_N)\cdots \underline{R_{04}(u,\xi_4)R_{01}(u,\xi_1)} R_{03}(u,\xi_3)R_{02}(u,\xi_2)
R_{13}(\xi_1,\xi_3)R_{12}(\xi_1,\xi_2),
\end{multline}
and so on.  Continuing this process we move all the $R$-matrices $R_{1k}(\xi_1,\xi_{k})$ to the right and eventually obtain
\begin{multline}\label{10-LISP-1-cycl-4}
\mathcal{T}(\xi_1) T_{0;1\dots N}(u|\{\xi\})=R_{01}(u,\xi_1)R_{0N}(u,\xi_N)\cdots R_{02}(u,\xi_2)\\
\times
R_{1N}(\xi_1,\xi_{N})\cdots R_{12}(\xi_1,\xi_2)=R_{01}(u,\xi_1)R_{0N}(u,\xi_N)\cdots R_{02}(u,\xi_2)
\mathcal{T}(\xi_1).
\end{multline}
Thus, equation \eqref{10-LISP-1-cycl-n}  is proved for  $m=1$.

To prove  \eqref{10-LISP-1-cycl-n} for $m=2$ we should check that
\begin{multline}\label{10-LISP-2-cycl}
\mathcal{T}(\xi_2) R_{01}(u,\xi_1)R_{0N}(u,\xi_N)\cdots R_{02}(u,\xi_2)\mathcal{T}^{-1}(\xi_2)\\
=R_{02}(u,\xi_2)R_{01}(u,\xi_1)R_{0N}(u,\xi_N)\cdots R_{03}(u,\xi_3).
\end{multline}
We could repeat all the above arguments related to the permutation of $R$-matrices
via the Yang--Baxter equation. However, this is not necessary, since a simple change of variables
reduces our problem to the case $m = 1$. Indeed, let us make a cyclic change of variables $\xi_k$ to $\xi'_k$
and the spaces $V_k$ to $V'_k$:
\begin{equation}\label{10-LISP-redenote}
\begin{array}{ccccc}
\xi_1&\xi_2&\dots&\xi_{N-1}&\xi_N\\
\downarrow&\downarrow&{}&\downarrow&\downarrow\\
\xi'_N&\xi'_1&\dots&\xi'_{N-2}&\xi'_{N-1}
\end{array}
\qquad\qquad
\begin{array}{ccccc}
1&2&\dots&{N-1}&N\\
\downarrow&\downarrow&{}&\downarrow&\downarrow\\
N'&1'&\dots&{N'-2'}&{N'-1'}
\end{array}
\end{equation}
Then, using the new notation we recast \eqref{10-LISP-2-cycl} in the form
\begin{multline}\label{10-LISP-2-cycl-1}
\mathcal{T}(\xi'_1) R_{0N'}(u,\xi'_{N'})R_{0N'-1'}(u,\xi'_{N'-1'})\cdots R_{0{1'}}(u,\xi'_1)\mathcal{T}^{-1}(\xi'_1)\\
=R_{01'}(u,\xi'_1)R_{0N'}(u,\xi'_{N'})R_{0N'-1'}(u,\xi'_{N'-1'})\cdots R_{02'}(u,\xi'_2).
\end{multline}
This form exactly coincides with already proved  \eqref{10-LISP-1-cycl-4}. It is clear that the process can be continued
for $m=3,4,\dots$, and we thus prove \eqref{10-LISP-1-cycl-n} for arbitrary $m$.

\section{Inverse problem\label{10-Sec2}}

The formulas given above are preparatory for our main goal: to express an arbitrary matrix $E_m$ that acts nontrivially in the space $V_m$
in terms of the matrix $E_0$ acting in the auxiliary space $V_0$,
and a transfer matrix. For this we consider expression
\begin{equation}\label{10-LISP-Calcul}
{\tr}_0\Bigl(E_0T_{0;1\dots N}(\xi_m|\{\xi\})\Bigr).
\end{equation}
Using equation \eqref{10-LISP-1-cycl-n} we can rewrite it as
\begin{multline}\label{10-LISP-Calcul-1}
{\tr}_0\Bigl(E_0T_{0;1\dots N}(\xi_m|\{\xi\})\Bigr)={\tr}_0\Bigl(T_{0;1\dots N}(\xi_m|\{\xi\})E_0\Bigr)
={\tr}_0\Bigl[\left(\prod_{k=1}^m\mathcal{T}^{-1}(\xi_k)\right)\\
%
P_{0m}R_{0m-1}(\xi_m,\xi_{m-1})\cdots R_{01}(\xi_m,\xi_1)R_{0N}(\xi_m,\xi_N)\cdots R_{0m+1}(\xi_m,\xi_{m+1})
\left(\prod_{k=1}^m\mathcal{T}(\xi_k)\right)E_0\Bigr],
\end{multline}
where we took into account that $R_{0m}(\xi_m,\xi_m)=P_{0m}$. Since the transfer matrices $\mathcal{T}(\xi_k)$ trivially
act in $V_0$, we can take them out of the trace ${\tr}_0$
\begin{multline}\label{10-LISP-Calcul-2}
{\tr}_0\Bigl(E_0T_{0;1\dots N}(\xi_m|\{\xi\})\Bigr)
=\left(\prod_{k=1}^m\mathcal{T}^{-1}(\xi_k)\right){\tr}_0\Bigl[P_{0m}R_{0m-1}(\xi_m,\xi_{m-1})\cdots R_{01}(\xi_m,\xi_1)\\
\times R_{0N}(\xi_m,\xi_N)\cdots R_{0m+1}(\xi_m,\xi_{m+1})
E_0P_{0m}P_{0m}\Bigr]\left(\prod_{k=1}^m\mathcal{T}(\xi_k)\right).
\end{multline}
Here, similarly to \eqref{10-LISP-res-tau2}, we have inserted the identity operator
in the form $\mathbf{1}=P_{0m}^2$. Using the basic property of the permutation matrix we obtain
\begin{multline}\label{10-LISP-Calcul-3}
{\tr}_0\Bigl(E_0T_{0;1\dots N}(\xi_m|\{\xi\})\Bigr)
=\left(\prod_{k=1}^m\mathcal{T}^{-1}(\xi_k)\right){\tr}_0\Bigl[R_{mm-1}(\xi_m,\xi_{m-1})\cdots R_{m1}(\xi_m,\xi_1)\\
\times R_{mN}(\xi_m,\xi_N)\cdots R_{mm+1}(\xi_m,\xi_{m+1})
E_mP_{0m}\Bigr]\left(\prod_{k=1}^m\mathcal{T}(\xi_k)\right).
\end{multline}
Now we can take the trace over the space $V_0$ as we already did above. This gives us
\begin{multline}\label{10-LISP-Calcul-4}
{\tr}_0\Bigl(E_0T_{0;1\dots N}(\xi_m|\{\xi\})\Bigr)=
\left(\prod_{k=1}^m\mathcal{T}^{-1}(\xi_k)\right)\cdot R_{mm-1}(\xi_m,\xi_{m-1})\cdots R_{m1}(\xi_m,\xi_1)\\
\times R_{mN}(\xi_m,\xi_N)\cdots R_{mm+1}(\xi_m,\xi_{m+1})
E_m\cdot\left(\prod_{k=1}^m\mathcal{T}(\xi_k)\right)\\
=\left(\prod_{k=1}^m\mathcal{T}^{-1}(\xi_k)\right)\cdot \mathcal{T}(\xi_m)E_m\cdot\left(\prod_{k=1}^m\mathcal{T}(\xi_k)\right)
=\left(\prod_{k=1}^{m-1}\mathcal{T}^{-1}(\xi_k)\right)\cdot E_m\cdot\left(\prod_{k=1}^m\mathcal{T}(\xi_k)\right).
\end{multline}
Hence, we finally obtain
\begin{equation}\label{10-LISP-ISP-sol}
E_m= \left(\prod_{k=1}^{m-1}\mathcal{T}(\xi_k)\right)\cdot{\tr}_0\Bigl(E_0T_{0}(\xi_m)\Bigr)
\cdot\left(\prod_{k=1}^m\mathcal{T}^{-1}(\xi_k)\right).
\end{equation}

The formula \eqref{10-LISP-ISP-sol} is the one we are looking for. In this formula, the operator $E_m$
acting in the local space $V_m$, is expressed in terms of the operator $E_0$ acting in the auxiliary space $V_0$,
and the transfer matrix.

{\sl Remark}. The derivation was done for the $XXZ$ chain. However, we did not use the explicit form of the $R$-matrix. We essentially used
two conditions:

\begin{itemize}
\item the monodromy matrix is the product of the $R$-matrices of the form \eqref{10-LISP-Def-T};

\item the $R$-matrix becomes the permutation matrix at some values of the arguments.
\end{itemize}

If these two conditions are satisfied, then the inverse problem is solvable, and its solution is still given by the formula \eqref{10-LISP-ISP-sol}.
In particular, all of the above is true for the $XXX$ chain and even for the $XYZ$ chain. Furthermore, we have never used the fact that the monodromy matrix was a $2\times 2$ matrix, therefore, the solution of the inverse problem {\it without any changes} can be obtained in the case of $n \times n$ matrices.
The possible generalizations of the inverse problem can be found in the work \cite{MaiT00}.

\section{Form factors of local operators\label{10-Sec3}}

The formulas obtained make it possible to easily obtain determinant representations for the matrix elements of local operators in the spin
chains \cite{KitMT99}. We call the matrix element of the operator $\sigma_m^\alpha$, $\alpha=\pm,z$
between two on-shell vectors a form factor of this operator. Consider the usual (homogeneous) $XXZ$ (or $XXX$) chain.
Then the monodromy matrix has the form
\be{10-T-XXZ}
T_0(u)=R_{0N}(u,\tfrac\eta2)\cdots R_{01}(u,\tfrac\eta2).
\ee
The inverse problem formula \eqref{10-LISP-ISP-sol} looks as follows:
\begin{equation}\label{10-solISP-XXZ}
\begin{aligned}
\sigma_m^-&=\mathcal{T}^{m-1}(\tfrac\eta2)\cdot B(\tfrac\eta2)
\cdot\mathcal{T}^{-m}(\tfrac\eta2),\\
\sigma_m^+&=\mathcal{T}^{m-1}(\tfrac\eta2)\cdot C(\tfrac\eta2)
\cdot\mathcal{T}^{-m}(\tfrac\eta2),\\
\sigma_m^z&=\mathcal{T}^{m-1}(\tfrac\eta2)\cdot \bigl(A(\tfrac\eta2)-D(\tfrac\eta2)\bigr)
\cdot\mathcal{T}^{-m}(\tfrac\eta2),\\
\mathbf{1}&=\mathcal{T}^{m-1}(\tfrac\eta2)\cdot \bigl(A(\tfrac\eta2)+D(\tfrac\eta2)\bigr)
\cdot\mathcal{T}^{-m}(\tfrac\eta2).
\end{aligned}
\end{equation}
The last of these equalities is an obvious identity, however the first three are quite nontrivial. With their help, we
can now calculate the form factors of local spin operators.

Let vector $B(\bu)|0\rangle$ be an on-shell vector and $\langle 0| C(\bv)$ be a dual on-shell vector.
Then we immediately obtain
\be{10-FF-sig+}
\langle 0| C(\bv)\sigma_m^+ B(\bu)|0\rangle = \frac{\tau^{m-1}(\tfrac\eta2|\bv)}{\tau^{m}(\tfrac\eta2|\bu)}\;
\langle 0| C(\bv)C(\tfrac\eta2)\cdot B(\bu)|0\rangle,
\ee
and
\be{10-FF-sig-}
\langle 0| C(\bv)\sigma_m^- B(\bu)|0\rangle = \frac{\tau^{m-1}(\tfrac\eta2|\bv)}{\tau^{m}(\tfrac\eta2|\bu)}\;
\langle 0| C(\bv)\cdot B(\tfrac\eta2)B(\bu)|0\rangle,
\ee
where $\tau$ are the transfer matrix eigenvalues.
Thus, the form factors of the operators $\sigma_m^\pm$ are reduced to the scalar products in which one of the vectors is on-shell.
Indeed, in \eqref{10-FF-sig+} the vector $B(\bu)|0\rangle$ remains on-shell, and in the formula \eqref{10-FF-sig-} the dual vector
$\langle 0| C(\bv)$ is on-shell. In both cases we know compact determinant representations for the scalar products.

The case of form factor of the operator $\sigma_m^z$ is a bit more complicated. It is clear that instead of this operator we can consider, for example, the
operator $\tfrac12(1-\sigma_m^z)$. Due to \eqref{10-solISP-XXZ} the form factor of the latter operator reduces to the form factor of the
operator $D$:
\be{10-FF-sigz}
\tfrac12\langle 0| C(\bv)(1-\sigma_m^z) B(\bu)|0\rangle = \frac{\tau^{m-1}(\tfrac\eta2|\bv)}{\tau^{m}(\tfrac\eta2|\bu)}\;
\langle 0| C(\bv)\cdot D(\tfrac\eta2)\cdot B(\bu)|0\rangle
\ee
In fact, the action of $D(\tfrac\eta2)$ on the vector $B(\bu)|0\rangle$ is known, therefore, it is not difficult to reduce this
form factor to a scalar products in which the dual vector remains on-shell. However, there is another way to
calculate the form factor of the operator $D(\tfrac\eta2)$, which we now demonstrate. This method may seem
a simple technical trick, but this trick has important consequences, which we will get acquainted with later.

Consider the twisted transfer matrix $\mathcal{T}_\kappa(z)=A(z)+\kappa D(z)$, where $\kappa$ is some complex parameter.
Let us introduce
\be{10-Qkap}
Q_\kappa= \langle 0| C(\bv)\bigl(\mathcal{T}_\kappa(\tfrac\eta2)-\mathcal{T}(\tfrac\eta2)\bigr)B(\bu)|0\rangle.
\ee
Here $\mathcal{T}(\tfrac\eta2)$ is the ordinary transfer matrix:  $\mathcal{T}(\tfrac\eta2)=A(\tfrac\eta2)+D(\tfrac\eta2)$. Let
$B(\bu)|0\rangle$ be an ordinary on-shell vector and $\langle 0| C(\bv)$ be a twisted on-shell vector.
Note that at $\kappa=1$, the twisted on-shell vector turns into
the ordinary on-shell vector.

On the one hand, $\mathcal{T}_\kappa(\tfrac\eta2)-\mathcal{T}(\tfrac\eta2)=(\kappa-1)D(\tfrac\eta2)$, therefore, we have
\be{10-Qkap-1c1}
Q_\kappa= (\kappa-1)\langle 0| C(\bv)D(\tfrac\eta2)B(\bu)|0\rangle.
\ee
Differentiating over $\kappa$ at $\kappa=1$ we obtain
\be{10-Qkap-1c2}
\frac{dQ_\kappa}{d\kappa}\Bigr|_{\kappa=1}= \langle 0| C(\bv)D(\tfrac\eta2)B(\bu)|0\rangle.
\ee
In this formula, we have already $\kappa=1$, hence, $ \langle 0| C(\bv)$ is an ordinary on-shell vector. Therefore, we have the form factor of the
operator $D(\tfrac\eta2)$ in the rhs.

On the other hand,
\be{10-Qkap2c1}
Q_\kappa= \bigl(\tau_\kappa(\tfrac\eta2|\bv)-\tau(\tfrac\eta2|\bu)\bigr)\langle 0| C(\bv)B(\bu)|0\rangle,
\ee
where $\tau_\kappa(\tfrac\eta2|\bv)$ is the twisted transfer matrix eigenvalue:
\be{10-Tw-ev}
\tau_\kappa(z|\bv)=a(z)f(\bv,z)+\kappa d(z)f(z,\bv).
\ee
The ordinary transfer matrix eigenvalue $\tau(\tfrac\eta2|\bu)$ is given by \eqref{10-Tw-ev}, where one should put  $\kappa=1$
and replace the set $\bv$ with the set $\bu$.

Thus,  we obtain
\be{10-FF-D1}
\langle 0| C(\bv)D(\tfrac\eta2)B(\bu)|0\rangle=\frac{d}{d\kappa}\bigl(\tau_\kappa(\tfrac\eta2|\bv)-\tau(\tfrac\eta2|\bu)\bigr)\langle 0| C(\bv)B(\bu)|0\rangle\Bigr|_{\kappa=1}.
\ee
Thus, to calculate the form factor of the operator $D(\tfrac\eta2)$, it suffices to calculate the scalar product of the ordinary and twisted
on-shell vectors, and then differentiate the result over $\kappa$ at $\kappa = 1$.

One can go a bit further. Consider the case when  $\bv=\bu$ at $\kappa=1$.
In another way, we can say that the parameters $\bv$
are $\kappa $-deformations of the parameters $\bu$. Then the difference $\tau_\kappa(\tfrac\eta2|\bv)-\tau(\tfrac\eta2|\bu)$
vanishes for $\kappa=1$. Therefore, the derivative over $\kappa$ in \eqref{10-FF-D1} must act only on this difference, but not on the scalar product:
\be{10-FF-D11}
\langle 0| C(\bv)D(\tfrac\eta2)B(\bv)|0\rangle=\frac{d}{d\kappa}\tau_\kappa(\tfrac\eta2|\bv)\Bigr|_{\kappa=1}\langle 0| C(\bv)B(\bv)|0\rangle.
\ee

Let now $\bv\ne\bu$ at $\kappa=1$. Then it is the scalar product $\langle 0| C(\bv)B(\bu)|0\rangle$ that vanishes at $\kappa=1$, because
it is the scalar product of two different on-shell vectors. Hence, in this case the $\kappa$-derivative  in \eqref{10-FF-D1} must act only on
the scalar product, but not on the difference of the transfer matrix eigenvalues. In the latter we can simply set $\kappa=1$:
\be{10-FF-D2}
\langle 0| C(\bv)D(\tfrac\eta2)B(\bu)|0\rangle=\bigl(\tau(\tfrac\eta2|\bv)-\tau(\tfrac\eta2|\bu)\bigr)
\frac{d}{d\kappa}\langle 0| C(\bv)B(\bu)|0\rangle\Bigr|_{\kappa=1}, \qquad \bv\ne\bu.
\ee
%


%
%

\chapter{Composite model\label{CHA-Cmod}}

The formulas of the quantum inverse problem make it possible to calculate the form factors of local operators in spin chains.
However, the solution of the quantum inverse problem is not known  for all quantum integrable models. How to calculate the form factors of local operators
in those models? To do this, we introduce a {\it composite model}. Within the framework of this model, one obtains an intermediate point of the interval,
on which the initial model was determined (for example, some point $x$ on the interval $[0,L]$ in the QNLS model). In this lecture we will get acquainted with the composite model. We also introduce the notion of a generalized model, that provides us with a convenient way of proving nontrivial identities.

\section{Generalized model\label{11-Sec1}}

The scalar product of two arbitrary Bethe vectors
\be{11-SP-SPdef}
S_n(\bv|\bu)=\frac{\langle0|C(\bv)B(\bu)|0\rangle}{d(\bv)d(\bu)}\,
\ee
is given by \eqref{07-SP-SPlm1}.
Assume that both vectors in this formula are on-shell.
Then due to the Bethe equations we have
\be{11-BE}
r(\bu_{\st})f(\bu_{\so},\bu_{\st})= f(\bu_{\st},\bu_{\so}),\qquad r(\bv_{\so})f(\bv_{\st},\bv_{\so}) = f(\bv_{\so},\bv_{\st}).
\ee
Substituting these expressions into \eqref{07-SP-SPlm1},
we obtain
\be{11-on-sh}
S_n(\bv|\bu)=\sum_{\substack{\bv\mapsto\{\bv_{\so},\bv_{\st}\}\\
\bu\mapsto\{\bu_{\so},\bu_{\st}\}}}
K_{k}( \bu_{\so}|\bv_{\so})K_{n-k}( \bv_{\st}|\bu_{\st})\,f(\bv_{\so},\bv_{\st})f(\bu_{\st},\bu_{\so}).
 \ee
Recall that here the sum is taken over partitions of the sets $\bu$ and $\bv$ into two subsets each. Both
partitions are independent except a restriction on the cardinalities of the subsets $\#\bu_{\so}=\#\bv_{\so}=k$, $k=0,1,\dots,n$.

The scalar product of two different on-shell vectors must vanish. Indeed, let us insert between the two vectors the transfer matrix
$\mathcal{T}(w)$. We can act with this operator
either to the left or to the right. Depending on this, we get
\be{11-SP-Tau}
\frac{\langle0|C(\bv)\mathcal{T}(w)B(\bu)|0\rangle}{d(\bv)d(\bu)}= \tau(w|\bv)S_n(\bv|\bu)=\tau(w|\bu)S_n(\bv|\bu).
\ee
If $\bu\ne\bv$, then $\tau(w|\bu)\ne \tau(w|\bv)$. It is appropriate to recall here that the eigenvalue of the transfer matrix is a
{\it function}  of $w$ (see \eqref{03-E-val}).
Therefore, if $\bv\ne\bu$, then coincidence  of the two functions $\tau(w|\bu)$ and $\tau(w|\bv)$ is possible only in some points $w_k$,
but not for any complex $w$. Thus, we arrive at $S_n(\bv|\bu)=0$.

There is a suspicion that the sum over the partitions in the formula \eqref{11-on-sh} vanishes (more precisely, $S_n(\bv|\bu)=\delta_{n0}$,
since for $n = 0$ we deal with the product of the vacuum vector and the dual vector).
However, there are several questions.

First, whether the equation
\be{11-Ident-2}
\sum_{\substack{\bv\mapsto\{\bv_{\so},\bv_{\st}\}\\
\bu\mapsto\{\bu_{\so},\bu_{\st}\}}}
K_{k}( \bu_{\so}|\bv_{\so})K_{n-k}( \bv_{\st}|\bu_{\st})\,f(\bv_{\so},\bv_{\st})f(\bu_{\st},\bu_{\so})=\delta_{n0}
 \ee
is valid identically? Deriving this formula we used Bethe equations, hence, it is possible
that \eqref{11-Ident-2} holds not for any $\bu$ and $\bv$, but only if these parameters satisfy certain conditions. Second, whether equation
\eqref{11-Ident-2} remains true, if $\bu=\bv$? It seems that in this case we should obtain the square of the norm of the on-shell vector.
The latter, generically, is not zero.

Let us consider the simplest example, when $n=1$. In this case, there are only two partitions: either the first set is empty or the second is empty.
Substituting  $K_1(x|y)=g(x,y)$ into \eqref{07-SP-SPlm1}
at $n=1$, we obtain
\be{11-n1-a}
S_1(v|u)=r(u)g(v,u)+r(v)g(u,v).
 \ee
Bethe equations for this particular case have the form $r(u)=r(v)=1$, what implies
\be{11-n1-b}
S_1(v|u)=g(v,u)+g(u,v)=0,
 \ee
due to the antisymmetry of the function $g(v,u)$. Certainly,  \eqref{11-n1-b} is the identity. More precisely, it is valid
for arbitrary $u\ne v$, while at $u=v$ we have a removable singularity. Defining $S_1(v|u)$ in this point by continuity we obtain that $S_1(v|u)$
is identically zero.

The above example suggests that the equality \eqref{11-Ident-2} is  the identity. It becomes clear also what is happening
in the case when $\bu = \bv$. As already discussed in section~\ref{07-Sec7},
in order to obtain the square of the norm of the on-shell vector, we must {\it first}  take the limit
$\bu\to \bv$ in   \eqref{07-SP-SPlm1},
and only after this we should impose the Bethe equations for the parameters $\bv$. Thus, if we consider the limit
$u\to v$ in the formula \eqref{11-n1-a}, then we obtain $S_1(v|v)=-cr'(v)$.  This expression is not necessarily zero, even if $v$
is a root of the Bethe equation. In the example above we did the opposite. We first imposed the Bethe equations on the parameters $\bu$ and $\bv$, and after
this we  began to study the behavior of the obtained function in the limit $\bu\to \bv$.

The identity \eqref{11-Ident-2} can be proved by direct calculation. Actually, we already did this in lecture~\ref{CHA-SP}.
There we calculated the scalar product under the condition that the vector $B(\bu)|0\rangle $ was an on-shell vector, and the second vector remained
arbitrary\footnote{Recall that all formulas of the lecture~\ref{CHA-SP}
are valid even if the value of the twist parameter is $\kappa = 1$.}. We can require that the second vector also
is on-shell. Then we obtain the sum \eqref{11-on-sh}. But, on the other hand, we can impose the Bethe equations on the parameters $\bv$ in the final
determinant representation  \eqref{07-SP-det-pres1}, \eqref{07-SP-det-pres0}.
Then we obtain the determinant of the matrix  \eqref{07-SP-cM}
with $\kappa = 1$. We know that this determinant vanishes.

There is, however, a simpler way of proving the identity of \eqref{11-Ident-2}, that is based on the concept of {\it generalized
model}. This model was introduced in \cite{Kor820}. In the early stages of the development of the algebraic Bethe ansatz it was actively used
\cite{Kor82,IzeK84,Kor84,Sla89},
but then was undeservedly forgotten.

The generalized model is in fact not a one model, but a class of models. The monodromy matrix of each representative of this class
satisfies the $RTT$-relation with a fixed $R$-matrix and has a vacuum vector (and also a dual vector). Different representatives of this class are
parameterized by different functions $r(u)$. We can say that the concrete representative of the generalized model is a concrete
representation of the operator algebra given by the $RTT$-relation.

We have seen already how one can build a model, in which the function $r(u)$ has the form
\be{11-ru}
r(u)=\prod_{k=1}^N\frac{a_k^{(1)}u+a_k^{(0)}}{d_k^{(1)}u+d_k^{(0)}},
\ee
where the only condition is imposed on the constants $a_k^{(j)}$ and $d_k^{(j)}$ ($j=0,1$):
$a_k^{(1)}\ne 0$, $d_k^{(1)}\ne 0$.
One can also consider all possible limits of the expression \eqref{11-ru}, when an arbitrary natural $N$ tends to infinity.
Thus, we have a very large reserve of functions $r(u)$. We can say that if the variable $u$ is fixed, then running through
different representatives of the generalized model, we can always find such of them that $r(u)$ takes any prescribed value.
In this sense, we can consider the variables $u$ and $r(u)$ as independent variables.
Then the scalar product of two generic Bethe vectors  becomes a function of variables of two types: on the one hand, they are variables
$u_k$ and $v_k$, on the other hand, the variables $r(u_k)$ and $r(v_k)$.
In the generalized model, both types of variables take arbitrary values independently of each other.

{\sl Remark.} If we want to avoid singularities in the scalar products at coinciding elements in the sets $\bu$ and $\bv$, then
one restriction is still required. In this case, it is reasonable to require that if $u_k = v_j$, then $r(u_k) = r(v_j)$. It is clear that
in this case, we should restrict ourselves with smooth functions $r(u)$.

If one or both vectors are on-shell, then the situation changes. Then a constraint arises for the two types of variables, which is specified by
Bethe equations. It is important to emphasize, however, that within the framework of the generalized model the Bethe equations are no longer equations.
Instead, they become conditions that connect two types of variables. If a representative of the generalized model is fixed (that is, a concrete function $r(u)$ is specified), then the Bethe equations again become equations: one should find possible values of the parameters $u_k$ for a given function
$r(u)$. But within the framework of the generalized model, we can look at these constraints from a different point of view, namely, we can consider the variables $u_k$ as arbitrary complex numbers and determine the values of the function $r(u)$  in the points $u_k$.

In the framework of the generalized model, the proof of the identity \eqref{11-Ident-2} becomes trivial. We first consider the scalar product \eqref{07-SP-SPlm1} in the general case, that is, the parameters $\bu$ and $\bv$ are arbitrary complex numbers with the only
restriction  $u_j\ne v_k$, $j,k=1,\dots,n$.
Then we impose the constraints (Bethe equations) {\it on the parameters $r(u_j)$ and $r(v_k)$}, still
considering $\bu $ and $\bv$ as arbitrary complex numbers. On the one hand, we get \eqref{11-on-sh}, on the other hand we have a scalar product
of two on-shell vectors, that should vanish. Thus, we prove the identity \eqref{11-Ident-2} for arbitrary
$u_j$ and $v_k$ provided $u_j\ne v_k$. In the points where some (or all) $u_j$ coincide with $v_k$, the identity is defined by continuity.
It is clear that if a rational function is equal to zero in the whole complex plane, except for a finite number of isolated points, then by continuity it is defined by zero everywhere.

\subsection{Two representations for scalar product\label{11-Sec11}}

Let us give another example that demonstrates the advantages of the generalized model. Consider the scalar product of two Bethe vectors
\eqref{11-SP-SPdef} where
the vector $B(\bu)|0\rangle$ is a twisted on-shell vector, and the vector $\langle0|C(\bv)$ is the ordinary on-shell vector.
This example is a special case of the scalar product considered in sections~\ref{07-Sec3}
and~\ref{07-Sec5}.
There the vector $B(\bu)|0\rangle$ was a twisted on-shell vector, and the vector $\langle0|C(\bv)$ was an arbitrary Bethe vector. In the particular case
it can also be an ordinary on-shell vector. The result for such the scalar product is given by formulas
 \eqref{07-SP-det-pres1} and \eqref{07-SP-det-pres0}.
However, now we can use  Bethe equations and can express functions $r(v_k)$ in terms of products of the $f$ functions. Then, after simple
transformations, we obtain
\begin{equation}\label{11-SP-det-pres1}
S_n(\bv|\bu)=\Delta'_n(\bu)\Delta_n(\bv)\; h(\bu,\bv)
\;\det_n\mathcal{M}_{jk}^{(1)}\,,
\end{equation}
where
\begin{equation}\label{11-SP-det-pres0}
\mathcal{M}_{jk}^{(1)}
=t(u_j,v_k) +\kappa\,t(v_k,u_j)\frac{h(v_k,\bu)h(\bv,v_k)}{h(\bu,v_k)h(v_k,\bv)}\,.
\end{equation}

This is one possible form of the result, but there exists another. Due to \eqref{07-SP-symm},
we can make the replacement of the vectors  $B(\bu)|0\rangle$ and $\langle0|C(\bv)$
\be{11-SP-SPdef0}
S_n(\bv|\bu)=\frac{\langle0|C(\bu)B(\bv)|0\rangle}{d(\bv)d(\bu)}\,.
\ee
Now we can say that the vector $B(\bv)|0\rangle$ is the twisted on-shell vector with $\kappa=1$,
while the vector $\langle0|C(\bu)$ is an arbitrary Bethe vector (which can be a twisted on-shell vector in the particular case).
Clearly, the result for this scalar product again is given by
\eqref{07-SP-det-pres1} and \eqref{07-SP-det-pres0},
where one should replace $\bu\leftrightarrow\bv$ and set $\kappa=1$:
\begin{equation}\label{11-SP-det-pres2}
S_n(\bv|\bu)=\Delta'_n(\bu)\Delta_n(\bv)\; h(\bu,\bv)
\;\det_n\mathcal{M}_{jk}^{(2)}\,,
\end{equation}
where
\begin{equation}\label{07-SP-det-pres3}
\mathcal{M}_{jk}^{(2)}
=t(u_k,v_j)
\left(1-r(u_k)\frac{f(\bv_j,u_k)}{f(u_k,\bv_j)}\right)\,,
\end{equation}
and we used $\Delta'_n(\bu)\Delta_n(\bv)=\Delta'_n(\bv)\Delta_n(\bu)$. Now we should express the functions $r(u_k)$ in terms of the products
of the $f$ functions via twisted Bethe equations, and we obtain
\begin{equation}\label{11-SP-det-pres4}
\mathcal{M}_{jk}^{(2)}
=t(u_k,v_j) +\kappa\,t(v_j,u_k)\frac{h(u_k,\bu)h(\bv,u_k)}{h(\bu,u_k)h(u_k,\bv)}\,.
\end{equation}
Thus, we have two different representations of the same scalar product
$S_n(\bv|\bu)$, and hence, there must be equivalence of the two determinants:
\be{11-det-det}
\det_n\left(t(u_j,v_k) +\kappa\,t(v_k,u_j)\frac{h(v_k,\bu)h(\bv,v_k)}{h(\bu,v_k)h(v_k,\bv)}\right)=
\det_n\left(t(u_k,v_j) +\kappa\,t(v_j,u_k)\frac{h(u_k,\bu)h(\bv,u_k)}{h(\bu,u_k)h(u_k,\bv)}\right).
\ee

From the point of view of a concrete model, this equality must be satisfied if the parameters $\bu$
are the roots of the twisted Bethe equations, and $\bv$ are the roots of the usual Bethe equations. However, from the point of view of the generalized model
the equality \eqref{11-det-det} must be the identity, because when we derive it, we impose restrictions only
on the parameters $r(u_k)$ and $r(v_k)$, while the parameters $\bv$ and $\bu$ remained arbitrary complex numbers.
Indeed, the two determinants in the formula \eqref{11-det-det} are identically equal to each other, that can be seen with a naked
eye at least for $n=1$. For $n> 1$ this is far from obvious, but the identity \eqref{11-det-det} can be proved
by induction. The method of proof is fairly standard. We give only the main idea, letting the reader check the details.

For definiteness, consider the case of the trigonometric $R$-matrix. The first step is to multiply both determinants by the product
$\Delta'_n(\bu)\Delta_n(\bv)\; h(\bu,\bv)$.
Then we need to pass from the variables $\bv$, $\bu$, and $\eta$ to the variables  $x_j=e^{2v_j}$, $y_j=e^{2u_j}$ and $q=e^\eta$.
After this we should consider both sides of equation
\eqref{11-det-det} as rational functions of $\bar x$ and $\bar y$ (up to a trivial product of radicals, as in the formula for DWPF
\eqref{08-K-defxy}).
One should check that the residues at the points $y_k = x_j$ are equal to the determinants of the $(n-1)\times(n-1)$ matrices, which must coincide by the inductive assumption. A separate problem is to prove that the determinant of the matrix $\mathcal{M}_{jk}^{(1)}$
has no poles at the points $x_jq^2=x_k$, and the determinant of the matrix $\mathcal{M}_{jk}^{(2)}$ has no poles
at the points $y_jq^2=y_k$. This is really so, because in these points
two columns of each of these matrices are proportional to each other. Having done all this, we will see that the difference
of two determinants (multiplied by $\Delta'_n(\bu)\Delta_n(\bv)\; h(\bu,\bv)$)
is an analytic function in the whole complex plane. In addition, it decreases when one of its arguments goes to infinity at other arguments fixed.
Hence, the identity \eqref{11-det-det} is true due to the Liouville theorem.

\section{Composite model\label{11-Sec2}}

\subsection{Definition of composite model\label{11-Sec21}}

Suppose that we have some lattice model. Then its monodromy matrix is a product of
$L$-operators  \eqref{02-T-LLL}.
If we deal with a continuous  model, then we can take the definition  \eqref{02-T-LLL}
as a lattice approximation of the monodromy matrix of this model. Local operators of the model are included in the $L$-operators.
At the same time, the monodromy matrix  is a global object: its matrix elements depend on the local operators in all
lattice sites. In particular, the Bethe vectors $B(\bu)|0\rangle$ are global objects. Therefore, in calculating the form factors of local operators
we need to establish commutation relations between local and global operators.

This problem can be solved in the framework of a composite model. Let us fix a site
$m$ and introduce {\it partial monodromy matrices} $T^{(1)}(u)$ and $T^{(2)}(u)$ as follows:
\be{11-T-TTL}
T^{(2)}(u)=L_N(u)\dots L_{m+1}(u), \qquad T^{(1)}(u)=L_m(u)\dots L_1(u).
\ee

Then the total monodromy matrix is equal to the product of the partial monodromy matrices
 \be{11-TTT}
 T(u)=T^{(2)}(u )T^{(1)}(u).
\ee
We call this model a generalized composite model\footnote{%
This model was introduced for the first time in the work \cite{IzeK84}. There it was called
a generalized two-site model. Later, in the work
\cite{Sla07}, the term two-component model was used. Unfortunately, both these terms
can lead to misunderstanding: the first one when considering
spin chains, the second when considering models of multicomponent gases. }.

The described method allows, on the one hand, to include in the consideration
operators that depend on some selected intermediate site $m$, and on the other hand, still to work with monodromy matrices
and the $R$-matrix. Let the matrix elements of the matrices $T^{(1)}(u)$ and $T^{(2)}(u)$
be equipped with the superscript
 \be{11-ABAT-2s}
 T^{(j)}(u)=\left(
\begin{array}{cc}
A^{(j)}(u)&B^{(j)}(u)\\
C^{(j)}(u)&D^{(j)}(u)
\end{array}\right), \qquad j=1,2.
\ee
Each of partial monodromy matrices satisfies the  $RTT$-relation
with the same  $R$-matrix, as the total monodromy matrix. This follows from the fact that they are equal to the product of $L$-operators,
where every $L$-operator satisfies the $RTT$-relation. The matrix elements of the different partial matrices
commute with each other. Each of the matrices
$T^{(j)}(u)$ possesses a vector $|0\rangle^{(j)}$ and a dual vector
$\langle0|^{(j)}$, where $|0\rangle=|0\rangle^{(2)}\otimes|0\rangle^{(1)}$.
The action formulas of the operators onto these vectors are similar to the action formulas for the total monodromy matrix
 \be{11-action-2s}
 \begin{array}{l}
 {\dis
 A^{(j)}(u)|0\rangle^{(j)}=a^{(j)}(u)|0\rangle^{(j)},\qquad
  D^{(j)}(u)|0\rangle^{(j)}=d^{(j)}(u)|0\rangle^{(j)},\qquad
   C^{(j)}(u)|0\rangle^{(j)}=0,}\non
{\dis \langle0|^{(j)}A^{(j)}(u)=a^{(j)}(u)\langle0|^{(j)},\qquad
  \langle0|^{(j)}D^{(j)}(u)=d^{(j)}(u)\langle0|^{(j)},\qquad
   \langle0|^{(j)}B^{(j)}(u)=0,}\end{array}
\ee
where $a^{(j)}(u)$ and $d^{(j)}(u)$ are some functions. We do not specify their explicit form.
It is easy to see that $a(u)=a^{(1)}(u)a^{(2)}(u)$ and
$d(u)=d^{(1)}(u)d^{(2)}(u)$.

We define partial Bethe vectors and their dual in the standard way
\be{11-PBV}
 B^{(j)}(\bu)|0\rangle^{(j)}, \qquad \langle0|^{(j)}C^{(j)}(\bu).
 \ee
The actions of the partial operators $A^{(j)}$, $B^{(j)}$, $C^{(j)}$ and $D^{(j)}$ on the corresponding Bethe vectors
are the same, as the actions of the total operators on the total Bethe vectors.
One should only replace everywhere the functions
$a(u)$ and $d(u)$ by $a^{(j)}(u)$ and $d^{(j)}(u)$. It is in this way, the dependence on a selected site arises in the formulas for the
scalar products and form factors.
If, for instance, we deal with a homogeneous model and the action of the diagonal matrix elements of the
$L$-operators on vacuum is given by
\be{11-actL}
L_n^{11}(u)|0\rangle=\alpha(u)|0\rangle,\qquad L_n^{22}(u)|0\rangle=\delta(u)|0\rangle,
\ee
then
\be{11-actT}
a(u)=\alpha^N(u),\qquad d(u)=\delta^N(u).
\ee
At the same time,
\be{11-actT12}
\begin{aligned}
&a^{(1)}(u)=\alpha^m(u),&\qquad & d^{(1)}(u)=\delta^m(u),\\
&a^{(2)}(u)=\alpha^{N-m}(u),&\qquad & d^{(2)}(u)=\delta^{N-m}(u).
\end{aligned}
\ee
As we shall see further, the final answers for the form factors of local operators somehow include the functions
$a^{(j)}(u)$ and $d^{(j)}(u)$,
and thus they depend on the intermediate $m$th site.

In order to obtain formulas for the action of partial operators on the total Bethe vector $B(\bu)|0\rangle$,
it is necessary to express the total vector in terms of the partial ones \eqref{11-PBV}.

\begin{prop}\label{11-PROP1}
A generic  total Bethe vector can be expressed in terms of the partial vectors as follows \cite{IzeK84}:
\be{11-IK}
B(\bu)|0\rangle=\sum_{\bu\mapsto\{\bu_{\so},\bu_{\st}\}}
   a^{(2)}(\bu_{\so})
 d^{(1)}(\bu_{\st})B^{(1)}(\bu_{\so})B^{(2)}(\bu_{\st})|0\rangle
\cdot  f(\bu_{\st},\bu_{\so}).
 \ee
Here the sum is taken over all possible partitions of the set $\bu$ into two subsets $\bu_{\so}$ and $\bu_{\st}$.
\end{prop}
{\sl Proof.}  For proving \eqref{11-IK}, we  use induction over $n=\#\bu$.
Obviously,  equation  \eqref{11-IK} is valid for $n=1$. Indeed, it follows from \eqref{11-TTT} that
\be{11-B-BB}
B(u)=A^{(2)}(u)B^{(1)}(u)+B^{(2)}(u) D^{(1)}(u),
\ee
what implies \eqref{11-IK} for $n=1$.
Let \eqref{11-IK} be valid for $n-1$:
\be{11-IKn}
B(\bu_n)|0\rangle=\sum_{\bu_n\mapsto\{\bu_{\so},\bu_{\st}\}}
   a^{(2)}(\bu_{\so})
 d^{(1)}(\bu_{\st})B^{(1)}(\bu_{\so})B^{(2)}(\bu_{\st})|0\rangle
\cdot  f(\bu_{\st},\bu_{\so}),
 \ee
where the sum is taken over partitions of the set $\bu_n=\bu\setminus u_n$. Then
\begin{multline}\label{11-step}
  B(\bu)|0\rangle=\bigl(A^{(2)}(u_n) B^{(1)}(u_n) + B^{(2)}(u_n) D^{(1)}(u_n)\bigr)\\
 \times\sum_{\bu_n\mapsto\{\bu_{\so},\bu_{\st}\}}
   a^{(2)}(\bu_{\so})d^{(1)}(\bu_{\st}) B^{(1)}(\bu_{\so})B^{(2)}(\bu_{\st})|0\rangle
\cdot f(\bu_{\st},\bu_{\so}).
 \end{multline}
Let us act with the operators
$A^{(2)}(u_n)$ and $D^{(1)}(u_n)$ onto obtained states. For this we should use
\eqref{03-2ABBB-act} and \eqref{03-2DBBB-act},
replacing everywhere  $a$, $d$, and $B$  respectively by $a^{(2)}$, $d^{(1)}$, and either $B^{(1)}$ or $B^{(2)}$.
It is convenient to write the action formulas in the following form:
\begin{multline}\label{11-A2B2}
A^{(2)}(v) B^{(2)}(\bu)|0\rangle^{(2)}=f(\bu,v)a^{(2)}(v) B^{(2)}(\bu)|0\rangle^{(2)}\\
+\sum_{\bu\mapsto\{u_{\rm i},\bu_{\rm ii}\}} a^{(2)}(u_{\rm i}) g(v,u_{\rm i})f(\bu_{\rm ii},u_{\rm i})
B^{(2)}(v) B^{(2)}(\bu_{\rm ii})|0\rangle^{(2)}.
\end{multline}
In the second line of this formula, the sum is taken over partitions of the set  $\bu$ into subsets $u_{\rm i}$ and $\bu_{\rm ii}$,
such that the subset $u_{\rm i}$ consists of one element. Similarly,
\begin{multline}\label{11-D1B1}
D^{(1)}(v) B^{(1)}(\bu)|0\rangle^{(1)}=f(v,\bu)d^{(1)}(v) B^{(1)}(\bu)|0\rangle^{(1)}\\
+\sum_{\bu\mapsto\{u_{\rm i},\bu_{\rm ii}\}} d^{(1)}(u_{\rm i}) g(u_{\rm i},v)f(u_{\rm i},\bu_{\rm ii})
B^{(1)}(v) B^{(1)}(\bu_{\rm ii})|0\rangle^{(1)}.
\end{multline}

First, we consider contributions arising in the case when the operators
$A^{(2)}$ and $D^{(1)}$ exchange their arguments with the arguments of the operators $B^{(2)}$ and
$B^{(1)}$. Then we should take the terms of the second line of \eqref{11-A2B2} for the action of $A^{(2)}(u_n)$. We obtain
 \begin{multline}\label{11-actA}
 \sum_{\bu_n\mapsto\{\bu_{\so},u_{\rm i},\bu_{\rm ii}\}}g(u_n,u_{\rm i})a^{(2)}(u_{\rm i})a^{(2)}(\bu_{\so})
 d^{(1)}(u_{\rm i})d^{(1)}(\bu_{\rm ii})B^{(1)}(u_n)B^{(2)}(u_n)\\
  \times B^{(1)}(\bu_{\so})B^{(2)}(\bu_{\rm ii})
 |0\rangle \cdot f(u_{\rm i},\bu_{\so})f(\bu_{\rm ii},\bu_{\so}) f(\bu_{\rm ii},u_{\rm i}).
 \end{multline}
Let us comment on this formula. In the equation \eqref{11-step}, the set  $\bu_{n}$ was already divided into subsets $\bu_{\so}$ and $\bu_{\st}$.
After the action of the operator $A^{(2)}(u_n)$ on the vector $B^{(2)}(\bu_{\st})|0\rangle$ according to \eqref{11-A2B2}, we obtain an additional
partition of the set $\bu_{\st}$ into subsets $u_{\rm i}$ and $\bu_{\rm ii}$. As a result, the set $\bu_n$ is divided into three subsets
$\bu_n\mapsto\{\bu_{\so},u_{\rm i},\bu_{\rm ii}\}$, where $u_{\rm i}$ consists of one element. It is also obvious that
$d^{(1)}(\bu_{\st})=d^{(1)}(u_{\rm i})d^{(1)}(\bu_{\rm ii})$ and
$f(\bu_{\st},\bu_{\so})=f(u_{\rm i},\bu_{\so})f(\bu_{\rm ii},\bu_{\so})$.

Similarly, using the second line of \eqref{11-D1B1}, we find a contribution
of the action of the operator $D^{(1)}(u_n)$
 \begin{multline}\label{11-actD}
 \sum_{\bu_n\mapsto\{\bu_{\rm ii},u_{\rm i},\bu_{\st}\}}g(u_{\rm i},u_n)a^{(2)}(u_{\rm i})a^{(2)}(\bu_{\rm ii})
 d^{(1)}(u_{\rm i})d^{(1)}(\bu_{\st})B^{(1)}(u_n)B^{(2)}(u_n)\\
  \times B^{(1)}(\bu_{\rm ii})B^{(2)}(\bu_{\st})
 |0\rangle \cdot f(u_{\rm i},\bu_{\rm ii})f(\bu_{\st},\bu_{\rm ii}) f(\bu_{\st},u_{\rm i}).
 \end{multline}

At the first sight it seems that the contributions of \eqref{11-actA} and \eqref{11-actD} do not have anything in common with each other. In reality
they differ from each other only in sign. In order to see this, it is enough to rename subsets in one
of these formulas so that the products of the operators $B^{(1)}$ and $B^{(2)}$ in both formulas would depend on the same subsets.
For example, we can first replace $\bu_{\rm ii}$ with $\bu_{\st}$ in the formula \eqref{11-actA}, and then replace
$\bu_{\so}$ with $\bu_{\rm ii}$. Then the contribution \eqref{11-actA} takes the form
 \begin{multline}\label{11-actA-d}
 \sum_{\bu_n\mapsto\{\bu_{\rm ii},u_{\rm i},\bu_{\st}\}}g(u_n,u_{\rm i})a^{(2)}(u_{\rm i})a^{(2)}(\bu_{\rm ii})
 d^{(1)}(u_{\rm i})d^{(1)}(\bu_{\st})B^{(1)}(u_n)B^{(2)}(u_n)\\
  \times B^{(1)}(\bu_{\rm ii})B^{(2)}(\bu_{\st})
 |0\rangle \cdot f(u_{\rm i},\bu_{\rm ii})f(\bu_{\st},\bu_{\rm ii}) f(\bu_{\st},u_{\rm i}).
 \end{multline}
We see now that due to $g(u_n,u_{\rm i})=-g(u_{\rm i},u_n)$, expression \eqref{11-actA-d} differs from \eqref{11-actD} in the sign only.
Thus, they cancel each other. We have only contributions in which the operators $A^{(2)}$ and $D^{(1)}$ keep  their arguments after the action
on the partial vectors  (i.e. the terms in the first lines in the formulas \eqref{11-A2B2} and \eqref{11-D1B1}). We obtain
\begin{multline}\label{11-step2}
  B(\bu)|0\rangle=\sum_{\bu_n\mapsto\{\bu_{\so},\bu_{\st}\}}a^{(2)}(u_n)
   a^{(2)}(\bu_{\so})d^{(1)}(\bu_{\st}) B^{(1)}(u_n)B^{(1)}(\bu_{\so})B^{(2)}(\bu_{\st})|0\rangle
\cdot f(\bu_{\st},u_n)f(\bu_{\st},\bu_{\so})\\
+\sum_{\bu_n\mapsto\{\bu_{\so},\bu_{\st}\}}
   a^{(2)}(\bu_{\so}) d^{(1)}(u_n)d^{(1)}(\bu_{\st}) B^{(1)}(\bu_{\so})B^{(2)}(u_n)B^{(2)}(\bu_{\st})|0\rangle
\cdot f(u_n,\bu_{\so})f(\bu_{\st},\bu_{\so}).
 \end{multline}
It is easy to see that this is exactly the formula \eqref{11-IK}. Indeed, if we have partitions of the set $\bu$ into two subsets, then the
parameter $u_n$ can belong either to the first subset or to the second one. Each of the two terms in \eqref{11-step2} corresponds to one of two possibilities mentioned above.

A similar formula exists for the dual vectors
\be{11-IK-d}
 \langle0| C(\bu)=\langle0|\sum_{\bu\mapsto\{\bu_{\so},\bu_{\st}\}}
a^{(1)}(\bu_{\st}) d^{(2)}(\bu_{\so})C^{(1)}(\bu_{\so})C^{(2)}(\bu_{\st})
\cdot  f(\bu_{\so},\bu_{\st}).
 \ee
The proof of this formula is completely analogous to the one given above.

\subsection{Multicomposite model\label{11-Sec22}}

To calculate the form factors of local operators, the composite model described in the previous section is already sufficient.
We will get acquainted with the details of this calculation in the following lectures. In the meantime, we note that we can go further and split the total monodromy matrix not by two, but by a larger number of partial matrices $T^{(j)}(u)$
\be{11-T-TTT}
T(u)=T^{(M)}(u)\dots T^{(1)}(u).
\ee
This model is called multicomposite.  We can define partial
Bethe vectors $B^{(j)}|0\rangle^{(j)}$ for each monodromy matrix $T^{(j)}(u)$.
Then the total Bethe vector expressed in terms of the partial ones as follows
\be{11-BV-mult-GL2}
B(\bu)|0\rangle=\sum_{\bu\mapsto\{\bu^{\bs{1}},\dots,\bu^{\bs{M}}\}} \prod_{1\le k<j\le M}\left\{ a^{(j)}(\bu^{\bs{k}}) d^{(k)}(\bu^{\bs{j}})
f(\bu^{\bs{j}},\bu^{\bs{k}})\right\}\;
\prod_{j=1}^M \left\{ B^{(j)}(\bu^{\bs{j}})\right\}\;|0\rangle.
\ee
In view of the large number of subsets, we do not use Roman numerals in this formula, but denote subsets by the superscript typed in bold.
The sum is taken over all partitions of the original set $\bu$ into $M$ subsets of $\bu^{\bs{1}},\dots,\bu^{\bs{M}}$.
The functions $a^{(j)}(u)$ and $d^{(j)}(u)$ respectively are the vacuum eigenvalues of the operators $A^{(j)}(u)$ and $D^{(j)}(u)$.
It is important to stress that the number of partial monodromy matrices $M$ is not related to the numbers of Bethe parameters $n=\#\bu$.
In particular, we can consider the case $M> n$. Then some subsets in the formula \eqref{11-BV-mult-GL2} necessarily are empty. However, we should not forget that for $M\le n$, we also can have empty subsets in the formula \eqref{11-BV-mult-GL2}.

The formula \eqref{11-BV-mult-GL2} can easily be proved by induction on $M$. In fact, assuming that it is true for $M-1$ partial monodromy matrices,
we apply the equality \eqref{11-IK} to the partial vector $B^{(M-1)}(\bu^{\bs{M-1}})|0\rangle^{(M-1)}$.
This immediately gives us \eqref{11-BV-mult-GL2} for $M$ partial monodromy matrices.

The introduction of the multicomposite model is a convenient way to express Bethe vectors in terms of local operators.
Consider, for example, an inhomogeneous $XXX$ chain consisting of $N$ sites.
The monodromy matrix is defined as the product of local $L$-operators
\be{11-T-Lxxx}
T(u)=L_N(u-\xi_M)\dots L_1(u-\xi_1),
\ee
where $\xi_k$ are inhomogeneities, and
\begin{equation}\label{11-L-opXXX}
L_n(u)=\frac1u\begin{pmatrix}
u+\frac c2(1+\sigma^z_n)& c\;\sigma^-_n \\
c\;\sigma^+_n &
u+\frac c2(1-\sigma^z_n)
\end{pmatrix}.
\end{equation}
The normalizing factor $1/u$ is chosen so as to have $d(u)=1$.
Recall that a vacuum vector is a state with all spins up
\be{11-vac-XXX}
|0\rangle = \left(\begin{smallmatrix} 1\\0\end{smallmatrix}\right)_N\otimes\dots\otimes\left(\begin{smallmatrix} 1\\0\end{smallmatrix}\right)_1.
\ee

Consider a multicomposite model consisting of $N$ partial monodromy matrices $T^{(j)}$. This means that each matrix $T^{(j)}$ coincides
with the $L$-operator $L_j(u- \xi_j)$, and the partial vacuum vector has the form $|0\rangle^{(j)}=\left(\begin{smallmatrix} 1\\0\end{smallmatrix}\right)_j$.
Then every  partial Bethe vector in the formula \eqref{11-BV-mult-GL2} corresponds to $j$th
site of the chain, therefore, by virtue of \eqref{11-L-opXXX}, we get
\be{11-PBV-xxx}
B^{(j)}(\bu^{\bs{j}})|0\rangle^{(j)}=g(\bu^{\bs{j}},\xi_j)     \left(\sigma^-_j\right)^{n_j}\left(\begin{smallmatrix} 1\\0\end{smallmatrix}\right)_j,
\ee
where $n_j$ is the cardinality of the subset $\bu^{\bs{j}}$.
Obviously, this vector vanishes for $n_j> 1$, because $\bigl(\sigma^-_j\bigr)^2=0$.
Thus, we come to the conclusion that
all $n_j\le 1$, and the subsets $\bu^{\bs{j}}$ are either empty or consist of one element. Let the total number of variables in the set $\bu$
be equal to $n$, and let the subsets $\bu^{\bs{j_k}}$ ($k=1,\dots,n$),
corresponding to lattice sites with numbers $j_1,\dots, j_n$
contain one element $u_k$, while the remaining subsets are empty. Then the sum
over partitions of the set $\bu$ turns into a sum over the permutations in the set $\bu$ and the sum over the lattice sites $j_1,\dots, j_n$
under the condition $j_n>\dots>j_1$.

It is easy to see that
\be{11-eig-XXX-vac}
\frac{u-\xi_j+\frac c2(1+\sigma^z_j)}{u-\xi_j}\left(\begin{smallmatrix} 1\\0\end{smallmatrix}\right)_j=
f(u,\xi_j)\left(\begin{smallmatrix} 1\\0\end{smallmatrix}\right)_j,
\qquad
\frac{u-\xi_j+\frac c2(1-\sigma^z_j)}{u-\xi_j}\left(\begin{smallmatrix} 1\\0\end{smallmatrix}\right)_j=
\left(\begin{smallmatrix} 1\\0\end{smallmatrix}\right)_j,
\ee
and hence,
\be{11-r1}
a^{(j)}(u)=f(u,\xi_j), \qquad d^{(j)}(u)=1.
\ee
Then equation \eqref{11-BV-mult-GL2} takes the form
\begin{equation}\label{11-BV-xxx-mult-2}
B(\bu)|0\rangle=\Sym_{\bar u} \prod_{1\le k<j\le n}f(u_j,u_k)
\sum_{N\ge j_n>\dots>j_1\ge 1}
\prod_{k=1}^n  \left[\Bigl(\prod\limits_{m=j_k+1}^N f(u_k,\xi_m)\Bigr) g(u_k,\xi_{j_k})\sigma^-_{j_k}\right]| 0\rangle,
\end{equation}
where symbol $\Sym$ means symmetrization (that is, the sum over permutations) over the set that is indicated in the subscript.
In the formula \eqref{11-BV-xxx-mult-2}, the symmetrization acts on the entire expression, that depends on $\bu$.
In the homogeneous limit $\xi_k=c/2$ this expression coincides with the formula for the eigenvector of the  $XXX$ Hamiltonian
found by H.~Bethe in 1931. (see \cite{Bet31, Gaud83}).

All the arguments above are completely applicable to the case of the $XXZ$ chain, and the formula for the Bethe vector \eqref{11-BV-xxx-mult-2}
remains the same. We present it in the detailed form for the homogeneous model
\begin{multline}\label{11-BV-xxz-mult-2}
B(\bu)|0\rangle=\prod_{j=1}^n\frac{\sinh\eta}{\sinh(u_j-\frac\eta2)}\Sym_{\bar u} \prod_{1\le k<j\le n}\frac{\sinh(u_j-u_k+\eta)}{\sinh(u_j-u_k)}\\
\times\sum_{N\ge j_n>\dots>j_1\ge 1}
\prod_{k=1}^n  \Biggl(\frac{\sinh(u_k+\frac\eta2)}{\sinh(u_k-\frac\eta2)}\Biggr)^{N-j_k} \sigma^-_{j_k}| 0\rangle.
\end{multline}
If we replace all hyperbolic functions by rational ones (that is, replace $u_k\to\epsilon u_k$, $\eta\to\epsilon c$
and consider the limit $\epsilon\to 0$) in this formula, then we obtain the Bethe vector of the $XXX$ chain.

The advantage of \eqref{11-BV-xxz-mult-2} is that there the Bethe vector is expressed in terms of local operators. Therefore, we can easily calculate the action of the local spin operators $\sigma_k^\alpha$, $\alpha=x,y,z$,
on this state. However, if we compare the rhs of \eqref{11-BV-xxz-mult-2} with the lhs, then it is clear that the latter looks somewhat simpler.

\section{Special on-shell Bethe vector\label{11-Sec3}}

In this section we finish the discussion of the example of the Bethe vector $B(u_1)B(u_2)|0\rangle$
in a spin chain of four sites.
We began our consideration of this example in section~\ref{04-Sec22}
and continued in section~\ref{05-Sec3}.
Let us recall briefly what we know by this moment. For definiteness, we will speak about the $XXZ$ chain.

The values $u_1=\eta/2$ and $u_2=-\eta/2$
formally satisfy the Bethe equations, but the Bethe vector corresponding to this solution,
becomes null-vector. If we normalize the original vector $B(u_1)B(u_2)|0\rangle$
so that for arbitrary $u_1$ and $u_2$ it would have a unit norm, then the result of the limit $u_1\to\eta/2$ and $u_2\to-\eta/2$
depends on how  $u_1$ and $u_2$ tend to their limiting values. If we do not take into account this solution of the Bethe equations at all, then we will have only five on-shell Bethe vectors, and we obviously will not be able to construct a basis in the $6$-dimensional space.

One can overcome this problem with by means of the twisted Bethe equations. In this case, for $\kappa \ne 1$, one can find $6$  admissible
solutions of the twisted Bethe equations and build on them $6$ twisted on-shell vectors. We already know that twisted on-shell vectors
corresponding to different solutions are mutually orthogonal. Therefore, this set of $6$ vectors forms a basis in the $6$-dimensional space.
It remains to find out what happens in the limit $\kappa \to 1$. We recall that in this limit one of the admissible solutions of the twisted Bethe equations
goes to the values $u_1=\eta/2$ and $u_2=-\eta/2$.

Let us write down representation \eqref{11-BV-xxz-mult-2}  for $N=4$ and $n=2$. For further convenience, we multiply this vector by a coefficient
\be{11-factor}
\frac{\sinh(u_1-u_2)\sinh(u_1+\frac\eta2)\sinh(u_2-\frac\eta2)}{{\sinh}^2\eta\sinh(u_1-u_2+\eta)}.
\ee
Then
\begin{multline}\label{11-BV0}
B(u_1)B(u_2)|0\rangle=\Biggl(\frac{\sinh(u_1+\frac\eta2)}{\sinh(u_k-\frac\eta2)}\Biggr)^{3}
\Biggl(\frac{\sinh(u_2+\frac\eta2)}{\sinh(u_2-\frac\eta2)}\Biggr)^{4}\\[2pt]
\times\Biggl[
\frac{\sinh(u_2-u_1+\eta)}{\sinh(u_2-u_1-\eta)} \sum_{4\ge j_2>j_1\ge 1}
\Biggl(\frac{\sinh(u_1-\frac\eta2)}{\sinh(u_1+\frac\eta2)}\Biggr)^{j_1}
\Biggl(\frac{\sinh(u_2-\frac\eta2)}{\sinh(u_2+\frac\eta2)}\Biggr)^{j_2}\sigma^-_{j_1}\sigma^-_{j_2}| 0\rangle\\[2pt]
+\sum_{4\ge j_2>j_1\ge 1}
\Biggl(\frac{\sinh(u_2-\frac\eta2)}{\sinh(u_2+\frac\eta2)}\Biggr)^{j_1}
\Biggl(\frac{\sinh(u_1-\frac\eta2)}{\sinh(u_1+\frac\eta2)}\Biggr)^{j_2}\sigma^-_{j_1}\sigma^-_{j_2}| 0\rangle\Biggr].
\end{multline}
It is already clear from this formula which problems arise in the limit $u_1\to\eta/2$ and $u_2\to-\eta/2$.
Indeed, if we put $u_1=\epsilon_1+\eta/2$ and $u_2=\epsilon_2-\eta/2$,
then we should answer the question  how the ratio  $\epsilon_1/\epsilon_2$ behaves when $\epsilon_k\to 0$, $k=1,2$.
Moreover, we also should say
how the difference $\epsilon_1- \epsilon_2$ behaves in this limit, because the formula \eqref{11-BV0} has terms proportional to
$u_2-u_1+\eta=\epsilon_2-\epsilon_1$. Apriori, we do not have any prescriptions on how to fix these ratio and difference\footnote{%
Generally speaking, we can substitute $u_1=\epsilon_1+\eta/2$ and $u_2=\epsilon_2-\eta/2$
into Bethe equations and require that there exist a limit of these equations for $\epsilon_k \to 0$. This will give additional restrictions on the parameters
$\epsilon_k \to 0$.}. However, with twisted Bethe equations the problem is easily solved.

In section~\ref{05-Sec3}
we parameterized the twist parameter as $\kappa=\theta^2$ and did a change of variables
\be{11-change}
\frac{\sinh(u_k-\frac\eta2)}{\sinh(u_k+\frac\eta2)}=w_k,\qquad k=1,2.
\ee
It also follows from the twisted Bethe equations that
\be{11-BE-part}
\frac{\sinh(u_2-u_1+\eta)}{\sinh(u_2-u_1-\eta)}=\theta^2\Biggl(\frac{\sinh(u_1-\frac\eta2)}{\sinh(u_1+\frac\eta2)}\Biggr)^{4}=\theta^2 w_1^4.
\ee
We are interested in a solution where $w_1\to 0$, $w_2\to\infty$,
and  the condition $w_1w_2 = -\theta$ is satisfied. At the same time, the parameter $\theta$ itself
goes to $1$.

We require that the Bethe vector in \eqref{11-BV0} corresponds to this solution of the twisted Bethe equations.
Let us write this vector in terms of
$w_1$ and $w_2$:
\be{11-BV1}
B(u_1)B(u_2)|0\rangle=\frac{w_1^3}{\theta^2}\sum_{4\ge j_2>j_1\ge 1}w_1^{j_1}w_2^{j_2}\sigma^-_{j_1}\sigma^-_{j_2}|0\rangle+
\frac{1}{w_1}\sum_{4\ge j_2>j_1\ge 1}w_2^{j_1}w_1^{j_2}\sigma^-_{j_1}\sigma^-_{j_2}|0\rangle.
\ee
Here we have used partly the condition $w_1w_2=-\theta$. Let us use it once more in order to get rid of the parameter $w_2$ from \eqref{11-BV1}:
\be{11-BV2}
B(u_1)B(u_2)|0\rangle=\sum_{4\ge j_2>j_1\ge 1}\Bigl((-\theta)^{j_2-2}w_1^{j_1-j_2+3}+(-\theta)^{j_1}w_1^{j_2-j_1-1}\Bigr)\sigma^-_{j_1}\sigma^-_{j_2}|0\rangle.
\ee
Now the limit $w_1\to0$  and $\theta\to 1$ can be easily computed. In the second term, we obtain the terms with
$j_2=j_1+1$, while in the first term, only the term with  $j_1=1$ and $j_2=4$ survives. We obtain
\be{11-BV-fin}
B(u_1)B(u_2)|0\rangle=\sum_{j=1}^4(-1)^{j}\sigma^-_{j}\sigma^-_{j+1}|0\rangle, \qquad u_1\to\tfrac\eta2,\quad u_2\to-\tfrac\eta2,
\ee
where we used the periodicity condition $\sigma^-_{5}\equiv\sigma^-_{1}$. We recommend that the reader to check
that the constructed vector is indeed an eigenvector of the $XXZ$ chain Hamiltonian, and thus, we found
a vector, which we lacked to build the complete set.

In conclusion we note that in the general case of a chain consisting of $N$ sites, a vector
\be{11-BV-fin1}
\sum_{j=1}^N(-1)^{j}\sigma^-_{j}\sigma^-_{j+1}|0\rangle
\ee
is an eigenvector of the $XXZ$ Hamiltonian (and the $XXX$ Hamiltonian, because this vector does not depend on $\Delta$). This vector
is the special limit of the Bethe vector $B(u_1)B(u_2)|0\rangle$, when  $u_1\to\tfrac\eta2$ and $u_2\to-\tfrac\eta2$.



%
%
\chapter{Zero modes\label{CHA-ZM}}

The content of this lecture concerns the models described by the rational $R$-matrix $R(u,v)=\mathbf{I}+g(u,v)\mathbf{P}$.
The generalization of the method described below to the case of models with trigonometric $R$-matrix remains an open question.

We introduce the notion of zero modes for the matrix elements of the monodromy matrix \cite{Dr88,Molevbook}. In this case, two goals are pursued. At first,
a method based on the use of zero modes allows one to solve the problem of calculating the form factors of local operators in models for which the solution of the quantum inverse problem is unknown. The idea of the calculation is formulated already in this lecture, but the details will be given later.
Secondly, with the help of zero modes, it is possible to establish additional relationships between the form factors of the operators $T^{ij}(z)$.
To say the truth,  in models with a $2\times 2$ monodromy matrix this  is not very relevant, because there are only four operators,
and some relations between form factors are obvious. For example, the form factor of the transfer matrix between two
different on-shell vectors should equal zero. Hence, it follows that the form factors of the operators $A(z)$ and $D(z)$
between different on-shell vectors differ from each other only in sign.

However, in models for which the $R$ matrix is still of the form $R(u,v)=\mathbf{I}+g(u,v)\mathbf{P}$,
while the monodromy matrix is an $N\times N$  matrix ($ N> 2 $), the benefits of additional relations between different form factors are already quite
essential. Using the method based on the application of zero modes, it is possible to prove that in such models it is sufficient
to calculate only one form factor of some diagonal element $T^{ii}(z)$. All the rest can be obtained
from this initial form factor by means of simple limits. Below we demonstrate this statement
in the example of models with a $2\times 2$ monodromy matrix.

\section{Zero modes\label{12-Sec1}}

Suppose that the monodromy matrix  $T(u)$ of some model has the following asymptotic expansion at $u\to\infty$:
\be{12-zero-modes}
T(u)=\mathbf{1}+ \sum_{n=0}^\infty T[n]\,\left(\tfrac cu\right)^{n+1}.
\ee
Here $T[n]$ are some operator-valued matrices which do not depend on $u$.
Such the expansion is consistent with the structure of the $ R $-matrix, which itself has exactly the same decomposition. It is not difficult to see, however, that the decomposition of \eqref{12-zero-modes} does not holds for all $ T $-matrices satisfying the $ RTT $-relation and having a vacuum vector.
Indeed, if \eqref{12-zero-modes} holds for some $ T (u) $, then for the twisted monodromy matrix $ \hat\kappa T (u) $ the expansion
does not begin with the identity matrix, but with some diagonal matrix $\hat\kappa $. Furthermore, it follows from the
asymptotic expansion \eqref{12-zero-modes} that similar asymptotic formulas hold for the functions $a(u)$, $d(u)$, and their ratio
$r(u)=a(u)/d(u)$:
\be{12-zero-modes-r}
\begin{aligned}
a(u)=1+ \sum_{n=0}^\infty a[n]\,\left(\tfrac cu\right)^{n+1},\\
d(u)=1+ \sum_{n=0}^\infty d[n]\,\left(\tfrac cu\right)^{n+1},\\
r(u)=1+ \sum_{n=0}^\infty r[n]\,\left(\tfrac cu\right)^{n+1}.
\end{aligned}
\ee
At the same time, we have seen already that in the QNLS model, the function $r(u)$ has the form $r(u)=e^{-iLu}$,
and thus, it has essential singularity at infinity.

Thus, the existence of the asymptotic expansion \eqref{12-zero-modes} is an assumption, which is not valid for all
the models with the rational $R$-matrix. Nevertheless, the method described below, is largely applicable to those models in which
the monodromy matrix has another asymptotic expansion. We will consider such models in more detail  later,
but for now we will assume that the formula \eqref{12-zero-modes} is valid.

The coefficients $T[0]$ are called zero modes. In fact, below we will need only these coefficients.

Using the $RTT$-relation, it is easy to obtain commutation relations between zero modes and the operators $A(u)$, $B(u)$, $C(u)$, and $D(u)$.
Consider, for example, the commutation relation for the operators $A(v)$ and $B(u)$  \eqref{03-CR-AB}.
Multiplying both sides by $u/c$ and letting $u$ go to infinity, we obtain
\begin{equation}
\lim_{u\to\infty}\tfrac uc A(v)B(u)=\lim_{u\to\infty}\tfrac uc \bigl(f(u,v)B(u)A(v)+g(v,u)B(v)A(u)\bigr).\label{12-AB-1}
\end{equation}
Taking into account that $f(u,v)\to 1$ as $u\to\infty$, we find
\begin{equation}
[B[0],A(v)]=B(v).\label{12-AB-2}
\end{equation}
We can continue and send  $v$ to infinity:
\begin{equation}
\lim_{v\to\infty}\tfrac vc [B[0],A(v)-1]=\lim_{v\to\infty}\tfrac vc B(v),\label{12-AB-3}
\end{equation}
what gives us
\begin{equation}
\bigl[B[0],A[0]\bigr]=B[0].\label{12-AB-4}
\end{equation}
Similarly, one can obtain other commutation relations.
We will give them without a detailed derivation as needed.

There are several reasons for introducing the zero modes. The main reason is that with the help of zero modes and the composite model we can calculate
form factors of local operators, without  an explicit solution of the quantum inverse problem. Suppose, as usual, that the mo\-no\-dro\-my matrix
is a product of local $L$-operators \eqref{02-T-LLL},
and each $L$-operator has an expansion analogous to  \eqref{12-zero-modes}
\be{12-zero-modesL}
L_k(u)=\mathbf{1}+ \sum_{n=0}^\infty L_k[n]\,\left(\tfrac cu\right)^{n+1}.
\ee
Observe that all the dependence of the monodromy matrix on local operators associated with the  $k$th site is hidden
in the $L$-operator $L_k (u)$, in particular, in its zero mode $L_k [0]$. On the other hand, it is obvious that the zero modes of the monodromy matrix
are sums of the zero modes of $L$-operators:
\be{12-T0L0}
T[0]=\sum_{k=1}^N L_k[0].
\ee
If we introduce now a composite model $T(u)=T^{(2)}(u)T^{(1)}(u)$ such that the partial matrix $T^{(1)}(u)$ corresponds to the
site $1,\dots, m$, where $1\le m\le N$, then
\be{12-T0L01}
T^{(1)}[0]=\sum_{k=1}^m L_k[0].
\ee
Thus, we obtain
\be{12-FF1}
\langle0|C(\bv)T^{(1)}_{ij}[0] B(\bu)|0\rangle  = \sum_{k=1}^m \langle0|C(\bv) (L_k)_{ij}[0]B(\bu)|0\rangle=\Phi_{ij}(m),
\ee
where $\Phi_{ij}(m)$ is a function depending on the number $m$ of the intermediate site. Then
\be{12-FF2}
\Phi_{ij}(m)-\Phi_{ij}(m-1)= \langle0|C(\bv) (L_m)_{ij}[0]B(\bu)|0\rangle.
\ee
Thus, if we learn how to calculate the form factors of the first partial zero modes $T^{(1)}_{ij}[0]$,
then we compute the form factors of the local operators $(L_m)_{ij}[0]$ associated with any lattice site with the number $m$.
It is also easy to understand,
that if the initial model is continuous, and the representation  \eqref{02-T-LLL}
is its lattice approximation, then in the continuous limit
instead of the finite difference (as in the formula \eqref{12-FF2}), a derivative of the form factor of the first partial zero mode will arise.

\section{Action of zero modes on Bethe vectors\label{12-Sec2}}

Actions of the zero modes on Bethe vectors can be easily obtained from the action formulas of the corresponding operators. We start with
an obvious equality
\be{12-actB}
B[0]\cdot B(\bu)|0\rangle=\lim_{z\to\infty}\tfrac zc B(z)B(\bu)|0\rangle.
\ee
In fact, the formula \eqref{12-actB} is a tautology: there is just the definition
of the zero mode $B[0]$ in the rhs. It is interesting to note, however, that if the initial vector $B(\bu)|0\rangle$
was on-shell, then the vector appearing in the rhs of  \eqref{12-actB} is also an on-shell. This follows from the fact that
by virtue of \eqref{12-zero-modes-r}, the Bethe equations admit infinite solutions. Indeed, the Bethe equations for the set $\{z,\bu\}$
have the form
\be{12-BEz}
\begin{aligned}
&r(z)=\frac{f(z,\bu)}{f(\bu,z)},\\
&r(u_k)=\frac{f(u_k,\bu_k)}{f(\bu_k,u_k)}\frac{f(u_k,z)}{f(z,u_k)},\qquad k=1,\dots,n.
\end{aligned}
\ee
Since $r(z)\to 1$ at  $z\to\infty$,
and the function $f(x,y)$ also goes to unity, when any of its arguments tend to infinity,
we see that in the limit $z\to\infty$ the first of the equations \eqref{12-BEz} is satisfied. From the remaining equations, the parameter $z$ simply
disappears in the limit $ z\to\infty $, and we obtain a system of Bethe equations for the set of variables $\bu$:
\be{12-BE}
r(u_k)=\frac{f(u_k,\bu_k)}{f(\bu_k,u_k)},\qquad k=1,\dots,n.
\ee
Thus, if the conditions \eqref{12-BE} are fulfilled, then \eqref{12-BEz} are also satisfied at $z\to\infty$.

The action of the zero mode $A[0]$ on the Bethe vector is already less trivial. To calculate it, we need to act with the operator $A(z)-1$
on the vector $B(\bu)|0\rangle$, then multiply the result by $z/c$ and let $z$ go to infinity. We have
\be{12-actA0}
A[0]\cdot B(\bu)|0\rangle=\lim_{z\to\infty}\tfrac zc \Bigl((a(z)f(\bu,z)-1)B(\bu)+
\sum_{k=1}^n a(u_k)g(z,u_k)f(\bu_k,u_k)B(z)B(\bu_k)\Bigr)|0\rangle.
\ee
Since $g(z,u_k)\sim c/z$ and $B(z)\sim c/z$ at $z\to\infty$,
all the terms in the sum over $k$ go to zero even after multiplication by $z/c$.
There remains the only contribution from the first term, and we obtain
\be{12-actA}
A[0]\cdot B(\bu)|0\rangle= (a[0]-n)B(\bu)|0\rangle,
\ee
where $n=\#\bu$. Similarly, we find
\be{12-actD}
D[0]\cdot B(\bu)|0\rangle= (d[0]+n)B(\bu)|0\rangle.
\ee
Thus, any Bethe vector is an eigenvector for the zero modes $A[0]$ and $D[0]$.

Perhaps the most curious is the action of the zero mode $C[0]$. Using  the action formula of $C(z)$ on the Bethe vector  \eqref{07-SP-ans-vid}
we find
\begin{multline}\label{12-SP-ans-vid}
C[0]\cdot B(\bu)|0\rangle=\lim_{z\to\infty}\tfrac zc\Biggl[\sum_{k=1}^n \bigl(d(z)a(u_k) X_k + a(z)d(u_k)\tilde X_k \bigr)
B(\bu_k)|0\rangle\\
+ \sum_{j<k}^n \bigl(a(u_j)d(u_k)Y_{jk}+a(u_k)d(u_j)Y_{kj}\bigr)
B(z)B(\bu_{jk})|0\rangle\Biggr]\,,
\end{multline}
and we recall that $\bu_{jk}=\bu\setminus\{u_j,u_k\}$. The coefficients in  \eqref{12-SP-ans-vid} are given by
formulas  \eqref{07-SP-ansLk}, \eqref{07-SP-anstLk}, \eqref{07-SP-Mjk}.
It is easy to see that for $z \to \infty$ the double sum in the formula \eqref{12-SP-ans-vid} disappears, since the coefficients $Y_ {jk}$ behave
as $z^{-2}$ at  $ z\to\infty $. The remaining terms give
\begin{equation}\label{12-C0BV}
C[0]\cdot B(\bu)|0\rangle=\sum_{k=1}^n \bigl(a(u_k) f(\bu_{k},u_k) - d(u_k)f(u_k,\bu_{k}) \bigr)
B(\bu_k)|0\rangle.
\end{equation}
We see that if the initial vector $B(\bu)|0\rangle$ was on-shell, then due to the Bethe equations, all the coefficients
in the rhs of \eqref{12-C0BV} turn to zero. That is, the zero mode $C[0]$ annihilates the on-shell vectors:
$C[0]B(\bu)|0\rangle=0$.

There is, however,  one subtlety. Calculating the limits of the coefficients $X_k$ and $\tilde X_k$ we assumed that all the parameters $u_k$
were finite. If the set $\bu$ contains an infinite parameter, then the result of the action $C[0]$ on the on-shell Bethe vector
changes. Indeed, according to \eqref{12-actB}, the on-shell Bethe vector with an infinite parameter can be interpreted as the action
of the zero mode $B[0]$ on an on-shell vector with finite parameters
\be{12-B-actB}
\lim_{z\to\infty}\tfrac zc B(z)B(\bu)|0\rangle=B[0] B(\bu)|0\rangle,
\ee
where the parameters $\bu$ are finite. Then due to the commutation relation
\be{12-CB-AD}
\bigl[C[0],B[0]\bigr]=A[0]-D[0],
\ee
and equations \eqref{12-zero-modes-r}, \eqref{12-actA}, \eqref{12-actD}, we find
\be{12-B-actB-inf}
C[0]\lim_{z\to\infty}\tfrac zc B(z)B(\bu)|0\rangle=C[0]B[0] B(\bu)|0\rangle=(A[0]-D[0]) B(\bu)|0\rangle=(r[0]-2n) B(\bu)|0\rangle.
\ee

The action of zero modes on dual vectors is completely analogous. In particular, all dual Bethe vectors are eigenvectors of
the zero modes $A[0]$ and $D[0]$, and the zero mode $B[0]$ annihilates the dual on-shell vectors,
if the latter depend on the finite parameters.

\section{Relationships between different form factors\label{12-Sec3}}

In this section we obtain new relations between the form factors of the matrix elements of the monodromy matrix.
It is convenient to normalize the original monodromy matrix $T(u)$ by the function $d(u)$. The matrix thus normalized
will be denoted by bold type $\bs{T}(u) = T(u)/d(u)$. We also denote in bold type normalized matrix
elements: $\bs{A}(u)=A(u)/d(u)$, and so on. Finally, we introduce a similar notation for the normalized eigenvalue of the transfer matrix
\be{12-btau}
\bs{\tau}(z|\bu)= \frac{\tau(z|\bu)}{d(z)}=r(z)f(\bu,z)+f(z,\bu).
\ee

As usual, we call a form factor of the opeartor $\bs{T}^{ij}(z)$ a matrix element of this operator between two  on-shell vectors
$\langle0|\bs{C}(\bv)$ and $\bs{B}(\bu)|0\rangle$. We assume that $\#\bu=n$, and denote this form factor of the matrix element
$\bs{T}^{ij}(z)$ by $\mathcal{F}_n^{(ij)}(z)$. Then
\be{12-defFF}
\mathcal{F}_n^{(ij)}(z)\equiv \mathcal{F}_n^{(ij)}(z|\bv;\bu)=\langle0|\bs{C}(\bv)\bs{T}^{ij}(z) \bs{B}(\bu)|0\rangle.
\ee
Observe that the number of parameters $\bv$ depends on the specific form factor $\mathcal{F}_n^{(ij)}(z)$:  if $i=j$, then $\#\bv=n$; if $i>j$, then $\#\bv=n-1$;
if $i<j$, then $\#\bv=n+1$.

Let us consider how different form factors  $\mathcal{F}_n^{(ij)}(z)$ are related with each other. For this we use the zero modes.

We have due to the commutation relation \eqref{12-AB-2}
\begin{equation}
\langle0|\bs{C}(\bv)\bs{B}(z) \bs{B}(\bu)|0\rangle=\langle0|\bs{C}(\bv)[B[0],\bs{A}(z)]\bs{B}(\bu)|0\rangle.\label{12-AB-FF}
\end{equation}
Assume that both vectors in \eqref{12-AB-FF} are on-shell, and there are no infinite parameters among the sets $\bv$ and $\bu$.
Then  we have in the lhs the form factor of the operator $\bs{B}(z)$.
Consider the rhs, that consists of two terms
\be{12-RHS}
\langle0|\bs{C}(\bv)B[0]\bs{A}(z)\bs{B}(\bu)|0\rangle-\langle0|\bs{C}(\bv)\bs{A}(z)B[0]\bs{B}(\bu)|0\rangle.
\ee
The action of the zero mode $B[0]$ on the dual on-shell vector yields zero, hence, the first term vanishes. In the second term,
the action of the zero mode $B[0]$ on the vector $\bs{B}(\bu)|0\rangle$ is given by the formula \eqref{12-actB}. We obtain
\begin{equation}
\langle0|\bs{C}(\bv)B(z) \bs{B}(\bu)|0\rangle=-\lim_{w\to\infty}\tfrac wc \langle0|\bs{C}(\bv)A(z) \bs{B}(w)\bs{B}(\bu)|0\rangle.\label{12-BA-FF0}
\end{equation}
Since the initial vector $\bs{B}(\bu)|0\rangle$ was on-shell, the resulting vector $\tfrac wc\bs{B}(w)\bs{B}(\bu)|0\rangle$
also is on-shell at $w\to\infty$. Therefore, in the rhs of \eqref{12-BA-FF0}, we have the form factor of the operator $\bs{A}(z)$, that is,
\begin{equation}
\mathcal{F}_n^{(12)}(z|\bv;\bu)=-\lim_{w\to\infty}\tfrac wc \mathcal{F}_{n+1}^{(11)}(z|\bv;\{\bu,w\}).\label{12-BA-FF}
\end{equation}
We arrive at a rather surprising conclusion: to calculate the form factor of the operator $\bs{B}(z)$ is sufficient to calculate the form factor
of the operator $\bs{A}(z)$ and to send one of the parameters to infinity in the obtained result. Similarly, one can also find relations between
other form factors. For example, using
\be{12-AC}
\bs{C}(z)=\bigl[\bs{A}(z),C[0]\bigr]
\ee
we obtain
\begin{equation}
\mathcal{F}_n^{(21)}(z|\bv;\bu)=-\lim_{w\to\infty}\tfrac wc \mathcal{F}_n^{(11)}(z|\{\bv,w\};\bu).\label{12-CA-FF}
\end{equation}

Let us now consider the case when some parameters of the on-shell vectors are infinite.
Suppose, for example, that a dual vector depends on one infinite parameter. Then
\begin{equation}
\lim_{w\to\infty}\tfrac wc \mathcal{F}_n^{(12)}(z|\{\bv,w\};\bu)=\langle0|\bs{C}(\bv)C[0]B(z) \bs{B}(\bu)|0\rangle.\label{12-FF-inf}
\end{equation}
Using commutation relations
\be{12-CB}
\bigl[C[0],\bs{B}(z)\bigr]=\bs{A}(z)-\bs{D}(z),\qquad C[0]\bs{B}(\bu)|0\rangle=0,
\ee
we immediately find
\begin{equation}
\lim_{w\to\infty}\tfrac wc \mathcal{F}_n^{(12)}(z|\{\bv,w\};\bu)=\mathcal{F}_n^{(11)}(z|\bv;\bu)-\mathcal{F}_n^{(22)}(z|\bv;\bu),\label{12-BAD-FF}
\end{equation}
where form factors in the rhs depend only on the finite parameters.

Thus, the form factors of different elements of the monodromy matrix are related with each other. It is interesting to verify these relations using
explicit formulas for form factors. This is done in the next section.

\section{Comparison of determinant representations\label{12-Sec4}}%

Let us check, for instance, equation \eqref{12-CA-FF}, that relates the form factors of the operators $\bs{C}(z)$ and $\bs{A}(z)$. For definiteness
we assume that the sets $\bv$ and $\bu$ are different and containing  finite parameters only.  Then the form factors of the operator $\bs{A}(z)$  differs from the one of $\bs{D}(z)$
only in the sign, because the form factor of the transfer matrix $\bs{A}(z)+\bs{D}(z)$ obviously vanishes. Hence, we have
\begin{equation}
\mathcal{F}_n^{(21)}(z|\bv;\bu)=\lim_{w\to\infty}\tfrac wc \mathcal{F}_n^{(22)}(z|\{\bv,w\};\bu).\label{12-CD-FF}
\end{equation}

\subsection{Transformation of the form factor $\mathcal{F}_n^{(21)}(z)$\label{12-Sec41}}

Consider the lhs of \eqref{12-CD-FF}. Form factor of $\bs{C}(z)$ is equal to the scalar product of the on-shell vector
$\bs{B}(\bu)|0\rangle$ and the dual vector $\langle 0| \bs{C}(\bv)\bs{C}(z)$. Let $\bxi=\{\bv,z\}$. Then
\be{12-FF21}
\mathcal{F}_n^{(21)}(z|\bv;\bu)=\langle 0| \bs{C}(\bxi)\bs{B}(\bu)|0\rangle.
\ee
According to the notation introduced above, the number of elements in the set $\bu$ is $n$. The number of elements in the set $\bxi$ is also $n$,
what implies $\#\bv=n-1$. In addition, the parameters $\bv$ satisfy Bethe equations
\be{12-BE-v}
r(v_k)=\frac{f(v_k,\bv_k)}{f(\bv_k,v_k)}=(-1)^n\frac{h(v_k,\bv)}{h(\bv,v_k)}.
\ee

The scalar product in \eqref{12-FF21} has the following representation:
\begin{equation}\label{12-SP-eig-arb}
\langle 0| \bs{C}(\bxi)\bs{B}(\bu)|0\rangle=\Delta'_n(\bu)
\Delta_n(\bxi)h(\bxi,\bu)\;\det_n\mathcal{M}_{jk} ,
\end{equation}
where
\begin{equation}\label{12-MCB}
\mathcal{M}_{jk}= t(\xi_k,u_j)+(-1)^{n-1}r(\xi_k)\;t(u_j,\xi_k)\frac{h(\bu,\xi_k)}{h(\xi_k,\bu)}.
\end{equation}
In the first $n-1$ columns of this matrix the parameters $\xi_k$ take the values $v_k$, while in the last column $\xi_n=z$.

We transform the matrix $\mathcal{M}_{jk}$ so that it  would be convenient to compare both sides of the equality \eqref{12-CD-FF}.
To do this, we add to the last row of this matrix all the remaining rows multiplied by the coefficients $\alpha_j/\alpha_n$, where
\be{12-alj}
\alpha_j=\frac{g(u_j,\bu_j)}{g(u_j,\bv)}=\prod_{m=1}^{n-1}(u_j-v_m)\prod_{\substack{m=1\\m\ne j}}^n(u_j-u_m)^{-1}\,.
\ee
It is clear that the determinant does not change under this transformation.

We denote the modified last row by $\widetilde{\mathcal{M}}_{nk}$.
In order to find its explicit form, we compute a sum
\be{12-sumalp}
H_k^+=\sum_{j=1}^n\alpha_jt(u_j,\xi_k).
\ee
We have already done a similar calculation in proving the orthogonality of on-shell vectors. Let us present this calculation once more,
because in this example there are some differences.
Consider the auxiliary integral over a circle of a large radius
\be{12-SP-Int-def}
I=\frac1{2\pi i}\oint_{|x|=R\to\infty} \frac{ c^2\,dx}{(x-\xi_k)(x-\xi_k+ c)}
\frac{\prod_{m=1}^{n-1}(x-v_m)}{\prod_{m=1}^{n}(x-u_m)}\,.
\ee
Since the integrand behaves as  $x^{-3}$ at $x\to\infty$,
we conclude that $I=0$. On the other hand, the integral \eqref{12-SP-Int-def} is equal to the sum of residues
in the poles within the integration contour. First, these are poles at $x=u_\ell$, $\ell=1,\dots,n$.
It is  easy to see that the sum of the residues in these poles gives exactly  $H_k^+$.  There are also two poles: at
$x=\xi_k$ and at $x=\xi_k-c$. Obviously,
\be{12-SP-res}
\begin{aligned}
&\Res \frac{ c^2}{(z-\xi_k)(z-\xi_k+ c)}
\frac{\prod_{m=1}^{n-1}(x-v_m)}{\prod_{m=1}^{n}(x-u_m)}\Bigr|_{x=\xi_k}=
c\frac{\prod_{m=1}^{n-1}(\xi_k-v_m)}{\prod_{m=1}^n(\xi_k-u_m)}=-\frac{g(\bu,\xi_k)}{g(\bv,\xi_k)}\,,\num
&\Res \frac{ c^2}{(z-\xi_k)(z-\xi_k+ c)}
\frac{\prod_{m=1}^{n-1}(x-v_m)}{\prod_{m=1}^{n}(x-u_m)}\Bigr|_{x=\xi_k- c}=
-c\frac{\prod_{m=1}^{n-1}(\xi_k-v_m- c)}{\prod_{m=1}^n(\xi_k-u_m- c)}=\frac{h(\bv,\xi_k)}{h(\bu,\xi_k)}\,.
\end{aligned}
\ee
Thus, we find
\be{12-Hp-res}
H_k^+=\frac{g(\bu,\xi_k)}{g(\bv,\xi_k)}-\frac{h(\bv,\xi_k)}{h(\bu,\xi_k)}.
\ee

Similarly, one can compute a sum
\be{12-Hm0}
H_k^-=\sum_{j=1}^n\alpha_jt(\xi_k,u_j).
\ee
However, we can avoid an additional calculation, if we notice that the sum  \eqref{12-Hm0} can be obtained from \eqref{12-sumalp}
via the replacement $c\to -c$. Thus, we immediately obtain
\be{12-Hm-res}
H_k^-=\frac{g(\xi_k,\bu)}{g(\xi_k,\bv)}-\frac{h(\xi_k,\bv)}{h(\xi_k,\bu)},
\ee
because the replacement $c\to -c$ leads  to the fact that the arguments for each of the functions $g$ and $h$
swap their positions. Now, taking into account
\eqref{12-Hp-res} and \eqref{12-Hm-res}, we find
\begin{multline}\label{12-modM}
\widetilde{\mathcal{M}}_{nk}=\mathcal{M}_{nk}+\frac1{\alpha_n}\sum_{j=1}^{n-1}\alpha_j \mathcal{M}_{jk}
=\frac1{\alpha_n}\sum_{j=1}^{n}\alpha_j \mathcal{M}_{jk}\\
=\frac1{\alpha_n}\left[(-1)^{n-1}r(\xi_k)\frac{h(\bu,\xi_k)}{h(\xi_k,\bu)}\left(\frac{g(\bu,\xi_k)}{g(\bv,\xi_k)}-\frac{h(\bv,\xi_k)}{h(\bu,\xi_k)}\right)
+\frac{g(\xi_k,\bu)}{g(\xi_k,\bv)}-\frac{h(\xi_k,\bv)}{h(\xi_k,\bu)}\right],
\end{multline}
and after simple algebra
\begin{equation}\label{12-modM0}
\widetilde{\mathcal{M}}_{nk}=\frac{\bs{\tau}(\xi_k|\bu)-\bs{\tau}(\xi_k|\bv)}{\alpha_n g(\xi_k,\bv)h(\xi_k,\bu)}.
\end{equation}

One can easily see that $\widetilde{\mathcal{M}}_{nk}=0$ for $k<n$. Indeed, if $k<n$, then $\xi_k=v_k$, and hence, $g^{-1}(\xi_k,\bv)=0$.
Of course, the eigenvalue $\bs{\tau}(\xi_k|\bv)$ contains products  $f(\xi_k,\bv)$ and $f(\bv,\xi_k)$, which compensate this zero.
However, due to the Bethe equations \eqref{12-BE-v}
\be{12-modMk}
-\frac{\bs{\tau}(\xi_k|\bv)}{g(\xi_k,\bv)}\Bigr|_{\xi_k=v_k}=(-1)^{n}r(v_k)h(\bv,v_k)-h(v_k,\bv)=0.
\ee
Thus, after the modification, all the last row vanishes, except the last element, where
\begin{equation}\label{12-modMn}
\widetilde{\mathcal{M}}_{nn}=\frac{g(u_n,\bv)(\bs{\tau}(z|\bu)-\bs{\tau}(z|\bv))}{ g(u_n,\bu_n) g(z,\bv)h(z,\bu)}.
\end{equation}
The determinant reduces to the product of this element by the corresponding minor, and we obtain
\be{12-FF21-res}
\mathcal{F}_n^{(21)}(z|\bv;\bu)=(\bs{\tau}(z|\bu)-\bs{\tau}(z|\bv))\frac{\Delta'_n(\bu)
\Delta_n(\bxi)h(\bxi,\bu) g(u_n,\bv)}{g(u_n,\bu_n) g(z,\bv)h(z,\bu)}\;\det_{n-1}\mathcal{M}_{jk}.
\ee
It remains to extract the explicit dependence on $z$ in the prefactor. Obviously,
\be{12-Delh}
\Delta_n(\bxi)=\Delta_{n-1}(\bv)g(z,\bv),\qquad  h(\bxi,\bu)=h(\bv,\bu)h(z,\bu).
\ee
Substituting this into \eqref{12-FF21-res}, we finally obtain
\be{12-FF21-res1}
\mathcal{F}_n^{(21)}(z|\bv;\bu)=(\bs{\tau}(z|\bu)-\bs{\tau}(z|\bv))\Delta'_n(\bu)
\Delta_{n-1}(\bv)h(\bv,\bu) \frac{g(u_n,\bv)}{g(u_n,\bu_n)}\;\det_{n-1}\mathcal{M}_{jk}^{(21)},
\ee
where
\begin{equation}\label{12-M21}
\mathcal{M}^{(21)}_{jk}= t(v_k,u_j)-t(u_j,v_k)\frac{h(\bu,v_k)h(v_k,\bv)}{h(v_k,\bu)h(\bv,v_k)}.
\end{equation}
Note that the matrix $\mathcal{M}^{(21)}_{jk}$ actually
coincides with the original matrix $\mathcal{M}_{jk}$, however we have
expressed functions $r(v_k)$ in terms of products of the $h$ functions using the Bethe equations.


\subsection{Transformation of form factor $\mathcal{F}_n^{(22)}(z)$\label{12-Sec42}}

We now consider the rhs of the equation \eqref{12-CD-FF}. We have already calculated the form factor of the operator $D(z)$ in terms of the derivative
of the scalar product of a twisted on-shell vector and the ordinary on-shell vectors\footnote{Strictly speaking, we calculated the form factor
of the operator $D(\tfrac\eta2)$ in the $XXZ$ spin chain. However, the calculation of the form factor of the operator $D(z)$
in models with the rational $R$-matrix is exactly the same.}.
Recall this representation:
\be{12-FF-D2}
\mathcal{F}_n^{(22)}(z|\bxi;\bu)=\bigl(\bs{\tau}(z|\bxi)-\bs{\tau}(z|\bu)\bigr)
\frac{d}{d\kappa}\langle 0| \bs{C}(\bxi)\bs{B}(\bu)|0\rangle\Bigr|_{\kappa=1}.
\ee
Here the parameters $\bu$ satisfy usual Bethe equations, while the parameters $\bxi$ satisfy twisted Bethe equations with the twist parameter $\kappa$.
The scalar product in the rhs of \eqref{12-FF-D2} has the form
\begin{equation}\label{12-SP-tw-us}
\langle 0| \bs{C}(\bxi)\bs{B}(\bu)|0\rangle=\Delta'_n(\bu)
\Delta_n(\bxi)h(\bxi,\bu)\;\det_n\mathcal{M}^{(\kappa)}_{jk} ,
\end{equation}
where
\begin{equation}\label{12-Mkap}
\mathcal{M}^{(\kappa)}_{jk}= t(\xi_k,u_j)+\kappa\;t(u_j,\xi_k)\frac{h(\bu,\xi_k)h(\xi_k,\bxi)}{h(\xi_k,\bu)h(\bxi,\xi_k)}.
\end{equation}

Let us calculate  explicitly the derivative over $\kappa$ in the formula \eqref{12-FF-D2}.
To do this, we transform the matrix $\mathcal{M}^{(\kappa)}_{jk}$, as before we transformed the matrix $\mathcal{M}_{jk}$.
Namely, we add to the last row of this matrix all the remaining rows multiplied by the coefficients $\nu_j/\nu_n$, where
\be{12-nuj}
\nu_j=\frac{g(u_j,\bu_j)}{g(u_j,\bxi)}=\frac1c\;\prod_{m=1}^n(u_j-\xi_m)\prod_{\substack{m=1\\m\ne j}}^n(u_j-u_m)^{-1}\,.
\ee
Calculation of the arising sums can be done again via an auxiliary contour integral. The result reads
\be{12-SP-Upm-res}
\sum_{j=1}^n \nu_j t(u_j,\xi_k)=\frac{h(\bxi,\xi_k)}{h(\bu,\xi_k)},
\qquad \sum_{j=1}^n \nu_j t(\xi_k,u_j)=-\frac{h(\xi_k,\bxi)}{h(\xi_k,\bu)}\,.
\ee
This implies
\begin{equation}\label{12-modMkap}
\widetilde{\mathcal{M}}^{(\kappa)}_{nk}=\mathcal{M}^{(\kappa)}_{nk}+\frac1{\nu_n}\sum_{j=1}^{n-1}\nu_j \mathcal{M}^{(\kappa)}_{jk}
=(\kappa-1)\frac{g(u_n,\bxi)}{g(u_n,\bu_n)}\frac{h(\xi_k,\bxi)}{h(\xi_k,\bu)}.
\end{equation}
Thus, the last row of the matrix $\mathcal{M}^{(\kappa)}_{jk}$ after the modification turned out to be proportional to $(\kappa-1)$.
Therefore, if we take the derivative with respect to $\kappa$ at $\kappa = 1$, we need to differentiate only the elements of this row, setting in the remaining
rows $\kappa = 1$. As a result, we get
\begin{equation}\label{12-FF22}
\langle 0| \bs{C}(\bxi)\bs{D}(z)\bs{B}(\bu)|0\rangle=\bigl(\bs{\tau}(z|\bxi)-\bs{\tau}(z|\bu)\bigr)\Delta'_n(\bu)
\Delta_n(\bxi)h(\bxi,\bu)\frac{g(u_n,\bxi)}{g(u_n,\bu_n)}\;\det_n\mathcal{M}^{(22)}_{jk} ,
\end{equation}
where
\be{12-Mat22}
\begin{aligned}
&\mathcal{M}^{(22)}_{jk}=t(\xi_k,u_j)+t(u_j,\xi_k)\frac{h(\bu,\xi_k)h(\xi_k,\bxi)}{h(\xi_k,\bu)h(\bxi,\xi_k)},\qquad j<n,\\
&\mathcal{M}^{(22)}_{nk}=\frac{h(\xi_k,\bxi)}{h(\xi_k,\bu)}.
\end{aligned}
\ee

Now we should put $\bxi=\{\bv,w\}$ and consider the limit $w\to\infty$. In doing this we use the following formulas
\be{12-beh}
g(w,x)\sim \frac cw,\qquad f(w,x)\sim 1,\qquad h(w,x)\sim \frac wc,\qquad t(w,x)\sim \frac {c^2}{w^2},\qquad w\to\infty.
\ee
It is easy to see that
\be{12-lim-tau}
\lim_{w\to\infty}\bs{\tau}(z|\bxi)=\bs{\tau}(z|\bv),
\ee
because $\bs{\tau}(z|\bxi)$ depends on $w$ only due to the functions $f(z,w)$ and $f(w,z)$, which go to one at   $w\to\infty$.

In the prefactor, we have
\begin{multline}\label{12-lim-pref}
\lim_{w\to\infty}
\Delta_n(\bxi)h(\bxi,\bu)g(u_n,\bxi)=
\lim_{w\to\infty}\Delta_n(\bv)h(\bv,\bu)g(u_n,\bv) g(w,\bv)h(w,\bu)g(u_n,w)\\
=-\Delta_n(\bv)h(\bv,\bu)g(u_n,\bv).
\end{multline}
Now we consider the limits of the matrix elements of the matrix $\mathcal{M}^{(22)}_{jk}$, preliminary multiplying  its last
column by the factor $w/c$, which appears in the formula \eqref{12-CD-FF}. Then in the last column for $j<n$ we have
\be{12-lim-Mjn}
\lim_{w\to\infty}\tfrac wc
\mathcal{M}^{(22)}_{jn}= \lim_{w\to\infty}\frac wc\left(t(w,u_j)+t(u_j,w)\frac{h(\bu,w)h(w,\bv)}{h(w,\bu)h(\bv,v)}\right)=0, \qquad j<n,
\ee
and for  $j=n$
\be{12-lim-Mnn}
\lim_{w\to\infty}\tfrac wc
\mathcal{M}^{(22)}_{nn}= \lim_{w\to\infty}\frac wc \frac{h(w,\bv)}{h(w,\bu)}=1.
\ee
It is important to recall that $\#\bu=n$ and   $\#\bv=n-1$.

Thus, the determinant of the matrix $\mathcal{M}^{(22)}_{jk}$ is reduced to its minor built on the first $n-1$ rows and columns. For
these matrix elements, we obtain
\be{12-lim-Mjk}
\lim_{w\to\infty}
\mathcal{M}^{(22)}_{jk}= \lim_{w\to\infty}\left(
t(v_k,u_j)+t(u_j,v_k)\frac{h(\bu,v_k)h(v_k,\bv)h(v_k,w)}{h(v_k,\bu)h(\bv,v_k)h(w,v_k)}\right)=\mathcal{M}^{(21)}_{jk},\qquad j,k<n,
\ee
where the matrix $\mathcal{M}^{(21)}_{jk}$ is given by \eqref{12-M21}. As a result, we have
\be{12-lim-FF22}
\lim_{w\to\infty}\tfrac wc \mathcal{F}_n^{(22)}(z|\{\bv,w\};\bu)=
(\bs{\tau}(z|\bu)-\bs{\tau}(z|\bv))\Delta'_n(\bu)
\Delta_{n-1}(\bv)h(\bv,\bu) \frac{g(u_n,\bv)}{g(u_n,\bu_n)}\;\det_{n-1}\mathcal{M}_{jk}^{(21)}.
\ee
Comparing \eqref{12-lim-FF22} and \eqref{12-FF21-res1} we see that they coincide.

\section{Universal form factor\label{12-Sec5}}

We have seen already that at least some form factors of the monodromy matrix entries can be presented in the form
\be{12-un-pred}
\mathcal{F}_n^{(ij)}(z|\bv;\bu)=\bigl(\bs{\tau}(z|\bv)-\bs{\tau}(z|\bu)\bigr)\mathfrak{F}_n^{(ij)}(\bv;\bu),
\ee
where a function $\mathfrak{F}_n^{(ij)}(\bv;\bu)$ does not depend on $z$. It is easy to show that such a representation is always true, if $\bv\ne\bu$.
For this we write the $RTT$-relation for arbitrary matrix elements $T^{ij}(z)$ and $T^{kl}(w)$
\be{12-RTT}
[T^{ij}(z),T^{kl}(w)]=g(z,w)\bigl(T^{kj}(w)T^{il}(z)-  T^{kj}(z)T^{il}(w)\bigr).
\ee
Observe that the rhs of \eqref{12-RTT} does not change under replacement $z$ by $w$. This implies that the commutator of any matrix element
$T^{ij}(z)$ with the transfer matrix $\mathcal{T}(w)$ is a symmetric function of $z$ and $w$:
\be{12-RTTau}
[T^{ij}(z),\mathcal{T}(w)]=[T^{ij}(w),\mathcal{T}(z)].
\ee
Let us divide \eqref{12-RTTau} by the product $d(z)d(w)$, and after this multiply it from the right by an on-shell vector $\bs{B}(\bu)|0\rangle$ and
form the left by a dual on-shell vector $\langle0|\bs{C}(\bv)$. We have
\be{12-RTTau-FF}
\langle0|\bs{C}(\bv)[\bs{T}^{ij}(z),\bs{\mathcal{T}}(w)]\bs{B}(\bu)|0\rangle=
\langle0|\bs{C}(\bv)[\bs{T}^{ij}(w),\bs{\mathcal{T}}(z)]\bs{B}(\bu)|0\rangle.
\ee
Calculating the action of the transfer matrices onto on-shell vectors we find
\be{12-tau-FF}
\bigl(\bs{\tau}(w|\bu)-\bs{\tau}(w|\bv)\bigr)\langle0|\bs{C}(\bv)\bs{T}^{ij}(z)\bs{B}(\bu)|0\rangle=
\bigl(\bs{\tau}(z|\bu)-\bs{\tau}(z|\bv)\bigr)\langle0|\bs{C}(\bv)\bs{T}^{ij}(w)\bs{B}(\bu)|0\rangle.
\ee
It remains to divide both sides of this equation by the differences of the transfer matrix eigen\-va\-lues. This gives us
\be{12-tau-FF-ind}
\frac{\mathcal{F}_n^{(ij)}(z|\bv;\bu)}{\bs{\tau}(z|\bv)-\bs{\tau}(z|\bu)}=
\frac{\mathcal{F}_n^{(ij)}(w|\bv;\bu)}{\bs{\tau}(w|\bv)-\bs{\tau}(w|\bu)}.
\ee
The lhs of \eqref{12-tau-FF-ind} depends only on  $z$, while the rhs depends only on  $w$.
This is possible if and only if
\be{12-FF-Univ}
\frac{\mathcal{F}_n^{(ij)}(z|\bv;\bu)}{\bs{\tau}(z|\bv)-\bs{\tau}(z|\bu)}=\mathfrak{F}_n^{(ij)}(\bv;\bu),
\ee
where $\mathfrak{F}_n^{(ij)}(\bv;\bu)$ does not depend on $z$.

The function $\mathfrak{F}_n^{(ij)}(\bv;\bu)$ is called the universal form factor of the operator $\bs{T}^{ij}(z)$ \cite{PakRS15}.
The meaning of this term is that $\mathfrak{F}_n^{(ij)}(\bv;\bu)$ does not depend on a particular model, if
there are no matching elements in the sets $\bv$ and $\bu$. Indeed, all information about a specific model is contained
in the function $r(w)=a(w)/d(w)$. In calculating form factors, we obtain functions $r$ that depend on the parameters $z$, $\bv$, and $\bu$.
However, the functions $ r(v_i)$ and $r(u_i)$ should be expressed in terms of products of the $f$ functions, according to the Bethe equations. Then
the only function that `remembers' the model in question is the function $r(z)$. But all dependence on
the parameter $z$ in the formula \eqref{12-FF-Univ} has already been extracted as the difference  of the transfer matrix eigenvalues.
Therefore, the function $\mathfrak{F}_n^{(ij)}(\bv;\bu)$ is independent of the model under consideration.

If, among the parameters $\bv$  and $\bu$, there are coincident elements, then, as we saw earlier, the form factor depends on
the derivatives $r'(v_i)$ or $r'(u_i)$. The latter are no longer expressed in terms of the Bethe equations. In this case, the universal
form factor still `remembers' a particular model and, strictly speaking, is not universal.

All relations between the ordinary form factors $\mathcal{F}_n^{(ij)}(z)$ are carried over to the universal form factors $\mathfrak{F}_n^{(ij)}$.
This is due to the fact that  the transfer matrix eigenvalues $\bs{\tau}(z|\bu)$ turn to new eigenvalues, when one
of the parameters $u_i$ goes to infinity
\be{12-tau-tau}
\lim_{u_i\to\infty}\bs{\tau}(z|\bu)=
\lim_{u_i\to\infty}\bigr(r(z)f(\bu,z)+f(z,\bu)\bigl)=r(z)f(\bu_i,z)+f(z,\bu_i)=\bs{\tau}(z|\bu_i).
\ee
Therefore, for example, equation \eqref{12-BA-FF} implies
\begin{equation}
-\lim_{w\to\infty}\tfrac wc\mathfrak{F}_{n+1}^{(11)}(\bv;\{\bu,w\})=-\lim_{w\to\infty}\frac{\tfrac wc \mathcal{F}_{n+1}^{(11)}(z|\bv;\{\bu,w\})}
{\bs{\tau}(z|\bv)-\bs{\tau}(z|\{\bu,w\})}
=\frac{\mathcal{F}_{n}^{(12)}(z|\bv;\bu)}
{\bs{\tau}(z|\bv)-\bs{\tau}(z|\bu)}=\mathfrak{F}_n^{(12)}(z|\bv;\bu).\label{12-BA-FFUN}
\end{equation}
%


%
%
\chapter{Form factors of local operators\label{CHA-FFLO}}

In this lecture, we obtain determinant representations for form factors of local operators using a method based on
the application of zero modes. In this case, the existence of a solution of the quantum inverse problem is not assumed, but we
still have to consider only such models that are described by the rational $R$-matrix.
A generalization of this method to models with trigonometric $R$-matrix has not yet been found.

\section{Necessary formulas\label{13-Sec1}}

In the previous lecture we introduced a `normalized' monodromy matrix $\bs{T}(u)=T(u)/d(u)$.
Therefore, we begin this lecture with giving once again familiar formulas, written in terms of the normalized matrix elements $\bs{T}(u)=T(u)/d(u)$.

A scalar product of two Bethe vectors given by \eqref{07-SP-SPdef},
now has the following form
\be{13-SP-SPdef}
S_n(\bv|\bu)=\langle0|\bs{C}(\bv)\bs{B}(\bu)|0\rangle\,.
\ee
If the parameters  $\bv$ and $\bu$ do not obey any additional constraints, then the scalar product is equal to the sum over partitions
\eqref{07-SP-SPlm1}.

Consider a twisted monodromy matrix with the twist matrix
$\hat\kappa=\diag(\kappa_1,\kappa_2)$ and let $\kappa_j=e^{\beta_j}$, $j=1,2$.
Suppose that the parameters $\bv$ and $\bu$ respectively satisfy twisted and ordinary Bethe equations. We write these
equations in the following form
\be{13-twBE-BE}
r(\bu_{\so})=\frac{f(\bu_{\so},\bu_{\st})}{f(\bu_{\st},\bu_{\so})},
\qquad
r(\bv_{\so})=e^{(\beta_2-\beta_1)n_{\so}}\frac{f(\bv_{\so},\bv_{\st})}{f(\bv_{\st},\bv_{\so})}.
\ee
Here we are dealing with two arbitrary partitions $\bu\mapsto\{\bu_{\so},\bu_{\st}\}$ and $\bv\mapsto\{\bv_{\so},\bv_{\st}\}$,
and the integer $n_{\so}$ is equal to the number of elements in the subset $\bv_{\so}$, that is
$n_{\so}=\#\bv_{\so}$.
It is easy to see that this form of writing is equivalent to (twisted) Bethe equations in their
standard form. Indeed, in the particular case when $\#\bv_{\so}=\#\bu_{\so}=1$  the system \eqref{13-twBE-BE} coincides with
(twisted) Bethe equations. If $\#\bv_{\so}>1$ or $\#\bu_{\so}>1$
then it is enough to take the product of (twisted)
Bethe equations with respect to the corresponding sets, and we arrive at the system \eqref{13-twBE-BE}.

Let us denote by $S_n^{(\kappa)}(\bv|\bu)$ a scalar product of twisted and ordinary on-shell vectors. Then, it follows from
\eqref{07-SP-SPlm1} that
\begin{equation}\label{13-SP-eigtw}
S_n^{(\kappa)}(\bv|\bu)=\sum_{\substack{\bv\mapsto\{\bv_{\so},\bv_{\st}\}\\
\bu\mapsto\{\bu_{\so},\bu_{\st}\}}} e^{\beta n_{\so}}
K_{n_{\so}}( \bu_{\so}|\bv_{\so})K_{n_{\st}}( \bv_{\st}|\bu_{\st})\,f(\bv_{\so},\bv_{\st})f(\bu_{\st},\bu_{\so})\,.
\end{equation}

Recall now formulas for the Bethe vectors in a composite model. It follows from \eqref{11-IK}
that
\be{13-IK}
\bs{B}(\bu)|0\rangle=\sum_{\bu\mapsto\{\bu_{\so},\bu_{\st}\}}
   r^{(2)}(\bu_{\so})f(\bu_{\st},\bu_{\so})\bs{B}^{(1)}(\bu_{\so})\bs{B}^{(2)}(\bu_{\st})|0\rangle.
 \ee
If the original vector $\bs{B}(\bu)|0\rangle$ is on-shell, then using Bethe equations \eqref{13-twBE-BE} we can express
the function $r^{(2)}(u)$ in terms of the function $r^{(1)}(u)$ and the product of the $f$ functions:
\be{13-BE-u2}
r^{(2)}(\bu_{\so})f(\bu_{\st},\bu_{\so})=\frac{r(\bu_{\so})f(\bu_{\st},\bu_{\so})}{r^{(1)}(\bu_{\so})}=
\frac{f(\bu_{\so},\bu_{\st})}{r^{(1)}(\bu_{\so})}.
\ee
Then equation \eqref{13-IK} takes the form
\be{13-IK-eig}
\bs{B}(\bu)|0\rangle=\sum_{\bu\mapsto\{\bu_{\so},\bu_{\st}\}}\frac{f(\bu_{\so},\bu_{\st})}{r^{(1)}(\bu_{\so})}
\bs{B}^{(1)}(\bu_{\so})\bs{B}^{(2)}(\bu_{\st})|0\rangle.
 \ee
Finally, we will need a formula for the dual Bethe vector \eqref{11-IK-d}.
Now it is written in the following form:
\be{13-IK-d}
 \langle0| \bs{C}(\bv)=\langle0|\sum_{\bv\mapsto\{\bv_{\so},\bv_{\st}\}}
r^{(1)}(\bv_{\st})f(\bv_{\so},\bv_{\st})\bs{C}^{(1)}(\bv_{\so})\bs{C}^{(2)}(\bv_{\st}).
 \ee

\section{Form factors of partial zero modes\label{13-Sec2}}

As we have learned in the previous lecture, to calculate the form factors of local operators it is sufficient
to consider the composite model and calculate the form factors of the zero modes of the first partial monodromy matrix
$T^{(1)}_{ij}[0]$. Let us introduce a notation
\begin{equation}
{\sf M}_n^{(ij)}(\bv;\bu)=\langle0| \bs{C}(\bv)\;T^{(1)}_{ij}[0]\;\bs{B}(\bu)|0\rangle.\label{13-defFF-ZM}
\end{equation}
Here both vectors $\langle0| \bs{C}(\bv)$ and $\bs{B}(\bu)|0\rangle$ are on-shell, and
$n=\#\bu$. As in the case of the ordinary form factors, the number of variables $\bv$ depends on the concrete operator
$T^{(1)}_{ij}[0]$.

We have already seen that the form factors of the matrix elements of the complete monodromy matrix $T(z)$ are related with each other. Actually,
it is sufficient to compute the form factor of a single operator, for example, the operator $A(z)$, and then the form factors of the remaining operators
can be obtained by taking the limits when one of the parameters in the sets $\bv$ or $\bu$ tends to infinity.
It is easy to show that the form factors of the zero modes of the first partial monodromy matrix possess exactly the same property.
For example, it follows from the relation for the total zero modes  \eqref{12-AC} that
\be{13-AC}
C[0]=\bigl[A[0],C[0]\bigr].
\ee
It is clear that the same relation exists for the partial zero modes
\be{13-AC1}
C^{(1)}[0]=\bigl[A^{(1)}[0],C^{(1)}[0]\bigr],
\ee
because the partial monodromy matrix satisfies the $RTT$-relation with the same $R$-matrix as the total monodromy matrix.
On the other hand, the total zero mode is the sum of the partial zero modes $C[0]=C^{(1)}[0]+C^{(2)}[0]$, and in addition
$\bigl[A^{(1)}[0],C^{(2)}[0]\bigr]=0$. Hence,
\be{13-AC10}
C^{(1)}[0]=\bigl[A^{(1)}[0],C[0]\bigr].
\ee
It remains to compute a matrix element of \eqref{13-AC10} between two on-shell Bethe vectors
\be{13-ACFF}
\langle0| \bs{C}(\bv)\;C^{(1)}[0]\;\bs{B}(\bu)|0\rangle=
\langle0| \bs{C}(\bv)\;\bigl[A^{(1)}[0],C[0]\bigr]\;\bs{B}(\bu)|0\rangle.
\ee
Thus, just as for ordinary form factors, we find
\begin{equation}
{\sf M}_n^{(21)}(\bv;\bu)=-\lim_{w\to\infty}\tfrac wc {\sf M}_n^{(11)}(\{\bv,w\};\bu).\label{13-CA-FF}
\end{equation}
Similarly, one can prove another relation
\begin{equation}
{\sf M}_n^{(12)}(\bv;\bu)=-\lim_{w\to\infty}\tfrac wc {\sf M}_n^{(11)}(\bv;\{\bu,w\}).\label{13-BA-FF}
\end{equation}

In the next section, we calculate the form factors of the diagonal operators $A^{(1)}[0]$ and $D^{(1)}[0]$.
In this case it is convenient to deal with twisted on-shell vectors, so we will consider a matrix matrix element of the form
\be{13-Mk-def}
{\sf M}^{(\kappa)}_n(\bv;\bu)=\langle0| \bs{C}(\bv) \;e^{\beta_1A^{(1)}[0]+\beta_2D^{(1)}[0]}\; \bs{B}(\bu)|0\rangle.
\ee
Here $\bs{B}(\bu)|0\rangle$ is an ordinary on-shell vector, and $\langle0| \bs{C}(\bv)$ is a twisted on-shell vector,
and the twist parameters are chosen, as in the previous section: $\kappa_j = e^{\beta_j}$. The matrix element \eqref{13-Mk-def}
is a generating functional for the form factors of the operators $A^{(1)}[0]$ and $D^{(1)}[0]$, that can be obtained by
differentiating ${\sf M}^{(\kappa)}_n$ with respect to $\beta_j$ at $\beta_j=0$

The reason we introduced the generating functional is simple enough. In section~\ref{10-Sec3}
we already dealt with the form factor of the operator $D$ and saw that in its calculation it is necessary to distinguish two cases:
$\bv=\bu$ and $\bv\ne\bu$.
The same is true for the form factors of the operators $A^{(1)}[0]$ and $D^{(1)}[0]$.
The introduction of the generating functional \eqref{13-Mk-def} allows us to consider these two cases simultaneously.

\section{Calculation of generating functional\label{13-Sec3}}

To calculate the action of the operator $\exp\{\beta_1A^{(1)} [0] + \beta_2D^{(1)} [0]\}$ on the Bethe vector, it is necessary to express
the total vector in terms of the partial ones. The same should be done with the total dual Bethe vector, if we want
to calculate the arising scalar products. Substituting the representations  \eqref{13-IK-eig} and \eqref{13-IK-d}
into the generating functional \eqref{13-Mk-def}, we obtain
\begin{multline}\label{13-Mk-rep}
{\sf M}^{(\kappa)}_n(\bv;\bu)=\sum_{\substack{\bu\mapsto\{\bu_{\so},\bu_{\st}\}  \\   \bv\mapsto\{\bv_{\so},\bv_{\st}\}}}
\frac{\ell(\bv_{\st})}{\ell(\bu_{\so})}f(\bv_{\so},\bv_{\st})f(\bu_{\so},\bu_{\st})\\
\times \langle0|^{(1)}\bs{C}^{(1)}(\bv_{\so}) \;e^{\beta_1A^{(1)}[0]+\beta_2D^{(1)}[0]}\;  \bs{B}^{(1)}(\bu_{\so})|0\rangle^{(1)}  \\
\times \langle0|^{(2)}\bs{C}^{(2)}(\bv_{\st}) \bs{B}^{(2)}(\bu_{\st})|0\rangle^{(2)},
\end{multline}
where we introduce a notation
\be{13-ell}
\ell(u)=r^{(1)}(u).
\ee
Now we can use the action formulas
\be{13-actAD}
\begin{aligned}
&A^{(1)}[0]\bs{B}^{(1)}(\bu_{\so})|0\rangle^{(1)}=(a^{(1)}[0]-n_{\so})\bs{B}^{(1)}(\bu_{\so})|0\rangle^{(1)},\\
&D^{(1)}[0]\bs{B}^{(1)}(\bu_{\so})|0\rangle^{(1)}=(d^{(1)}[0]+n_{\so})\bs{B}^{(1)}(\bu_{\so})|0\rangle^{(1)},
\end{aligned}
\ee
where $n_{\so}=\#\bu_{\so}$. This implies
\be{13-exp-actAD}
e^{\beta_1A^{(1)}[0]+\beta_2D^{(1)}[0]}\;  \bs{B}^{(1)}(\bu_{\so})|0\rangle^{(1)}=
e^{\beta_1a^{(1)}[0]+\beta_2d^{(1)}[0]+n_{\so}(\beta_2-\beta_1)}\;  \bs{B}^{(1)}(\bu_{\so})|0\rangle^{(1)},
\ee
and the generating functional ${\sf M}^{(\kappa)}_n(\bv;\bu)$ takes the form
\begin{equation}\label{13-Mk-act}
{\sf M}^{(\kappa)}_n(\bv;\bu)=\sum_{\substack{\bu\mapsto\{\bu_{\so},\bu_{\st}\}  \\   \bv\mapsto\{\bv_{\so},\bv_{\st}\}}}
 \;e^{\beta_1a^{(1)}[0]+\beta_2d^{(1)}[0]+n_{\so}(\beta_2-\beta_1)}\frac{\ell(\bv_{\st})}{\ell(\bu_{\so})}f(\bv_{\so},\bv_{\st})f(\bu_{\so},\bu_{\st})
 S_{n_{\so}}^{(1)}(\bv_{\so}|\bu_{\so})S_{n_{\st}}^{(2)}(\bv_{\st}|\bu_{\st}).
%
%
\end{equation}
Here $S_m^{(j)}(\bv|\bu)$ are  scalar products of the partial Bethe vectors:
\be{13-SP-k}
S_m^{(j)}(\bv|\bu)= \langle0|^{(j)}\bs{C}^{(j)}(\bv) \bs{B}^{(j)}(\bu)|0\rangle^{(j)},\qquad j=1,2,\qquad m=n_{\so},n_{\st}.
\ee
For their calculation, we can use the formula  \eqref{07-SP-SPlm1},
replacing in it the functions $r$ by the functions $r^{(1)} = \ell$ in the case of the
scalar product $S_ {n_ {\so}}^{(1)}$, and $r$ by $r^{(2)}$ for the scalar product $S_ {n_ {\st}}^{(2)}$. Clearly, we will have new partitions
into new subsets. In view of the large number of such subsets, we will for a time reject the use of Roman numerals and will mark
the subsets by bold superscripts. Then we obtain for the scalar product $S_ {n_ {\so}}^{(1)}$
\be{13-SP1}
 S_{n_{\so}}^{(1)}(\bv_{\so}|\bu_{\so})=
 \sum_{\substack{\bu_{\so}\mapsto\{\bu^{\bs{1}},\bu^{\bs{3}}\}\\
\bv_{\so}\mapsto\{\bv^{\bs{1}},\bv^{\bs{3}}\}}}
\ell(\bu^{\bs{3}})\ell(\bv^{\bs{1}})\,
K_{n_{1}}( \bu^{\bs{1}}|\bv^{\bs{1}})K_{n_{3}}( \bv^{\bs{3}}|\bu^{\bs{3}})\,f(\bv^{\bs{3}},\bv^{\bs{1}})f(\bu^{\bs{1}},\bu^{\bs{3}})\,.
\ee
Here the sum is taken over all possible partitions $\bu_{\so}\mapsto\{\bu^{\bs{1}},\bu^{\bs{3}}\}$ and $\bv_{\so}\mapsto\{\bv^{\bs{1}},\bv^{\bs{3}}\}$,
such that $\#\bu^{\bs{1}}=\#\bv^{\bs{1}}=n_{1}$, $\#\bu^{\bs{3}}=\#\bv^{\bs{3}}=n_{3}$ and $n_{1}+n_{3}=n_{\so}$.

Similarly,
\begin{equation}\label{13-SP2}
 S_{n_{\st}}^{(2)}(\bv_{\st}|\bu_{\st})=
 \sum_{\substack{\bu_{\st}\mapsto\{\bu^{\bs{2}},\bu^{\bs{4}}\}\\
\bv_{\st}\mapsto\{\bv^{\bs{2}},\bv^{\bs{4}}\}}}
r^{(2)}(\bu^{\bs{4}})r^{(2)}(\bv^{\bs{2}})\,
K_{n_{2}}( \bu^{\bs{2}}|\bv^{\bs{2}})K_{n_{4}}( \bv^{\bs{4}}|\bu^{\bs{4}})
f(\bv^{\bs{4}},\bv^{\bs{2}})f(\bu^{\bs{2}},\bu^{\bs{4}})\,.
\end{equation}
Here the notation is the same as in  \eqref{13-SP1}.

We already took into account that the original vector $\bs{B}(\bu)|0\rangle$ was on-shell, when we substi\-tu\-ted into the generating
functional representation \eqref{13-IK-eig}, rather than \eqref{13-IK}. Now we use the fact that the original dual vector $\langle0|\bs{C}(\bv)$ was
a twisted on-shell vector, and express the functions $r^{(2)}$ in terms of
the function $\ell$ in the formula \eqref{13-SP2}. Due to the (twisted) Bethe equations, we have
\be{13-BE-u4}
r^{(2)}(\bu^{\bs{4}})=\frac{r(\bu^{\bs{4}})}{\ell(\bu^{\bs{4}})}=\frac{1}{\ell(\bu^{\bs{4}})}\;
\frac{f(\bu^{\bs{4}},\bu^{\bs{1}})f(\bu^{\bs{4}},\bu^{\bs{2}})f(\bu^{\bs{4}},\bu^{\bs{3}})}
{f(\bu^{\bs{1}},\bu^{\bs{4}})f(\bu^{\bs{2}},\bu^{\bs{4}})f(\bu^{\bs{3}},\bu^{\bs{4}})},
\ee
and
\be{13-BE-v2}
r^{(2)}(\bv^{\bs{2}})=\frac{r(\bv^{\bs{2}})}{\ell(\bv^{\bs{2}})}=\frac{e^{n_2(\beta_2-\beta_1)}}{\ell(\bv^{\bs{2}})}\;
\frac{f(\bv^{\bs{2}},\bv^{\bs{1}})f(\bv^{\bs{2}},\bv^{\bs{3}})f(\bv^{\bs{2}},\bv^{\bs{4}})}
{f(\bv^{\bs{1}},\bv^{\bs{2}})f(\bv^{\bs{3}},\bv^{\bs{2}})f(\bv^{\bs{4}},\bv^{\bs{2}})}.
\ee
Then the scalar product $ S_{n_{\st}}^{(2)}(\bv_{\st}|\bu_{\st})$ takes the form
\begin{multline}\label{13-SP2-mod}
 S_{n_{\st}}^{(2)}(\bv_{\st}|\bu_{\st})=
 \sum_{\substack{\bu_{\st}\mapsto\{\bu^{\bs{2}},\bu^{\bs{4}}\}\\
\bv_{\st}\mapsto\{\bv^{\bs{2}},\bv^{\bs{4}}\}}}
\frac{e^{n_2(\beta_2-\beta_1)}}{\ell(\bu^{\bs{4}})\ell(\bv^{\bs{2}})}
\,K_{n_{2}}( \bu^{\bs{2}}|\bv^{\bs{2}})K_{n_{4}}( \bv^{\bs{4}}|\bu^{\bs{4}})\\
\times \frac{f(\bu^{\bs{4}},\bu^{\bs{1}})f(\bu^{\bs{4}},\bu^{\bs{2}})f(\bu^{\bs{4}},\bu^{\bs{3}})f(\bv^{\bs{2}},\bv^{\bs{1}})f(\bv^{\bs{2}},\bv^{\bs{3}})f(\bv^{\bs{2}},\bv^{\bs{4}})}
{f(\bu^{\bs{1}},\bu^{\bs{4}})f(\bu^{\bs{3}},\bu^{\bs{4}})f(\bv^{\bs{1}},\bv^{\bs{2}})f(\bv^{\bs{3}},\bv^{\bs{2}})}\,.
\end{multline}
The last thing that remains for us is to express the products
 $\ell(\bv_{\st})$, $\ell(\bu_{\so})$,  $f(\bv_{\so},\bv_{\st})$, and  $f(\bu_{\so},\bu_{\st})$
in \eqref{13-Mk-act} in terms of products over new subsets:
\begin{multline}\label{13-fact}
e^{n_{\so}(\beta_2-\beta_1)}\frac{\ell(\bv_{\st})}{\ell(\bu_{\so})}f(\bv_{\so},\bv_{\st})f(\bu_{\so},\bu_{\st})=
e^{(n_{1}+n_{3})(\beta_2-\beta_1)}\frac{\ell(\bv^{\bs{2}})\ell(\bv^{\bs{4}})}{\ell(\bu^{\bs{1}})\ell(\bu^{\bs{3}})}f(\bv^{\bs{1}},\bv^{\bs{2}}) f(\bv^{\bs{1}},\bv^{\bs{4}})\\
\times f(\bv^{\bs{3}},\bv^{\bs{2}})f(\bv^{\bs{3}},\bv^{\bs{4}})
f(\bu^{\bs{1}},\bu^{\bs{2}}) f(\bu^{\bs{1}},\bu^{\bs{4}})f(\bu^{\bs{3}},\bu^{\bs{2}})f(\bu^{\bs{3}},\bu^{\bs{4}}).
\end{multline}

Now we need to substitute the formulas \eqref{13-SP1}, \eqref{13-SP2-mod}, and \eqref{13-fact} into  representation \eqref{13-Mk-act}.
All of these formulas are long enough, therefore, one should  be careful. However, nothing but accuracy here
is not required, and as a result we obtain
\begin{multline}\label{13-Mk-act2}
{\sf M}^{(\kappa)}_n(\bv;\bu)=\sum
 \;e^{\beta_1a^{(1)}[0]+\beta_2d^{(1)}[0]+(n_{1}+n_{2}+n_{3})(\beta_2-\beta_1)}
 \frac{\ell(\bv^{\bs{1}})\ell(\bv^{\bs{4}})}{\ell(\bu^{\bs{1}})\ell(\bu^{\bs{4}})}\;F_u\, F_v\,\\
 \times
K_{n_{1}}( \bu^{\bs{1}}|\bv^{\bs{1}})K_{n_{3}}( \bv^{\bs{3}}|\bu^{\bs{3}})K_{n_{2}}( \bu^{\bs{2}}|\bv^{\bs{2}})K_{n_{4}}( \bv^{\bs{4}}|\bu^{\bs{4}}),
\end{multline}
where
\be{13-Fuv}
\begin{aligned}
&F_u=f(\bu^{\bs{1}},\bu^{\bs{2}}) f(\bu^{\bs{1}},\bu^{\bs{3}})f(\bu^{\bs{3}},\bu^{\bs{2}})f(\bu^{\bs{4}},\bu^{\bs{1}})f(\bu^{\bs{4}},\bu^{\bs{2}}) f(\bu^{\bs{4}},\bu^{\bs{3}}),\\
&F_v=f(\bv^{\bs{2}},\bv^{\bs{1}})f(\bv^{\bs{2}},\bv^{\bs{4}}) f(\bv^{\bs{1}},\bv^{\bs{4}})f(\bv^{\bs{2}},\bv^{\bs{3}})f(\bv^{\bs{3}},\bv^{\bs{4}})f(\bv^{\bs{3}},\bv^{\bs{1}}).
\end{aligned}
\ee
The sum in \eqref{13-Mk-act2}  is taken over partitions of the original sets $\bv$ and $\bu$ into four subsets each:
\be{13-part}
\bu\mapsto\{\bu^{\bs{1}},\bu^{\bs{2}},\bu^{\bs{3}},\bu^{\bs{4}}\},\qquad
\bv\mapsto\{\bv^{\bs{1}},\bv^{\bs{2}},\bv^{\bs{3}},\bv^{\bs{4}}\}.
\ee
The restrictions are imposed only on the cardinalities of the subsets, namely, $\#\bv^{\bs{j}}=\#\bu^{\bs{j}}=n_{j}$, $j=1,2,3,4$.
By the way, let us recall once again that these restrictions are implicitly indicated in the formula \eqref{13-Mk-act2} in the form of subscripts of the DWPF, which are defined only under the condition that the number of elements to the left of the vertical line is equal to the number of elements to the right of the vertical line.

Actually, all the calculations are already done, and we only need to see the answer.
For this we introduce new subsets:
\be{13-new-part}
\begin{aligned}
&\bu_{\rm i}=\{\bu^{\bs{1}},\bu^{\bs{4}}\},\qquad & \bu_{\rm ii}=\{\bu^{\bs{2}},\bu^{\bs{3}}\},\\
&\bv_{\rm i}=\{\bv^{\bs{1}},\bv^{\bs{4}}\},\qquad & \bv_{\rm ii}=\{\bv^{\bs{2}},\bv^{\bs{3}}\} .
\end{aligned}
\ee
Then, it is easy to see that
\be{13-Fuv-2}
\begin{aligned}
&F_u=f(\bu_{\rm i},\bu_{\rm ii})f(\bu^{\bs{3}},\bu^{\bs{2}})f(\bu^{\bs{4}},\bu^{\bs{1}}),\\
&F_v=f(\bv_{\rm ii},\bv_{\rm i}) f(\bv^{\bs{1}},\bv^{\bs{4}})f(\bv^{\bs{2}},\bv^{\bs{3}}).
\end{aligned}
\ee
As a result, we obtain
\begin{equation}\label{13-Mk-act3}
{\sf M}^{(\kappa)}_n(\bv;\bu)=e^{\beta_1a^{(1)}[0]+\beta_2d^{(1)}[0]}
\sum_{\substack{ \bu\mapsto\{\bu_{\rm i},\bu_{\rm ii}\} \\ \bv\mapsto\{\bv_{\rm i},\bv_{\rm ii}\} } }
 \; \frac{\ell(\bv_{\rm i})}{\ell(\bu_{\rm i})}\;f(\bu_{\rm i},\bu_{\rm ii})f(\bv_{\rm ii},\bv_{\rm i})
 G_1(\bv_{\rm i}|\bu_{\rm i})\,  G_2(\bv_{\rm ii}|\bu_{\rm ii}) ,
\end{equation}
and functions $G_1(\bv_{\rm i}|\bu_{\rm i})$ and $G_2(\bv_{\rm ii}|\bu_{\rm ii})$ are given as sums over partitions
into additional subsets:
\be{13-G1}
 G_1(\bv_{\rm i}|\bu_{\rm i})=
 \sum_{\substack{ \bu_{\rm i}\mapsto\{\bu^{\bs{1}},\bu^{\bs{4}}\} \\ \bv_{\rm i}\mapsto\{\bv^{\bs{1}},\bv^{\bs{4}}\} } }
 e^{(n-n_{4})(\beta_2-\beta_1)}K_{n_{1}}( \bu^{\bs{1}}|\bv^{\bs{1}})K_{n_{4}}( \bv^{\bs{4}}|\bu^{\bs{4}})\,
 f(\bv^{\bs{1}},\bv^{\bs{4}})f(\bu^{\bs{4}},\bu^{\bs{1}}),
 \ee
and
\be{13-G2}
 G_2(\bv_{\rm ii}|\bu_{\rm ii})=
 \sum_{\substack{ \bu_{\rm ii}\mapsto\{\bu^{\bs{2}},\bu^{\bs{3}}\} \\ \bv_{\rm ii}\mapsto\{\bv^{\bs{2}},\bv^{\bs{3}}\} } }
K_{n_{3}}( \bv^{\bs{3}}|\bu^{\bs{3}})K_{n_{2}}( \bu^{\bs{2}}|\bv^{\bs{2}})\,
f(\bv^{\bs{2}},\bv^{\bs{3}})f(\bu^{\bs{3}},\bu^{\bs{2}}).
\ee
In other words, the initial summation over  partitions \eqref{13-part} is now organized into several stages: first
we have partitions $\bu\mapsto\{\bu_{\rm i},\bu_{\rm ii}\}$ and $\bv\mapsto\{\bv_{\rm i},\bv_{\rm ii}\}$
and then each of the resulting subsets is divided once again into two subsets.

Looking carefully at the sum \eqref{13-G2}, one can observe that it differs from the sum \eqref{11-Ident-2}
only by the labels of the subsets.
Therefore, $G_2(\bv_{\rm ii}|\bu_{\rm ii})=\delta_{0,n_2+n_3}$, that is,   $\bv_{\rm ii}=\bu_{\rm ii}=\emptyset$. It follows from this that
$\bv_{\rm i}=\bv$, $\bu_{\rm i}=\bu$ and $n-n_4=n_1$. Then, comparing the sum \eqref{13-G2} with the sum \eqref{13-SP-eigtw}, we obtain
\be{13-G1-res}
G_1(\bv|\bu)=  S_n^{(\kappa)}(\bv|\bu),
\ee
and hence,
\begin{equation}\label{13-Mk-res}
{\sf M}^{(\kappa)}_n(\bv;\bu)=e^{\beta_1a^{(1)}[0]+\beta_2d^{(1)}[0]}\frac{\ell(\bv)}{\ell(\bu)}
S_n^{(\kappa)}(\bv|\bu).
\end{equation}

Thus, all the numerous and very cumbersome sums over the partitions eventually reduced to the scalar product of the twisted
and the ordinary on-shell vectors, for which we have a compact representation in the form of a determinant. To calculate the form factors, we
need only to take the derivatives with respect to the parameters $\beta_j$. Let us calculate, for example, the form factor of the operator $A^{(1)}[0]$.

Differentiating definition \eqref{13-Mk-def} over $\beta_1$, we obtain
\begin{multline}\label{13-Mk-derl}
\frac{d{\sf M}^{(\kappa)}_n(\bv;\bu)}{d\beta_1}\Bigr|_{\beta_j=0}=\frac{d}{d\beta_1}
\langle0| \bs{C}(\bv) \;e^{\beta_1A^{(1)}[0]+\beta_2D^{(1)}[0]}\; \bs{B}(\bu)|0\rangle\Bigr|_{\beta_j=0}\\
=\langle0| \bs{C}(\bv) \;A^{(1)}[0]\; \bs{B}(\bu)|0\rangle+
\frac{d}{d\beta_1}S_n^{(\kappa)}(\bv|\bu)\Bigr|_{\beta_j=0}.
\end{multline}
Here $ \beta_j = 0 $ means that both the parameters $\beta_1$ and $\beta_2$ are zero. Note that the parameters $\beta_j$ are already equal to zero
in the term $\langle0| \bs{C}(\bv) \;A^{(1)}[0]\; \bs{B}(\bu)|0\rangle$.
This means that the dual vector $ \langle0|\bs{C}(\bv)$ is the ordinary on-shell vector, not twisted. In other words,
this is exactly the form factor of the partial zero mode $A^{(1)}[0]$ that we are looking for:
\be{13-FFAdef}
\langle0| \bs{C}(\bv) \;A^{(1)}[0]\; \bs{B}(\bu)|0\rangle={\sf M}^{(11)}_n(\bv;\bu).
\ee

On the other hand, differentiating \eqref{13-Mk-res}, we have
\begin{equation}\label{13-Mk-derr}
\frac{d{\sf M}^{(\kappa)}_n(\bv;\bu)}{d\beta_1}\Bigr|_{\beta_j=0}=
\frac{\ell(\bv)}{\ell(\bu)}\frac{d}{d\beta_1}S_n^{(\kappa)}(\bv|\bu)\Bigr|_{\beta_j=0}
+\delta_{\bu,\bv}\left(a^{(1)}[0]+\frac{d}{d\beta_1}\ln \ell(\bv)\Bigr|_{\beta_j=0}\right)
\langle0| \bs{C}(\bu) \bs{B}(\bu)|0\rangle.
\end{equation}
Here $\delta_{\bu,\bv} =1$, if $\bu=\bv$, and $ \delta_{\bu,\bv} = 0$ for $\bu \ne \bv$. In the latter case, as was shown earlier,
the derivative of the scalar product with respect to the twist parameter\footnote{Earlier we calculated the derivatives  over the parameters
$\kappa_j = e^{\beta_j}$, but not over the parameters
$\beta_j$. It is easy to see, however, that this is the same at the point $\beta_j=0$.} is equal to the universal form factor of the operator $A(z)$
\be{13-der-kappa}
\frac{d}{d\beta_1}S_n^{(\kappa)}(\bv|\bu)\Bigr|_{\beta_j=0}=\mathfrak{F}_n^{(11)}(\bv;\bu),\qquad \bu\ne\bv.
\ee
Thus, comparing \eqref{13-Mk-derl} with \eqref{13-Mk-derr} and taking into account \eqref{13-der-kappa} we finally find
\begin{multline}\label{13-FF11}
{\sf M}^{(11)}_n(\bv;\bu)=\langle0| \bs{C}(\bv) \;A^{(1)}[0]\; \bs{B}(\bu)|0\rangle=\left(\frac{\ell(\bv)}{\ell(\bu)}-1\right)
\mathfrak{F}_n^{(11)}(\bv;\bu)\\
+\delta_{\bu,\bv}\left(a^{(1)}[0]+\frac{d}{d\beta_1}\ln \ell(\bv)\Bigr|_{\beta_j=0}\right)
\langle0| \bs{C}(\bu) \bs{B}(\bu)|0\rangle.
\end{multline}

The form factor of the operator $D^{(1)}[0]$ is computed in a completely analogous way
\begin{multline}\label{13-FF22}
{\sf M}^{(22)}_n(\bv;\bu)=\langle0| \bs{C}(\bv) \;D^{(1)}[0]\; \bs{B}(\bu)|0\rangle=\left(\frac{\ell(\bv)}{\ell(\bu)}-1\right)
\mathfrak{F}_n^{(22)}(\bv;\bu)\\
+\delta_{\bu,\bv}\left(d^{(1)}[0]+\frac{d}{d\beta_2}\ln \ell(\bv)\Bigr|_{\beta_j=0}\right)
\langle0| \bs{C}(\bu) \bs{B}(\bu)|0\rangle.
\end{multline}

The remaining two form factors (of the operators $C^{(1)}[0]$ and $B^{(1)}[0]$) can be found via equations \eqref{13-CA-FF}, \eqref{13-BA-FF}.
It is easy to see that
\begin{equation}\label{13-FF21}
{\sf M}^{(21)}_n(\bv;\bu)=\langle0| \bs{C}(\bv) \;C^{(1)}[0]\; \bs{B}(\bu)|0\rangle=\left(\frac{\ell(\bv)}{\ell(\bu)}-1\right)
\mathfrak{F}_n^{(21)}(\bv;\bu),
\end{equation}
and
\begin{equation}\label{13-FF12}
{\sf M}^{(12)}_n(\bv;\bu)=\langle0| \bs{C}(\bv) \;B^{(1)}[0]\; \bs{B}(\bu)|0\rangle=\left(\frac{\ell(\bv)}{\ell(\bu)}-1\right)
\mathfrak{F}_n^{(12)}(\bv;\bu).
\end{equation}
In the derivation of these formulas, we used the relations between universal form factors, and also the property
of the functions $\ell(w)\to 1$ at $w\to\infty$. We recall that this property follows from the asymptotic expansion over $u^{-1}$
of the monodromy matrix $T(u)$. In addition, in the case of the form factors of the operators $C^{(1)}[0]$ and $B^{(1)}[0]$
the number of the elements in the sets $\bu$ and $\bv$ is different, therefore we always have $\bu \ne \bv$.

If the first partial monodromy matrix $ T^{(1)}(u)$ corresponds to the lattice sites $1,\dots,m$, then all the dependence
of the form factor ${\sf M}^{(jk)}_n$ of $m$ is in the functions $\ell$. In models describing physical systems,
the function $\ell (u)$ is representable in the form $\ell(u)=\ell_0^m(u)$, where $\ell_0(u)$ is some function corresponding to one
lattice site. Then for the form factors of local operators we obtain
\begin{equation}\label{13-Ljk}
\langle0| \bs{C}(\bv) \;(L_m[0])_{jk}\; \bs{B}(\bu)|0\rangle=\frac{\ell_0^{m-1}(\bv)}{\ell_0^{m-1}(\bu)}
\left(\frac{\ell_0(\bv)}{\ell_0(\bu)}-1\right)
\mathfrak{F}_n^{(jk)}(\bv;\bu), \qquad \bu\ne \bv.
\end{equation}
In the case $\bv=\bu$, we have
\begin{equation}\label{13-Ljj}
\langle0| \bs{C}(\bu) \;(L_m[0])_{jj}\; \bs{B}(\bu)|0\rangle=\frac{d}{d\beta_j}\ln \ell_0(\bv)\Bigr|_{\beta_j=0}
\langle0| \bs{C}(\bu) \bs{B}(\bu)|0\rangle.
\end{equation}
Thus, the average value of the local operator $(L_m[0])_{jj}$ does not depend on the lattice site $m$.

\subsection{Differentiating over twist parameters}

We have already pointed out that before using the Bethe equations in formulas for scalar products, it is necessary
to make sure that there are no coinciding elements in the sets $\bu$ and $\bv$, that is $\bu\cap\bv=\emptyset$. However, we did not do this,
when expressing the function $r^{(2)}(v)$ in terms of $\ell(v)$ in the formula \eqref{13-BE-v2}. At the same time, we managed to calculate the form factors
of local operators even in the case when the sets $\bu$ and $\bv$ are the same \eqref{13-Ljj}.

The reason is that in the formula \eqref{13-BE-v2} we used not the usual Bethe equations, but twisted ones. Then the variables $v_i$ become functions of the twist parameters $\beta_j$: $v_i=v_i(\beta_j)$. Therefore, we can always choose values $\beta_j$ such that  $\bu\cap\bv(\beta_j)=\emptyset$.
Then we can perform  all the calculations and, in the final answer, consider any special values of $\beta_j$ when the sets $\bu$ and $\bv$
are partially or even completely coincide. Actually, this is exactly what is done in the formula \eqref{13-Ljj}.

Here a question arises on the calculating the derivatives with respect to twist parameters. Formally, one can find the derivatives
$dv_i/d\beta_j$ at $\beta_j=0$, differentiating the twisted Bethe equations. However, in most cases this is not required.
For example, as follows from the formula \eqref{13-der-kappa},  for the  calculating the universal form factor $\mathfrak{F}_n^{(11)}$,
we need simply to take the derivative of the scalar product of the twisted and ordinary on-shell vectors over $\beta_1$  at $\beta_j = 0 $.
The scalar product $S_n^{(\kappa)}(\bv | \bu)$ is given by the formulas  \eqref{11-SP-det-pres2}, \eqref{11-SP-det-pres4},
in which it is necessary to make a replacement $ \bu \leftrightarrow \bv $:
\begin{equation}\label{13-SP-det-pres2}
S_n^{(\kappa)}(\bv|\bu)=\Delta'_n(\bu)\Delta_n(\bv)\; h(\bv,\bu)
\;\det_n\mathcal{M}_{jk}^{(2)}\,,
\end{equation}
where
\begin{equation}\label{13-SP-det-pres4}
\mathcal{M}_{jk}^{(2)}
=t(v_k,u_j) +e^{\beta_2-\beta_1}\,t(u_j,v_k)\frac{h(v_k,\bv)h(\bu,v_k)}{h(\bv,v_k)h(v_k,\bu)}\,.
\end{equation}
Strictly speaking, taking the derivative with respect to $\beta_1$, we should differentiate all the parameters $\bv$, which enter both into the prefactor \eqref{13-SP-det-pres2}, and in the matrix elements \eqref{13-SP-det-pres4}. However, it is not difficult to show that the determinant of the matrix
$\mathcal{M}_{jk}^{(2)}$ is proportional to $(e^{\beta_2- \beta_1} -1)$. To do this, we need to add to the last (or any other) row of the matrix
$\mathcal{M}_{jk}^{(2)}$ all the other
rows multiplied by the coefficients $\nu_j^{(n)}/\nu_n^{(n)}$, where $\nu_j^{(n)}$ are defined by the formula  \eqref{07-SP-nuk}.
Repeating the calculations made in the proof of proposition~\ref{07-SP-prop-ortho},
we obtain that the new last row is proportional to $(e^{\beta_2- \beta_1}-1) $. And if so, then when taking the derivative with respect to $\beta_1$
at $\beta_j = 0$ it suffices to differentiate only this exponent. At the same time, we eliminate the need to differentiate the parameters $v_i$, which depend on $\beta_1$ implicitly through the twisted Bethe equations.

The described method does not work when $\bu = \bv$ for $\beta_j = 0$ (in this case $\nu_j^{(n)} = 0$). However, here one can also make a calculation
of the derivatives with respect to the parameters $\beta_j$ more explicit. Indeed, the logarithmic derivative
of the product of functions $\ell_0$ over $ \beta_1 $ in formula \eqref{13-Ljj} has the form
\be{13-diff-ell}
\frac{d}{d\beta_1}\ln \ell_0(\bv)\Bigr|_{\beta_j=0}=
\frac{d\ln \ell_0(v)}{d\beta_1}\sum_{k=1}^n\frac{dv_k}{d\beta_1}\Bigr|_{\beta_j=0}.
\ee
On the other hand, differentiating the  system of twisted Bethe equations written in terms of functions $\Psi_j$ \eqref{07-SP-Psi},
we find
\be{13-diff-tBE}
\sum_{k=1}^n\frac{\partial\Psi_j}{\partial u_k}\frac{dv_k}{d\beta_1}\Bigr|_{\beta_j=0}=-1,
\ee
where we took into account that $\bv=\bu$ at $\beta_j=0$.
After this, it remains to use the formula  \eqref{07-SP-Jak-norm},
which states that the norm of an on-shell vector is proportional to the Jacobian of the matrix $\partial\Psi_j/\partial u_k$
\be{13-norm-J}
\langle0| \bs{C}(\bu) \bs{B}(\bu)|0\rangle=(-c)^n\prod_{\substack{j,k=1\\j\ne k}}^nf(u_j,u_k)\;\det_n\left(\frac{\partial\Psi_j}{\partial u_k}\right).
\ee
It is easy to see that due to \eqref{13-diff-tBE}, one has an identity
\begin{multline}\label{13-vynosPsi}
\frac{d}{d\alpha}\det_n\left(\frac{\partial\Psi_j}{\partial u_k}-\alpha\right)\Bigr|_{\alpha=0}=\det_n\left(\frac{\partial\Psi_j}{\partial u_k}\right)\cdot
\frac{d}{d\alpha}\det_n\left(\delta_{jk}+\alpha\frac{dv_k}{d\beta_1}\Bigr|_{\beta_j=0}\right)\Bigr|_{\alpha=0}\\[2pt]
=\det_n\left(\frac{\partial\Psi_j}{\partial u_k}\right)\cdot\sum_{i=1}^n\frac{dv_i}{d\beta_1}\Bigr|_{\beta_j=0}.
\end{multline}
From this we find
\be{13-sum-det}
\frac{d}{d\beta_j}\ln \ell_0(\bv)\Bigr|_{\beta_j=0}
\langle0| \bs{C}(\bu) \bs{B}(\bu)|0\rangle
=(-c)^n\prod_{\substack{j,k=1\\j\ne k}}^nf(u_j,u_k)\;\frac{d}{d\alpha}\det_n\left(\frac{\partial\Psi_j}{\partial u_k}-\alpha\right)\Bigr|_{\alpha=0}.
\ee
Taking the derivative with respect to the auxiliary parameter $\alpha$ in the equality \eqref{13-sum-det} is no more difficult. One can, for example,
subtract the last column from all the others, after which the dependence on $\alpha$ will remain only in this last column. Hence,
the derivative of the determinant reduces to the differentiation of the last column.

Such a method of taking derivatives with respect to the parameters $\beta_j$ is also convenient in the case when the sets $\bv$ and $\bu$ intersect only partially, and we need to solve the indeterminateness  arising in the matrix elements $\mathcal{M}_{jk}^{(2)}$ because of the poles of the functions
$t(v_k,u_j)$ and $t(u_j,v_k)$.
However, it is simpler to express the ratios $h(v_k,\bv)/h(\bv,v_k)$ in terms of the functions $r(v_k)$ and rewrite the matrix elements in the form
\begin{equation}\label{13-SP-det-pres5}
\mathcal{M}_{jk}^{(2)}
=t(v_k,u_j) +(-1)^{n-1}r(v_k)\,t(u_j,v_k)\frac{h(\bu,v_k)}{h(v_k,\bu)}\,.
\end{equation}
Now we can solve the indeterminateness by simply sending $v_k\to u_j$ (that is, it is not necessary to assume that $v_k=v_k(\beta_1,\beta_2)$ and
the condition $v_k =u_j$ is attained in the limit $\beta_ {1,2}\to 0$). Indeed, using  $t(u,v)=g(u,v)/h(u,v)$ and
$g(u,v)=-g(v,u)$ we obtain
\begin{equation}\label{13-SP-det-pres6}
\mathcal{M}_{jk}^{(2)}
=g(v_k,u_j)\left(\frac1{h(v_k,u_j)} -(-1)^{n-1}\frac{r(v_k)}{h(u_j,v_k)}\frac{h(\bu,v_k)}{h(v_k,\bu)}\right)\,.
\end{equation}
It is easy to see that the expression in the parenthesis vanishes at $v_k=u_j$ provided the Bethe equations for the set
$\bu$ are fulfilled.

The representation \eqref{13-SP-det-pres5} is also more convenient for
calculating form factors of the off-diagonal elements  ${\sf M}_n^{(21)}(\bv;\bu)$ and ${\sf M}_n^{(12)}(\bv;\bu)$
for $\bu\cap\bv\ne\emptyset$, using the formulas \eqref{13-CA-FF}, \eqref{13-BA-FF}.

\section{Calculating form factors of  off-diagonal elements\label{13-Sec4}}

In this section, we obtain a representation for the form factor of the partial zero mode $C^{(1)}[0]$ by direct calculation. We do this despite the fact that such the representation has already been obtained (see \eqref{13-FF21}). However, direct calculation will allow us to establish new  identities that
will be useful later. For definiteness, we assume that $\bu\cap\bv =\emptyset$.

Thus, we should calculate
\begin{equation}
{\sf M}_n^{(21)}(\bv;\bu)=\langle0| \bs{C}(\bv)\;C^{(1)}[0]\;\bs{B}(\bu)|0\rangle,\label{13-defFF-ZM21}
\end{equation}
where both vectors are on-shell. Presenting them in terms of the partial Bethe vectors, we obtain
\begin{multline}\label{13-M21-rep}
{\sf M}^{(21)}_n(\bv;\bu)=\sum_{\substack{\bu\mapsto\{\bu_{\so},\bu_{\st}\}  \\   \bv\mapsto\{\bv_{\so},\bv_{\st}\}}}
\frac{\ell(\bv_{\st})}{\ell(\bu_{\so})}f(\bv_{\so},\bv_{\st})f(\bu_{\so},\bu_{\st})
\;\langle0|^{(1)}\bs{C}^{(1)}(\bv_{\so}) \;C^{(1)}[0]\;  \bs{B}^{(1)}(\bu_{\so})|0\rangle^{(1)}  \\
\times \langle0|^{(2)}\bs{C}^{(2)}(\bv_{\st}) \bs{B}^{(2)}(\bu_{\st})|0\rangle^{(2)}.
\end{multline}
The action of $C^{(1)}[0]$ onto the dual vector is given by
\be{13-actC}
\langle0|^{(1)}\bs{C}^{(1)}(\bv_{\so})C^{(1)}[0]=\lim_{w\to\infty}\tfrac wc \langle0|^{(1)}\bs{C}^{(1)}(\bv_{\so})\bs{C}^{(1)}(w).
\ee
Thus, we reduce the matrix element of the partial zero mode $C^{(1)}[0]$ to a sum of scalar products of the partial Bethe vectors:
\begin{equation}\label{13-M21-act}
{\sf M}^{(21)}_n(\bv;\bu)=\lim_{w\to\infty}\frac wc\sum_{\substack{\bu\mapsto\{\bu_{\so},\bu_{\st}\}  \\   \bv\mapsto\{\bv_{\so},\bv_{\st}\}}}
\frac{\ell(\bv_{\st})}{\ell(\bu_{\so})}f(\bv_{\so},\bv_{\st})f(\bu_{\so},\bu_{\st})
 S_{n_{\so}}^{(1)}(\{\bv_{\so},w\}|\bu_{\so})S_{n_{\st}}^{(2)}(\bv_{\st}|\bu_{\st}).
\end{equation}

So far, everything was completely analogous to the calculation of the generating functional  ${\sf M}^{(\kappa)}_n$ \eqref{13-Mk-def}.
Now small differences appear. We introduce a set of variables $\bet=\{\bv,w\}$.
Then the formula \eqref{13-M21-act} can be rewritten as:
\begin{equation}\label{13-M21-act2}
{\sf M}^{(21)}_n(\bv;\bu)=\lim_{w\to\infty}\frac wc\sum_{\substack{\bu\mapsto\{\bu_{\so},\bu_{\st}\}  \\   \bet\mapsto\{\bet_{\so},\bet_{\st}\}
\\ w\in \bet_{\so}}}
\frac{\ell(\bet_{\st})}{\ell(\bu_{\so})}f(\bet_{\so},\bet_{\st})f(\bu_{\so},\bu_{\st})
 S_{n_{\so}}^{(1)}(\bet_{\so}|\bu_{\so})S_{n_{\st}}^{(2)}(\bet_{\st}|\bu_{\st}).
\end{equation}
Here the sum is taken over partitions $\bu\mapsto\{\bu_{\so},\bu_{\st}\}$ and $\bet\mapsto\{\bet_{\so},\bet_{\st}\}$,
but the partitions are restricted with an additional condition $w\in \bet_{\so}$. Indeed, under such the constraint, we can put
$\bet_{\so}=\{\bv_{\so},w\}$ and  $\bet_{\st}=\bv_{\st}$.
One can see that the formulas \eqref{13-M21-act2} and \eqref{13-M21-act} coincide if we take into account that $f(w,\bet_{\st})=1$ at $w\to\infty$.

Now we are basically repeating the calculations that have already been done. First, we should explicitly write the scalar products that include
in the formula \eqref{13-M21-act2}. This will give us a sum over partitions
\be{13-part21}
\bu\mapsto\{\bu^{\bs{1}},\bu^{\bs{2}},\bu^{\bs{3}},\bu^{\bs{4}}\},\qquad
\bet\mapsto\{\bet^{\bs{1}},\bet^{\bs{2}},\bet^{\bs{3}},\bet^{\bs{4}}\},
\ee
and the initial restriction $w\in \bet_{\so}$ is transformed into $w\in \{\bet_{1},\bet_{3}\}$, or, which is the same, $w\notin \{\bet_{2},\bet_{4}\}$.
In the calculation of the scalar products, the functions $r^{(2)}$ arise. In analogy with the formulas \eqref{13-BE-u4}, \eqref{13-BE-v2},
they should be expressed  in terms of the functions $\ell$. Here there is only one small difference, namely, the analog of formula
\eqref{13-BE-v2} now has the form
\be{13-BE-v2-21}
r^{(2)}(\bet^{\bs{2}})=\frac{r(\bet^{\bs{2}})}{\ell(\bet^{\bs{2}})}=\frac{1}{\ell(\bet^{\bs{2}})}\;
\frac{f(\bet^{\bs{2}},\bet^{\bs{1}})f(\bet^{\bs{2}},\bet^{\bs{3}})f(\bet^{\bs{2}},\bet^{\bs{4}})}
{f(\bet^{\bs{1}},\bet^{\bs{2}})f(\bet^{\bs{3}},\bet^{\bs{2}})f(\bet^{\bs{4}},\bet^{\bs{2}})}.
\ee
First, here we do not use  the twisted Bethe equations, but the usual ones. Secondly, the rhs of \eqref{13-BE-v2-21} includes the $f$ functions
depending on the parameter $w$, while this parameter does not appear in Bethe equations for the parameters $\bv$. However, for $w\to\infty $
the equations \eqref{13-BE-v2-21} actually turn into the Bethe equations, since the $f$ function goes to unity when any of its arguments tends to infinity.

Thus, we eventually arrive at an analog of the formula \eqref{13-Mk-act3}
\begin{equation}\label{13-M21-act3}
{\sf M}^{(21)}_n(\bv;\bu)=\lim_{w\to\infty}\frac wc
\sum_{\substack{ \bu\mapsto\{\bu_{\rm i},\bu_{\rm ii}\} \\ \bet\mapsto\{\bet_{\rm i},\bet_{\rm ii}\} } }
 \; \frac{\ell(\bet_{\rm i})}{\ell(\bu_{\rm i})}\;f(\bu_{\rm i},\bu_{\rm ii})f(\bet_{\rm ii},\bet_{\rm i})
 G_1(\bet_{\rm i}|\bu_{\rm i})\,  G_2(\bet_{\rm ii}|\bu_{\rm ii}) ,
\end{equation}
where functions $G_1(\bet_{\rm i}|\bu_{\rm i})$ and $G_2(\bet_{\rm ii}|\bu_{\rm ii})$ are given as sums over partitions into additional
subsets:
\be{13-G1-21}
 G_1(\bet_{\rm i}|\bu_{\rm i})=
 \sum_{\substack{ \bu_{\rm i}\mapsto\{\bu^{\bs{1}},\bu^{\bs{4}}\} \\ \bet_{\rm i}\mapsto\{\bet^{\bs{1}},\bet^{\bs{4}}\}\\ w\notin \bet^{\bs{4}} } }
 K_{n_{1}}( \bu^{\bs{1}}|\bet^{\bs{1}})K_{n_{4}}( \bet^{\bs{4}}|\bu^{\bs{4}})\,
 f(\bet^{\bs{1}},\bet^{\bs{4}})f(\bu^{\bs{4}},\bu^{\bs{1}}),
 \ee
and
\be{13-G2-21}
 G_2(\bet_{\rm ii}|\bu_{\rm ii})=
 \sum_{\substack{ \bu_{\rm ii}\mapsto\{\bu^{\bs{2}},\bu^{\bs{3}}\} \\ \bet_{\rm ii}\mapsto\{\bet^{\bs{2}},\bet^{\bs{3}}\} \\ w\notin \bet^{\bs{2}}} }
K_{n_{3}}( \bet^{\bs{3}}|\bu^{\bs{3}})K_{n_{2}}( \bu^{\bs{2}}|\bet^{\bs{2}})\,
f(\bet^{\bs{2}},\bet^{\bs{3}})f(\bu^{\bs{3}},\bu^{\bs{2}}).
\ee
In both cases we deal with the sums with additional restrictions, which follow from the condition $w\in\{\bet^{\bs{1}},\bet^{\bs{3}}\}$.

Assume that $w\in\bet^{\bs{1}}$. Then $G_2=\delta_{0,n_2+n_3}$.
Indeed, in this case there are no restrictions in the sum \eqref{13-G2-21},
and then, up to the notation, it coincides with the identity  \eqref{11-Ident-2}.
Therefore, $\bet_{\rm ii}=\bu_{\rm ii}=\emptyset$, and respectively
$\bet_{\rm i}=\bet$, $\bu_{\rm i}=\bu$.

Similarly, if $w\in\bet^{\bs{3}}$, then $G_1=\delta_{0,n_1+n_4}$. In this case $\bet_{\rm i}=\bu_{\rm i}=\emptyset$,
and $\bet_{\rm ii}=\bet$, $\bu_{\rm ii}=\bu$. We come to the conclusion that the form factor ${\sf M}^{(21)}_n(\bv;\bu)$ has the form
\be{13-Om1Om2}
{\sf M}^{(21)}_n(\bv;\bu)=\lim_{w\to\infty}\frac wc\left(\frac{\ell(\bet)}{\ell(\bu)}\Omega_1(\bet|\bu)+\Omega_2(\bet|\bu)\right),
\ee
where
\be{13-Om1-1}
 \Omega_1(\bet|\bu)=
 \sum_{\substack{ \bu\mapsto\{\bu^{\bs{1}},\bu^{\bs{4}}\} \\ \bet\mapsto\{\bet^{\bs{1}},\bet^{\bs{4}}\}\\ w\in \bet^{\bs{1}} } }
 K_{n_{1}}( \bu^{\bs{1}}|\bet^{\bs{1}})K_{n_{4}}( \bet^{\bs{4}}|\bu^{\bs{4}})\,
 f(\bet^{\bs{1}},\bet^{\bs{4}})f(\bu^{\bs{4}},\bu^{\bs{1}}),
 \ee
and
\be{13-Om2-1}
 \Omega_2(\bet|\bu)=
 \sum_{\substack{ \bu\mapsto\{\bu^{\bs{2}},\bu^{\bs{3}}\} \\ \bet\mapsto\{\bet^{\bs{2}},\bet^{\bs{3}}\} \\ w\in \bet^{\bs{3}}} }
K_{n_{3}}( \bet^{\bs{3}}|\bu^{\bs{3}})K_{n_{2}}( \bu^{\bs{2}}|\bet^{\bs{2}})\,
f(\bet^{\bs{2}},\bet^{\bs{3}})f(\bu^{\bs{3}},\bu^{\bs{2}}).
\ee
Let us make a change of indices in the formulas \eqref{13-Om1-1}, \eqref{13-Om2-1}. Namely, we relabel subsets with superscripts
${}^{\bs{1}}$ and ${}^{\bs{3}}$  to the subsets with subscript $\so$. We also relabel the subsets with superscripts ${}^{\bs{2}}$ and  ${}^{\bs{4}}$
to the subsets with subscript $\st$. Then
\be{13-Om1-2}
 \Omega_1(\bet|\bu)=
 \sum_{\substack{ \bu\mapsto\{\bu_{\so},\bu_{\st}\} \\ \bet\mapsto\{\bet_{\so},\bet_{\st}\}\\ w\in \bet_{\so} } }
 K_{n_{\so}}( \bu_{\so}|\bet_{\so})K_{n_{\st}}( \bet_{\st}|\bu_{\st})\,
 f(\bet_{\so},\bet_{\st})f(\bu_{\st},\bu_{\so}),
 \ee
\be{13-Om2-2}
 \Omega_2(\bet|\bu)=
 \sum_{\substack{ \bu\mapsto\{\bu_{\so},\bu_{\st}\} \\ \bet\mapsto\{\bet_{\so},\bet_{\st}\}\\ w\in \bet_{\st} } }
 K_{n_{\so}}( \bu_{\so}|\bet_{\so})K_{n_{\st}}( \bet_{\st}|\bu_{\st})\,
 f(\bet_{\so},\bet_{\st})f(\bu_{\st},\bu_{\so}).
 \ee
We see that the functions $\Omega_1$ and $\Omega_2$ are given by the same sum over partitions, but in the first case $w\in \bet_{\so}$,
while in the second one $w\in \bet_{\st}$. Then the sum $\Omega_1 + \Omega_2$ is given by the sum over partitions without additional constraints that coincides
with the identity  \eqref{11-Ident-2}.
Therefore, if the original sets $\bv$ and $\bu$ are not empty, then $\Omega_1+\Omega_2=0$. In this way,
we find
\be{13-FF-Om1}
{\sf M}^{(21)}_n(\bv;\bu)=\left(\frac{\ell(\bv)}{\ell(\bu)}-1\right)
\lim_{w\to\infty}\frac wc\; \Omega_1(\bet|\bu).
\ee
Here we partly have taken the limit when replaced  $\ell(\bet)$ by $\ell(\bv)$, because  $\ell(w)\to 1$ at $w\to\infty$. Comparing equations \eqref{13-FF-Om1} and \eqref{13-FF21}, we see that
\be{13-Om1-FF}
\lim_{w\to\infty}\frac wc\; \Omega_1(\bet|\bu)=\mathfrak{F}^{(21)}_n(\bv;\bu).
\ee
Thus, we obtain an identity.
\begin{prop}\label{13-PROP1}
Let $\bv$ and $\bu$ be two sets of parameters, such that  $\#\bv=n-1$,  $\#\bu=n$ ($n>0$), and $\bv\cap\bu=\emptyset$. Let
$\bet=\{\bv,w\}$. Then
\be{13-sum-FF21}
 \lim_{w\to\infty}\frac wc\;
 \sum_{\substack{ \bu\mapsto\{\bu_{\so},\bu_{\st}\} \\ \bet\mapsto\{\bet_{\so},\bet_{\st}\}\\ w\in \bet_{\so} } }
 K_{n_{\so}}( \bu_{\so}|\bet_{\so})K_{n_{\st}}( \bet_{\st}|\bu_{\st})\,
 f(\bet_{\so},\bet_{\st})f(\bu_{\st},\bu_{\so})=\mathfrak{F}^{(21)}_n(\bv;\bu).
 \ee
The sum is taken over partitions with the restriction $w\in \bet_{\so}$.
\end{prop}

The equality \eqref{13-sum-FF21} is an identity, since it was obtained within the generalized model, and no
conditions on the parameters $\bv$ and $\bu$ were imposed, except $\bv\cap\bu=\emptyset$.

Similarly, one can prove one more identity.
\begin{prop}\label{13-PROP2}
Let $\bv$ and $\bu$ be two sets of parameters, such that  $\#\bv=n+1$,  $\#\bu=n$ ($n\ge 0$), and $\bv\cap\bu=\emptyset$. Let
$\bet=\{\bu,w\}$. Then
\be{13-sum-FF12}
 \lim_{w\to\infty}\frac wc\;
 \sum_{\substack{ \bu\mapsto\{\bv_{\so},\bv_{\st}\} \\ \bet\mapsto\{\bet_{\so},\bet_{\st}\}\\ w\in \bet_{\so} } }
 K_{n_{\so}}( \bet_{\so}|\bv_{\so})K_{n_{\st}}( \bv_{\st}|\bet_{\st})\,
 f(\bv_{\so},\bv_{\st})f(\bet_{\st},\bet_{\so})=\mathfrak{F}^{(12)}_n(\bv;\bu).
 \ee
The sum is taken over partitions with the restriction $w\in \bet_{\so}$.
\end{prop}

In the case when there are coinciding elements in the sets $\bu$ and $\bv$, the calculations are much more complicated.
Therefore, the easiest way to act is to use the formulas \eqref{13-CA-FF}, \eqref{13-BA-FF}.

\section{Proofs of new identities\label{13-Sec5}}

For the sake of completeness, we give a direct proof of the identity \eqref{13-sum-FF21}. In the previous lecture, we already calculated
the universal form factor of the operator $C(z)$. Recall the explicit formula:
\be{13-FF21-res1}
\mathfrak{F}_n^{(21)}(\bv;\bu)=-\Delta'_n(\bu)
\Delta_{n-1}(\bv)h(\bv,\bu) \frac{g(u_n,\bv)}{g(u_n,\bu_n)}\;\det_{n-1}\mathcal{M}_{jk}^{(21)},
\ee
where the matrix $\mathcal{M}^{(21)}_{jk}$ is given by \eqref{12-M21}.

We reproduce this result by computing the sum over the partitions in the formula \eqref{13-Om1-2}. The summation identity  \eqref{06-MCR-Part-old1}
for DWPF allows us to calculate the sum over partitions of the set $\bu$:
\be{13-SSDS}
 \sum_{\bu\mapsto\{\bu_{\so},\bu_{\st}\} }
 K_{n_{\so}}( \bu_{\so}|\bet_{\so})K_{n_{\st}}( \bet_{\st}|\bu_{\st})\,
 f(\bu_{\st},\bu_{\so})=(-1)^{n_{\so}}f(\bu,\bet_{\so})K_{n}( \{\bet_{\so}-c,\bet_{\st}\}|\bu).
 \ee
Then equation \eqref{13-Om1-2} turns into
\be{13-Om1-sum1}
 \Omega_1(\bet|\bu)=
 \sum_{\substack{  \bet\mapsto\{\bet_{\so},\bet_{\st}\}\\ w\in \bet_{\so} } } (-1)^{n_{\so}}f(\bu,\bet_{\so}) f(\bet_{\so},\bet_{\st})
 K_{n}( \{\bet_{\so}-c,\bet_{\st}\}|\bu)\,.
 \ee
Summation over partitions of the set $\bet$ is carried out with the help of the lemma, that was proved in the derivation of the determinant representation
for the scalar product. Recall this lemma.
\begin{lemma}\label{13-Long-Det}
Let $\bet$ and $\bu$ be two sets of complex numbers, such that $\#\bet=\#\bu=n$. Let also
$\phi_1(\eta)$ and $\phi_2(\eta)$ be two arbitrary functions of the complex variable $\eta$. Then
\begin{multline}\label{13-SumDet1}
\sum_{\bet\mapsto\{\bet_{\so},\bet_{\st}\}} K_n(\{\bet_{\so}-c, \bet_{\st}\}|\bu)f(\bu, \bet_{\so})f(\bet_{\st},\bet_{\so})
\phi_1(\bet_{\so})\phi_2(\bet_{\st})\num
=\Delta'_n(\bu)\Delta_n(\bet)
\det_n\Bigl(\phi_2(\eta_k)t(\eta_k,u_j)h(\eta_k,\bu)+(-1)^n \phi_1(\eta_k)t(u_j,\eta_k)h(\bu,\eta_k)\Bigr).
\end{multline}
Here we have used the shorthand notation for the products of the functions  $\phi_1$ and $\phi_2$.
\end{lemma}

We need to find functions $\phi_1$ and $\phi_2$ such that the sum in  \eqref{13-SumDet1} would turn into the sum \eqref{13-Om1-sum1}.
It is easy to see that the functions
\be{13-f1f2}
\begin{aligned}
&\phi_1(\eta_i)=-1,\\
&\phi_2(\eta_i)=\frac{f(\bet_i,\eta_i)}{f(\eta_i,\bet_i)}=(-1)^{n-1}\frac{h(\bet,\eta_i)}{h(\eta_i,\bet)},\qquad \eta_i\in\bv,\\
&\phi_2(w)=0,
\end{aligned}
\ee
satisfy this condition. Indeed, in this case,
\be{13-prof2-ff}
f(\bet_{\st},\bet_{\so})\phi_2(\bet_{\st})=f(\bet_{\st},\bet_{\so})\frac{f(\bet_{\so},\bet_{\st})}{f(\bet_{\st},\bet_{\so})}=f(\bet_{\so},\bet_{\st}),
\ee
and the condition $\phi_2(w)=0$ automatically selects only such the partitions, for which $w\in\bet_{\so}$. Then we have
\be{13-Om1-sum2}
 \Omega_1(\bet|\bu)=\Delta'_n(\bu)\Delta_n(\bet)
\det_n\mathcal{N}^{(21)}_{jk},
\ee
where
\be{13-N21}
\begin{aligned}
&\mathcal{N}^{(21)}_{jk}=(-1)^{n-1}t(\eta_k,u_j)\frac{h(\eta_k,\bu)h(\bet,\eta_k)}{h(\eta_k,\bet)}+(-1)^{n-1}t(u_j,\eta_k)h(\bu,\eta_k),
\qquad k<n,\\
&\mathcal{N}^{(21)}_{jn}=(-1)^{n-1}t(u_j,\eta_n)h(\bu,\eta_n), \qquad \eta_n=w.
\end{aligned}
\ee
Let us extract from each column the following factor
\be{13-factor}
(-1)^{n-1}\frac{h(\eta_k,\bu)h(\bet,\eta_k)}{h(\eta_k,\bet)}.
\ee
Then we obtain
\be{13-Om1-sum3}
 \Omega_1(\bet|\bu)=\Delta'_n(\bu)\Delta_n(\bet)h(\bet,\bu)
\det_n\widetilde{\mathcal{N}}^{(21)}_{jk},
\ee
where
\be{13-wtN21}
\begin{aligned}
&\widetilde{\mathcal{N}}^{(21)}_{jk}=t(v_k,u_j)+t(u_j,v_k)\frac{h(v_k,\bet)h(\bu,v_k)}{h(v_k,\bu)h(\bet,v_k)},
\qquad k<n,\\
&\widetilde{\mathcal{N}}^{(21)}_{jn}=t(u_j,w)\frac{h(w,\bet)h(\bu,w)}{h(w,\bu)h(\bet,w)}.
\end{aligned}
\ee
Here we have shown explicitly that $\eta_k=v_k$ for $k<n$, and  $\eta_n=w$.

The result should be multiplied by $w/c$ and then one should send $w$ to infinity. It is convenient to present $w/c$ as a product
$w^2/c^2$ and $c/w$ and multiply the last column of the matrix $\widetilde{\mathcal{N}}^{(21)}$ by the factor $w^2/c^2$. Then
\be{13-lim1}
\begin{aligned}
&\lim_{w\to\infty} \widetilde{\mathcal{N}}^{(21)}_{jk}=\mathcal{M}^{(21)}_{jk},\qquad k<n\\
&\lim_{w\to\infty} \frac{w^2}{c^2}\widetilde{\mathcal{N}}^{(21)}_{jn}=-1,
\end{aligned}
\ee
where $\mathcal{M}^{(21)}_{jk}$ is given by \eqref{12-M21},
and
\be{13-lim2}
\lim_{w\to\infty}\frac cw \Delta'_n(\bu)\Delta_n(\bet)h(\bet,\bu)= \Delta'_n(\bu)\Delta_{n-1}(\bv)h(\bv,\bu).
\ee
We almost reproduced the result \eqref{13-FF21-res1}. It remains only to get rid of the last column in the matrix  $\widetilde{\mathcal{N}}^{(21)}$.
This is done in a standard way. We add to the last row of the matrix $\widetilde{\mathcal{N}}^{(21)}_{jk}$ all other rows multiplied by
the coefficients $\alpha_j/\alpha_n$, where
\be{13-alj}
\alpha_j=\frac{g(u_j,\bu_j)}{g(u_j,\bv)}=\prod_{m=1}^{n-1}(u_j-v_m)\prod_{\substack{m=1\\m\ne j}}^n(u_j-u_m)^{-1}\,.
\ee
The resulting sums are calculated explicitly via auxiliary contour integrals. Actually, we have already proved that
\be{13-sumM1}
\sum_{j=1}^n\alpha_j\mathcal{M}^{(21)}_{jk}=0,\qquad k<n,
\ee
therefore, after this transformation,  we obtain zeros everywhere in the last row except the last element. The last element is $-1/\alpha_n$
due to an identity
\be{13-sumM2}
\sum_{j=1}^n\alpha_j=1.
\ee
Thus, the determinant of the $n\times n$ matrix reduces to the product of the coefficient $-1/\alpha_n$ and the minor constructed on
first $n-1$ rows and columns. The matrix elements of this minor coincide with $\mathcal{M}^{(21)}_{jk}$, and we reproduce the
result \eqref{13-FF21-res1}.


%
%
\chapter{Form factors of local operators in QNLS model\label{CHA-FFLONS}}

This lecture is devoted to the calculation of the form factors of local operators in the QNLS  model.
The matter is that the asymptotic expansion of the monodromy matrix $T(u)$ at $u\to\infty$ in this model has a structure that
differs from that considered in previous lectures. Because of this, the definition of zero modes changes. We will see, however, that despite
these changes, the basic actions for calculating the form factors of local operators, remain almost the same as in
lecture~\ref{CHA-FFLO}.

\section{QNLS model\label{14-Sec1}}

Description of the QNLS model and its formulation within the framework of QISM was given in section~\ref{03-Sec12}.
Therefore, here we only recall basic formulas referring the reader for details to the works \cite{KBIr,Gaud83,LiebL63,Lieb63,YanY69}.

The Hamiltonian of the QNLS model has the form
 \be{14-HamQ}
 H=\int_0^L\left(\partial_x\Psi^\dagger(x)\partial_x\Psi(x)+\varkappa\Psi^\dagger(x)\Psi^\dagger(x)\Psi(x)\Psi(x)\right)\,dx,
 \ee
where  $\Psi(x)$ and $\Psi^\dagger(x)$  are canonical Bose fields, and  $\varkappa$ is a coupling constant, that is related with
the $R$-matrix constant $c$ by $\varkappa=ic$.
We consider the model on a finite interval  $[0,L]$ with periodic boundary condition.

To formulate the QNLS model in the language of the QISM we
introduce a lattice appro\-xi\-ma\-tion. Namely, we chose $N$ points $x_1,\dots,x_N$ in the interval $[0, L]$, so that $x_n=\Delta n$, where
$\Delta=L/N$ is the lattice step. With each lattice site, we associate the $L$-operator  \eqref{03-L-op22},
in which the operators $\psi_n$ and $\psi^\dagger_n$ are lattice approximations of the original Bose fields and satisfy the relations
\be{14-com-rel-psin}
[\psi_n, \psi^\dagger_m]=\frac{\delta_{nm}}\Delta.
\ee
The monodromy matrix appears in the continuous limit $\Delta\to0$ in the standard expression
\be{14-T-LL}
T(u)=L_N(u)\dots L_1(u).
\ee
Our goal is to obtain explicit representations of the
monodromy matrix entries in terms of the Bose fields. Knowing such representations, we can obtain the asymptotic expansion of $T(u)$ for $u\to\infty$ and determine the zero modes.

\subsection{Representation for monodromy matrix in terms of Bose fields\label{14-Sec11}}
For calculating the monodromy matrix in the continuous limit we need the  $L$-operator up to terms linear over $\Delta$
\begin{equation}\label{14-L-op-sm}
 L_n(u)=\frac1{1-\frac{iu\Delta}2}\begin{pmatrix}
1-\frac{iu\Delta}2& -i\sqrt{\varkappa}\Delta\psi^\dagger_n \\
i\sqrt{\varkappa}\Delta \psi_n&1+\frac{iu\Delta}2
\end{pmatrix}+O(\Delta^2),
\end{equation}
where the normalization factor is taken so as to provide $a(u)=1$.

{\sl Remark}. Generally speaking, one should be careful when writing such an expansion, if terms of order $O(\Delta^2)$ depend
on unbounded operators. The point is that when commuting operators $\psi_n$ and $\psi^\dagger_n$, the factor
of order $1/\Delta$ appears. However, in our case we will multiply the $L$-operators corresponding to different lattice sites,
therefore, the representation \eqref{14-L-op-sm} is justified.

We present the infinitesimal $L$-operator \eqref{14-L-op-sm} as the sum of the diagonal and antidiagonal matrices\footnote{%
Here and below, we omit the terms $ O(\Delta^2)$, since they do not contribute to the continuous limit.}:
\begin{equation}\label{14-L-op-LW}
L_n(u)=\frac{\Lambda(u) +W_n}{1-\frac{iu\Delta}2},
\end{equation}
where
\begin{equation}\label{14-L-op-L-W}
\Lambda(u)=\begin{pmatrix}1-\frac{iu\Delta}2&0\\0&1+\frac{iu\Delta}2\end{pmatrix},
\qquad W_n=\begin{pmatrix}0&b_n\\c_n&0\end{pmatrix}=\begin{pmatrix}0&-i\sqrt{\varkappa}\Delta\psi^\dagger_n\\
i\sqrt{\varkappa}\Delta\psi_n&0\end{pmatrix}.
\end{equation}

Let us substitute  representation \eqref{14-L-op-LW} in \eqref{14-T-LL} and expand the resulting expression over degrees of $W_n$.
We have
\be{14-T-power}
T(u)=\sum_{n=0}^N  T_n(u),
\ee
where
\be{14-Tnu}
T_n(u)=\left(1-\frac{iu\Delta}2\right)^{-N}\sum_{N\ge k_n>\dots>k_1\ge 1}\Lambda^{N-k_n}W_{k_n}\Lambda^{k_n-k_{n-1}-1}
\cdots\Lambda^{k_2-k_1-1}W_{k_1}\Lambda^{k_1-1}.
\ee
Let
\be{14-tW}
\widetilde{W}_{k_i}=\Lambda^{-k_i}W_{k_i}\Lambda^{k_i-1}=\begin{pmatrix}0&\tilde b_{k_i}\\ \tilde c_{k_i}&0\end{pmatrix},
\ee
where
\be{14-tb-tc}
\tilde b_{k_i} =\frac{b_{k_i}}{1+\frac{iu\Delta}2}\left(\frac{1+\frac{iu\Delta}2}{1-\frac{iu\Delta}2} \right)^{k_i},
\qquad
\tilde c_{k_i} =\frac{c_{k_i}}{1-\frac{iu\Delta}2}\left(\frac{1-\frac{iu\Delta}2}{1+\frac{iu\Delta}2} \right)^{k_i}.
\ee
Then  equation \eqref{14-Tnu} takes the form
\be{14-Tnu-1}
T_n(u)=\left(1-\frac{iu\Delta}2\right)^{-N}\Lambda^{N}\sum_{N\ge k_n>\dots>k_1\ge 1}\widetilde{W}_{k_n}\widetilde{W}_{k_{n-1}}\cdots
\widetilde{W}_{k_1}.
\ee
Since the product of two matrices $\widetilde{W}_{k_i}$ and $\widetilde{W}_{k_{i-1}}$ gives a diagonal matrix
\be{14-prod-two}
\widetilde{W}_{k_i}\widetilde{W}_{k_i-1}=\begin{pmatrix} \tilde b_{k_i}\tilde c_{k_{i-1}} &0\\0&\tilde c_{k_i}\tilde b_{k_{i-1}}\end{pmatrix},
\ee
we easily find
\be{14-T-ell}
\begin{aligned}
&T_{0}(u)= \begin{pmatrix}1&0\\ 0&\left(\frac{1+\frac{iu\Delta}2}{1-\frac{iu\Delta}2} \right)^{N}\end{pmatrix},\\
&T_{2\ell}(u)= \begin{pmatrix}A_\ell(u)&0\\ 0&D_\ell(u)\end{pmatrix}, \qquad \ell=1,2,\dots,\\
&T_{2\ell+1}(u)= \begin{pmatrix}0&B_\ell(u)\\ C_\ell(u)&0\end{pmatrix}, \qquad \ell=0,1,\dots,
\end{aligned}
\ee
where
\be{14-A-ell}
\begin{aligned}
& A_\ell(u)=\sum_{N\ge k_{2\ell}>\dots>k_1\ge 1} \prod_{i=1}^\ell\tilde b_{k_{2i}}\tilde c_{k_{2i-1}},\\
& B_\ell(u)=\sum_{N\ge k_{2\ell+1}>\dots>k_1\ge 1} \tilde b_{k_{2\ell+1}}\prod_{i=1}^\ell\tilde c_{k_{2i}}\tilde b_{k_{2i-1}},\\
& C_\ell(u)=\left(\frac{1+\frac{iu\Delta}2}{1-\frac{iu\Delta}2} \right)^{N}
\sum_{N\ge k_{2\ell+1}>\dots>k_1\ge 1} \tilde c_{k_{2\ell+1}}\prod_{i=1}^\ell\tilde b_{k_{2i}}\tilde c_{k_{2i-1}},\\
& D_\ell(u)=\left(\frac{1+\frac{iu\Delta}2}{1-\frac{iu\Delta}2} \right)^{N}
\sum_{N\ge k_{2\ell}>\dots>k_1\ge 1} \prod_{i=1}^\ell\tilde c_{k_{2i}}\tilde b_{k_{2i-1}}.
\end{aligned}
 \ee
Note that in the operator products in \eqref{14-A-ell}, all the operators commute, because they correspond to different lattice sites.

Now everything is ready for taking the continuous limit $\Delta\to0$. First of all, note that $N\to\infty$ in this limit, therefore,
the finite sum  \eqref{14-T-power} turns into an infinite series. It is also easy to see that
\be{14-r0-n}
\lim_{\Delta\to 0}\left(\frac{1+\frac{iu\Delta}2}{1-\frac{iu\Delta}2} \right)^{k_i}=\lim_{\Delta\to 0}\left(\frac{1+\frac{iu\Delta}2}{1-\frac{iu\Delta}2}\right)^{x_{k_i}/\Delta}=e^{iux_{k_i}}.
\ee
This implies
\be{14-LN}
\lim_{\Delta\to0}\left(\frac{1+\frac{iu\Delta}2}{1-\frac{iu\Delta}2} \right)^{N}= e^{iuL} ,
\ee
and
\be{14-tb-tc-cont}
\tilde b_{k_i} \to -i\sqrt{\varkappa}\Delta e^{iux_{k_i}}\psi^\dagger_{k_i},
\qquad
\tilde c_{k_i} \to i\sqrt{\varkappa}\Delta e^{-iux_{k_i}}\psi_{k_i}, \qquad \Delta\to0.
\ee
Substituting this into \eqref{14-A-ell}, we obtain
\be{14-ABCD-ell}
\begin{aligned}
& A_\ell(u)=\varkappa^\ell\Delta^{2\ell}\sum_{k_{2\ell}>\dots>k_1\ge 1}\prod_{i=1}^\ell  e^{iux_{2i}}\psi^\dagger_{2i}
  \prod_{i=1}^\ell  e^{-iux_{2i-1}}\psi_{2i-1},\\
& B_\ell(u)=-i\varkappa^{\ell+1/2}\Delta^{2\ell+1}\sum_{k_{2\ell+1}>\dots>k_1\ge 1}
\prod_{i=1}^{\ell+1}  e^{iux_{2i-1}}\psi^\dagger_{2i-1}  \prod_{i=1}^\ell  e^{-iux_{2i}}\psi_{2i},\\
& C_\ell(u)=i\varkappa^{\ell+1/2}\Delta^{2\ell+1}e^{iuL}
\sum_{k_{2\ell+1}>\dots>k_1\ge 1} \prod_{i=1}^\ell  e^{iux_{2i}}\psi^\dagger_{2i}
  \prod_{i=1}^{\ell+1}  e^{-iux_{2i-1}}\psi_{2i-1},\\
& D_\ell(u)=\varkappa^\ell\Delta^{2\ell}e^{iuL}
\sum_{k_{2\ell}>\dots>k_1\ge 1}  \prod_{i=1}^\ell  e^{iux_{2i-1}}\psi^\dagger_{2i-1}
  \prod_{i=1}^\ell  e^{-iux_{2i}}\psi_{2i}.
\end{aligned}
 \ee

It remains to go from the discrete operators $\psi_{i}$ and $\psi^\dagger_{i}$ to the fields $\Psi(x)$ and $\Psi^\dagger(x)$. We agree upon that
here and below all limits of operator-valued expressions should be understood in the weak sense. This is all the more justified,
that in the final analysis we are interested in the form factors of local operators, not the operators themselves. Then it is clear,
that in the continuous limit the discrete sums over the lattice sites become integrals. Indeed,
let $\Phi(x)$ be an integrable function on the interval $[0, L]$. Then, due to \eqref{03-lat-con} we have
\be{14-sum-int}
\Delta\sum_{j=1}^N\Phi(x_j)\psi^\dagger(j)=
\sum_{j=1}^N\Phi(x_j)\int_{x_{j-1}}^{x_j}\Psi^\dagger_k(x)\,dx \longrightarrow \int_{0}^{L}\Phi(x)\Psi^\dagger_k(x)\,dx,\qquad \Delta\to0.
\ee
Thus, the sum over the lattice sites multiplied by $\Delta$  goes to the integral in the continuous limit. In the same way, one can prove that
an $m$-fold sum over the lattice sites multiplied by $\Delta^m$  becomes an $m$-fold
integral in the continuous limit. As a result, we obtain for the diagonal part of the monodromy matrix
\begin{equation}\label{14-A-ell-cont}
 A_\ell(u)=\varkappa^\ell\int_{\mathcal{D}_{2\ell}(\bar x)} \prod_{i=1}^\ell  e^{iux_{2i}}\Psi^\dagger (x_{2i})\,dx_{2i}
  \prod_{i=1}^\ell  e^{-iux_{2i-1}}\Psi(x_{2i-1})\,dx_{2i-1}
   \end{equation}
and
\begin{equation}\label{14-D-ell-cont}
 D_\ell(u)=e^{iuL}\varkappa^\ell\int_{\mathcal{D}_{2\ell}(\bar x)} \prod_{i=1}^\ell  e^{iux_{2i-1}}\Psi^\dagger (x_{2i-1})\,dx_{2i-1}
  \prod_{i=1}^\ell  e^{-iux_{2i}}\Psi(x_{2i})\,dx_{2i},
 \end{equation}
where $\mathcal{D}_{2\ell}(\bar x)=\{L>x_{2\ell}>\dots,>x_1>0\}$. Similarly, for the antidiagonal part, we obtain
\begin{equation}\label{14-B-ell-cont}
 B_\ell(u)=-i\varkappa^{\ell+1/2}\int_{\mathcal{D}_{2\ell+1}(\bar x)} \prod_{i=1}^{\ell+1}  e^{iux_{2i-1}}\Psi^\dagger (x_{2i-1})\,dx_{2i-1}
  \prod_{i=1}^\ell  e^{-iux_{2i}}\Psi(x_{2i})\,dx_{2i}
 \end{equation}
and
\begin{equation}\label{14-C-ell-cont}
 C_\ell(u)=i\varkappa^{\ell+1/2}e^{iuL}\int_{\mathcal{D}_{2\ell+1}(\bar x)}\prod_{i=1}^\ell  e^{iux_{2i}}\Psi^\dagger (x_{2i})\,dx_{2i}
  \prod_{i=1}^{\ell+1}  e^{-iux_{2i-1}}\Psi(x_{2i-1})\,dx_{2i-1},
   \end{equation}
where $\mathcal{D}_{2\ell+1}(\bar x)=\{L>x_{2\ell+1}>\dots,>x_1>0\}$. Let us stress once more that within the integration domains  $\mathcal{D}_{2\ell}$ and
$\mathcal{D}_{2\ell+1}$ all the Bose fields commute with each other, therefore, their ordering actually is not important.

Thus, we obtained a representation for the matrix elements of the monodromy matrix $T(u)$ in the form of a series whose terms depend explicitly on the local
Bose fields. This series is formal, and we do not study the question of its convergence. However, it is easy to see that if we introduce a vector
\be{14-Phi}
|\Phi_{n}\rangle=\int_0^L \,dx_1,\dots,\,dx_n \; \Phi_{n}(x_1,\dots,x_n)\prod_{i=1}^n\Psi^\dagger(x_i)|0\rangle,
\ee
where $\Phi_{n}(x_1,\dots,x_n)$ is a smooth function in the integration domain, then the action of any
$T_{ij}(u)$  on $|\Phi_{n}\rangle$ turns into a finite sum.

\section{Zero modes\label{14-S-ZM}}

For $u\to\infty$, the expansion for the monodromy matrix contains multiple integrals of rapidly oscillating functions.
The methods for calculating rapidly oscillating integrals are well known. In our case
one of the simplest ways to obtain an asymptotic expansion for
$T(u)$ is integration by parts.

{\sl Remark}. As well as the passage from discrete operators to boson fields, the asymptotic expansion of $T(u)$ in a power series
over $u^{-1}$ should be understood in the weak sense. For example, when evaluating the asymptotic behavior of the operator $A_\ell(u)$, we should
study the matrix elements of
this operator between vectors of the type \eqref{14-Phi}. This leads to an asymptotic analysis of integrals of the form
\begin{equation}\label{14-A-ell-est}
\mathcal{A}_\ell=\int_{\mathcal{D}_{2\ell}(\bar x)}\left( \prod_{i=1}^\ell  e^{iu(x_{2i}-x_{2i-1})}\right)\Phi (x_{1},\dots x_{2\ell})
\,dx_{1},\dots \,dx_{2\ell},
   \end{equation}
where $\Phi (x_{1},\dots x_{2\ell})$ is a smooth function in the integration domain.

It is easy to show (see section~\ref{14-A-OI}) that integrals over a single integration variable in the formulas \eqref{14-B-ell-cont}, \eqref {14-C-ell-cont} and the double integrals in the formulas \eqref{14-A-ell-cont}, \eqref{14-D-ell-cont} behave as $1/u$ at $u\to\infty$. All other
terms in which the multiplicity of integrals is greater than two, give contributions of order $o(u^{-1})$. Therefore, in
order to find zero modes, it suffices to take only the first nontrivial
terms of the expansion of $T(u)$. Then we have
\be{14-Tij}
A(u)= 1+\varkappa\int_0^L e^{iu(z-y)}\theta(z-y)\Psi^\dagger(z)\Psi(y)\,dz\,dy+O(\varkappa^2),
\ee
\begin{equation}\label{14-T33}
D(u)=e^{iuL}+\varkappa\,e^{iuL}\int_0^L e^{iu(y-z)}\theta(z-y)
\Psi^\dagger(y)\Psi(z)\,dz\,dy+O(\varkappa^2).
 \end{equation}
\begin{equation}\label{14-Ti3}
B(u)=-i\sqrt{\varkappa}\int_0^L e^{iuy}\Psi^\dagger(y)\,dy  +O(\varkappa^{3/2}),
 \end{equation}
\begin{equation}\label{14-T3j}
C(u)=i\sqrt{\varkappa}e^{iuL}\int_0^L e^{-iuy}\Psi(y)\,dy  +O(\varkappa^{3/2}).
 \end{equation}
All the terms denoted by $O(\varkappa^2)$ or $O(\varkappa^{3/2})$ give contributions of order  $O(u^{-2})$ as $u\to\infty$, therefore, they are not
essential. Integrating by parts we obtain
\be{14-Tij-1}
A(u)= 1+\frac{i\varkappa}{u}\int_0^L \Psi^\dagger(y)\Psi(y)\,dy+O(u^{-2}),
\ee
\begin{equation}\label{14-T33-1}
 D(u)=e^{iuL}-\frac{i\varkappa}{u}\,e^{iuL}\int_0^L
\Psi^\dagger(y)\Psi(y)\,dy+O(u^{-2}).
 \end{equation}
\begin{equation}\label{14-Ti3-1}
B(u)=-\frac{\sqrt{\varkappa}}{u}\left(e^{iuL}\Psi^\dagger(L)-\Psi^\dagger(0)\right)+O(u^{-2}),
 \end{equation}
\begin{equation}\label{14-T3j-1}
C(u)=-\frac{\sqrt{\varkappa}}{u}\left(\Psi(L)-e^{iuL}\Psi(0)\right)+O(u^{-2}).
 \end{equation}

Now we define the zero modes as follows:
\be{14-0Tij}
A[0]=\lim_{u\to\infty}\tfrac uc(A(u)-1)=-\int_0^L \Psi^\dagger(y)\Psi(y)\,dy,
\ee
and we recall that $c=-i\varkappa$. This is the same definition as we used earlier. The zero mode
$D[0]$ is defined somewhat differently:
\be{14-0T33}
D[0]=\lim_{u\to\infty}\tfrac uc\left(e^{-iuL} D(u)-1\right)=\int_0^L
\Psi^\dagger(y)\Psi(y)\,dy,
\ee
and hence, $D[0]=-A[0]$.

Looking at the expressions \eqref{14-Ti3-1}, \eqref{14-T3j-1}, we see that there exist two types of zero modes for these operators:
\be{14-0Ti3}
\begin{aligned}
&B_{(\text{\tiny R})}[0]=\lim_{u\to-i\infty}e^{-iuL}\tfrac uc \;B(u)=\frac1{i\sqrt{\varkappa}}\;\Psi^\dagger(L),\\
&B_{(\text{\tiny L})}[0]=\lim_{u\to+i\infty}\tfrac uc \;B(u)=-\frac1{i\sqrt{\varkappa}}\;\Psi^\dagger(0),
\end{aligned}
\ee
and
\be{14-0T3j}
\begin{aligned}
&C_{(\text{\tiny R})}[0]=\lim_{u\to+i\infty}\tfrac uc \;C(u)=\frac1{i\sqrt{\varkappa}}\;\Psi(L), \\
&C_{(\text{\tiny L})}[0]=\lim_{u\to-i\infty}e^{-iuL}\tfrac uc \;C(u)=-\frac1{i\sqrt{\varkappa}}\;\Psi(0).
\end{aligned}
\ee

The formulas obtained for the zero modes make it possible to investigate the form factors of local operators within the framework of the composite model.
Indeed, if the partial monodromy matrix $T^{(1)}(u)$ corresponds to an interval $[0,x]$, where
$x$ is some fixed point of the interval $[0, L]$, then the partial zero modes are given by the expressions \eqref{14-0Tij}--\eqref{14-0T3j}, in which
one should replace everywhere  $L$ by $x$. In particular, we obtain
\be{14-ZM-TCBG}
\begin{aligned}
&\Psi^\dagger(x)\Psi(x)=-\frac{d}{dx}A^{(1)}[0]=\frac1{i\varkappa}\frac{d}{dx}\lim_{u\to\infty} u(A^{(1)}(u)-1), \\
&\Psi^\dagger(x)=i\sqrt{\varkappa}\;B^{(1)}_{(\text{\tiny R})}[0]=\frac1{\sqrt{\varkappa}}\lim_{u\to-i\infty}e^{-iux}  u \;B^{(1)}(u)\\
 &\Psi(x)=i\sqrt{\varkappa}\;C^{(1)}_{(\text{\tiny R})}[0]=\frac1{\sqrt{\varkappa}}\lim_{u\to+i\infty} u \;C^{(1)}(u).
\end{aligned}
\ee
Thus, we again reduced the problem of calculating form factors of local operators to the cal\-cu\-la\-ti\-on of the form factors of partial zero modes.

\section{Form factors of local operators\label{14-Sec3}}

\subsection{Form factor of operator $\Psi^\dagger(x)\Psi(x)$\label{14-Sec31}}

Since the definition of the partial zero mode  $A^{(1)}[0]$ is the same as the one considered in
lectures~\ref{CHA-ZM}  and~\ref{CHA-FFLO},
we can immediately write formula \eqref{13-FF11}
for the corresponding form factor.  We need only to substitute concrete functions in this expression.
In the QNLS model, for given normalization of the  $L$-operator, we have
\be{14-adr}
a(u)=1,\qquad d(u)= e^{iLu}, \qquad r(u)= e^{-iLu}, \qquad \ell(u)= e^{-ixu}.
\ee
From this we find $a[0]=0$.

Twisted Bethe equations with the twist parameters $\kappa_1=e^{\beta_1}$ and $\kappa_2=1$ have the form
\be{14-Betw}
e^{-iLv_j}=e^{-\beta_1}\frac{f(v_j,\bv_j)}{f(\bv_j,v_j)}.
\ee
We see that the dependence of $v_j$ on $\beta_1$ is very simple:
$v_j=u_j-i\beta_1/L$, where parameters  $u_j$ solve ordinary Bethe equations. This implies
\be{14-der-ell}
\frac{d}{d\beta_1}\ln \ell(v_j)=-\frac xL.
\ee

Thus, if $\bv\ne\bu$, then taking into account the first equation \eqref{14-ZM-TCBG} we obtain
\begin{equation}\label{14-FF11-diff}
\langle0| \bs{C}(\bv) \;\Psi^\dagger(x)\Psi(x)\; \bs{B}(\bu)|0\rangle=-i\mathcal{P}_{11}(\bu;\bv)e^{ix\mathcal{P}_{11}(\bu;\bv)}\;
\mathfrak{F}_n^{(11)}(\bv|\bu),
\end{equation}
where
\be{14-mom}
\mathcal{P}_{11}(\bu;\bv)=\sum_{i=1}^n(u_i-v_i).
\ee
On the other hand, if $\bv=\bu$, then
\be{14-FF11-equi}
\langle0| \bs{C}(\bu) \;\Psi^\dagger(x)\Psi(x)\; \bs{B}(\bu)|0\rangle= \frac nL\langle0| \bs{C}(\bu)  \bs{B}(\bu)|0\rangle,
\ee
where $n=\#\bu$.

\section{Form factors of fields $\Psi(x)$  and $\Psi^\dagger(x)$\label{14-Sec4}}

According to equations \eqref{14-ZM-TCBG}, for calculating form factors of the fields $\Psi(x)$  and $\Psi^\dagger(x)$, one should
compute form factors of the partial zero modes $C^{(1)}_{(\text{\tiny R})}[0]$ and $B^{(1)}_{(\text{\tiny R})}[0]$.
Consider an example of the first of these two operators
\begin{equation}
{\sf M}_n^{(21)}(\bv;\bu)=\langle0| \bs{C}(\bv)\;C^{(1)}_{(\text{\tiny R})}[0]\;\bs{B}(\bu)|0\rangle.\label{14-defFF-ZM21}
\end{equation}
Recall the here both vectors are on-shell. Substituting their expressions in terms of the partial vectors we obtain
\begin{multline}\label{14-M21-rep}
{\sf M}^{(21)}_n(\bv;\bu)=\sum_{\substack{\bu\mapsto\{\bu_{\so},\bu_{\st}\}  \\   \bv\mapsto\{\bv_{\so},\bv_{\st}\}}}
\frac{\ell(\bv_{\st})}{\ell(\bu_{\so})}f(\bv_{\so},\bv_{\st})f(\bu_{\so},\bu_{\st})
\;\langle0|^{(1)}\bs{C}^{(1)}(\bv_{\so}) \;C^{(1)}[0]\;  \bs{B}^{(1)}(\bu_{\so})|0\rangle^{(1)}  \\
\times \langle0|^{(2)}\bs{C}^{(2)}(\bv_{\st}) \bs{B}^{(2)}(\bu_{\st})|0\rangle^{(2)},
\end{multline}
The action of the operator $C^{(1)}_{(\text{\tiny R})}[0]$ onto the dual vector is given by
\be{14-actC}
\langle0|^{(1)}\bs{C}^{(1)}(\bv_{\so})C^{(1)}_{(\text{\tiny R})}[0]=\lim_{w\to+i\infty}\tfrac wc \langle0|^{(1)}\bs{C}^{(1)}(\bv_{\so})C^{(1)}(w)
=\lim_{w\to+i\infty}\tfrac wc\; e^{iwx}\langle0|^{(1)}\bs{C}^{(1)}(\bv_{\so})\bs{C}^{(1)}(w).
\ee
Actually, the main difference from the case considered in the previous lecture is precisely in the formula \eqref{14-actC}. Recall that
$\bs{C}^{(1)}(w)=C^{(1)}(w)/d^{(1)}(w)$. Earlier we considered models in which the function $d^{(1)}(w)$ tended to unity as $w\to\infty$. Therefore,
the actions of the operators $\bs{C}^{(1)}(w)$ and $C^{(1)}(w)$ on dual vectors  were the same in this limit. Now, $d^{(1)}(w)=e^{iwx}$, therefore,
we must take this factor into account when replacing $C^{(1)}(w)$ by $\bs{C}^{(1)}(w)$. In addition, one  should keep in mind that now
$w$ approaches the  limit in a specific direction, namely $w\to+i\infty$. However, despite these differences, we  still
reduce the matrix element of the partial zero mode $C^{(1)}_{(\text{\tiny R})}[0]$
to the sum of the scalar products of the partial Bethe vectors:
\begin{equation}\label{14-M21-act}
{\sf M}^{(21)}_n(\bv;\bu)=\lim_{w\to+i\infty}\tfrac wc\; e^{iwx}\sum_{\substack{\bu\mapsto\{\bu_{\so},\bu_{\st}\}  \\   \bv\mapsto\{\bv_{\so},\bv_{\st}\}}}
\frac{\ell(\bv_{\st})}{\ell(\bu_{\so})}f(\bv_{\so},\bv_{\st})f(\bu_{\so},\bu_{\st})
 S_{n_{\so}}^{(1)}(\{\bv_{\so},w\}|\bu_{\so})S_{n_{\st}}^{(2)}(\bv_{\st}|\bu_{\st}).
\end{equation}

As before, we introduce a set of variables $\bet=\{\bv,w\}$. Then we can rewrite equation \eqref{14-M21-act} as
\begin{equation}\label{14-M21-act2}
{\sf M}^{(21)}_n(\bv;\bu)=\lim_{w\to\infty}\tfrac wc\; e^{iwx}\sum_{\substack{\bu\mapsto\{\bu_{\so},\bu_{\st}\}  \\   \bet\mapsto\{\bet_{\so},\bet_{\st}\}
\\ w\in \bet_{\so}}}
\frac{\ell(\bet_{\st})}{\ell(\bu_{\so})}f(\bet_{\so},\bet_{\st})f(\bu_{\so},\bu_{\st})
 S_{n_{\so}}^{(1)}(\bet_{\so}|\bu_{\so})S_{n_{\st}}^{(2)}(\bet_{\st}|\bu_{\st}),
\end{equation}
where the sum is taken over partitions with an additional restriction $w\in \bet_{\so}$. Further calculations completely repeat the ones that
we did in section~\ref{13-Sec4},
and eventually we arrive at the following result:
\be{14-FF-Om1}
{\sf M}^{(21)}_n(\bv;\bu)=\lim_{w\to+i\infty}\tfrac wc\; e^{iwx}\left(\frac{\ell(\bet)}{\ell(\bu)}-1\right)
 \Omega_1(\bet|\bu),
\ee
where
\be{14-Om1-2}
 \Omega_1(\bet|\bu)=
 \sum_{\substack{ \bu\mapsto\{\bu_{\so},\bu_{\st}\} \\ \bet\mapsto\{\bet_{\so},\bet_{\st}\}\\ w\in \bet_{\so} } }
 K_{n_{\so}}( \bu_{\so}|\bet_{\so})K_{n_{\st}}( \bet_{\st}|\bu_{\st})\,
 f(\bet_{\so},\bet_{\st})f(\bu_{\st},\bu_{\so}).
 \ee
It remains to take the limit $w\to+i\infty$. Obviously,
\be{14-lim-ell}
\lim_{w\to+i\infty} e^{iwx}\left(\frac{\ell(\bet)}{\ell(\bu)}-1\right)=\frac{\ell(\bv)}{\ell(\bu)}=e^{ix\mathcal{P}_{21}(\bu;\bv)},
\ee
where
\be{14-mom21}
\mathcal{P}_{21}(\bu;\bv)=\sum_{i=1}^nu_i-\sum_{i=1}^{n-1}v_i.
\ee
The remaining limit has been already calculated
\be{14-lim-OM}
\lim_{w\to+i\infty}\tfrac wc\;  \Omega_1(\bet|\bu)=\mathfrak{F}^{(21)}(\bv;\bu),
\ee
since $\Omega_1(\bet|\bu)$ is a rational function of  $w$, and hence, its limit does not depend on how $w$
approaches its limiting value. Finally, taking into account \eqref{14-ZM-TCBG}, we obtain
\begin{equation}\label{14-FF21}
\langle0| \bs{C}(\bv) \;\Psi(x)\; \bs{B}(\bu)|0\rangle=i\sqrt{\varkappa}\;e^{ix\mathcal{P}_{21}(\bu;\bv)}\;
\mathfrak{F}_n^{(21)}(\bv|\bu).
\end{equation}

The form factor of the field $\Psi^\dagger(x)$ is computed in the same way. It is not difficult to guess that the final result has the form
\begin{equation}\label{14-FF12}
\langle0| \bs{C}(\bv) \;\Psi^\dagger(x)\; \bs{B}(\bu)|0\rangle=i\sqrt{\varkappa}\;e^{ix\mathcal{P}_{12}(\bu;\bv)}\;
\mathfrak{F}_n^{(12)}(\bv|\bu),
\end{equation}
where now
\be{14-mom12}
\mathcal{P}_{12}(\bu;\bv)=\sum_{i=1}^nu_i-\sum_{i=1}^{n+1}v_i.
\ee

Let us summarize. We have seen that the definition of the zero modes of certain operators in the QNLS model
differs (and very significantly) from the definition of zero modes in models in which the monodromy matrix goes to the identity operator
with an infinite value of its argument. Nevertheless, these differences do not in any way affect the basic calculation: the summation over
partitions. It is for this reason that we did not give details of these calculations. The differences in the definitions of zero modes are affected
only at the final stage, when we take the limit.


\section{Integrals of oscillating functions\label{14-A-OI}}

Let us consider the simplest example of an integral of a rapidly oscillating function with respect to a final interval
 \be{14-J1}
 J_1(u)=\int_{0}^Le^{iu x}\Phi(x)\,dx.
 \ee
Assume that the parameter $u$ is real and large enough, and the function $\Phi(x)$ is holomorphic within some domain containing the interval $[0,L]$.
Integrating twice by parts we find
 \be{14-J1-parts}
 J_1(u)=
 \frac{\Phi(L)}{iu}\;e^{iL u}-\frac{\Phi(0)}{iu}-\frac{\Phi'(L)}{(iu)^2}\;e^{iL u}+\frac{\Phi'(0)}{(iu)^2}
 +\frac1{(iu)^2}\int_{0}^Le^{iu x}\Phi''(x)\,dx.
 \ee
It is clear that
\be{14-est-int}
\left|\int_{0}^Le^{iu x}\Phi''(x)\,dx\right|\le \int_{0}^L\left|\Phi''(x)\right|\,dx\le L \max_{x\in[0,L]}\left|\Phi''(x)\right|.
\ee
Therefore,
 \be{14-J1-est}
 J_1(u)=
 \frac{\Phi(L)}{iu}\;e^{iL u}-\frac{\Phi(0)}{iu}+O(u^{-2}).
 \ee
It is clear that the integration by parts can be continued to obtain a more precise estimate.
In this case, the asymptotics is expressed in terms of the values of the function
$\Phi(x)$ and its derivatives at the endpoints of the interval. One can understand this
in the following way. For large $u$, the contributions from the integrals over
half-period $[x,x+\frac{\pi i}u]$ almost
compensated by contributions from integrals over neighboring half-periods,
since the function $\Phi(x)$ does not change significantly at such small intervals. The absence of
compensation is possible only at the ends of the interval $[0, L]$. This qualitative consideration
explains the form of the asymptotic formula.

Thus, the integral of an oscillating function of the form \eqref{14-J1} behaves at least as\footnote{%
If the function $\Phi(x)$ vanishes at the endpoints of the interval, then the integral \eqref{14-J1} decreases faster than $u^{-1}$.}
$u^{-1}$ at $u\to\infty$.
It is also clear that an $m$-fold integral of the form
 \be{14-Jm}
 J_m(u)=\int_{0}^L\prod_{j=1}^m e^{iu x_j}\;\Phi(x_1,\dots,x_m)\,dx_1\dots\,dx_m
 \ee
should behave like $u^{-m}$, since the integration over each variable should give $u^{-1}$.
This is true, however, provided that the function $\Phi(x_1,\dots,x_m)$ has no singularities or other special points within the domain of integration.
In some cases the situation is slightly more complicated. Let us consider an example of a double integral
\be{14-J2}
J_2=\varkappa\int_0^L e^{iu(x_2-x_1)}\theta(x_2-x_1)\Phi(x_1,x_2)\,dx_1\,dx_2,
\ee
where the function $\Phi(x_1,x_2)$ is smooth in the square $[0,L]\times[0,L]$. The Heaviside function $\theta(x_2-x_1)$ reduces the
integration domain to the subdomain  $x_2> x_1$. It is this kind of integrals that arise in the expressions for  the monodromy matrix entries
in terms of the Bose fields.

Integrating by parts over  $x_2$, we obtain
\begin{multline}\label{14-J2-part1}
J_2(u)=\frac{\varkappa}{iu}\Bigl[e^{iuL}\int_0^L e^{-iux_1}\Phi(x_1,L)\,dx_1\\
-\int_0^L e^{iu(x_2-x_1)}\Bigl(\delta(x_2-x_1)\Phi+\theta(x_2-x_1)
\partial_{x_2}\Phi\Bigr)\,dx_1\,dx_2\Bigr].\\
\end{multline}
The  integral over $x_1$ in the first line of \eqref{14-J2-part1} gives an additional factor $u^{-1}$, thus, this term behaves as $u^{-2}$.
However, in the second line of \eqref{14-J2-part1}, there is an integral containing $\delta(x_2-x_1)$. This $\delta$-function kills
the oscillating exponent, and we get
\be{14-J2-part2}
J_2(u)=-\frac{\varkappa}{iu}\int_0^L \Phi(x,x)\,dx +O(u^{-2}).\\
\ee
Thus, in spite of the fact that $J_2(u)$ is given as the double integral of the oscillating function, it behaves
as $u^{-1}$. The reason for this behavior is the presence of the nonanalytic function $\theta(x_2-x_1)$ in the original integral.

At the same time, triple integrals over the domain $x_3>x_2>x_1$ behave as $u^{-2}$. Let us consider an example of such an integral that
arises in the expansion of the operator $C(u)$:
\be{14-J3}
J_3(u)=i\varkappa^{3/2}e^{iuL}\int_0^L e^{iu(x_2-x_1-x_3)}\theta(x_3-x_2)\theta(x_2-x_1)\Phi(x_1,x_2,x_3)\,dx_1\,dx_2\,dx_3.
\ee
Here the function $\Phi(x_1,x_2,x_3)$ is smooth within the integration domain. Integrating over $x_3$ by parts, we obtain
\begin{multline}\label{14-J3-part1}
J_3(u)=-\frac{\varkappa^{3/2}e^{iuL}}{u}\Bigl[e^{-iuL}\int_0^L e^{iu(x_2-x_1)}\theta(x_2-x_1)\Phi(x_1,x_2,L)\,dx_1\,dx_2\\
-\int_0^L e^{iu(x_2-x_1-x_3)}\theta(x_2-x_1)\Bigl(\delta(x_3-x_2)\Phi+\theta(x_3-x_2)
\partial_{x_3}\Phi\Bigr)\,dx_1\,dx_2\,dx_3\Bigr].\\
\end{multline}
We see that even the term with $\delta$-function does not kill the  oscillating exponent completely, so the remaining integrals give at
least one additional factor of $u^{-1}$. Therefore $J_3(u)$ behaves as  $u^{-2}$. It is also clear that the integrals
of higher multiplicities  decrease faster than $u^{-1}$.


%
%

\chapter{Correlation functions\label{CHA-CF}}

In this lecture we consider a question of application of the results obtained earlier  to the calculation of correlation functions.
In this field many issues are still open.

The methods for calculating the correlation functions are quite diverse. Often the appli\-ca\-bi\-li\-ty of one or another method depends not only on
a specific model, but also on a specific correlation function. We consider an example of a generating functional,
which allows us to calculate the two-point correlator of the third components of  spins in the $XXZ$ Heisenberg chain at zero temperature.
In a special limit, this generating functional gives one more correlation function, which is called the emptiness formation probability.
We will use the method of form factor expansion, although in this particular example we can apply
purely algebraic methods as well. However, the approach described in this lecture clearly demonstrates the benefit of
determinant representations for the scalar products of Bethe vectors, and also the advantages of the twisted Bethe equations
in comparison with the ordinary ones.

\section{Definition of correlation functions\label{15-Sec1}}

By definition, a correlation function
of an operator $\mathcal{O}$ is
 \begin{equation}\label{15-def-cf}
 \langle\mathcal{O}\rangle_T=\frac{\tr\left(\mathcal{O}e^{-\frac HT}\right)}
 {\tr e^{-\frac HT}}.
 \end{equation}
Here $H$ is the Hamiltonian of the system, $T$ is the temperature, and the trace is taken over the space of states.
If we chose the system of the Hamiltonian eigenfunctions $|\omega_n\rangle$ as a basis in this space, then we can
rewrite equation \eqref{15-def-cf} in the form
 \begin{equation}\label{15-def-cf-1}
 \langle\mathcal{O}\rangle_T=\frac{\sum\limits_n\langle\omega_n|\mathcal{O}|\omega_n\rangle e^{-\frac {E_n}T}}
 {\sum\limits_n\langle\omega_n|\omega_n\rangle e^{-\frac {E_n}T}},
 \end{equation}
where $E_n$ is the Hamiltonian eigenvalue corresponding to the eigenfunction $|\omega_n\rangle$.
At zero tem\-pe\-ra\-tu\-re, in the sums \eqref{15-def-cf-1},
only one term survives, that corresponds to the minimal eigenvalue. Thus, we deal with the average
value in the ground state $|\omega_0\rangle$ (assuming that the ground state is not degenerated)
 \begin{equation}\label{15-def-cf-2}
 \langle\mathcal{O}\rangle=\frac{\langle\omega_0|\mathcal{O}|\omega_0\rangle} {\langle\omega_0|\omega_0\rangle}.
 \end{equation}

We restrict ourselves with the case of zero temperature. Then in the  algebraic Bethe ansatz  solvable models, the correlation
functions have the form of the average of the operator $\mathcal{O} $ with respect to the normalized eigenstate (on-shell vector) $B(\bu)|0\rangle$
 \begin{equation}\label{15-def-cf-ABA}
 \langle\mathcal{O}\rangle=\frac{\langle0|C(\bu)\;\mathcal{O}\;B(\bu)|0\rangle} {\langle 0|C(\bu)B(\bu)|0\rangle},
 \end{equation}
where $\bu$ are the Bethe equations roots  corresponding to the minimal value of the energy.

{\sl Remark.}  In some models (in particular, in the $XXZ$ Heisenberg chain) the algebraic Bethe ansatz is well suited
for calculating correlation functions at finite temperature as well. We do not consider this question in our lectures, but we refer the
the interested reader to the works \cite{GohKS04,GohBKS07,Klu93}.

We already know how to calculate some correlation functions, for example, the average value of the operator $\sigma^z_k$ in the $XXX$ and $XXZ$ chains, or the average of $\Psi^\dagger(x)\Psi(x)$  in the QNLS model. In this case, the vector $ B(\bu)|0\rangle$ can be an arbitrary on-shell
vector, and not necessarily corresponding to the lowest value of the energy. However, the most interesting are two-point correlation functions,
for example, $\langle\sigma^z_{k_1}\sigma^z_{k_2}\rangle $  or $\langle\Psi^\dagger(x_1)\Psi(x_2)\rangle $.
An effective way to calculate such functions is their expansion with respect to the complete set of the intermediate states. If the operator $\mathcal{O}$
depends on the space variable $x$ (continuous or discrete) $\mathcal{O}=\mathcal{O}(x)$, then
\be{15-FF-sum}
\langle\mathcal{O}(x_2)\mathcal{O}(x_1)\rangle=\sum_{\bv}\frac{\langle0|C(\bu)\;\mathcal{O}(x_2)\; B(\bv)|0\rangle
\langle0|C(\bv)\;\mathcal{O}(x_1)\; B(\bu)|0\rangle}
{\langle0|C(\bu) B(\bu)|0\rangle   \langle0|C(\bv) B(\bv)|0\rangle}.
\ee
Here the summation is over the complete set of the intermediate states $B(\bv)|0\rangle$, that is, in fact, over all possible
solutions of the Bethe equations $\bv$. Every term of the sum \eqref{15-FF-sum} is expressed in terms of the norms of on-shell vectors and the form factors of the operator $\mathcal{O}(x)$ of the form $\langle0|C(\bv)\;\mathcal{O}(x_k)\; B(\bu)|0\rangle$.  For all these objects, we have already obtained compact representations in the form of determinants.
It is this circumstance that makes the formula \eqref{15-FF-sum} especially attractive for computer calculations. However, it allows one to obtain
some analytical results as well. This will be demonstrated below.

The expansion \eqref{15-FF-sum} is also convenient for considering correlation functions that depend on time. If the evolution of
the operator $\mathcal{O}(x)$ is defined in the standard way
\be{15-evolut}
\mathcal{O}(x,t)= e^{iHt}\mathcal{O}(x)\;e^{-iHt},
\ee
then a new phase factor appears in the form factor expansion \eqref{15-FF-sum}:
\be{15-FF-sumt}
\langle\mathcal{O}(x_2,t_2)\mathcal{O}(x_1,t_1)\rangle=\sum_{\bv}e^{i(t_2-t_1)(E_u-E_v)}\frac{\langle0|C(\bu)\;\mathcal{O}(x_2)\; B(\bv)|0\rangle
\langle0|C(\bv)\;\mathcal{O}(x_1)\; B(\bu)|0\rangle}
{\langle0|C(\bu) B(\bu)|0\rangle   \langle0|C(\bv) B(\bv)|0\rangle}.
\ee
Here $E_u$ and $E_v$ respectively are the energies of the states $B(\bu)|0\rangle$ and $B(\bv)|0\rangle$.

Strictly speaking, it is the time-dependent correlation functions that are of greatest interest, since they are measured experimentally.
From the point of view of numeric calculations, such time-dependent correlations \eqref{15-FF-sumt} are no more complex than time-independent \eqref{15-FF-sum} ones. However, analytical calculations in this case are encountered with serious technical difficulties, therefore, we will not
add the time dependence and confine ourselves to the correlation functions of the form \eqref{15-FF-sum}. Even in this relatively simple case, there are a lot of questions to solve.

\section{Thermodynamic limit\label{15-Sec2}}

The majority of results on the correlation functions are obtained in the thermodynamic limit. In order to describe
this notion, it is necessary to fix the model. Let this be the $XXZ$ Heisenberg chain with the anisotropy parameter $\Delta$, such
that $|\Delta|<1$. In the thermodynamic limit, the number of sites of the chain $N$ goes to infinity.

{\sl Remark}. A more accurate approach is that the number $N$ is large, but finite. This is all the more justified, since there is no chains
of infinite length. Further, assuming $N$ to be  large enough, all the results for physical quantities need to be expanded in an asymptotic series
over $1/N$, and leave only those terms that remain finite for $N\to\infty$. From the formal point of view, this is equivalent to the fact that
we consider the limit of an infinite chain. However, with this approach, we get rid of many troubles. For example, we do not have
to define the Hamiltonian, which acts in the infinite tensor product of the spaces $\mathbb{C}^2$.
On the contrary, as long as we are dealing with operators, we are always within the framework of linear algebra. And only after
having calculated some matrix element of the operator, we consider its limit at $N\to\infty$.

In the thermodynamic limit, it is necessary to fix the state $B(\bu)|0\rangle$, with respect to which we compute the
correlation functions. As already mentioned, we restrict ourselves with the case of the ground state, that is, the eigenstate of the Hamiltonian corresponding to the lowest eigenvalue of the energy. More details about the construction of the ground state in the $XXZ$ chain  can be found in the
books \cite{Gaud83,KBIr}. We give only the final results without proof.

We consider the limit of a large $N$, therefore, for simplicity we assume that this number is even, since the results for physical observables
for a long length of the chain  should not depend on the parity of $N$. Then in the $XXZ$ model for $|\Delta|<1$ the ground state $B(\bu)|0\rangle$
has vanishing third component of the total spin. This means that in this state, there are $N/2$ spins up and $N/2$ spins down.
Since the number of spins down is equal to the number $n$ of the parameters $u_j$ in the vector $B(\bu)|0\rangle$, we conclude that $n=N/2$.
The parameters $\bu$, as usual, satisfy the system of Bethe equations, that we write in the form
\be{15-BE}
\left(\frac{\sinh(u_j-\tfrac{i\zeta}2)}{\sinh(u_j+\tfrac{i\zeta}2)}\right)^N
\prod_{k=1}^n\frac{\sinh(i\zeta+u_j-u_k)}{\sinh(i\zeta-u_j+u_k)}=(-1)^{n-1}.
\ee
Here $0<\zeta<\pi$, and $\cos\zeta=\Delta$. We recall that in the framework of the algebraic Bethe ansatz, the $XXZ$ chain is described by the trigonometric
$R$-matrix. When describing the trigonometric $R$-matrix, we usually used a parameter $\eta$, so as $\cosh\eta=\Delta$. Since we consider the
case $|\Delta|<1$, then $\eta$ turns out to be purely imaginary. Therefore, it is convenient to make a change of variables $i\eta=\zeta$, taking
$\zeta$ as a real parameter belonging to the interval $0<\zeta<\pi$. It is easy to see that under such an agreement, the familiar form for writing the Bethe equations
\be{15-BE0}
r(u_j)=\frac{f(u_j,\bu_j)}{f(\bu_j,u_j)}
\ee
turns to  \eqref{15-BE}.

Taking the logarithm of the both sides of \eqref{15-BE}, we obtain
\be{15-BE-log}
N\log\frac{\sinh(u_j-\tfrac{i\zeta}2)}{\sinh(u_j+\tfrac{i\zeta}2)}+
\sum_{k=1}^n\log\frac{\sinh(i\zeta+u_j-u_k)}{\sinh(i\zeta-u_j+u_k)}=2\pi i I_j,
\ee
where $I_j$ are integer or half-integer numbers, depending on the parity of $n$. The ground state
corresponds to a set of (half)integers
\be{15-intger}
I_j=j-\frac{n-1}2, \qquad j=1,\dots,n.
\ee
In this case $I_{j+1}-I_j=1$. Moreover, the roots of the Bethe equations $u_j$ turn out to be real
and fill the real axis\footnote{More precisely, in the limit $N\to\infty$, the roots $u_j$ belong to the interval $[-\frac\zeta\pi\log N, \frac\zeta\pi\log N]$.
At the same time, near the ends of the interval, the distance between adjacent roots is not small, but of order $\log N$. However, in the bulk of the interval,
the difference between neighboring roots is indeed of the order of $1/N$. },  so that  $u_{j+1}-u_j\sim1/N$. In this case, instead of
a discrete set of roots, it is natural to introduce a distribution density
\be{15-rho}
\rho(u_j)=\frac1{N(u_{j+1}-u_j)}.
\ee
It is easy to see that the introduced object does have the meaning of the distribution density. Indeed,
calculating the average of a function $\phi(u)$ with respect to the roots of the Bethe equations, we have
\be{15-sum-rho1}
\frac1N\sum_{j=1}^n\phi(u_j)=\sum_{j=1}^n\phi(u_j)\frac{u_{j+1}-u_j}{N(u_{j+1}-u_j)}=
\sum_{j=1}^n\phi(u_j)\rho(u_j)(u_{j+1}-u_j).
\ee
The latter sum is nothing but the integral sum, therefore, in the limit $N\to\infty$
\be{15-sum-rho2}
\frac1N\sum_{j=1}^n\phi(u_j)\longrightarrow\int_{-\infty}^\infty
\phi(u)\rho(u)\,du, \qquad N\to\infty.
\ee
Thus, $\rho(u)$ does have the meaning of the distribution density\footnote{Traditionally, in the Bethe ansatz method, the distribution density is  normalized by the average density, not by $1$. The average density in this case is equal to $n/N=1/2$.}.

It is easy to obtain an integral equation for the distribution density, starting from Bethe equations \eqref{15-BE-log}. To do this,
we consider the difference of the equations for
parameters $u_{j+1}$ and $u_{j}$. Using the fact that the difference $u_{j+1}-u_{j}$ is small, we replace the finite differences by the derivatives. Then,
up to terms of order $O(N^{-1})$
\be{15-BE-log-dif}
N(u_{j+1}-u_j)\frac{d}{du_j}\log\frac{\sinh(u_j-\tfrac{i\zeta}2)}{\sinh(u_j+\tfrac{i\zeta}2)}+
(u_{j+1}-u_j)\frac{d}{du_j}\sum_{k=1}^n\log\frac{\sinh(i\zeta+u_j-u_k)}{\sinh(i\zeta-u_j+u_k)}=2\pi i.
\ee
Let us introduce functions
\be{15-mom-K}
p'_0(u)=\frac{\sin\zeta}{\sinh(u-\tfrac{i\zeta}2)\sinh(u+\tfrac{i\zeta}2)},\qquad
\mathcal{K}(u)=\frac{\sin2\zeta}{\sinh(u-i\zeta)\sinh(u+i\zeta)}.
\ee
Notation $p'_0(u)$ is due to the fact the in the $XXZ$ model, the function $p_0(u)$ plays the role of the naked momentum of the elementary excitation.
Dividing both sides of \eqref{15-BE-log-dif} by $(u_{j+1}-u_j)$, we obtain
\be{15-BE-mom-K}
p'_0(u_j)-\frac1N\sum_{k=1}^n\mathcal{K}(u_j-u_k)=2\pi \rho(u_j).
\ee
It remains to replace the discrete sum with an integral according to  \eqref{15-sum-rho2}:
\be{15-Inteq-rho}
\rho(u)+\frac1{2\pi}\int_{-\infty}^\infty\mathcal{K}(u-v)\rho(v)\,dv=\frac1{2\pi}p'_0(u).
\ee

We obtained a linear integral equation with an integral operator acting on the whole real axis, and with a kernel, depending on the difference of the arguments. Such equations are solved by means of the Fourier transform. Let us demonstrate this with the example of equation
\eqref{15-Inteq-rho}. Let
\be{15-fourier}
\begin{aligned}
&\hat\rho(\lambda)=\int_{-\infty}^\infty e^{iu\lambda}\rho(u)\,du,\\
&\hat{\mathcal{K}}(\lambda)=\frac1{2\pi}\int_{-\infty}^\infty e^{iu\lambda}\mathcal{K}(u)\,du\\
&\hat p'_0(\lambda)=\frac1{2\pi}\int_{-\infty}^\infty e^{iu\lambda}p'_0(u)\,du.
\end{aligned}
\ee
Multiplying equation \eqref{15-Inteq-rho} by  $e^{iu\lambda}$ and integrating over $u$, we have
\be{15-Inteq-rho-sol}
\int_{-\infty}^\infty e^{iu\lambda}\rho(u)\,du+\frac1{2\pi}\int_{-\infty}^\infty e^{i(u-v)\lambda}e^{iv\lambda}\mathcal{K}(u-v)\rho(v)\,du\,dv
=\frac1{2\pi}\int_{-\infty}^\infty e^{iu\lambda}p'_0(u)\,du.
\ee
Changing variable $u-v=u'$ and $v=v'$ in the double integral, we immediately find
\be{15-Inteq-rho-four}
\hat\rho(\lambda)\bigl(1+\hat{\mathcal{K}}(\lambda)\bigr)=\hat p'_0(\lambda),
\ee
what implies
\be{15-Inteq-rho-four1}
\hat\rho(\lambda)=\frac{\hat p'_0(\lambda)}{1+\hat{\mathcal{K}}(\lambda)}.
\ee

Thus, to solve the equation \eqref{15-Inteq-rho}, one should find Fourier images
$\hat{\mathcal{K}}(\lambda)$ and $\hat p'_0(\lambda)$. They can be easily computed. Consider an example of calculating $\hat{\mathcal{K}}(\lambda)$:
\be{15-calc-K1}
\hat{\mathcal{K}}(\lambda)=\frac1{2\pi}\int_{-\infty}^\infty \frac{e^{iu\lambda}\sin2\zeta }{\sinh(u-i\zeta)\sinh(u+i\zeta)}\,du.
\ee
Let $\Lambda$ be some positive number. Let us draw a contour $\Gamma$, which is a rectangle with vertices in points $-\Lambda$, $\Lambda$, $\Lambda+i\pi$,
and $-\Lambda+i\pi$. Consider an integral over this contour in the positive direction
\be{15-calc-K10}
J(\lambda,\Lambda)=\frac1{2\pi}\oint_{\Gamma} \frac{e^{iu\lambda}\sin2\zeta }{\sinh(u-i\zeta)\sinh(u+i\zeta)}\,du.
\ee
Since the integrand exponentially decreases at $u\to\infty$, the integrals over the vertical intervals $[\Lambda,\Lambda+i\pi]$ and
$[-\Lambda+i\pi,-\Lambda]$ go to zero at $\Lambda\to\infty$. In the same limit, the integral over $[-\Lambda,\Lambda]$ goes to
$\hat{\mathcal{K}}(\lambda)$. Finally, due to  $\sinh(u)=-\sinh(u+i\pi)$, the integral over $[\Lambda+i\pi,-\Lambda+i\pi]$
reduces to the integral over  $[-\Lambda,\Lambda]$:
\be{15-calc-K2}
\frac1{2\pi}\int_{\Lambda+i\pi}^{-\Lambda+i\pi} \frac{e^{iu\lambda}\sin2\zeta }{\sinh(u-i\zeta)\sinh(u+i\zeta)}\,du=
-e^{-\pi\lambda}\frac1{2\pi}\int_{-\Lambda}^\Lambda \frac{e^{iu\lambda}\sin2\zeta }{\sinh(u-i\zeta)\sinh(u+i\zeta)}\,du.
\ee
Hence,
\be{15-KJ}
\lim_{\Lambda\to\infty}J(\lambda,\Lambda)=\left(1-e^{-\pi\lambda}\right)\hat{\mathcal{K}}(\lambda),
\ee
while the integral over the closed contour $\Gamma$ is equal to the sum of residues within the contour
\be{15-calc-K3}
\hat{\mathcal{K}}(\lambda)\left(1-e^{-\pi\lambda}\right)=i\Res\frac{e^{iu\lambda}\sin2\zeta}{\sinh(u-i\zeta)\sinh(u+i\zeta)}\Bigr|_{\substack{u=i\zeta\\ u=i\pi -i\zeta}}=
e^{-\zeta\lambda}-e^{-(\pi-\zeta)\lambda} .
\ee
From this we find
\be{15-calc-K4}
\hat{\mathcal{K}}(\lambda)=\frac{\sinh\lambda(\tfrac\pi2-\zeta)}{\sinh\tfrac{\pi\lambda}2}.
\ee

Similarly, we can compute
\be{15-calc-p0}
\hat p'_0(\lambda)=\frac{\sinh\tfrac\lambda2(\pi-\zeta)}{\sinh\tfrac{\pi\lambda}2},
\ee
and then, due to \eqref{15-Inteq-rho-four1}
\be{15-calc-rho}
\hat \rho(\lambda)=\frac{1}{2\cosh\tfrac{\zeta\lambda}2}\;.
\ee
Making the inverse Fourier transform we finally obtain
\be{15-rho-res}
\rho(u)=\frac{1}{2\zeta\cosh\tfrac{\pi u}\zeta}\;.
\ee

\subsection{Resolvent of integral operator\label{15-Sec21}}

We introduce a kernel $\mathcal{R}(u-v)$ of the resolvent by an integral equation
\be{15-inteqRes}
\mathcal{R}(u-v)+\frac1{2\pi}\int_{-\infty}^\infty\mathcal{K}(u-w)\mathcal{R}(w-v)\,dw=\frac1{2\pi}\mathcal{K}(u-v).
\ee
The operator $\delta(u-v)-\mathcal{R}(u-v)$ is inverse to the operator  $\delta(u-v)+\frac1{2\pi}\mathcal{K}(u-v)$,
hence,
\be{15-inteqInv}
p'_0(u)-\int_{-\infty}^\infty\mathcal{R}(u-v)p'_0(v)\,dv=2\pi\rho(u).
\ee

Fourier image of the kernel $\mathcal{R}(u)$ is expressed in terms of the Fourier image of $\mathcal{K}(u)$ via
\be{15-Fur-Res1}
1-\hat{\mathcal{R}}(\lambda)= \frac1{1+ \hat{\mathcal{K}}(\lambda)},
\ee
Knowing the explicit expression for $\hat{\mathcal{K}}(\lambda)$, we easily find
\be{15-Fur-Res2}
\hat{\mathcal{R}}(\lambda)= \frac{\hat{\mathcal{K}}(\lambda)}{1+ \hat{\mathcal{K}}(\lambda)}=
\frac{\sinh\lambda(\tfrac\pi2-\zeta)  }{2\cosh\tfrac{\zeta\lambda}2\sinh\lambda(\tfrac{\pi-\zeta}2)}
\ee
Now, in order to find an explicit formula for $\mathcal{R}(u-v)$, it is sufficient to make the inverse Fourier transform. However,
the result can be expressed in terms of the elementary functions only if $\pi/\zeta=q/p$, where  $q$ and $p$ are integers.

Let us give one more equation that will be used later for calculating correlation functions.
Since\footnote{We recall the definition of the function $t(u,v)$ in formula \eqref{15-gfht} on page~\pageref{15-Page1}.}
\be{15-p0-t}
p'_0(u)=-\frac1{\sin\zeta}\;t(u,-\tfrac{i\zeta}2),
\ee
one can rewrite equation \eqref{15-inteqInv} in the form
\be{15-Inteq-rhot01}
t(u,-\tfrac{i\zeta}2)-\int_{-\infty}^\infty\mathcal{R}(u-v)t(v,-\tfrac{i\zeta}2)\,dv=-2\pi\sin\zeta\;\rho(u)
\ee
Shifting  $u$ here by some $w$, we obtain
\be{15-Inteq-rhot02}
t(u-w,-\tfrac{i\zeta}2)-\int_{-\infty}^\infty\mathcal{R}(u-v)t(v-w,-\tfrac{i\zeta}2)\,dv=-2\pi\sin\zeta\;\rho(u-w).
\ee
This transformation is possible, because the kernel of the integral operator  $\mathcal{R}(u-v)$ depends on the difference of the arguments and
acts on the real axis. The shift $w$ also can be imaginary, but in this case one should take care about condition $-\tfrac{\zeta}2<\Im(w)<\tfrac{\zeta}2$.
This is because we should not cross the poles of the function $t(u,-\tfrac{i\zeta}2)$ when shifting the variable $u$.

Using the fact that $t(u-w,-\tfrac{i\zeta}2)=t(u,w-\tfrac{i\zeta}2)$, and setting  $w-\tfrac{i\zeta}2=z$, such that $-\zeta<\Im(z)<0$, we obtain
\be{15-inteqtuz}
t(u,z)-\int_{-\infty}^\infty\mathcal{R}(u-v)t(v,z)\,dv=-2\pi\sin\zeta\;\rho(u-z-\tfrac{i\zeta}2)=
\frac{-i\pi\sin\zeta}{\zeta\sinh\frac\pi\zeta(u-z)}.
\ee
This equation will be used later for calculating the thermodynamic limit of correlation functions.

\section{Solutions of twisted Bethe equations\label{15-Sec3}}

\subsection{Preliminary remarks\label{15-Sec31}}

Considering the $XXZ$ model, we are permanently dealing with hyperbolic functions that have the property of periodicity.
For example, the Bethe equations \eqref{15-BE} are $i\pi$-periodic in any variable $u_j$. Solutions that differ from each other only
by shifting one or more $u_j$ to $i\pi$, describe the same on-shell vector, thus, it is natural  to identify such
solutions. In order to avoid the ambiguity associated with the periodicity, we  assume that the imaginary part of the parameters $u_j$ belongs to the strip
$-\pi/2<\Im(u_j)\le \pi/2$. We also agree upon that an analogous condition will be satisfied for other arguments of hyperbolic functions that
will be used later (for example, the inhomogeneities or solutions to the twisted Bethe equations).

We will still use the standard functions $g(u,v)$, $f(u,v)$, $h(u,v)$, and $t(u,v)$, but sometimes
we will write them in an explicit \label{15-Page1} form:
\be{15-gfht}
\begin{aligned}
g(u,v)&=\frac{-i\sin\zeta}{\sinh(u-v)},\qquad\qquad &f(u,v)=\frac{\sinh(u-v-i\zeta)}{\sinh(u-v)},\\[2mm]
h(u,v)&=\frac{\sinh(u-v-i\zeta)}{-i\sin\zeta},\qquad\qquad &t(u,v)=\frac{-\sin^2\zeta}{\sinh(u-v)\sinh(u-v-i\zeta)}.
\end{aligned}
\ee
We will also use the convention on the shorthand notation for the products, if it does not lead to misunderstanding. However, sometimes
we will have to abandon this form of writing and write down all the products explicitly.

\subsection{Inhomogeneous model\label{15-Sec32}}

As already noted, summing the form factors over the intermediate states, we actually have to calculate the sums over the solutions of the Bethe equations.
At the same time, we need to solve two problems. First, we need to make sure that the roots of the Bethe equations  really
parameterize a complete set of intermediate states. Secondly, we need to get rid of spurious solutions, because, as we have already seen,
not every solution of the Bethe equations corresponds to an eigenvector of the transfer matrix.

To solve these problems, it is convenient to consider an inhomogeneous $XXZ$ chain\footnote{Of course,  there is no the notion of the ground state
in this model, because the eigenvalues of the Hamiltonian, generally speaking, are not real, and hence, there is no the lowest among them. However, up to a certain point we shall consider the average values of the operators with respect to an arbitrary on-shell vector, and only at the final stage we will
pass to the homogeneous limit. Only at this stage we require that the on-shell vector under consideration be the ground state of the Hamiltonian.} with inhomogeneities  $\xi_1,\dots,\xi_N$.  Then
\be{15-adr}
a(u)=\prod_{j=1}^N\sinh(u-\xi_j-i\zeta),\qquad d(u)=\prod_{j=1}^N\sinh(u-\xi_j),\qquad
r(u)=\prod_{j=1}^N\frac{\sinh(u-\xi_j-i\zeta)}{\sinh(u-\xi_j)}.
\ee
In the end, we will be interested in the homogeneous limit $\xi_k\to-i\zeta/2$, however, at the initial stage it is useful to
consider all $\xi_k$ different. Nevertheless,  we can require from the very beginning that $\xi_k$ be in a small neighborhood of the point $-i\zeta/2$,
so that $\xi_j-\xi_k\pm i\zeta\ne 0$ for all $j$ and $k$.

Instead of the usual Bethe equations, we will consider their twisted version. Indeed, in the formula \eqref{15-FF-sum}, we are not obliged to
take the sum over the basis of the eigenstates of the Hamiltonian, and instead we can insert any other complete system of vectors\footnote{%
This argument does not work for time-dependent correlation functions. However, in this case it is possible to deform the initial Hamiltonian so as to make
twisted on-shell vectors its eigenstates (see \cite{KitMST05b}). }. In particular, it can be twisted on-shell vectors.
As we shall see, it is for the system of twisted on-shell vectors that it is easy to prove the completeness, while
there are serious problems to prove the completeness of the system of the ordinary on-shell vectors.

Recall that the eigenvalue $\tau_\kappa(x|\bar y)$ of the twisted transfer matrix $\mathcal{T}_\kappa(x)$ on the twisted
on-shell vector  $B(\bar y)|0\rangle$ has the form
\be{15-deftau-k}
\tau_\kappa(x|\bar y)= a(x)f(\bar y,x)+\kappa\;d(x)f(x,\bar y).
\ee
Let us introduce a function
\be{15-defY}
Y_\kappa(x|\bar y)=\frac{\tau_\kappa(x|\bar y)}{g(x,\bar y)}=(-1)^n a(x)h(\bar y,x)+\kappa\;d(x)h(x,\bar y).
\ee
Then we can write the twisted Bethe equations in the form
\be{15-BE-Yv}
Y_\kappa(v_j|\bv)=0,\qquad j=1,\dots n,
\ee
or, more detailed
\be{15-BE-Yv-expl}
\prod_{k=1}^N\sinh(v_j-\xi_k-i\zeta)\prod_{\ell=1}^n\sinh(v_\ell-v_j-i\zeta)=\kappa\!
\prod_{k=1}^N\sinh(v_j-\xi_k)\prod_{\ell=1}^n\sinh(v_\ell-v_j+i\zeta),\quad j=1,\dots n.
\ee
Recall that according to the results of the work \cite{Tarv95b}, the eigenvectors of the twisted transfer matrix correspond to such solutions of the system
\eqref{15-BE-Yv-expl}, where all the parameters $v_j$ are pairwise distinct, and  both sides of the equations do not vanish for all $j=1,\dots n$
(admissible solutions). Spurious (inadmissible) solutions, therefore, are those for which both sides of the equations \eqref{15-BE-Yv-expl} vanish at least for one $j$. It is easy to describe such solutions. If, for example, $v_1=\xi_s$ and $v_2=\xi_s+i\zeta$ for some $s\in\{1,\dots,N\}$,
then the equations for $j=1,2$ are certainly satisfied, but both sides of each of these equations  being equal to zero.

Another type of inadmissible solution arises with special values of the parameter $\zeta=\pi p/q$, where  $p$ and $q$ are positive integers, with $p<q$
and $q\le n$. It is easy to check that in this case the subset of the parameters $v_1,\dots,v_q$, such that $v_j=w+i\zeta(j-1)$,
satisfies the equations with $j=1,\dots,q$, and
again, both sides of these equations vanish. It is remarkable that the constant $w$ is not fixed, and thus, these solutions are
not isolated from each other.

We will see below that when summing the form factors, the inadmissible solutions described above are automatically cut off. We now turn
to the question of the completeness of the system of intermediate states. If both sides of the Bethe equations do not vanish, then
the roots of the system \eqref{15-BE-Yv-expl} depend on the parameter $\kappa$, that is, $v_j=v_j(\kappa)$. Therefore, we can
treat the twisted Bethe equations as the specification of implicit functions $v_j(\kappa)$. For $\kappa=0$, the system \eqref{15-BE-Yv-expl} turns into
\be{15-BE-Yvk0}
\prod_{k=1}^N\sinh(v_j-\xi_k-i\zeta)\prod_{\ell=1}^n\sinh(v_\ell-v_j-i\zeta)=0,\qquad j=1,\dots n,
\ee
and has obvious admissible solutions $v_j=\xi_{s_j}+i\zeta$, where  $\xi_{s_j}$ is any from the set $\bxi$. In total, thus,
there are $\binom{N}{n}$  sets of the roots $\bv=\{v_1,\dots,v_n\}$, consisting of pairwise distinct numbers. That is, the number of solutions
to the twisted Bethe equations for $\kappa =0$ coincides with the dimension of the subspace with $n$ spins down and $N-n$ spins up.

The matrix of the first derivatives of the system \eqref{15-BE-Yvk0} in the point $v_j=\xi_{s_j}+i\zeta$, ($j=1,\dots,n$) has the form
\be{15-BE-Jacob}
\frac{\partial}{\partial v_m}\prod_{k=1}^N\sinh(v_j-\xi_k-i\zeta)\prod_{\ell=1}^n\sinh(v_\ell-v_j-i\zeta)=
\delta_{jm}\prod_{\substack{k=1\\k\ne s_j}}^N\sinh(\xi_{s_j}-\xi_k)\prod_{\ell=1}^n\sinh(\xi_{s_\ell}-\xi_{s_j}-i\zeta).
\ee
It is obvious that the Jacobian of the system is non-degenerate. It was for this purpose that we considered the inhomogeneous model and imposed the condition
$\xi_j-\xi_k\pm i\zeta\ne 0$. Then all the conditions of the implicit function theorem are satisfied, and we can formulate

\begin{prop}\label{15-PROP1}
There exists $\kappa_0>0$ such that for $|\kappa|<\kappa_0$ the system \eqref{15-BE-Yv-expl} has solutions $v_j=v_j(\kappa)$,
and $v_j(\kappa)$ are analytic functions of $\kappa$, such that $v_j(0)=\xi_{s_j}+i\zeta$.
\end{prop}

Thus, for $|\kappa|$ small enough, the system \eqref{15-BE-Yv-expl} has exactly $\binom{N}{n}$ admissible solutions. We also know
that the twisted on-shell Bethe vectors constructed from these solutions are pairwise orthogonal. Hence, the system of these vectors forms a basis
in the subspace of states with $n$ spins down and $N-n$ spins up.

In the case of a homogeneous chain $\xi_k=-i\zeta/2$, the Jacobian of the system of twisted Bethe equations   becomes degenerate for $\kappa=0$, and
the implicit function theorem becomes inapplicable in its classical formulation. However, there are many generalizations of this theorem,
which allow us to consider this case \cite{KraP13}. Then the assertion about the existence of the necessary number of admissible solutions is made for
a punctured neighborhood $0<|\kappa|<\kappa_0$. Thus, the condition of inhomogeneity of the chain is not fundamental, but only facilitates
the proof of the completeness of the system of twisted on-shell vectors. More important is the presence of the twist parameter $\kappa$.
The implicit function theorem says
nothing  about the size of the neighborhood $|\kappa|<\kappa_0$, and we can not say that the roots $v_j(\kappa)$ can be analytically continued
to the point $\kappa =1$. Consequently, the completeness of the admissible solutions of the twisted Bethe equations does not yet imply the completeness of the admissible solutions of the ordinary Bethe equations. Moreover, we have already seen that for $\kappa=1$ such an effect as the coincidence
of admissible  and inadmissible solutions takes place. This further complicates the task.

The inability to reach the value of $\kappa=1$ may seem like a serious obstacle. Indeed, in most of the examples in which
we used twisted on-shell vectors, to obtain the final result, we needed to take the $\kappa$-derivative at the point $\kappa =1$.
Below we will see how one can solve this problem.

\section{Generating functional\label{15-Sec4}}

As an example, we consider a two-point correlation function of the third components of spin
\be{15-sigsig}
S_{zz}(m)=\frac{\langle0|C(\bu)\; \sigma^z_{1}\sigma^z_{m+1}\; B(\bu)|0\rangle}
{\langle0|C(\bu) B(\bu)|0\rangle},
\ee
where vector  $B(\bu)|0\rangle$ is on-shell. We assume that in the homogeneous limit $\xi_k=-i\zeta/2$ this vector becomes
the ground state of the Hamiltonian of the $XXZ$ chain. However, for the time being we do not specify the roots of the Bethe equations $\bu$.

As in the case of form factors, instead of the correlation function \eqref{15-sigsig}, it is more convenient to deal with a special generating functional.
For this we introduce an operator
\be{15-GenFun}
\mathcal{Q}_{1,m}^{(\kappa)}=\prod_{k=1}^m e^{\frac\beta2(1-\sigma^z_k)}, \qquad  e^\beta= \kappa,
\ee
where $\beta$ is a complex parameter, and, as usual, we denote $e^\beta= \kappa$.

{\sl Remark}. It is easy to check that in the case of the $XXX$ chain the operator $\mathcal{Q}_{1,m}^{(\kappa)}$
is directly related to the first partial zero mode of the operator $D(u)$: $\mathcal{Q}_{1,m}^{(\kappa)}=e^{\beta D^{(1)}[0]}$.

The operator \eqref{15-GenFun}
has a very simple representation in terms of solution of the quantum inverse problem. Indeed, it is easy to see that
\be{15-power}
\left(\frac{1-\sigma^z_k}2\right)^p = \frac{1-\sigma^z_k}2,\qquad p>0.
\ee
Hence,
\be{15-exp-pres}
e^{\frac\beta2(1-\sigma^z_k)}=1+\sum_{p=1}^\infty \frac{\beta^p}{p\,!} \frac{1-\sigma^z_k}2= 1+\left(e^\beta-1\right)\frac{1-\sigma^z_k}2.
\ee
In terms of elementary units $E_k^{11}=\frac{1+\sigma^z_k}2$ and $E_k^{22}=\frac{1-\sigma^z_k}2$,  equation \eqref{15-exp-pres} takes the form
\be{15-exp-pres1}
e^{\frac\beta2(1-\sigma^z_k)}=E_k^{11}+\kappa E_k^{22}.
\ee
Now, using the solution of the quantum inverse problem we obtain
\be{15-exp-pres-tau}
e^{\frac\beta2(1-\sigma^z_k)}=\prod_{j=1}^{k-1}\mathcal{T}(\xi_j)\cdot \bigl((A(\xi_k)+\kappa\;D(\xi_k)\bigr)\cdot\prod_{j=1}^k\mathcal{T}^{-1}(\xi_j)
=\prod_{j=1}^{k-1}\mathcal{T}(\xi_j)\cdot \mathcal{T}_\kappa(\xi_k)\cdot\prod_{j=1}^k\mathcal{T}^{-1}(\xi_j),
\ee
and thus,
\be{15-GenFun-T}
\mathcal{Q}_{1,m}^{(\kappa)}=\prod_{k=1}^m\mathcal{T}_\kappa(\xi_k)\prod_{k=1}^m\mathcal{T}^{-1}(\xi_k).
\ee

On the other hand, we have
\be{15-lat-der}
\mathcal{Q}_{1,m+1}^{(\kappa)}-\mathcal{Q}_{1,m}^{(\kappa)}-\mathcal{Q}_{2,m+1}^{(\kappa)}+\mathcal{Q}_{2,m}^{(\kappa)}=  \left(e^{\frac\beta2(1-\sigma^z_{1})}-1\right)\cdot
\prod_{k=2}^m e^{\frac\beta2(1-\sigma^z_k)}\cdot \left(e^{\frac\beta2(1-\sigma^z_{m+1})}-1\right),
\ee
and hence,
\be{15-lat-der-def}
2\frac{d^2}{d\beta^2}\Bigl[\mathcal{Q}_{1,m+1}^{(\kappa)}-\mathcal{Q}_{1,m}^{(\kappa)}-\mathcal{Q}_{2,m+1}^{(\kappa)}+\mathcal{Q}_{2,m}^{(\kappa)}\Bigr]_{\beta=0}
=  (1-\sigma^z_{1})(1-\sigma^z_{m+1}).
\ee

Let us introduce a generating functional
\be{15-Qm-def0}
Q_m^{(\kappa)}=\frac{\langle0|C(\bu)\mathcal{Q}_{1,m}^{(\kappa)} B(\bu)|0\rangle}{\langle0|C(\bu) B(\bu)|0\rangle}.
\ee
Then it follows from \eqref{15-lat-der-def} that
\be{15-Szz-Q}
S_{zz}(m)=2\frac{d^2}{d\beta^2}\Bigl[Q_{m+1}^{(\kappa)}-2Q_m^{(\kappa)}+Q_{m-1}^{(\kappa)}\Bigr]_{\beta=0}+2\langle \sigma^z\rangle-1,
\ee
where we took into account the translation invariance of $Q_m^{(\kappa)}$. The term $\langle \sigma^z\rangle$ is the average value
of the operator $\sigma^z_k$
\be{15-sz-def0}
\langle \sigma^z\rangle=\frac{\langle0|C(\bu)\;\sigma^z_k\; B(\bu)|0\rangle}{\langle0|C(\bu) B(\bu)|0\rangle}.
\ee
Due to the translation invariance it does not depend on the site $k$ and can be obtained via taking the $\beta$-derivative of the difference $Q_m^{(\kappa)}-Q_{m-1}^{(\kappa)}$ in the point\footnote{Using the quantum inverse problem formulas one can directly reduce
the average value $\langle \sigma^z\rangle$ to the form factors of the operators $A$ and $D$. Besides,
from the symmetry of the ground state with respect to the reversal of the directions of all the spins, it follows that this average value must vanish. Direct calculation confirms this conclusion. In the general case,  $\langle \sigma^z\rangle=1-\tfrac{2n}N$. } $\beta=0$.\label{15-pageSZ}
Thus, we associated the required two-point correlation function with the generating functional $Q_m^{(\kappa)}$.

\section{Summation of form factors\label{15-Sec5}}

We should calculate the following average value:
\be{15-Qm-def}
Q_m^{(\kappa)}=\frac{\langle0|C(\bu)\prod_{k=1}^m\mathcal{T}_\kappa(\xi_k)\prod_{k=1}^m\mathcal{T}^{-1}(\xi_k) B(\bu)|0\rangle}
{\langle0|C(\bu) B(\bu)|0\rangle}.
\ee
Pay attention that here and below we do not use a `normalized' monodromy matrix $\bs{T}(u)=T(u)/d(u)$. The matter is that we will often
deal with such arguments of the function $d(u)$ that it turns to zero. Therefore,  formulas for the scalar products given below differ
from the previous ones by normalization factors.

The action of the operators $\mathcal{T}^{-1}(\xi_k)$ on the on-shell vector $B(\bu)|0\rangle$ is obvious, however, the action of the twisted
transfer matrices $\mathcal{T}_\kappa(\xi_k)$ onto the ordinary on-shell vectors is more complex. In fact, one can compute this action in a relatively
compact form (see \cite{KitMST05}), but we use another method. Let $\kappa$ be small enough.
Then we know that the system of the twisted on-shell Bethe vectors is complete, and we can insert this system between the products of the operators
$\mathcal{T}_\kappa(\xi_k)$ and $\mathcal{T}^{-1}(\xi_k)$:
\be{15-Qm-sum}
Q_m^{(\kappa)}=\sum_{\bv}\frac{\langle0|C(\bu)\prod_{k=1}^m\mathcal{T}_\kappa(\xi_k) B(\bv)|0\rangle
\langle0|C(\bv)\prod_{k=1}^m\mathcal{T}^{-1}(\xi_k) B(\bu)|0\rangle}
{\langle0|C(\bu) B(\bu)|0\rangle   \langle0|C(\bv) B(\bv)|0\rangle}.
\ee
Here the sum is taken over all admissible solutions of the twisted Bethe equations. Now the action of the operators $\mathcal{T}_\kappa(\xi_k)$
on the twisted on-shell Bethe vectors is obvious, and we obtain
\be{15-Qm-sum-eig}
Q_m^{(\kappa)}=\sum_{\bv}\prod_{k=1}^m\frac{\tau_\kappa(\xi_k|\bv)}{\tau(\xi_k|\bu)}\;\frac{\langle0|C(\bu) B(\bv)|0\rangle
\langle0|C(\bv) B(\bu)|0\rangle}{\langle0|C(\bu) B(\bu)|0\rangle   \langle0|C(\bv) B(\bv)|0\rangle}.
\ee
We have an expression that contains scalar products of the twisted and ordinary on-shell vectors, their norms, and
eigenvalues of the usual and twisted transfer matrices. Since $d(\xi_k)=0$ for any  $\xi_k$, we obtain
\be{15-prod-tau}
\prod_{k=1}^m\frac{\tau_\kappa(\xi_k|\bv)}{\tau(\xi_k|\bu)}=\prod_{k=1}^m\prod_{j=1}^n\frac{f(v_j,\xi_k)}{f(u_j,\xi_k)}.
\ee
As for the scalar products in the formula \eqref{15-Qm-sum-eig}, we have determinant representations for them.
Moreover, for the scalar product of the twisted and ordinary on-shell vectors, we have even two representations
(see \eqref{11-SP-det-pres1}, \eqref{11-SP-det-pres2}).
Recall that this is due to the fact that in this type of the scalar product, we have
a choice which of the two vectors is considered to be a twisted on-shell vector, because
the ordinary on-shell vector is a special case of the twisted one at $\kappa=1$.

We need both of these formulas. In order to write them uniformly, we introduce an $n\times n$ matrix $\Omega_\kappa$ with matrix elements
\be{15-Omk-def}
(\Omega_\kappa)_{jk}=\Omega_\kappa(x_j,y_k|\bar x)=(-1)^{n-1}a(y_k)t(x_j,y_k)h(\bar x,y_k)+\kappa\;d(y_k)t(y_k,x_j)h(y_k,\bar x).
\ee
This matrix is defined for any two sets of variables $\bar x$ and $\bar y$, such that $\#\bar x=\#\bar y=n$.
We agree upon to denote the same matrix for $\kappa= 1$ by the symbol $\Omega(x_j,y_k|\bar x)$
(that is, simply omitting the subscript $\kappa$).
Then, if the parameters $\bv$ satisfy the twisted Bethe equations, while $\bu$ are roots of the usual Bethe equations, we have
\be{15-SP1}
\begin{aligned}
\langle0|C(\bu) B(\bv)|0\rangle=\langle0|C(\bv) B(\bu)|0\rangle&\\
&= d(\bv)\Delta'(\bu)\Delta(\bv)\det_n \Omega_\kappa(v_j,u_k|\bv)\\
&=d(\bu)\Delta'(\bv)\Delta(\bu)\det_n \Omega(u_j,v_k|\bu).
\end{aligned}
\ee

The norms of the usual and the twisted on-shell Bethe vectors are the particular cases of the scalar products described above and they can be expressed in terms of the determinants of the same matrices:
\be{15-norm-Om}
\begin{aligned}
&\langle0|C(\bu) B(\bu)|0\rangle=d(\bu)\Delta'(\bu)\Delta(\bu)\det_n \Omega(u_j,u_k|\bu),\\
&\langle0|C(\bv) B(\bv)|0\rangle=d(\bv)\Delta'(\bv)\Delta(\bv)\det_n \Omega_\kappa(v_j,v_k|\bv).
\end{aligned}
\ee
Explicit expressions for the matrices $\Omega$ and $\Omega_\kappa$ in these particular cases will be given in the next section.

Thus, substituting \eqref{15-prod-tau},  \eqref{15-SP1}, and \eqref{15-norm-Om}  into \eqref{15-Qm-sum-eig},
we obtain
\be{15-Qm-sum-Om}
Q_m^{(\kappa)}=\sum_{\bv}\prod_{k=1}^m\prod_{j=1}^n\frac{f(v_j,\xi_k)}{f(u_j,\xi_k)}\;
\frac{\det_n \Omega_\kappa(v_j,u_k|\bv)\;\det_n \Omega(u_j,v_k|\bu)}{\det_n \Omega_\kappa(v_j,v_k|\bv) \; \det_n \Omega(u_j,u_k|\bu)}.
\ee
Note that we used both representations \eqref{15-SP1} for the scalar product  $\langle0|C(\bu) B(\bv)|0\rangle$. In the next section we will see why this was done.

\subsection{Cut off of spurious solutions\label{15-Sec51}}

A question arises: how to distinguish the inadmissible solutions of the twisted Bethe equations and how to get rid of them  in the sum
\eqref{15-Qm-sum-Om}. It turns out that such a cut off of unwanted solutions occurs automatically.
Recall once again the formulas for matrices whose determinants enter in \eqref{15-Qm-sum-Om}. In the numerator we have  determinants of the matrices
\be{15-Omk-vu0}
\Omega_\kappa(v_j,u_k|\bar v)=(-1)^{n-1}a(u_k)t(v_j,u_k)h(\bar v,u_k)+\kappa\;d(u_k)t(u_k,v_j)h(u_k,\bar v),
\ee
and
\be{15-Omk-uv0}
\Omega(u_j,v_k|\bu)=(-1)^{n-1}a(v_k)t(u_j,v_k)h(\bu,v_k)+d(v_k)t(v_k,u_j)h(v_k,\bu).
\ee
The matrices in the denominator are (see \eqref{07-SP-Matel-all}, \eqref{07-SP-Norm})
\be{15-Om-uu}
\Omega(u_j,u_k|\bu)=(-1)^{n} a(u_k)h(\bu,u_k)\sin\zeta\left[\delta_{jk}\left(-i\log'r(u_j)-\sum_{\ell=1}^n\mathcal{K}(u_j-u_\ell)\right)
+\mathcal{K}(u_j-u_k)\right],
\ee
and
\be{15-Om-vv}
\Omega_\kappa(v_j,v_k|\bv)=(-1)^{n} a(v_k)h(\bv,v_k)\sin\zeta\left[\delta_{jk}\left(-i\log'r(v_j)-\sum_{\ell=1}^n\mathcal{K}(v_j-v_\ell)\right)
+\mathcal{K}(v_j-v_k)\right],
\ee
where we used the relation $a(\bv)=\kappa^nd(\bv)$, that follows from the twisted Bethe equations. Note that  $\kappa$  is not explicitly included
in the formula \eqref{15-Om-vv}. It enters only implicitly, through the variables $\bv $.

One can convince himself by a direct calculation that
\be{15-Y-Om}
\Omega_\kappa(v_j,v_k|\bar v)=-i\sin\zeta \frac{\partial Y_\kappa(v_j|\bar v)}{\partial v_k},
\ee
where the function $Y_\kappa(v_j|\bar v)$ is defined by  \eqref{15-defY}. It is important to emphasize that the equality \eqref{15-Y-Om} takes
place on the twisted Bethe equations, that is, after calculating the derivatives, it is necessary to take into account the equation \eqref{15-BE-Yv}.
The relation \eqref{15-Y-Om} will play a very important role in further transformations.

It is immediately evident that if there are coinciding variables among $\bv$, then matrices \eqref{15-Omk-vu0} and \eqref{15-Omk-uv0} have matching
rows or columns. At the same time, the determinant of the matrix $\Omega_\kappa(v_j,v_k|\bv)$ does not vanish because of the nontrivial diagonal
parts. Hence, the summation in the formula \eqref{15-Qm-sum-Om} occurs only with respect to solutions in which all $v_j$ are pairwise distinct.

Now let $v_1=\xi_s$ and $v_2=\xi_s+i\zeta$. Let us prove that the determinant of the matrix $\Omega(u_j,v_k|\bu)$ then vanishes.
Indeed, with this choice, we have $d(v_1)=0$ and $a(v_2)=0$. Then the first and second columns of the matrix $\Omega(u_j,v_k|\bu)$ have the following
form:
\be{15-Omv1v2}
\begin{aligned}
&\Omega(u_j,v_1|\bu)=(-1)^{n-1}a(\xi_s)t(u_j,\xi_s)h(\bu,\xi_s),\\
&\Omega(u_j,v_2|\bu)=d(\xi_s+i\zeta)t(\xi_s+i\zeta,u_j)h(\xi_s+i\zeta,\bu).
\end{aligned}
\ee
But $t(\xi_s+i\zeta,u_j)=t(u_j,\xi_s)$, and we see that these two columns are proportional to each other. Hence, the determinant vanishes.

Finally, let $\zeta=\pi p/q$, and we have the parameters $v_1,\dots,v_q$, such that $v_j=w+i\zeta(j-1)$. In this case
it is easier to prove that  there are linearly dependent rows in the matrix $\Omega_\kappa(v_j,u_k|\bar v)$. Indeed, it is easy
see that in this case
\be{15-sum-t}
\sum_{j=1}^q t(v_j,u_k)=\sum_{j=1}^q t(u_k,v_j)=0.
\ee
To prove this statement it enough to present the function $t(x,y)$ in the form
\be{15-t-cth-cth}
t(x,y)=i\sin\zeta\bigl(\coth(x-y-i\zeta)-\coth(x-y)\bigr),
\ee
and substitute this expression into the sums \eqref{15-sum-t} with $v_j=w+i\zeta(j-1)$. Since the entries of the matrix
$\Omega_\kappa$ have the structure
\be{15-Omk-vu-short}
\Omega_\kappa(v_j,u_k|\bar v)=\alpha_k t(v_j,u_k)+\beta_k t(u_k,v_j),
\ee
where $\alpha_k$ and $\beta_k$ only depend on the number of the column, we conclude that
\be{15-sum-O}
\sum_{j=1}^q \Omega_\kappa(v_j,u_k|\bar v)=0.
\ee
Hence,
the determinant of $\Omega_\kappa(v_j,u_k|\bar v)$ vanishes for these solutions.

Thus, the formulas for scalar products turn out to be sufficiently `smart' and turn to zero for inadmissible solutions of the twisted Bethe equations.
Therefore, the terms corresponding to inadmissible solutions do not contribute to the sum of \eqref{15-Qm-sum-Om}. Consequently, the
summation automatically occurs only on the admissible roots.

{\sl Remark}. We have just seen the usefulness of different representations for scalar products. Despite the fact that
the determinants of the matrices $\Omega(u_j,v_k|\bu)$ and $\Omega_\kappa(v_j,u_k|\bar v)$
describe the same scalar product, the properties of the matrix elements are different. Therefore, for one type of inadmissible solution, we could easily prove the linear dependence of rows in the matrix $\Omega(u_j,v_k|\bu)$, and for another type the linear dependence
of columns in the matrix $\Omega_\kappa(v_j,u_k|\bar v)$.
Below, after we express the sum \eqref{15-Qm-sum-Om} through a multidimensional contour integral,
we will see that the properties  of these two matrices differ even more.

\subsection{Summation over roots of equations\label{15-Sec52}}

Thus, we reduced the generating functional $Q_m^{(\kappa)}$ to the sum over the solutions of the twisted Bethe equations.
We cannot  find explicitly the roots $v_j$, but it is not necessary. The sum over the roots of a system of equations can be reduced to a multiple contour integral of the Cauchy type.

Consider an example of an algebraic equation. Let we have a polynomial $P_M(z)$ of degree $M$, and let $x_1,\dots,x_M$ be the roots of this polynomial:
\be{15-polyn}
P_M(x_j)=0, \qquad j=1,\dots,M.
\ee
Suppose that a function $\phi(x)$ is analytic in some domain containing the roots $x_j$. We want to sum up this function with respect to the roots $x_j$
\be{15-sum}
S_\phi=\sum_{j=1}^M \phi(x_j).
\ee
It is easy to see that the sum \eqref{15-sum} can be written as a contour integral
\be{15-sum-int}
S_\phi=\oint_{\Gamma(\bar x)}\frac{dz}{2\pi i}\frac{\phi(z)P'_M(z)}{P_M(z)}.
\ee
Here $\Gamma(\bar x)$ is a positively oriented closed contour lying in the region of analyticity of the function $\phi(z)$,
surrounding the roots $\bar x$ and not containing any other singularities of the integrand within it. The calculation of the integral \eqref{15-sum-int} by
the residues in the poles within the contour returns us to the sum \eqref{15-sum}.
However, we can take the integral also by residues outside the contour. Singular points of the integrand outside the contour are special points
of the function $\phi(z)$. Thus, we pass from the  contour integral
around the unknown poles $x_j$ to the integral around the singularities of the known function $\phi(z)$.

Let, for example, $\phi(z)=1/(z-a)$, where $a$ is such that $P_M(a)\ne 0$. Then
\be{15-sum-res1}
S_\phi=\oint_{\Gamma(\bar x)}\frac{dz}{2\pi i}\frac{P'_M(z)}{(z-a)P_M(z)}.
\ee
The integrand behaves as  $z^{-2}$ at $z\to\infty$, thus, the residue at infinity is zero. Therefore, the integral over
the original contour reduces to an integral over a closed contour around the point $a$ with the opposite sign:
\be{15-sum-res2}
S_\phi=
-\oint_{\Gamma(a)}\frac{dz}{2\pi i}\frac{P'_M(z)}{(z-a)P_M(z)} = -\frac{P'_M(a)}{P_M(a)}.
\ee
Thus, we have calculated the sum $S_\phi$ without solving the equation $P_M(z)=0$.

The fact that we considered $P_M(z)$ as a polynomial is not a fundamental one. It is clear that the representation of \eqref{15-sum-int}
remains valid even if $P_M(z)$ is not a polynomial. Therefore, we can always go from integrating around the domain,
containing the roots of the equation, to integration around the outer region. However, if outside the contour the integrand
has a complex analytic structure (essential singularities, cuts, etc.), it can
turn out that horseradish is not sweeter than radishes, and such a transition does not give anything.

The method described above can be naturally generalized to a sum over roots of a system of equations. Let a system of equations be given
\be{15-sys}
F_j(\bar x)=0,\qquad j=1,\dots,n.
\ee
Here $F_j$ are some $n$-variables functions, and $\bar x =\{x_1,\dots,x_n\}$ are solutions to the system \eqref{15-sys}.
We can formulate a problem of calculating the sum of some function $\phi(\bar z)$ over all possible solutions $\bar x$.
Then
\be{15-sum-sys}
\sum_{\bar x} \phi(\bar x)=\oint_{\Gamma(\bar x)}\phi(\bar z)
\det_n\left(\frac{\partial F_j(\bar z)}{\partial z_k}\right)\;\prod_{j=1}^n\frac{dz_j}{2\pi i F_j(\bar z)}.
\ee
Here  $\Gamma(\bar x)$ is a multidimensional closed contour (the boundary of a polydisc) that encompasses all solutions of the system \eqref{15-sys}
and does not contain any other singularities of the integrand within it. Now the denominator is the product
of the functions $F_j(\bar z)$, and the Jacobian of the system \eqref{15-sys} appears in the numerator instead of the derivative. Similarly, as before, we
can try to calculate this integral by residues outside the contour, that is, at the singular points of $\phi(\bar z)$. Success of
this enterprise again depends on the properties of the integrand beyond the integration contour.

\subsection{Integral representation for generating functional\label{15-Sec53}}

Let us turn back to the sum over intermediate states \eqref{15-Qm-sum-Om}
\be{15-Qm-sum-Om0}
Q_m^{(\kappa)}=\left(\frac{i}{\sin\zeta}\right)^n\sum_{\bv}\prod_{k=1}^m\prod_{j=1}^n\frac{f(v_j,\xi_k)}{f(u_j,\xi_k)}\;
\frac{\det_n \Omega_\kappa(v_j,u_k|\bv)\;\det_n \Omega(u_j,v_k|\bu)}{\det_n(\partial Y_\kappa(v_j|\bar v)/\partial v_k) \; \det_n \Omega(u_j,u_k|\bu)},
\ee
where we used equation \eqref{15-Y-Om} and replaced the matrix
$\Omega_\kappa(v_j,v_k|\bar v)$ by the matrix of derivatives of the function $Y_\kappa(v_j|\bar v)$.
According to the results of the previous section, we can replace the sum by
an $n$-fold contour integral. Then the Jacobian  $\det_n(\partial Y_\kappa(v_j|\bar v)/\partial v_k)$ cancels, and we obtain
\be{15-Qm-int1}
Q_m^{(\kappa)}=\frac{i^n}{(\sin^n\zeta) n!}\oint_{\Gamma(\bv)}\prod_{j=1}^n\frac{dz_j}{2\pi i}\prod_{k=1}^m\prod_{j=1}^n\frac{f(z_j,\xi_k)}{f(u_j,\xi_k)}\;
\frac{\det_n \Omega_\kappa(z_j,u_k|\bar z)\;\det_n \Omega(u_j,z_k|\bu)}{\prod_{j=1}^nY_\kappa(z_j|\bar z) \; \det_n \Omega(u_j,u_k|\bu)}.
\ee
Here the (multidimensional) integration contour $\Gamma(\bv)$ surrounds all admissible solutions of the twisted Bethe equations. We know that for
$\kappa $ small enough these solutions are close to the values $\xi_k+i\zeta$. On the other hand, we from the very beginning agreed that the
 inhomogeneities $\xi_k$ can be chosen arbitrarily close to their limiting value $-i\zeta/2$. Thus, all admissible solutions turn out to be in a small
neighborhood of the point $i\zeta/2$. Therefore, as the contour $\Gamma(\bv)$, we can choose a system of circles of small radius
around the point $i\zeta/2$ (for each integration variable $z_j$).

Compared with the formula \eqref{15-sum-sys}, an additional factor of $1/n!$ appeared in the integral representation \eqref{15-Qm-int1}.
This is due to the fact that formally the integral \eqref{15-Qm-int1} is equal to the sum over {\it all} the solutions of the twisted Bethe equations,
including such that differ from each other only by a permutation in the set $\bv$. But these solutions correspond to the same
vector, therefore, we introduced an additional combinatorial factor.  Formally, the integral \eqref{15-Qm-int1} also takes into account inadmissible solutions.
However, as we have already seen, the corresponding contributions vanish.

Now we need to go from integrating over the contour $\Gamma(\bv)$ to integration over a new contour that surrounds the poles of the integrand
outside $\Gamma(\bv)$. What are these `outer' singularities? First, the integrand contains poles for $z_j=\xi_k$  due to the product
of the unctions $f(z_j,\xi_k)$. Secondly, there are poles at the points $z_j=u_k$. To clearly see these poles, let us recall once again the expressions for
the matrices $\Omega_\kappa$ and $\Omega$, in which the variables $\bv$ are replaced by $\bar z$:
\be{15-Omk-zu}
\Omega_\kappa(z_j,u_k|\bar z)=(-1)^{n-1}a(u_k)t(z_j,u_k)h(\bar z,u_k)+\kappa\;d(u_k)t(u_k,z_j)h(u_k,\bar z),
\ee
\be{15-Omk-uz}
\Omega(u_j,z_k|\bu)=(-1)^{n-1}a(z_k)t(u_j,z_k)h(\bu,z_k)+d(z_k)t(z_k,u_j)h(z_k,\bu).
\ee
We see that the matrix element $\Omega_\kappa(z_j,u_k|\bar z)$  has a simple pole for $z_j=u_k$ due to the functions $t(z_j,u_k)$ and $t(u_k,z_j)$.
On the other hand, the matrix element $\Omega(u_j,z_k|\bu)$ turns out to be regular at the same points, because the parameters $\bu$ satisfy the
Bethe equations. Indeed,
\be{15-ResOm}
\Res\Omega(u_j,z_k|\bu)\Bigr|_{z_k=u_j}=i\sin\zeta\Bigl((-1)^{n-1}a(u_j)h(\bu,u_j)-d(u_j)h(u_j,\bu)\Bigr),
\ee
and we see that the residue vanishes.

At first glance, this may seem strange, because the determinants of the matrices $\Omega_\kappa(z_j,u_k|\bar z)$ and $\Omega(u_j,z_k|\bu)$
describe the same scalar product (see the formulas \eqref{15-SP1}). However, in the formulas \eqref{15-SP1} the parameters $\bv$
satisfy twisted Bethe equations, and the parameters $\bu$ satisfy ordinary Bethe equations. At the same time, in the formulas \eqref{15-Omk-zu}, \eqref{15-Omk-uz}, only the parameters $\bu$ satisfy Bethe equations, while $z_j$ are integration variables. Therefore, the matrices
$\Omega_\kappa(z_j,u_k|\bar z)$ and  $\Omega(u_j,z_k|\bu)$
have completely different analytical properties.

Thus, we found that the integrand has simple poles when $z_j$ coincides  with either $\xi_k$ or $u_k$.
There are no other singularities outside the contour $\Gamma(\bv)$. We also note that the integrand is $i\pi$-periodic in each $z_j$
and decreases as $e^{-2|z_j|}$ at  $z_j\to\pm\infty$.
Therefore, the integral over each $z_j$ along the boundary of any horizontal strip of width $i\pi$ is equal to zero. Hence we find
\be{15-int-int}
\oint_{\Gamma(\bv)}=\oint_{\text{Strip}}-\oint_{\Gamma(\bxi)\cup\Gamma(\bu)} = -\oint_{\Gamma(\bxi)\cup\Gamma(\bu)}.
\ee
Here the contour $\Gamma(\bxi)\cup\Gamma(\bu)$  encloses the points $\bxi$  and $\bu$, and the integral with the subscript `$\scriptstyle\text{Strip}$'
means the integral over the above-mentioned strip of width $i\pi$, which is equal to zero. Thus, the $n$-fold integral \eqref{15-Qm-int1}
turns into
\be{15-Qm-int3}
Q_m^{(\kappa)}=\left(\frac{-i}{\sin\zeta}\right)^n \frac1{n!}\oint_{\Gamma(\bxi)\cup\Gamma(\bu)}\prod_{j=1}^n\frac{dz_j}{2\pi i}\prod_{k=1}^m\prod_{j=1}^n\frac{f(z_j,\xi_k)}{f(u_j,\xi_k)}\;
\frac{\det_n \Omega_\kappa(z_j,u_k|\bar z)\;\det_n \Omega(u_j,z_k|\bu)}{\prod_{j=1}^nY_\kappa(z_j|\bar z) \; \det_n \Omega(u_j,u_k|\bu)}.
\ee
Here it is appropriate to emphasize once again the importance of the fact that the sum in the formula \eqref{15-Qm-sum-Om0} was taken
over the solutions of twisted, rather than ordinary Bethe equations. Due to this circumstance, the initial contour $\Gamma(\bv)$ exists.
After all, according to the formula \eqref{15-sum-sys}, the integration contour should  enclose
only the roots of the system of equations. In our case, the integrand has singularities both at the points $\bu$
and  $\bv$. If the parameters $\bv$ were solutions of the ordinary Bethe equations, then we could not construct a contour that would enclose $\bv$, but would not enclose $\bu$, since $\bu$ is one of the solutions of the ordinary Bethe equations. But the introduction of the twist allowed us to move all solutions
$\bv$ to a neighborhood of the point $i\zeta/2$ and thereby separate them from the solutions of the ordinary Bethe equations.

\subsection{Calculation of the integral\label{15-Sec54}}

Now we can calculate the resulting integral \eqref{15-Qm-int3} by the residues at the points where the integration variables $z_j$ coincide either with the inhomogeneities $\xi_k$,  or with the roots of the Bethe  equations $u_k$. Just before the calculation it is convenient to go from the matrices
$\Omega_\kappa(x_j,y_k|\bar x)$ to the new matrices $\widetilde{\Omega}_\kappa(x_j,y_k|\bar x)$, defined as follows:
\be{15-Om-tOm0}
\widetilde{\Omega}_\kappa(x_j,y_k|\bar x)=\frac{\Omega_\kappa(x_j,y_k|\bar x)}{(-1)^{n-1}a(y_k)h(\bar x,y_k)} .
\ee
Similarly, we introduce a function $\widetilde{Y}_\kappa(z_j|\bar z)$ by
\be{15-Y-tY1}
\widetilde{Y}_\kappa(z_j|\bar z)=\frac{Y_\kappa(z_j|\bar z)}{(-1)^{n}a(z_k)h(\bar z,z_j)} =1+(-1)^n\kappa \frac{d(z_j)h(z_j,\bar z)}{a(z_j)h(\bar z,z_j)}.
\ee
As a result of such transformations, the matrices $\widetilde{\Omega}$ depend on the ratio of the functions $a$ and $d$, what allows us to
use Bethe equations in those cases when these functions depend on the parameters $u_j$. We arrive at the following representation:
\begin{multline}\label{15-Qm-int2}
Q_m^{(\kappa)}=\frac{i^n}{(\sin^n\zeta) n!}\oint_{\Gamma(\bxi)\cup\Gamma(\bu)}\prod_{j=1}^n\frac{dz_j}{2\pi i}\prod_{k=1}^m\prod_{j=1}^n\frac{f(z_j,\xi_k)}{f(u_j,\xi_k)}\;W_n(\bar z;\bu)\\
\times
\frac{\det_n \widetilde{\Omega}_\kappa(z_j,u_k|\bar z)\;\det_n \widetilde{\Omega}(u_j,z_k|\bu)}
{\prod_{j=1}^n\widetilde{Y}_\kappa(z_j|\bar z) \; \det_n \widetilde{\Omega}(u_j,u_k|\bu)},
\end{multline}
where
\be{15-W}
W_n(\bar z;\bu)=\frac{h(\bar z,\bu) h(\bu, \bar z)}{h(\bar z,\bar z)h(\bu,\bu)}.
\ee
Let us also give the explicit form of all the matrices in  \eqref{15-Qm-int2}:
\be{15-tOm2}
\widetilde{\Omega}(u_j,z_k|\bar u)=t(u_j,z_k)+(-1)^{n-1}t(z_k,u_j)\frac{d(z_k)h(z_k,\bu)}{a(z_k)h(\bu,z_k)},
\ee
\be{15-tOm1}
\widetilde{\Omega}_\kappa(z_j,u_k|\bar z)=t(z_j,u_k)+\kappa\;t(u_k,z_j)\frac{h(u_k,\bar z)h(\bu,u_k)}
{h(\bar z,u_k)h(u_k,\bu)},
\ee
\be{15-Omuu}
 \widetilde{\Omega}(u_j,u_k|\bu)=-\sin\zeta\left[\delta_{jk}\left(-i\frac d{du_j}\log r(u_j)-\sum_{\ell=1}^n\mathcal{K}(u_j-u_\ell)\right)+\mathcal{K}(u_j-u_k)\right].
\ee
In the formulas \eqref{15-tOm1} and \eqref{15-Omuu}, we have used Bethe equations for the set $\bu$.

For each integration variable  $z_j$, the integral over the contour $\Gamma(\bxi)\cup\Gamma(\bu)$  can be written as the sum of two integrals:
over the contour $\Gamma(\bxi)$ (around the points $\bxi$) and over the contour $\Gamma(\bu)$ (around the points $\bu$). Since the integrand in
\eqref{15-Qm-int2} is symmetric with respect to the variables $\bar z$, the total $n$-fold integral is then written as
\be{15-int-2int}
\frac{1}{n!}\oint_{\Gamma(\bxi)\cup\Gamma(\bu)}\prod_{j=1}^n\frac{dz_j}{2\pi i}=\sum_{s=0}^n
\frac{1}{s!(n-s)!}\;\oint_{\Gamma(\bxi)}\prod_{j=1}^s\frac{dz_j}{2\pi i}\quad\oint_{\Gamma(\bu)}\prod_{j=1}^{n-s}\frac{dz_j}{2\pi i}\;.
\ee
Here, for the integration variables $z_1,\dots,z_s$, the contours encircle the points $\bxi$, and for the remaining variables $z_{s+1},\dots,z_n$
the integration is carried out along contours enclosing the points $\bu$. Note that the number of poles whithin the contour $\Gamma(\bxi)$ is equal to $m$
(these are the points $\xi_1,\dots,\xi_m$). In addition, due to two determinants, the integrand has a zero of the second order if two variables $z_j$
and $z_k$ coincide. Therefore, calculating the residues at the points $\xi_1,\dots,\xi_m$, we cannot put two different $z_j$
and $z_k$ equal to the same inhomogeneity $\xi_\ell$. It follows that in the integrals over the contours $\Gamma(\bxi)$,
the number of integration variables
$z_1,\dots,z_s$  can not exceed the number of the poles $\xi_1,\dots,\xi_m$, that is, $s \le m$. Otherwise, the $s$-fold integral over the contours
$\xi_1,\dots,\xi_m$ vanishes. Thus, the sum over $s$ in the formula \eqref{15-int-2int} is actually limited to $s= m$.

For the time being, let us leave the integrals over the contours $\Gamma(\bxi)$ and pass to the integrals over the contours $\Gamma(\bu)$.
In these integrals we also cannot put two different $z_j$ and $z_k$ equal the same value $u_\ell$. Therefore, the integral over the contours
$\Gamma(\bu)$ is equal to the sum of the residues
at the points $z_{s+k}=u_{j_k}$, $k=1,\dots,n-s$, where the sum is taken over all possible sets $u_{j_1},\dots,u_{j_{n-s}}$. In other words,
this integral generates a sum over partitions of the complete set $\bu$ into two subsets:  $\bu^{\so}$ and  $\bu^{\st}$,
where $\#\bu^{\so}=s$ and$\#\bu^{\st}=n-s$.  The residues are taken at the points $u_j\in\bu^{\st}$. Taking into account the fact that
we can do all possible permutations of the elements in the subset $\bu^{\st}$, the sum over the partitions $\bu\mapsto\{\bu^{\so},\bu^{\st}\}$
should be multiplied by $(n-s)!$.

Let us now see what happens with the integrand, if we put $\{z_{s+1},\dots,z_n\}=\bu^{\st}$. Before doing this it is convenient to reorder the columns and rows in the matrices $\widetilde{\Omega}_\kappa(z_j,u_k|\bar z)$ and $\widetilde{\Omega}(u_j,z_k|\bu)$. Namely, we move all the columns of
$\widetilde{\Omega}_\kappa$ with $u_k\in\bu^{\st}$ to the right, and in the matrix $\widetilde{\Omega}$, we move all the rows
 with $u_j\in\bu^{\st}$ down.  It is clear that with such a reordering of rows and columns, the common sign of the integrand will not change.

All the poles at $z_j=u_k$ are collected in the matrix $\widetilde{\Omega}_\kappa(z_j,u_k|\bar z)$. It is not difficult to check that
\be{15-ResOmY}
\Res\frac{\widetilde{\Omega}_\kappa(z_j,u_k|\bar z)}{\widetilde{Y}_\kappa(z_j|\bar z)}\Bigr|_{z_j=u_k}=-i\sin\zeta.
\ee
From this we easily find
\be{15-ResdetOmY}
\Res\frac{\det_n\widetilde{\Omega}_\kappa(z_j,u_k|\bar z)}{\prod_{j=1}^n\widetilde{Y}_\kappa(z_j|\bar z)}
\Bigr|_{\{z_{s+1},\dots,z_n\}=\bu^{\st}}=(-i\sin\zeta)^{n-s}
\frac{M_s(\{z_{1},\dots,z_s\};\bu^{\so})}{\prod_{j=1}^s\widetilde{Y}_\kappa(z_j|\{z_{1},\dots,z_s\},\bu^{\st})}.
\ee
where
\be{15-Ms}
M_s(\{z_{1},\dots,z_s\};\bu^{\so})=\det_{s}\left(t(z_j,u^{\so}_k)+\kappa\;t(u^{\so}_k,z_j)\prod_{\ell=1}^s
\frac{h(u^{\so}_k,z_\ell)}{h(z_\ell,u^{\so}_k)}  \prod_{u_\ell\in\bu^{\so}} \frac{h(u_\ell,u^{\so}_k)}
{h(u^{\so}_k,u_\ell)}\right),
\ee
and $u^{\so}_k$ means the $k$th element of the set  $\bu^{\so}$. Taking into account obvious relations
\be{15-W-W}
W_n(\bar z;\bu)\Bigr|_{\{z_{s+1},\dots,z_n\}=\bu^{\st}}=W_s(\{z_{1},\dots,z_s\};\bu^{\so}),
\ee
and
\be{15-BP-BP}
\prod_{k=1}^m\prod_{j=1}^n\frac{f(z_j,\xi_k)}{f(u_j,\xi_k)}\Bigr|_{\{z_{s+1},\dots,z_n\}=\bu^{\st}}=
\prod_{k=1}^m\frac{\prod_{j=1}^s f(z_j,\xi_k)}{\prod_{u_j\in\bu^{\so}}f(u_j,\xi_k)},
\ee
we obtain new representation for the generating functional
\begin{multline}\label{15-Qm-int3a}
Q_m^{(\kappa)}=\sum_{s=0}^m\frac{1}{(-i\sin\zeta)^s s!}\sum_{\bu\mapsto\{\bu^{\so},\bu^{\st}\}}\oint_{\Gamma(\bxi)}\prod_{j=1}^s\frac{dz_j}{2\pi i}
\prod_{k=1}^m\frac{\prod_{j=1}^s f(z_j,\xi_k)}{\prod_{u_j\in\bu^{\so}}f(u_j,\xi_k)}
\;W_s(\{z_{1},\dots,z_s\};\bu^{\so})
\\
\times \frac{M_s(\{z_{1},\dots,z_s\};\bu^{\so})}{\prod_{j=1}^s\widetilde{Y}_\kappa(z_j|\{z_{1},\dots,z_s\},\bu^{\st})}
\frac{\det_n\Bigl[ \widetilde{\Omega}(u_j,z_k|\bu);\widetilde{\Omega}(u_j,u^{\st}_k|\bu)\Bigr]}
{ \det_n \widetilde{\Omega}(u_j,u_k|\bu)}.
\end{multline}
Here in the determinant $\det_n\Bigl[ \widetilde{\Omega}(u_j,z_k|\bu);\widetilde{\Omega}(u_j,u^{\st}_k|\bu)\Bigr]$,
we have the matrix $\widetilde{\Omega}(u_j,z_k|\bu)$ in the columns $1,\dots,s$, while in the remaining $n-s$ columns,
we have the matrix $\widetilde{\Omega}(u_j,u^{\st}_k|\bu)$.
The sum over partitions of the set $\bu$ is taken under condition $\#\bu^{\so}=s$. For $s=0$ , the corresponding term in the sum over $s$
is equal to $1$.

In the remaining integral, we can put all the functions $d(z_j)=0$. Indeed, after calculating this integral, all the variables $z_j$  become equal to the inhomogeneities  $z_j=\xi_\ell$. At the same time, $d(\xi_\ell)=0$ for any $\xi_\ell$. Therefore,  we can make replacements in the integrand
\be{15-change}
\begin{aligned}
\widetilde{Y}_\kappa(z_j|\{z_{1},\dots,z_s\},\bu^{\st})\longrightarrow 1,\\
\widetilde{\Omega}(u_j,z_k|\bu)\longrightarrow t(u_j,z_k).
\end{aligned}
\ee
Such replacements are equivalent to discarding a function that is holomorphic within the  integ\-ra\-ti\-on contour.
After this, representation \eqref{15-Qm-int3a} takes the form
\begin{multline}\label{15-Qm-int4}
Q_m^{(\kappa)}=\sum_{s=0}^m\frac{1}{(-i\sin\zeta)^s s!}\sum_{\bu\mapsto\{\bu^{\so},\bu^{\st}\}}\oint_{\Gamma(\bxi)}\prod_{j=1}^s\frac{dz_j}{2\pi i}
\prod_{k=1}^m\frac{\prod_{j=1}^s f(z_j,\xi_k)}{\prod_{u_j\in\bu^{\so}}f(u_j,\xi_k)}
\;W_s(\{z_{1},\dots,z_s\};\bu^{\so})
\\
\times M_s(\{z_{1},\dots,z_s\};\bu^{\so})
\frac{\det_n\Bigl[ t(u_j,z_k);\widetilde{\Omega}(u_j,u^{\st}_k|\bu)\Bigr]}
{ \det_n \widetilde{\Omega}(u_j,u_k|\bu)}.
\end{multline}

Observe that after all transformations, the parameter $\kappa$ is now included only in the function $M_s(\{z_{1},\dots,z_s\};\bu^{\so})$
(see \eqref{15-Ms}). That is, the dependence on $\kappa$ turns out to be polynomial. Recall that initially we considered $|\kappa| $ small enough
to ensure the completeness of the system of twisted on-shell vectors. Now we can get rid of this restriction. Indeed,
if some polynomial in $\kappa$ is known in a small vicinity of $\kappa =0$, then it is analytically continued to any point of the complex plane.
Thus, differentiation with respect to $\kappa$ at $\kappa =1$ is possible in the representation \eqref{15-Qm-int4}.

\subsection{Thermodynamic limit of the generating functional\label{15-Sec55}}

It is clear that if we calculate the remaining integrals in the formula \eqref{15-Qm-int4}, then we obtain an additional sum over the partitions, this time
these are partitions of the set of inhomogeneities $\{\xi_1,\dots,\xi_m\}$. Such a representation for the generating function $Q_m^{(\kappa)}$
can also be obtained directly by algebraic methods, without summing the form factors. For this, we need to
calculate the action of the product of twisted transfer matrices $\mathcal{T}_\kappa(\xi_k)$
on the ordinary on-shell vectors $\langle0|C(\bu)$ or $B(\bu)|0\rangle$ in the formula \eqref{15-Qm-def}.
Details can be found in \cite{KitMST05}.

We leave the integrals over the contours $\Gamma(\bxi)$ as they are. Then we can easily take the homogeneous limit, simply putting
$\xi_k=-i\zeta/2$ for all $k=1,\dots,m$ in the formula \eqref{15-Qm-int4}.
Moreover, the transition to the homogeneous model must be done if we
want to consider the thermodynamic limit, otherwise we will have to deal with the function $r(u)=a(u)/d(u)$,
depending on the infinite set of arbitrary  inhomogeneities.

Up to now, all calculations have been performed under the assumption that the variables $\bu$ were some roots of the ordinary Bethe equations. Now we  assume,
that the set $\bu$ corresponds to the ground state of the Hamiltonian. This will enable us to calculate the ratio of the determinants in the formula \eqref{15-Qm-int4} in the thermodynamic limit.

The matrix elements of the matrix $\widetilde{\Omega}(u_j,u_k|\bu)$
are given by  \eqref{15-Omuu}. Let us find the inverse matrix in the limit of a large $n$. For the homogeneous model, we have the equality
\be{15-r-p0}
-i\frac{r'(u_j)}{r(u_j)}=Np'_0(u),
\ee
where $p'_0(u)$ is given by \eqref{15-mom-K}.
Then, due to the condition \eqref{15-BE-mom-K} we have
\begin{equation}\label{15-sum-diag}
-i\frac{r'(u_j)}{r(u_j)}-\sum_{\ell=1}^n\mathcal{K}(u_j-u_\ell)=N\left(p'_0(u_j)-\frac1N\sum_{\ell=1}^n\mathcal{K}(u_j-u_\ell)\right)
%
\to 2\pi N\rho(u_j),\qquad N\to\infty,
\end{equation}
what implies
\be{15-Om-lim}
\widetilde{\Omega}(u_j,u_k|\bu)\to-2\pi N\sin\zeta\rho(u_k)\left(\delta_{jk}+\frac{\mathcal{K}(u_j-u_k)}{2\pi N\rho(u_k)}\right),
\qquad N\to\infty.
\ee
Recall that the equality \eqref{15-BE-mom-K} was derived under the assumption that $n$ and $N$ are large enough. Therefore, the formula \eqref{15-Om-lim}
is approximate. It discards terms that disappear in the limit $n,N\to\infty$.

Let us denote by $\Xi(u_j,u_k)$ a matrix that is inverse to $\widetilde{\Omega}(u_j,u_k|\bu)$:
\be{15-invMat}
\sum_{\ell=1}^n\widetilde{\Omega}(u_j,u_\ell|\bu)\Xi(u_\ell,u_k)=\delta_{jk}.
\ee
One can see from \eqref{15-Om-lim} that  $\Xi(u_j,u_k)$ is expressed in terms of the resolvent $\mathcal{R}(u-v)$ \eqref{15-inteqRes}
\be{15-invMat-res}
\Xi(u_j,u_k)=\frac{-1}{2\pi N\sin\zeta\rho(u_j)}\left(\delta_{jk}-\frac{\mathcal{R}(u_j-u_k)}{N\rho(u_k)}\right).
\ee
This can be easily verified by substituting \eqref{15-invMat-res} in the formula \eqref{15-invMat}. Then, in the limit of large $n$ and $N$,
we can replace the summation by the integration and use the integral equation for the resolvent \eqref{15-inteqRes}.

Now we can easily calculate the ratio of determinants in the formula \eqref{15-Qm-int4}.
We rearrange the rows and columns in the matrix $\Xi(u_j,u_k)$ so that in the first
$s$ rows and columns would contain elements from the subset $\bu^{\so}$: $\Xi(u^{\so}_j,u^{\so}_k)$. The resulting matrix is denoted by
$\widetilde{\Xi}(u_j,u_k)$. Then
\begin{multline}\label{15-Rat-beg}
\frac{\det_n\Bigl[ t(u_j,z_k);\widetilde{\Omega}(u_j,u^{\st}_k|\bu)\Bigr]}
{ \det_n \widetilde{\Omega}(u_j,u_k|\bu)}=
\det_n \Bigl[\widetilde{\Xi}(u_j,u_k)\Bigr]\cdot\det_n\Bigl[ t(u_j,z_k);\widetilde{\Omega}(u_j,u^{\st}_k|\bu)\Bigr]\\
=\det_n\Bigl[ \sum_{\ell=1}^n \widetilde{\Xi}(u_j,u_\ell)t(u_\ell,z_k)\;;\;\sum_{\ell=1}^n \widetilde{\Xi}(u_j,u_\ell)\widetilde{\Omega}(u_\ell,u^{\st}_k|\bu)\Bigr].
\end{multline}
Obviously,
\be{15-ochevid}
\sum_{\ell=1}^n \widetilde{\Xi}(u_j,u_\ell)\widetilde{\Omega}(u_\ell,u^{\st}_k|\bu)=\delta_{jk},
\ee
therefore the determinant of the $n\times n$ matrix reduces to its diagonal minor of the size $s\times s$
\begin{equation}\label{15-Rat-prod}
\frac{\det_n\Bigl[ t(u_j,z_k);\widetilde{\Omega}(u_j,u^{\st}_k|\bu)\Bigr]}
{ \det_n \widetilde{\Omega}(u_j,u_k|\bu)}=\det_s\Bigl[ \sum_{\ell=1}^n \widetilde{\Xi}(u^{\so}_j,u_\ell)t(u_\ell,z_k)\Bigr].
\end{equation}
Substituting here \eqref{15-invMat-res}, we obtain for $n$ and $N$ large enough
\begin{multline}\label{15-actRes}
 \sum_{\ell=1}^n\widetilde{\Xi}(u^{\so}_j,u_\ell)t(u_\ell,z_k)=\frac{-1}{2\pi N\sin\zeta\rho(u^{\so}_j)}\left(t(u^{\so}_j,z_k)-
 \sum_{\ell=1}^n\frac{\mathcal{R}(u^{\so}_j-u_\ell)t(u_\ell,z_k)}{N\rho(u_\ell)}\right)\\
 \to \frac{-1}{2\pi N\sin\zeta\rho(u^{\so}_j)}\left(t(u^{\so}_j,z_k)-
 \int_{-\infty}^\infty\mathcal{R}(u^{\so}_j-v)t(v,z_k)\,dv\right) = \frac{i}{N\rho(u^{\so}_j)\;2\zeta\sinh\frac\pi\zeta(u^{\so}_j-z_k)},
\end{multline}
where we used equation \eqref{15-inteqtuz}. Thus, we finally arrive at
\begin{equation}\label{15-Rat-fin}
\frac{\det_n\Bigl[ t(u_j,z_k);\widetilde{\Omega}(u_j,u^{\st}_k|\bu)\Bigr]}
{ \det_n \widetilde{\Omega}(u_j,u_k|\bu)}=\prod_{u_j\in\bu^{\so}}\frac1{N\rho(u_j)}\;
\det_s\Bigl[\frac{i}{2\zeta\sinh\frac\pi\zeta(u^{\so}_j-z_k)}\Bigr].
\end{equation}
Substituting this expression to the formula for the generating functional
\eqref{15-Qm-int4}, we obtain
\begin{multline}\label{15-Qm-int5}
Q_m^{(\kappa)}=\sum_{s=0}^m\frac{1}{(-i\sin\zeta)^s s!}\sum_{\bu\mapsto\{\bu^{\so},\bu^{\st}\}}\oint_{\Gamma(-\tfrac{i\zeta}2)}\prod_{j=1}^s\frac{dz_j}{2\pi i}
\prod_{j=1}^s \varphi^m(z_j)\prod_{u_j\in\bu^{\so}}\varphi^{-m}(u_j)
\\
\times \;W_s(\{z_{1},\dots,z_s\};\bu^{\so})\;M_s(\{z_{1},\dots,z_s\};\bu^{\so})
\prod_{u_j\in\bu^{\so}}\frac1{N\rho(u_j)}\;
\det_s\Bigl[\frac{i}{2\zeta\sinh\frac\pi\zeta(u^{\so}_j-z_k)}\Bigr],
\end{multline}
where
\be{15-varphi}
\varphi(z)=\frac{\sinh(z-\tfrac{i\zeta}2)}{\sinh(z+\tfrac{i\zeta}2)}.
\ee
Here we already consider the $XXZ$ model in the homogeneous limit, therefore we put all $\xi_\ell=-i\zeta/2$,
and the contour integrals over the variables $z_j$ are taken around the point $-i\zeta/2$.

It remains for us to take the last step and replace the sum over the partitions of the set $\bu$ by the $s$-fold integral. Observe that all the terms in the sum over the partitions in the formula \eqref{15-Qm-int5} evidently depend only on the subset $\bu^{\so}$.
Consider a sum of the form
\be{15-sum-part-sum0}
\sum_{\bu\mapsto\{\bu^{\so},\bu^{\st}\}} F(\bu^{\so})\prod_{u_j\in\bu^{\so}}\frac1{N\rho(u_j)},
\ee
where $F$ is a symmetric function of  $s$ variables, integrable in $\mathbb{R}^s$. We also assume that $F$ vanishes as soon as any two arguments coincide.
Then it is easy to see that
\be{15-sum-part-sum1}
\sum_{\bu\mapsto\{\bu^{\so},\bu^{\st}\}} F(\bu^{\so})\prod_{u_j\in\bu^{\so}}\frac1{N\rho(u_j)}=
\frac1{s!}\sum_{u_1\in\bu}\dots \sum_{u_s\in\bu} F(\{u_1,\dots,u_s\})\prod_{j=1}^s\frac1{N\rho(u_j)}.
\ee
Here in the rhs, each variable $u_k$ independently runs through the set $\bu$. Indeed, due to the condition imposed on the function $F$,
in the rhs of the formula \eqref{15-sum-part-sum1}, there remain only those terms in which all $u_k$ are distinct.
Those terms that differ from each other only by permuting the arguments are killed by the factor $1/s!$. There remains a sum over partitions
of the set $\bu$ into subsets $\bu^{\so}$ and $\bu^{\st}$.

In its turn, the sum in the rhs of \eqref{15-sum-part-sum1} becomes an $s$-fold integral in the thermodynamic limit, therefore,
\be{15-sum-part-sum2}
\sum_{\bu\mapsto\{\bu^{\so},\bu^{\st}\}} F(\bu^{\so})\prod_{u_j\in\bu^{\so}}\frac1{N\rho(u_j)}\to
\frac1{s!}\int_{-\infty}^\infty \,du_1\dots \int_{-\infty}^\infty \,du_s \;F(\{u_1,\dots,u_s\}).
\ee

It only remains to apply the result to the formula \eqref{15-Qm-int5}. To write down the final result, it is convenient
slightly to change the meaning of the notation that we have used so far. Now let the set
$\bar z$ denotes the set of integration variables $\{z_1,\dots,z_s\}$. The corresponding
integrals  are taken around the point $-i\zeta/2$.
Similarly, by the symbol $\bu$, we now denote the set of integration variables $\{u_1,\dots,u_s\}$. The
integrals over these variables are taken along the real axis. Then
\begin{multline}\label{15-Qm-TD-lim}
Q_m^{(\kappa)}=\sum_{s=0}^m\frac{1}{(-i\sin\zeta)^s (s!)^2}\oint_{\Gamma(-\frac{i\zeta}2)}\prod_{j=1}^s\frac{dz_j}{2\pi i}
\int_{-\infty}^\infty\prod_{j=1}^s du_j
\\
\times \prod_{j=1}^s \varphi^m(z_j)\varphi^{-m}(u_j)
\;W_s(\bar z;\bu)\;M_s(\bar z;\bu)\;
\det_s\Bigl[\frac{i}{2\zeta\sinh\frac\pi\zeta(u_j-z_k)}\Bigr].
\end{multline}
This is the required integral representation for the generating functional $Q_m^{(\kappa)}$ in the thermo\-dy\-na\-mic limit.
We recall that the functions $\varphi(z)$
are given by the formula \eqref{15-varphi}, and the functions $W_s(\bar z;\bu)$ and $M_s(\bar z;\bu)$
have the form
\be{15-Ws-new}
W_s(\bar z;\bu)=\prod_{j=1}^s\prod_{k=1}^s
\frac{\sinh(u_j-z_k-i\zeta)\sinh(z_j-u_k-i\zeta)}{\sinh(z_j-z_k-i\zeta)\sinh(u_j-u_k-i\zeta)}\;,
\ee
\be{15-Ms-new}
M_s(\bar z;\bu)=\det_{s}\left(t(z_j,u_k)+\kappa\;t(u_k,z_j)\prod_{\ell=1}^s
\frac{\sinh(u_k-z_\ell-i\zeta)\sinh(u_\ell-u_k-i\zeta)}{\sinh(z_\ell-u_k-i\zeta)\sinh(u_k-u_\ell-i\zeta)}  \right),
\ee
where the function $t(u,v)$ is given by one of the formulas \eqref{15-gfht}.

\section{Several simple particular cases\label{15-Sec7}}

It is interesting to see how equations \eqref{15-Qm-TD-lim} works in simple particular cases, where the result is known in advance.

\subsection{The case $\kappa=1$\label{15-Sec71}}

The first obvious case arises when $\kappa=1$, or equivalently, for $\beta=0$ (recall that $\kappa=e^\beta$).
Then the operator $\mathcal{Q}_{1,m}$ (see \eqref{15-GenFun}) turns into the identity operator. The average of the latter should be equal to
$1$. The formula \eqref{15-Qm-TD-lim} does give such an answer, because
\be{15-kappa0}
M_s(\bar z;\bu)\Bigr|_{\kappa=1}=\delta_{s0}.
\ee
Indeed, if we consider the scalar product of a twisted on-shell vector $\langle0|C(\bar z)$
and the ordinary on-shell vector
$B(u)|0\rangle$, then the answer will be proportional
to the determinant  \eqref{15-Ms-new} (see, for example,  \eqref{11-SP-det-pres0}).
For $\kappa=1$ such the scalar product vanishes. One can also prove this directly
observing a linear dependence of the rows in the determinant \eqref{15-Ms-new}. We have already done this for the case of the rational $R$-matrix.
The proof for the case of the trigonometric $R$-matrix is completely analogous, and we suggest that the reader do this himself.

Thus, for  $\kappa=1$, all the terms in the sum \eqref{15-Qm-TD-lim} vanish, except the term at $s=0$, which is equal to $1$.

\subsection{The case $m=1$\label{15-Sec72}}

For $m=1$ the operator $\mathcal{Q}_{1,1}^{(\kappa)}$ has the form
\begin{equation}\label{15-Qm-1-exp}
\mathcal{Q}_{1,1}^{(\kappa)}=e^{\frac\beta2(1-\sigma_1^z)}=1+\frac{\kappa-1}{2}(1-\sigma_1^z).
\end{equation}
Recall that the average value  of the operator $\sigma_k^z$ in the ground state
vanishes for any $k$ (see footnote on page~\pageref{15-pageSZ}). Therefore,
taking into account the equality \eqref{15-Qm-1-exp}, we find
\begin{equation}\label{15-Qm-14}
Q_1^{(\kappa)}=\frac{\langle0|C(\bu)\left(1+\frac{\kappa-1}{2}(1-\sigma_1^z)\right)B(u)|0\rangle}
{\langle0|C(\bu)B(u)|0\rangle}=\frac{\kappa+1}{2}.
\end{equation}

Let us try to reproduce this result via representation \eqref{15-Qm-TD-lim}. For $m=1$ we have
\begin{equation}\label{15-Qm-1}
Q_1^{(\kappa)}=1+\frac{\kappa-1}{-i\sin\zeta}\oint_{\Gamma(-\frac{i\zeta}2)}\frac{dz}{2\pi i}
\int_{-\infty}^\infty
\frac{\sinh(z-\tfrac{i\zeta}2)\sinh(u+\tfrac{i\zeta}2)}{\sinh(z+\tfrac{i\zeta}2)\sinh(u-\tfrac{i\zeta}2)}
\;\frac{\sinh(u-z-i\zeta)}{\sinh(u-z)}\;\frac{i\, du}{2\zeta\sinh\frac\pi\zeta(u-z)}.
\end{equation}
The contour integral over $z$ is equal to the residue in the point $z=-i\zeta/2$. Taking this residue, we obtain
\begin{equation}\label{15-Qm-12}
Q_1^{(\kappa)}=1+\frac{\kappa-1}{2\zeta}
\int_{-\infty}^\infty\frac{\, du}{\cosh\frac{\pi u}\zeta}.
\end{equation}
The remaining integral over $u$ is trivially computed, and we reproduce the desirable result.
\begin{equation}\label{15-Qm-13}
Q_1^{(\kappa)}=\frac{\kappa+1}{2}.
\end{equation}

\subsection{Free fermions \label{15-Sec73}}

If the anisotropy parameter $\Delta$ is zero, then the $XXZ$ model becomes equivalent to the free fermions \cite{JorW28,CreTW80}.
This is the simplest case of the
$XXZ$ chains. In particular, the Bethe equations \eqref{15-BE} are solved explicitly if we take into account that the point $\Delta=0$
corresponds to the value $\zeta=\pi/2$.
In the integral representation for the generating functional $Q_m^{(\kappa)}$, there are also significant simplifications, which allows us to
calculate explicitly the two-point function of the third  components of spin \eqref{15-sigsig}.
The simplification of the generating functional is based on formula  \eqref{ASM-detCauchy-S}.

We proceed to a direct calculation of the generating functional for $\zeta=\pi/2$.
Substituting this value in \eqref{15-Ms-new}, we get
\begin{multline}\label{15-Msff}
M_s(\bar z;\bu)=\det_s\Bigl[\frac{i(\kappa-1)}{\sinh2(z_j-u_k)}\Bigr]\\
=\bigl(2i(\kappa-1)\bigr)^s\frac{\prod_{j>k}^s \cosh(z_j-z_k)\cosh(u_k-u_j)}{\prod_{j,k=1}^s\cosh(z_j-u_k)}
\det_s\Bigl[\frac{1}{\sinh(z_j-u_k)}\Bigr].
\end{multline}
The second determinant in the integral representation \eqref{15-Qm-TD-lim} is computed exactly in the same way
\be{15-detff}
\det_s\Bigl[\frac{-i}{\pi\sinh2(z_j-u_k)}\Bigr]=
\left(\frac1{ 2\pi i}\right)^s\frac{\prod_{j>k}^s \cosh(z_j-z_k)\cosh(u_k-u_j)}{\prod_{j,k=1}^s\cosh(z_j-u_k)}
\det_s\Bigl[\frac{1}{\sinh(z_j-u_k)}\Bigr].
\ee
Finally,
\be{15-Wsff}
W_s(\bar z;\bu)=\prod_{j=1}^s\prod_{k=1}^s \frac{\cosh^2(z_j-u_k)}{\cosh(z_j-z_k)\cosh(u_j-u_k)}.
\ee
Thus, we obtain
\begin{equation}\label{15-Qm-ff}
Q_m^{(\kappa)}=\sum_{s=0}^m\frac{i^s(\kappa-1)^s}{(2\pi)^s (s!)^2}\oint_{\Gamma(-\frac{i\pi}4)}\prod_{j=1}^s\frac{dz_j}{2\pi i}
\int_{-\infty}^\infty\prod_{j=1}^s du_j
 \prod_{j=1}^s \left(\frac{\varphi(z_j)}{\varphi(u_j)}\right)^m
\left(\det_s\Bigl[\frac{1}{\sinh(u_j-z_k)}\Bigr]\right)^2.
\end{equation}

We recall that the two-point correlation function of the third spin components is extracted from the generating functional $Q_m^{(\kappa)}$ by the formula
\eqref{15-Szz-Q}
\be{15-Szz-Qff}
S_{zz}(m)=2\frac{d^2}{d\beta^2}D^2_m Q_m^{(\kappa)}\Bigr|_{\beta=0}-1,
\ee
where the symbol $D^2_m$ denotes the second derivative on the lattice, which acts on  functions of the discrete argument $\phi(m)$  as
follows:
\be{15-LatDir}
D^2_m \phi(m)=\phi(m+1)-2\phi(m)+\phi(m-1).
\ee
We also took into account that the average value of the operator $\sigma^z_k$ with respect to the ground state vanishes.

We see that the $s$th term in the sum \eqref{15-Qm-ff} is proportional to $(\kappa-1)^s$. Since for calculating the correlation function we need to take the second derivative with respect to the parameter $\beta$ at the point $\beta=0$ (that is, for $\kappa=1$), then it is sufficient to consider only terms with $s=1$ and $s=2$ (the term corresponding to $s=0$ obviously disappears after taking the derivatives).

Let us consider the contribution to the generating functional for $s=1$:
\begin{equation}\label{15-L1-ff}
\Lambda_1(m)=\frac{i(\kappa-1)}{2\pi}\int_{-\infty}^\infty du
\oint_{\Gamma(-\frac{i\pi}4)}\frac{dz}{2\pi i}
  \left(\frac{\sinh(z-\tfrac{i\pi}4)\sinh(u+\tfrac{i\pi}4)}{\sinh(z+\tfrac{i\pi}4)\sinh(u-\tfrac{i\pi}4)}\right)^m
\frac{1}{\sinh^2(u-z)}.
\end{equation}
Instead of calculating the integral over $z$ by taking the residue at the point $z=-i\pi/4$,
where there is a pole of order $m$, we can calculate this integral
by the residue at the point $z=u$, where there is a pole of only the second order. Here the arguments are quite analogous to those that we used when
replacing  the integration contour around the solutions of the twisted Bethe equations by a new contour around the inhomogeneities and the ground state parameters (see \eqref{15-int-int}). Indeed, the integrand in \eqref{15-L1-ff} is $i\pi$-periodic over $z$ and decreases exponentially at infinity.
Hence, the integral over the boundary of an infinite horizontal strip of width $i\pi$ vanishes. On the other hand, this integral is equal to
the sum of the residues at the points $z=-i\pi/4$ and $z=u$, since there are no other singularities in the integrand. Therefore, the integral along the contour around the point $ i\pi/4$ can be replaced by an integral along the contour around the point $z=u$ with the opposite sign. Calculating the residue at the second-order pole, we immediately find
\begin{equation}\label{15-L1-ffres}
\Lambda_1(m)=\frac{im(\kappa-1)}{2\pi}
\int_{-\infty}^\infty \frac{d}{du}\log
  \left(\frac{\sinh(u+\tfrac{i\pi}4)}{\sinh(u-\tfrac{i\pi}4)}\right)\,du= \frac{m(\kappa-1)}{2} .
\end{equation}
The result appears to be linear over $m$, therefore, after taking the second lattice derivative this contribution vanishes.

For $s=2$ we have two contributions. The first one has the form
\begin{equation}\label{15-L21-ff}
\Lambda_2^{(1)}(m)=-\frac{(\kappa-1)^2}{8\pi^2}\left\{\int_{-\infty}^\infty du
\oint_{\Gamma(-\frac{i\pi}4)}\frac{dz}{2\pi i}
  \left(\frac{\sinh(z-\tfrac{i\pi}4)\sinh(u+\tfrac{i\pi}4)}{\sinh(z+\tfrac{i\pi}4)\sinh(u-\tfrac{i\pi}4)}\right)^m
\frac{1}{\sinh^2(u-z)}\right\}^2.
\end{equation}
In fact, up to a numerical coefficient, this is the square of the integral considered above. Therefore, we easily find
\begin{equation}\label{15-L21-ff1}
\Lambda_2^{(1)}(m)=\frac{m^2}{8}(\kappa-1)^2.
\ee

The second contribution at $s=2$ has a bit more complex structure:
\begin{equation}\label{15-L22-ff}
\Lambda_2^{(2)}(m)=\frac{(\kappa-1)^2}{8\pi^2}
\int_{-\infty}^\infty \frac{\,du_1\,du_2}{\varphi^m(u_1)\varphi^m(u_2)}     \left\{ \oint_{\Gamma(-\frac{i\pi}4)}\frac{dz}{2\pi i}
\frac{\varphi^m (z)}{\sinh(z-u_1)\sinh(z-u_2)}\right\}^2
\end{equation}
Again, moving the $z$-integration contour over $z$ to the points $u_1$ and $u_2$, we obtain
\begin{equation}\label{15-L22-ff1}
\Lambda_2^{(2)}(m)=\frac{(\kappa-1)^2}{8\pi^2}
\int_{-\infty}^\infty \frac{\left( \varphi^m(u_1)-\varphi^m(u_2)\right)^2}{\sinh^2(u_1-u_2)\varphi^m(u_1)\varphi^m(u_2)}
\,du_1\,du_2 .
\end{equation}
The integral with respect to the variables $u_1$ and $u_2$ is more difficult to calculate, but it can be simplified if we immediately  proceed to
the second lattice derivative:
\begin{equation}\label{15-L22-ff2}
D^2_m\Lambda_2^{(2)}(m)=\frac{(\kappa-1)^2}{4\pi^2}
\int_{-\infty}^\infty \frac{ \varphi^{m-1}(u_1)}{\varphi^{m-1}(u_2)}
\left(\frac{ \varphi(u_1)}{\varphi(u_2)}-1\right)^2
\frac{du_1\,du_2}{\sinh^2(u_1-u_2)} .
\end{equation}
It is easy to check that
\be{15-diff-phi}
\frac{ \varphi(u_1)}{\varphi(u_2)}-1=\frac{ \sinh(u_1-\tfrac{i\pi}4)\sinh(u_2+\tfrac{i\pi}4)}{\sinh(u_1+\tfrac{i\pi}4)\sinh(u_2-\tfrac{i\pi}4)}-1
=\frac{ i\sinh(u_1-u_2)}{\sinh(u_1+\tfrac{i\pi}4)\sinh(u_2-\tfrac{i\pi}4)}.
\ee
Substituting this expression into \eqref{15-L22-ff2} we find
\begin{equation}\label{15-L22-ff3}
D^2_m\Lambda_2^{(2)}(m)=-\frac{(\kappa-1)^2}{4\pi^2}
\left|\int_{-\infty}^\infty \frac{ \sinh^{m-1}(u-\tfrac{i\pi}4)}{\sinh^{m+1}(u+\tfrac{i\pi}4)}\,du\right|^2 .
\end{equation}
We only need to use a relation
\be{15-deriv}
\frac{d}{du}\left(\frac{ \sinh(u-\tfrac{i\pi}4)}{\sinh(u+\tfrac{i\pi}4)}\right)=\frac{i}{\sinh^2(u+\tfrac{i\pi}4)},
\ee
and after this the integral \eqref{15-L22-ff3} is easily computed
\be{15-L22-ff4}
D^2_m\Lambda_2^{(2)}(m)=\frac{(\kappa-1)^2}{4\pi^2m^2}\left(1-(-1)^m\right).
\ee
Substituting the contributions $\Lambda_2^{(1)}$ and $\Lambda_2^{(2)}$ into \eqref{15-Szz-Qff}, we finally arrive at
\be{15-ansSz-ff}
S_{zz}(m)=\frac{1-(-1)^m}{\pi^2m^2}.
\ee

Thus, we have obtained an explicit and fairly simple result for the correlation function of the third  components of spin in the $XXZ$ chain
for $\Delta=0$. The simplicity of this answer is mainly due to the fact that with this value of the anisotropy parameter the model
is equivalent to free fermions.

\subsection{Several general remarks\label{15-Sec74}}

\subsubsection{On the periodicity\label{15-Sec741}}

Calculating the correlation function for $\zeta=\pi/2$, we explicitly used the $i\pi$-periodicity of the integrands. Generically,
however, this is not the case, because the functions $\sinh\frac\pi\zeta(u_j-z_k)$
in the formula \eqref{15-Qm-int5} are periodic with period $2i\zeta$.
Such a violation of $i\pi$-periodicity may seem strange. And indeed, even in the representation  \eqref{15-Qm-int4}
there was no violation.

In fact, there is no miracle here. The point is that the solution of the integral equation
\be{15-intrhotuz}
\rho(u-z-\tfrac{i\zeta}2)+\frac1{2\pi}\int_{-\infty}^\infty\mathcal{K}(u-v)\rho(v-z-\tfrac{i\zeta}2) \,dv=-\frac1{2\pi\sin\zeta}\;t(u,z)
\ee
depends on the parameter $z$ in a nonanalytic way. This is due to the presence of poles of the function $t(u,z)$ at $u=z$ and $u=z+i\zeta$.
Because of this, the Fourier transform of the function $t(u,z)$ strongly depends on  the imaginary part of the parameter $z$ The reader can easily
verify this fact independently. If $-\zeta<\Im(z)<0$, then
\be{15-ro-t}
\rho(u-z-\tfrac{i\zeta}2)=\frac{i}{2\zeta\sinh\frac\pi\zeta(u-z)}.
\ee
However, if $\Im(z)$ lies outside this strip, then the integral equation \eqref{15-intrhotuz} has another solution that
cannot be expressed in terms of elementary functions for generic $\zeta$. In general, the solution of the equation \eqref{15-intrhotuz}
is an $i\pi$-periodic function, but this function has cuts on the axes $\Im(z)=0$ and $\Im(z)=-\zeta$.
In our case the parameter $z$ was an integration variable that varied in a small neighborhood of the point $-i\zeta/2$.
That is why we have chosen the function \eqref{15-ro-t} as a solution to the equation \eqref{15-intrhotuz} .

\subsubsection{Asymptotic analysis\label{15-Sec742}}

The integral representation \eqref{15-Qm-int5} looks quite attractive from the point of view of asymptotic analysis of the
correlation function $S_{zz}(m)$ for  $m$ large. Indeed,  all the dependence on the distance $m$ is in the functions $\varphi$.
Naturally, an idea arises to deform the integration contours  so as to make $|\varphi(z)|<1$ on them.
For example, it is possible to blow up small circles around the point $-i\zeta/2$ into infinite horizontal strips, at the boundaries of which
$|\varphi(z)|<1$ for $|\Re(z)|<\infty$. However, serious combinatorial complications arise on this path. The point is,
that moving the initial integration contour  to the contour of steepest descent, we inevitably cross the singularities of the integrand,
in particular at the points $z_j=u_k$. In this case, the functions $\varphi^m(z_j)$ cancel with the functions $\varphi^{-m}(u_k)$
until they disappear completely in the integrand. Calculating these contributions is extremely difficult, therefore the question of the
asymptotic analysis of the integral representation  \eqref{15-Qm-int5} has not been resolved to date.

\subsubsection{The $XXX$ limit\label{15-Sec743}}

Completely analogously, we can obtain an integral pre\-sen\-ta\-tion for the generating functional $Q_m^{(\kappa)}$ in the $XXX$ chain.
However, the same result can be extracted directly from the re\-pre\-sen\-ta\-tion \eqref{15-Qm-int5} in the limit $\zeta\to 0$.
To do this, we need to make the change of variables $u_j\to \zeta u_j$, $z_j\to \zeta z_j$ and let $\zeta$ go to zero. The function
$\sinh\frac\pi\zeta(u-z)$ turns into $\sinh\pi(u-z)$, and all other hyperbolic functions go to the rational ones.
This is the required answer.

 \section{Emptiness formation probability \label{15-Sec8}}

In conclusion of this lecture we  consider one more correlation function that was called {\it emptiness formation probability} (EFP) \cite{KorIEU94}.
This term is used because in the QNLS model an analogous correlation function does
have the sense of the probability of the absence of particles in an interval of a fixed length.
A similar meaning can be given to this correlation in
$XXZ$ chain, if we introduce the notion of {\it quasiparticle}.

We assume that there is a  quasiparticle in the $k$th site, if the spin is directed downward at this site. If the spin is directed upwards, then
there is no quasiparticle in this site. Then the vacuum vector $|0\rangle$ is a state without quasiparticles.
On-shell vector $B(\bu)|0\rangle$ with $\#\bu=n$ is a state with $n$ quasiparticles. It should only be remembered that the position of
these quasiparticles is not fixed. The vector $B(\bu)|0\rangle$
is a linear combination of states, in each of which the quasiparticles occupy fixed sites $j_1,\dots,j_n$.
The sum in this linear
combination runs through all possible sites (see  \eqref{11-BV-xxz-mult-2}).

Consider the limit of the operator $\mathcal{Q}_{1,m}^{(\kappa)}$ at $\kappa\to 0$ (or equivalently, at $\beta\to-\infty$). Due to
\eqref{15-exp-pres1} we obtain
\be{15-lim-k0}
\lim_{\kappa\to 0}\mathcal{Q}_{1,m}^{(\kappa)}=\mathcal{Q}_{1,m}^{(0)}=\prod_{k=1}^mE_k^{11}.
\ee
In other words, the operator $\mathcal{Q}_{1,m}^{(0)}$ is a projector onto the states in which there are no quasi\-par\-ticles
at sites with numbers $1,\dots,m$. Therefore, if $B(\bu)|0\rangle$ is a ground state, then the quantity
\be{15-EFP-def0}
P_m=\frac{\langle0|C(\bu)\mathcal{Q}_{1,m}^{(0)} B(\bu)|0\rangle}{\langle0|C(\bu) B(\bu)|0\rangle}
\ee
has the meaning of the emptiness formation probability (in the sense of the absence of quasiparticles) in the ground state in
an interval of length $m$.

In the standard language, $P_m$ is the probability of finding a cluster of $m$ particles with spins up in the ground state.
Since the ground state is symmetric with respect to replacing the spins up by the spins down, the probability of detecting a cluster of
$m$ particles with spins down\footnote{In the language of quasiparticles, such a probability could be called the completeness formation probability.}
also is equal to $P_m$.
Therefore, apart from the limit $\kappa\to 0$, we can also consider the limit $\kappa\to\infty$
\be{15-lim-kinf}
\lim_{\kappa\to \infty}\kappa^{-m}\mathcal{Q}_{1,m}^{(\kappa)}=\prod_{k=1}^mE_k^{22},
\ee
when the operator $\mathcal{Q}_{1,m}^{(\kappa)}$ turns into a projector into states with spins down at sites with numbers $1,\dots,m$.
Hence, we conclude that the EFP arises in two limiting cases of the generating functional $Q_{m}^{(\kappa)}$:
\be{15-P-Q}
P_m= \lim_{\kappa\to 0}Q_{m}^{(\kappa)} = \lim_{\kappa\to \infty}\kappa^{-m}Q_{m}^{(\kappa)}.
\ee

We now turn to the integral representation \eqref{15-Qm-TD-lim} for the generating functional $Q_{m}^{(\kappa)}$. In this integral
representation, the parameter $\kappa$ enters only the determinant $M_s(\bar z;\bu)$ \eqref{15-Ms-new}. It is obvious that for a fixed $s$ this determinant
is a polynomial in $\kappa$ of degree $s$. Therefore, if we use the second formula \eqref{15-P-Q} (where $\kappa\to\infty$), then in the whole sum over
$s$ there will be only one term with $s = m$. It is easy to see that
\be{15-Ms-lim}
\lim_{\kappa\to \infty}\kappa^{-m}M_m(\bar z;\bu)=\prod_{j=1}^m\prod_{k=1}^m \frac{\sinh(u_k-z_j-i\zeta)}{\sinh(z_j-u_k-i\zeta)}\det_{m}t(u_k,z_j).
\ee
Then we obtain
\begin{multline}\label{15-Pm-lim}
P_m=\frac{1}{(-i\sin\zeta)^m (m!)^2}\oint_{\Gamma(-\frac{i\zeta}2)}\prod_{j=1}^m\frac{dz_j}{2\pi i}
\int_{-\infty}^\infty\prod_{j=1}^m du_j\;\prod_{j=1}^m \varphi^m(z_j)\varphi^{-m}(u_j)
\\
\times
\;\prod_{j=1}^s\prod_{k=1}^s
\frac{\sinh^2(u_j-z_k-i\zeta)}{\sinh(z_j-z_k-i\zeta)\sinh(u_j-u_k-i\zeta)}\;\det_{m}t(u_k,z_j)\;
\det_m\Bigl[\frac{i}{2\zeta\sinh\frac\pi\zeta(u_j-z_k)}\Bigr].
\end{multline}
In the resulting representation, we can get rid of the integrals with respect to the variables $z_k$ if we artificially introduce inhomogeneities. Namely, we make the following change of functions $\varphi^m(z)$ and $\varphi^{-m}(u)$:
\be{15-varphi-inh}
\begin{aligned}
&\varphi^m(z)=\left(\frac{\sinh(z-\tfrac{i\zeta}2)}{\sinh(z+\tfrac{i\zeta}2)}\right)^m\to\prod_{k=1}^m
\frac{\sinh(z-\xi_k-i\zeta)}{\sinh(z-\xi_k)},\non
&\varphi^{-m}(u)=\left(\frac{\sinh(u+\tfrac{i\zeta}2)}{\sinh(u-\tfrac{i\zeta}2)}\right)^m\to\prod_{k=1}^m
\frac{\sinh(u-\xi_k)}{\sinh(u-\xi_k-i\zeta)}.
\end{aligned}
\ee
Here$\{\xi_1,\dots,\xi_m\}$ is a set of  parameters belonging to a neighborhood of the point $-i\zeta/2$.
It is clear that the original functions
$\varphi^m(z)$ and $\varphi^{-m}(u)$ appear in the limit $\xi_k\to-i\zeta/2$.
We denote by $P_m(\bxi)$ the EFP modified in this way. Then
\begin{multline}\label{15-Pm-inh}
P_m(\bxi)=\frac{1}{(-i\sin\zeta)^m (m!)^2}\oint_{\Gamma(\bxi)}\prod_{j=1}^m\frac{dz_j}{2\pi i}
\int_{-\infty}^\infty\prod_{j=1}^m du_j\;\prod_{j=1}^m \prod_{k=1}^m
\frac{\sinh(z_j-\xi_k-i\zeta)\sinh(u_j-\xi_k)}{\sinh(z_j-\xi_k)\sinh(u_j-\xi_k-i\zeta)}\\
\times
\;\prod_{j=1}^m\prod_{k=1}^m
\frac{\sinh^2(u_j-z_k-i\zeta)}{\sinh(z_j-z_k-i\zeta)\sinh(u_j-u_k-i\zeta)}\;\det_{m}t(u_k,z_j)\;
\det_m\Bigl[\frac{i}{2\zeta\sinh\frac\pi\zeta(u_j-z_k)}\Bigr].
\end{multline}
Now, in the integrals with respect to the $z_j$-variables, instead of poles of the order $m$, we have only simple poles at the points $\xi_k$.
Due to the fact that the number of integration variables is equal to the number of poles, the integral over all $z_j$ is easily taken.
It is equal to the sum of the residues at the points
$z_j=\xi_{P(j)}$, where $P$ is any permutation in the set $\{\xi_1,\dots,\xi_m\}$. However, because of the symmetry of the integrand
with respect to the variables $z_j$, we can take the residues only at the points $z_j=\xi_j$ and multiply the resulting contribution by $m!$.
As a result, we obtain a new representation that no longer contains contour integrals
\begin{multline}\label{15-Pm-inh1}
P_m(\bxi)=\frac{1}{(-i\sin\zeta)^m m!}\prod_{\substack{j,k=1\\j\ne k}}^m\frac{1}{\sinh(\xi_j-\xi_k)}
\int_{-\infty}^\infty\prod_{j=1}^m du_j
\;\det_m\Bigl[\frac{i}{2\zeta\sinh\frac\pi\zeta(u_j-\xi_k)}\Bigr]\\
\times\;\det_{m}\left[\frac{-\sin^2\zeta}{\sinh(u_j-\xi_k)\sinh(u_j-\xi_k-i\zeta)}\right]\;
\prod_{j=1}^m\prod_{k=1}^m
\frac{\sinh(u_j-\xi_k-i\zeta)\sinh(u_j-\xi_k)}{\sinh(u_j-u_k-i\zeta)}.
\end{multline}
Thus, we computed the integrals over the $z_j$-variables, but this did not happen for free. Now, to calculate the EFP
we will have to take the limit $\xi_k\to-i\zeta/2$, and this procedure is far from trivial.

{\sl Remark 1.} One can also introduce the inhomogeneities $\xi_k$ in the integral representation for the generating functional $Q_{m}^{(\kappa)}$. This
allows us to compute the integrals over the variables $z_j$. However, in this case, in the $s$th term of the sum, the number of integration variables is $s$,
and the number poles is equal to $m$. Therefore, the result of integration will give a sum over partitions of the set $\bxi$, but not one term, as in
the case of the EFP.

{\sl Remark 2.} The representation for the EFP appeared in the limit of the generating functional $Q_{m}^{(\kappa)}$ at
$\kappa\to\infty$, and the generating functional itself was obtained by summing the form factors. However, the integral representation  \eqref{15-Pm-inh1}
can be derived in a much simpler way. Taking into account the formula \eqref{15-lim-kinf} and the formulas of the quantum inverse problem,
one can easily convince himself that
\be{15-P-simpl}
P_m(\bxi)=\prod_{k=1}^m \frac1{\tau(\xi_k|\bu)}\;\frac{\langle0|C(\bu)\Bigl(\prod_{k=1}^m D(\xi_k)\Bigr) B(\bu)|0\rangle}{\langle0|C(\bu) B(\bu)|0\rangle}.
\ee
We know the action of the operators $ D(\xi_k)$ on a vector Bethe $ B(\bu)|0\rangle$ \eqref{06-MCR-DB}.
It is given by the sum over the partitions of the set
$\{\bxi,\bu\}$. However, in the inhomogeneous model $d(\xi_k)=0$, therefore, in this sum only those terms survive that correspond to partitions
only of the set $\bu$. It is this sum over the partitions  that turns into an integral over the variables $u_j$ in the thermodynamic limit,
similarly to how this was with the generating functional $Q_{m}^{(\kappa)}$. An interested reader can see this for himself.

\subsection{Emptiness formation probability at $\Delta=1/2$\label{15-Sec81}}

For the value of the anisotropy parameter $\Delta=1/2$ (that is, $\zeta=\pi/3$), one can explicitly calculate all the integrals
over $u_j$ in the representation  \eqref{15-Pm-inh1} \cite{KitMST02cc}. Let us rewrite equation \eqref{15-Pm-inh1} for $\zeta=\pi/3$:
\begin{multline}\label{15-Pm-inh2}
P_m(\bxi)=\frac{1}{(-i\sin\tfrac{\pi}3)^m m!}\prod_{\substack{j,k=1\\j\ne k}}^m\frac{1}{\sinh(\xi_j-\xi_k)}
\int_{-\infty}^\infty\prod_{j=1}^m du_j\;\det_m\Bigl[\frac{3i}{2\pi\sinh3(u_j-\xi_k)}\Bigr]
\\
\times\;\det_{m}\left[\frac{-\sin^2\tfrac{\pi}3}{\sinh(u_j-\xi_k)\sinh(u_j-\xi_k-\tfrac{i\pi}3)}\right]\;
\;\prod_{j=1}^m\prod_{k=1}^m
\frac{\sinh(u_j-\xi_k-\tfrac{i\pi}3)\sinh(u_j-\xi_k)}{\sinh(u_j-u_k-\tfrac{i\pi}3)}.
\end{multline}

The main simplification of the integrand at $\zeta=\pi/3$ is based on an elementary formula
\be{15-formula}
\sinh 3x=4\sinh x\sinh(x+\tfrac{i\pi}3)\sinh(x-\tfrac{i\pi}3),
\ee
that was already used in lecture~\ref{CHA-ASM}.

Let us make a series of preliminary transforms. First, we rewrite one of the double products as follows:
\be{15-prod-prod}
\prod_{j=1}^m\prod_{k=1}^m\sinh(u_j-u_k-\tfrac{i\pi}3) =(-i\sin \tfrac{\pi}3)^m (-1)^{m(m-1)/2}
\prod_{j>k}^m \sinh(u_j-u_k-\tfrac{i\pi}3)\sinh(u_j-u_k+\tfrac{i\pi}3)
\ee
Then, using \eqref{15-formula} we have
\begin{multline}\label{15-prod-prod1}
\prod_{j=1}^m\prod_{k=1}^m
\frac{\sinh(u_j-\xi_k-\tfrac{i\pi}3)\sinh(u_j-\xi_k)}{\sinh(u_j-u_k-\tfrac{i\pi}3)}\\
=\frac{(-1)^{m(m-1)/2}}{2^{m^2+m}(-i\sin \tfrac{\pi}3)^m}
\prod_{j=1}^m\prod_{k=1}^m
\frac{\sinh3(u_j-\xi_k)}{\sinh(u_j-\xi_k+\tfrac{i\pi}3)}
\prod_{j>k}^m \frac{\sinh(u_j-u_k)}{\sinh3(u_j-u_k)}.
\end{multline}
Secondly, one of the determinants in \eqref{15-Pm-inh1} is explicitly computed via \eqref{ASM-detCauchy-S}
\be{15-det-prod}
\det_m\Bigl[\frac{3i}{2\pi\sinh 3(u_j-\xi_k)}\Bigr]=\left(\frac{3i}{2\pi}\right)^m
\frac{\prod_{j>k}^m \sinh3(u_j-u_k)\sinh3(\xi_k-\xi_j)}{\prod_{j,k=1}^m\sinh3(u_j-\xi_k)}.
\ee
Combining \eqref{15-prod-prod1} and \eqref{15-det-prod} we obtain
\begin{multline}\label{15-prod-det}
\prod_{j=1}^m\prod_{k=1}^m
\frac{\sinh(u_j-\xi_k-\tfrac{i\pi}3)\sinh(u_j-\xi_k)}{\sinh(u_j-u_k-\tfrac{i\pi}3)}\;\det_m\Bigl[\frac{3i}{2\pi\sinh 3(u_j-\xi_k)}\Bigr]\\[2mm]
=\left(\frac{3i}{2\pi}\right)^m\frac{(-1)^{m(m-1)/2}}{2^{m^2+m}(-i\sin \tfrac{\pi}3)^m}
\frac{\prod_{j>k}^m \sinh(u_j-u_k)\sinh3(\xi_k-\xi_j)}{\prod_{j,k=1}^m\sinh(u_j-\xi_k+\tfrac{i\pi}3)}\\[2mm]
=\left(\frac{3i}{2\pi}\right)^m\frac{(-1)^{m(m-1)/2}}{2^{m^2+m}(-i\sin \tfrac{\pi}3)^m}
\prod_{j>k}^m \frac{\sinh3(\xi_k-\xi_j)}{\sinh(\xi_k-\xi_j)}\;\det_m\left[\frac1{\sinh(u_j-\xi_k+\tfrac{i\pi}3)}\right].
\end{multline}
As a result of these transformations, the integral representation for $P_m(\bxi)$  takes the form
\begin{multline}
P_m(\bxi)=\frac{(-1)^{m(m-1)/2}}{ m!\;2^{m^2}}\left(\frac{3i}{4\pi}\right)^m\prod_{\substack{j,k=1\\j\ne k}}^m\frac{1}{\sinh(\xi_j-\xi_k)}
\prod_{j>k}^m \frac{\sinh3(\xi_k-\xi_j)}{\sinh(\xi_k-\xi_j)}\\[2mm]
\times \int_{-\infty}^\infty\,du_1\dots\,du_m\;\det_{m}\left[\frac{1}{\sinh(u_j-\xi_k)\sinh(u_j-\xi_k-\tfrac{i\pi}3)}\right]\;
\det_m \Biggl[\frac1{\sinh(u_j-\xi_k+\tfrac{i\pi}3)}\Biggr].
\label{15-Pm-preob}
\end{multline}
Now the integrals over $u_j$ can be taken using the following lemma.

\begin{lemma}\label{15-two-det} Let $F_j(u)$ and $G_j(u)$ be a set of functions that are square integrable on an interval
$[a,b]$  ($a$ and $b$ can be infinite). Then
 \be{15-lemma-1}
 \int_a^b \det_m F_k(u_j)
 \det_mG_k(u_j)\,du_1\dots\,du_m=m!\det_m\left(\int_a^bF_j(u)G_k(u)\,du\right).
 \ee
\end{lemma}

{\sl Proof}. We have
 \ba{15-lemma-2}
 \int_a^b \det_m F_k(u_j)
 \det_mG_k(u_j)\,du_1\dots\,du_m=\sum_P(-1)^{[P]}\int_a^b \prod_{j=1}^m
 F_j(u_{P(j)})\cdot  \det_mG_k(u_j)\,du_1\dots\,du_m\\
 \hspace{27mm} =\sum_P(-1)^{[P]}\int_a^b \prod_{j=1}^m
 F_j(u_j)\cdot  \det_mG_k(u_{P^{-1}(j)})\,du_1\dots\,du_m\\
 \hspace{32mm}=\sum_P(-1)^{[P]}\int_a^b \prod_{j=1}^m
 F_j(u_j)\cdot  \det_mG_k(u_j)\cdot(-1)^{[P]}\,du_1\dots\,du_m\\
 \hspace{-7cm}=m!\int_a^b \prod_{j=1}^m
 F_j(u_j)\cdot  \det_mG_k(u_j)\,du_1\dots\,du_m.
 \end{multline}
Let us explain the transformations. First we wrote down the first of the two
determinants according to its definition. Further, in each term of the sum
over permutations, the integration variables were changed
$u_{P(j)}\to u_j$. With such a replacement,
the rows of the second determinant are rearranged. Returning these rows to the initial positions
we get the additional sign $(-1)^{[P]}$, which
cancels with the original $(-1)^{[P]}$. Since after this nothing
depends on the permutation, the sum of these permutations gives
$m!$. In fact, it is proved that since the original integrand
was symmetric with respect to the variables $u_1,\dots,u_m$,
then each term of the sum over the permutations gives the same contribution as the product
of the diagonal terms of the first matrix.

Let us continue our transformations.
 \begin{align}
 m!\int_a^b \prod_{j=1}^m
 F_j(u_j)\cdot  \det_mG_k(u_j)\,du_1\dots\,du_m&=
 m!\int_a^b
   \det_m\Bigl(F_j(u_j)G_k(u_j)\Bigr)\,du_1\dots\,du_m\nonumber\\
 &= m!\det_m\left(\int_a^bF_j(u_j)G_k(u_j)\,du_j\right).\label{15-lemma-3}
 \end{align}
Again, we make an explanation. Each of the factors  $F_j(u_j)$
is entered in the appropriate row of the second determinant. Now
each integration variable $u_j$ enters only in the $j$th row.
Therefore, one can perform the integration separately in every row.
It remains to remove the subscript $j$ from each integration variable, and we
arrive at the assertion of the lemma.\qed

In our case
\be{15-FG}
\begin{aligned}
&F_k(u_j)=\frac{1}{\sinh(u_j-\xi_k)\sinh(u_j-\xi_k-\tfrac{i\pi}3)},\\
&G_k(u_j)=\frac1{\sinh(u_j-\xi_k+\tfrac{i\pi}3)},
\end{aligned}
\ee
and the integration contour $[a,b]$ is the real axis $\mathbb{R}$. Thus, we should calculate the integral
\be{15-IntJ}
J=\int_{-\infty}^\infty \frac{du}{\sinh(u-\xi_j)\sinh(u-\xi_j-\tfrac{i\pi}3)\sinh(u-\xi_k+\tfrac{i\pi}3)}.
\ee
The value of this integral depends on the imaginary parts of the inhomogeneities $\xi_j$. Therefore it is convenient to put
$\xi_j=\epsilon_j-\tfrac{i\pi}6$, and then we can assume that $\Im(\epsilon_j)=0$, and in the homogeneous limit $\epsilon_j\to 0$.
Then the integral takes the form
\be{15-IntJe}
J=\int_{-\infty}^\infty \frac{du}{\sinh(u-\epsilon_j+\tfrac{i\pi}6)\sinh(u-\epsilon_j-\tfrac{i\pi}6)\sinh(u-\epsilon_k+\tfrac{i\pi}2)}.
\ee
This integral is easily calculated by residues. Indeed, we can replace the integral along the real axis by an integral over the closed
contour consisting of $\mathbb{R}$,  $\mathbb{R}+i\pi$ (in the negative direction), and two vertical segments $[\Lambda,\Lambda+i\pi]$
and $[-\Lambda,-\Lambda+i\pi]$ at $\Lambda\to\infty$.
It is clear that the contribution from the vertical segments tends to zero, and the integral
in the negative direction over $\mathbb{R}+i\pi$ is equal to $J$ due to the antiperiodicity of the integrand. Therefore,
\be{15-IntJres}
J= \sum\Res\frac{i\pi}{\sinh(u-\epsilon_j+\tfrac{i\pi}6)\sinh(u-\epsilon_j-\tfrac{i\pi}6)\sinh(u-\epsilon_j+\tfrac{i\pi}2)}
\Bigr|_{u=\epsilon_j+\tfrac{i\pi}6;\; u=\epsilon_j+\tfrac{5i\pi}6;\; u=\epsilon_k+ \tfrac{i\pi}2}\;.
\ee
After simple algebra we find
\be{15-IntJresf}
J=-4\pi i\;\frac{\sinh\frac12(\epsilon_j-\epsilon_k)}{\sinh\frac32(\epsilon_j-\epsilon_k)},
\ee
and hence,
\begin{equation}\label{15-Pm-noint}
P_m(\bar\epsilon)=\frac{(-1)^{m(m-1)/2}3^m}{2^{m^2}}
\prod_{j>k}^m \frac{\sinh3(\epsilon_k-\epsilon_j)}{\sinh(\epsilon_k-\epsilon_j)}
\prod_{\substack{j,k=1\\j\ne k}}^m\frac{1}{\sinh(\epsilon_j-\epsilon_k)}
\; \det_m \left[\frac{\sinh\frac12(\epsilon_j-\epsilon_k)}{\sinh\frac32(\epsilon_j-\epsilon_k)}\right].
\end{equation}
In the resulting representation, all the integrals are already taken, and we only need to proceed to the homogeneous limit $\epsilon_j\to 0$.
Note that the determinant in the formula \eqref{15-Pm-noint} is very similar to that, which was studied in the lecture on the
alternating sign matrices. It is easy to make this determinant not just similar, but exactly the same. For this we introduce an additional
set of variables  $\{\epsilon'_1,\dots,\epsilon'_m\}$.  Then
\begin{multline}\label{15-lim2det}
\lim_{\epsilon_j\to 0}\prod_{\substack{j,k=1\\j\ne k}}^m\frac{1}{\sinh(\epsilon_j-\epsilon_k)}
\; \det_m \left[\frac{\sinh\frac12(\epsilon_j-\epsilon_k)}{\sinh\frac32(\epsilon_j-\epsilon_k)}\right]\\
=\lim_{\substack{\epsilon_j\to 0\\ \epsilon'_k\to 0}}
\prod_{j>k}^m\frac{1}{\sinh(\epsilon_j-\epsilon_k)\sinh(\epsilon'_j-\epsilon'_k)}
\; \det_m \left[\frac{\sinh\frac12(\epsilon_j-\epsilon'_k)}{\sinh\frac32(\epsilon_j-\epsilon'_k)}\right].
\end{multline}
Indeed, due to proposition~\ref{ASM-Pred-limdet}
the limit on the rhs of \eqref{15-lim2det} exists and does not depend on how the parameters $\epsilon_j$ and $\epsilon'_j$ approach their
limiting values. Since they all go to zero, we can put $\epsilon_j=\epsilon'_j$ before taking the limit,
what gives us the lhs of \eqref{15-lim2det}.

Using the auxiliary set of variables $\{\epsilon'_1,\dots,\epsilon'_m\}$, we can write the representation for the EFP
as follows:
\begin{equation}\label{15-Pm-new}
P_m=\frac{
3^m}{2^{m^2}}
 \lim_{\substack{\epsilon_j\to 0\\ \epsilon'_k\to 0}}\frac{\sinh3(\epsilon_k-\epsilon_j)}{\sinh(\epsilon_k-\epsilon_j)}
\prod_{j>k}^m\frac{1}{\sinh(\epsilon_j-\epsilon_k)\sinh(\epsilon'_j-\epsilon'_k)}
\; \det_m \left[\frac{\sinh\frac12(\epsilon_j-\epsilon'_k)}{\sinh\frac32(\epsilon_j-\epsilon'_k)}\right].
\end{equation}
Now we can set
$\epsilon_j=2\epsilon j$, $\epsilon'_k=2\epsilon(1-k)$ and consider the limit $\epsilon\to 0$. Then we find
\begin{equation}\label{15-Pm-fin-lim}
P_m=(-1)^{m(m-1)/2}3^{m(m+1)/2}2^{-m^2}\lim_{\epsilon\to 0}
\prod_{j>k}^m\frac{1}{\sinh^22\epsilon(j-k)}
\; \det_m \left[\frac{\sinh\epsilon(j+k-1)}{\sinh3\epsilon(j+k-1)}\right].
\end{equation}
And this is exactly the determinant  \eqref{ASM-Izer3},
with which we met in the study of  the alternating sign matrices. We know that
this determinant is
\be{15-det-Kup}
\det_m \left[\frac{\sinh\epsilon(j+k-1)}{\sinh3\epsilon(j+k-1)}\right]=2^{m^2-m}\prod_{j>k}^m{\sinh}^23\epsilon(j-k)
\prod_{j=1}^m\prod_{k=1}^m\frac{\sinh\epsilon(1+3(j-k))}{\sinh3\epsilon(j+k-1)}\;.
\ee
Substituting this into \eqref{15-Pm-fin-lim} and taking the elementary limit $\epsilon\to 0$, we find
\begin{equation}\label{15-Pm-aft-lim}
P_m=(-1)^{m(m-1)/2}3^{(3m^2-m)/2}2^{-m^2}\prod_{j=1}^m\prod_{k=1}^m\frac{j-k+1/3}{j+k-1}.
\end{equation}
It remains to repeat the calculations that we already did before, and we eventually obtain
\be{15-fin-answ}
P_m=\frac{\mathcal{A}_m}{2^{m^2}},
\ee
where $\mathcal{A}_m$  is the number of the alternating sign matrices of the size $m\times m$:
\be{15-ASM-numb}
\mathcal{A}_m=\prod_{k=0}^{m-1}\frac{(3k+1)!}{(m+k)!}.
\ee

Thus, we have obtained an exact answer for the EFP at the value of the anisotropy parameter $\Delta=1/2$.
This was due to the complete factorization of the $m$-fold integral with respect to the variables $u_j$ in the representation \eqref{15-Pm-inh1}.
Similar factorization takes place in two more cases. The first of them is $\Delta=0$ (that is, $\zeta=\pi/2$). But
this case corresponds to free fermions, and therefore the factorization of the $m$-fold integral is quite expected.
On the contrary, it would be surprising if it did not happen. We sugest the reader to consider this case.

The second case is the value $\Delta=1/\sqrt{2}$  (that is, $\zeta=\pi/4$). In this case, too, there is a complete factorization,
but the analogue of the representation \eqref{15-Pm-noint} in the case $\Delta=1/\sqrt{2}$ contains a determinant of a matrix,
whose entries  have a much more complex structure than the matrix
in the formula \eqref{15-Pm-noint}. The question of calculating its determinant remains open.

Generally speaking, the factorization of multiple integrals in representations for correlation functions in the $XXZ$ Heisenberg chain is also possible in the case of an arbitrary $\Delta$ (see \cite{BGKS06}). However, all the results of this type were obtained within the framework of the inhomogeneous model. For final results it is also necessary to calculate the homogeneous limit. This is a far from simple task, which for today has been solved only in a number of special cases.


%
%

\end{document}